\RequirePackage{lineno}
\documentclass[iop,apj]{emulateapj} 
\usepackage{times}

\usepackage[section]{placeins}

\usepackage{amssymb,amsmath}
\usepackage{graphicx,subfigure}
\usepackage{color}
\usepackage{rotating}
\usepackage{ulem} 
\usepackage{float}
\newcommand{\be}{\begin{equation}}
\newcommand{\ee}{\end{equation}}

\newcommand{\gray}{$\gamma$-ray}

\newcommand{\hi}{H~{\sc i}}

\newcommand{\Xco}{$X_{\rm CO}$}

\newcommand{\fermilat}{{\it Fermi}--LAT}
\newcommand{\fermi}{{\it Fermi}}

\newcommand{\GP}{\mbox{GALPROP}}

\newcommand{\hess}{H.E.S.S.}
\newcommand{\model}[5]{$^{\rm S}$#1$^{\rm Z}#2^{\rm R}#3^{\rm T}#4^{\rm C}#5$}

\shorttitle{GeV Observations of the Galactic Centre}
\shortauthors{\fermilat{} Collaboration}

\begin{document}



\title{\fermilat\ Observations of High-Energy $\gamma$-Ray Emission Toward the 
Galactic Centre}

\author{
M.~Ajello\altaffilmark{1}, 
A.~Albert\altaffilmark{2}, 
W.~B.~Atwood\altaffilmark{3}, 
G.~Barbiellini\altaffilmark{4,5}, 
D.~Bastieri\altaffilmark{6,7}, 
K.~Bechtol\altaffilmark{8}, 
R.~Bellazzini\altaffilmark{9}, 
E.~Bissaldi\altaffilmark{10}, 
R.~D.~Blandford\altaffilmark{2}, 
E.~D.~Bloom\altaffilmark{2}
% ...
}
% ...
\altaffiltext{1}{Department of Physics and Astronomy, Clemson University, Kinard Lab of Physics, Clemson, SC 29634-0978, USA}
\altaffiltext{2}{W. W. Hansen Experimental Physics Laboratory, Kavli Institute for Particle Astrophysics and Cosmology, Department of Physics and SLAC National Accelerator Laboratory, Stanford University, Stanford, CA 94305, USA}
\altaffiltext{3}{Santa Cruz Institute for Particle Physics, Department of Physics and Department of Astronomy and Astrophysics, University of California at Santa Cruz, Santa Cruz, CA 95064, USA}
\altaffiltext{4}{Istituto Nazionale di Fisica Nucleare, Sezione di Trieste, I-34127 Trieste, Italy}
\altaffiltext{5}{Dipartimento di Fisica, Universit\`a di Trieste, I-34127 Trieste, Italy}
\altaffiltext{6}{Istituto Nazionale di Fisica Nucleare, Sezione di Padova, I-35131 Padova, Italy}
\altaffiltext{7}{Dipartimento di Fisica e Astronomia ``G. Galilei'', Universit\`a di Padova, I-35131 Padova, Italy}
\altaffiltext{8}{Dept.  of  Physics  and  Wisconsin  IceCube  Particle  Astrophysics  Center, University  of  Wisconsin, Madison,  WI  53706, USA}
\altaffiltext{9}{Istituto Nazionale di Fisica Nucleare, Sezione di Pisa, I-56127 Pisa, Italy}
\altaffiltext{10}{Istituto Nazionale di Fisica Nucleare, Sezione di Bari, I-70126 Bari, Italy}
% ...

\begin{abstract}
The \fermi{} Large Area Telescope (LAT)
has provided the most detailed view to date of the emission towards the 
Galactic centre (GC) in high-energy \gray{s}. 
This paper describes the analysis of 
data taken during the first 62~months of the mission in the energy 
range $1-100$~GeV from a $15^\circ \times 15^\circ$ region about 
the direction of the GC, and  implications  for the interstellar emissions produced by cosmic ray 
(CR) particles interacting with the gas and radiation fields in the 
inner Galaxy and for the
point sources detected.
Specialised interstellar emission models (IEMs) are constructed that enable
separation of the \gray{} emission from the inner $\sim 1$~kpc about the 
GC from the fore- and background emission from the Galaxy. 
Based on these models, the interstellar emission from 
CR electrons interacting with the interstellar radiation field via 
the inverse Compton (IC) process and CR nuclei inelastically scattering
off the gas producing \gray{s} via $\pi^0$ decays from the 
inner $\sim 1$~kpc is determined. 
The IC contribution is found to be dominant in the region and 
strongly enhanced compared to previous studies.
A catalog of point sources for the $15^\circ \times 15^\circ$ 
region is self-consistently constructed using these 
IEMs: the First \fermilat\ Inner Galaxy point source Catalog (1FIG).
The spatial locations, fluxes, and spectral properties of the 1FIG 
sources are presented, and compared with \gray{} point sources over the 
same region taken from existing catalogs, including 
the Third \fermilat\ Source Catalog (3FGL). 
In general, the spatial density of 1FIG sources 
differs from those in the 3FGL, which is
attributed to the different treatments of the interstellar emission and energy
ranges used by the respective analyses.
Three 1FIG sources are found to spatially overlap with supernova remnants (SNRs)
listed in Green's SNR catalog; these SNRs have not 
previously been associated with high-energy \gray{} sources.
Most 3FGL sources with known multi-wavelength counterparts are also found.
However, the majority of 1FIG point sources are unassociated.
After subtracting the interstellar emission and point-source contributions 
from the data a residual is found that is a sub-dominant fraction of the 
total flux. 
But, it is brighter than the \gray{} emission associated with interstellar
gas in the inner $\sim 1$~kpc derived for the IEMs used in this paper,  
and comparable to the integrated brightness 
of the point sources in the region for energies $\gtrsim 3$~GeV.
If spatial templates that peak toward the GC are used to model the positive 
residual and included in the total model for the $15^\circ \times 15^\circ$ 
region, the agreement with the data improves, but they do not account for 
all the residual structure.
The spectrum of the positive residual modelled with these templates 
has a strong dependence on the choice of IEM.
\end{abstract}



\section{Introduction}
\label{intro}
\setcounter{footnote}{0}

The region surrounding the Galactic centre (GC) is among the brightest and 
most complex in high-energy \gray{s}, with on-going massive star 
formation providing all types of known or suspected 
cosmic ray (CR) and \gray{} sources. 
The GC also houses a $\sim 10^6$ M$_\odot$ black hole 
\citep[e.g.,][]{2010RvMP...82.3121G} 
and the region is predicted to 
be the brightest source of \gray{s} associated with 
annihilation or decay of massive weakly-interacting particles \citep[see the reviews by, e.g.,][]{1996PhR...267..195J, 2000RPPh...63..793B, 2010ARA&A..48..495F}.  
Despite detection in the 100~MeV to GeV range by the EGRET instrument 
on the {\it Compton Gamma-Ray Observatory} \citep{EGRETGCref} and 
at higher energies by the \hess\ Cherenkov array 
\citep{2006Natur.439..695A,2006ApJ...636..777A} the 
characterisation of the \gray{} emission for $<100$~GeV energies 
in the region surrounding the GC has remained elusive.

The \gray{} emission in the Galaxy is predominantly due to the 
interactions of CR particles with the interstellar gas and 
radiation fields.
This interstellar emission is a fore-/background against which \gray{} 
point sources are detected.
In the Galactic plane the intensity of this emission makes disentangling 
the contributions by \gray{} point sources and truly diffuse processes 
challenging.
Particularly toward the GC, where the intensity of the interstellar emission 
and number of point sources is maximised, self-consistent modelling is 
necessary to deal with the strong confusion.

Since 2008 the Large Area Telescope instrument on the \fermi{} Gamma-Ray 
Space Telescope (\fermilat) has been taking data in the range 20~MeV to more 
than 300~GeV energies. 
Analyses of the data toward the region surrounding the GC
have been made by various authors 
\citep{2009arXiv0910.2998G,2011PhLB..697..412H,2012PhRvD..86h3511A,2013PDU.....2..118H,2013PhRvD..88h3521G,2013arXiv1307.6862H,2014PhRvD..90b3526A,2014arXiv1402.6703D,2015PhRvD..91f3003C}. 
The results of these works have been interpreted as evidence for 
an unresolved point source population or annihilating dark matter (DM).
Versions of the  
interstellar emission models (IEMs) distributed by the Fermi Science 
Support Center (FSSC)\footnote{http://fermi.gsfc.nasa.gov/ssc/data/access/lat/BackgroundModels.html} 
have typically been employed in these analyses, although some works have 
used IEMs \citep[e.g.,][]{2015PhRvD..91f3003C} 
that are based on CR propagation calculations using the \GP\ code\footnote{For a detailed description of the \GP\ code the reader is referred to the dedicated website: http://galprop.stanford.edu}~\citep[e.g.,][]{1998ApJ...493..694M,2012ApJ...752...68V}.  

The FSSC IEMs are optimised to flatten residuals
over large regions of the sky in support of the generation of the \fermilat\ 
source catalogs. 
The optimisations vary according to the version of the FSSC IEM. 
The most widely used by the analyses cited above 
\citep[supporting the generation of the Second
\fermilat\ Source Catalog; ][]{2FGLref} 
includes patches with spatially uniform spectral intensity 
to account for positive residuals.
Some of these are in and about the GC, which makes interpretation 
of positive residuals after fitting additional templates and subtracting the
IEM and point sources uncertain.

In this paper, an analysis is described of the \gray{} emission observed by 
the \fermilat\ during the first 62~months of the mission toward the inner 
Milky Way that characterises 
the $15^\circ\times15^\circ$ region in Galactic coordinates centred on the GC.
This encompasses the innermost $\sim 1$~kpc where
the CR intensities, interstellar gas and radiation field densities are 
highest but most uncertain, and signatures of new physics may be detectable.
The analysis uses multiple IEMs together with an iterative fitting procedure 
to determine the contributions by diffuse and discrete sources 
of high-energy \gray{} emission. 
The \GP\ CR
propagation code
is used to calculate components of IEMs that are fit to the \fermilat\ data to 
predict the 
interstellar emission fore-/background toward the $15^\circ\times15^\circ$ 
region.
Candidate locations of point sources are found using a 
wavelet-based algorithm \citep[][]{1997ApJ...483..350D,Ciprini:2007zz}.
These are used together 
with the IEMs to define a model for the emission of the region, 
which is then optimised in a maximum-likelihood fit to determine
the contribution by CR-induced diffuse emission from the innermost 
$\sim 1$~kpc and \gray{} point sources.
This is the first self-consistent modelling balancing the various sources
of \gray{s} toward the inner Galaxy.
The point sources found as a result 
are presented as the  First \fermilat\ Inner Galaxy
point source catalogue (1FIG), which is compared with the sources in the recent 
Third \fermilat\ Source Catalog \citep[3FGL;][]{2015ApJS..218...23A}
for the same region
\footnote{The optimised IEMs, 1FIG, and sub-threshold source candidate lists are available at: https://www-glast.stanford.edu/pub\_data.}.

\section{Large Area Telescope and Data Selection}

The \fermi\ Gamma-ray Space Telescope was launched on 
11~June~2008.
The LAT, which is the main instrument on \fermi, is a pair-conversion 
telescope composed of a $4 \times 4$ grid of 
towers, with each tower consisting of a silicon micro-strip tracker with 
interleaved tungsten foils for conversion of incident \gray{s} into 
electron-positron pairs, mated with a hodoscopic cesium-iodide calorimeter. 
This grid of towers is covered by a segmented plastic scintillator 
anti-coincidence detector.   
The tracker is divided in two sections, ``front'' 
(4.1\% radiation lengths [R.L.] per layer, first 12 layers 
below the anti-coincidence detector where a layer comprises 
the tungsten converter foil with two silicon detector planes and 
associated support structures) and ``back'' 
(19.3\% R.L. per layer, next 4 layers) where the 
last two layers do not have conversion foils.  
The effective collecting areas of both sections are comparable, but 
the angular resolution for \gray{s} that convert in the front section
is approximately a factor of two better than for back-converting \gray{s}.
For the former the 68/95\% containment 
radii of the point-spread function (PSF) are 
$0.4^\circ/1.5^\circ$ at 1~GeV decreasing to $0.1^\circ/0.3^\circ$ 
at 10~GeV.  
The LAT is sensitive to \gray{s} with energies in the range from 20~MeV to 
over 300~GeV, and its on-axis effective area is $\sim 8000$~cm$^2$ for 
energies $>1$~GeV.
The LAT is described in detail 
in \citet{Atwood:2009}, with specifics related to its on-orbit performance 
reported in \citet{2009APh....32..193A} and \citet{2012ApJS..203....4A}.

The analysis described in this paper employs events with 
reconstructed energy in the range $1-100$~GeV, where the effective area of the 
LAT is largest and not strongly dependent on energy.
To allow the best separation between point sources and the 
structured interstellar emission in the analysis 
procedure (described below), only front-converting events are used.

Events and instrument response functions (IRFs) for
the standard low-residual CR background ``Clean'' events
from the Pass~7 event selections \citep[][]{2012ApJS..203....4A} 
\footnote{The reprocessed data and instrument response functions P7REP\_CLEAN\_V15 are employed. See http://fermi.gsfc.nasa.gov/ssc/data/analysis/documentation/Pass7\_usage.html} are used.
To minimise the contribution from the very bright Earth limb,
the event selection and exposure calculation is restricted to zenith angles
less than $100^\circ$.

Events are selected from approximately 62~months of data from 
2008-08-11 until 2013-10-15.
Exposure maps and the PSF for the pointing
history of the observations were generated using the standard
\fermilat{} ScienceTools package (version 09-34-02) available 
from the FSSC\footnote{http://fermi.gsfc.nasa.gov/ssc/data/analysis/}.
For the $15^\circ \times 15^\circ$ region about the direction toward 
the GC the resulting exposure is $7 \times 10^{10}$ cm$^2$~s at 1~GeV.

\section{Methodology}
\label{methodology}

\subsection{Interstellar Emission Models}
\label{sec:interstellaremission}
The diffuse \gray{} emission is produced by the interaction of high-energy 
CRs with the interstellar gas and radiation fields.
The limited angular resolution and statistics that are a characteristic of 
high-energy \gray{} data, coupled with the relatively intense 
interstellar emission at low latitudes, make accurate modelling of the 
latter important
for characterising all but the brightest point sources there
\citep{2010ApJS..188..405A,2FGLref}.

\begin{figure}[ht]
\includegraphics[scale=0.8]{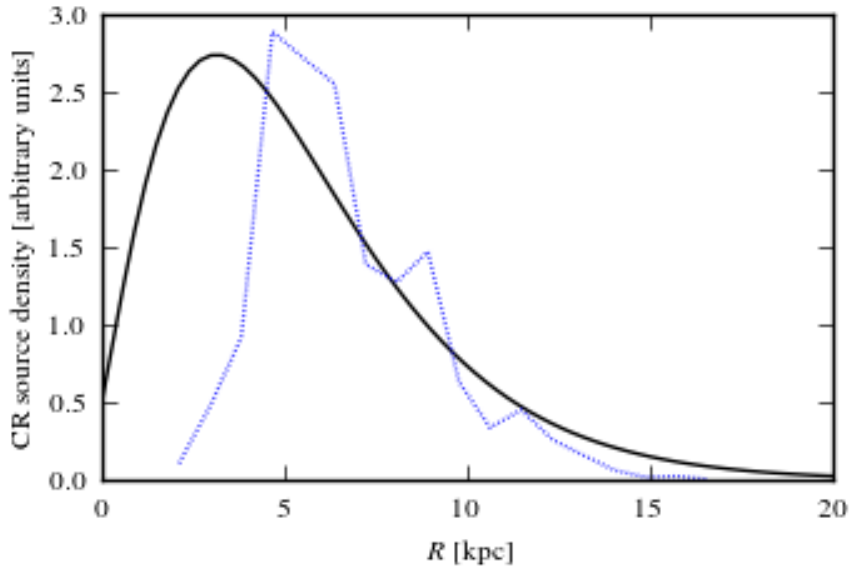}
\caption{Galactocentric radial dependence of the spatial 
distribution of CR sources per unit volume. 
Line styles/colours: solid/black, Pulsars; dotted/blue, OB-stars.}
\label{fig:source_distribution}
\end{figure}

\begin{deluxetable*}{ccccc}[htb]
\tablecolumns{5}
\tablecaption{Galactocentric Annular Boundaries.
\label{table:rings}}
\tablehead{
\colhead{Annulus} &
\colhead{$R_{\rm min}$} &
\colhead{$R_{\rm max}$} &
\colhead{Longitude} &
\colhead{Longitude}\\
\colhead{\#} &
\colhead{[kpc]} &
\colhead{[kpc]} &
\colhead{Range (Full)} &
\colhead{Range (Tangent)}
}
\startdata
1 & 0   & 1.5 & $-10^\circ \leq l \leq 10^\circ$ & \\
2 & 1.5 & 2.5 & $-17^\circ \leq l \leq 17^\circ$ & $10^\circ \leq |l| \leq 17^\circ$\\
3 & 2.5 & 3.5 & $-24^\circ \leq l \leq 24^\circ$ & $17^\circ \leq |l| \leq 24^\circ$\\
4 & 3.5 & 8.0 & $-70^\circ \leq l \leq 70^\circ$ & $24^\circ \leq |l| \leq 70^\circ$\\
5 & 8.0 & 10.0 & $-180 \leq l \leq 180^\circ$ & \\
6 & 10.0 & 50.0 & $-180 \leq l \leq 180^\circ$ &
\enddata
\end{deluxetable*}

\begin{figure*}[ht]
\subfigure{
\includegraphics{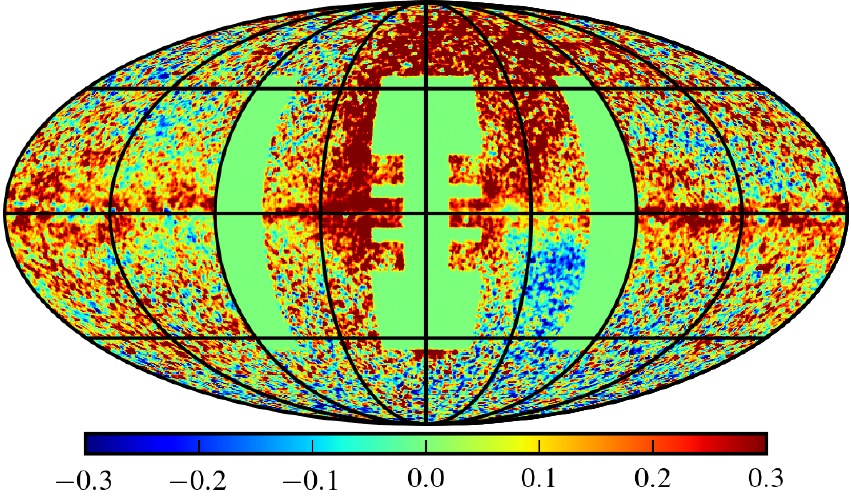}
\includegraphics{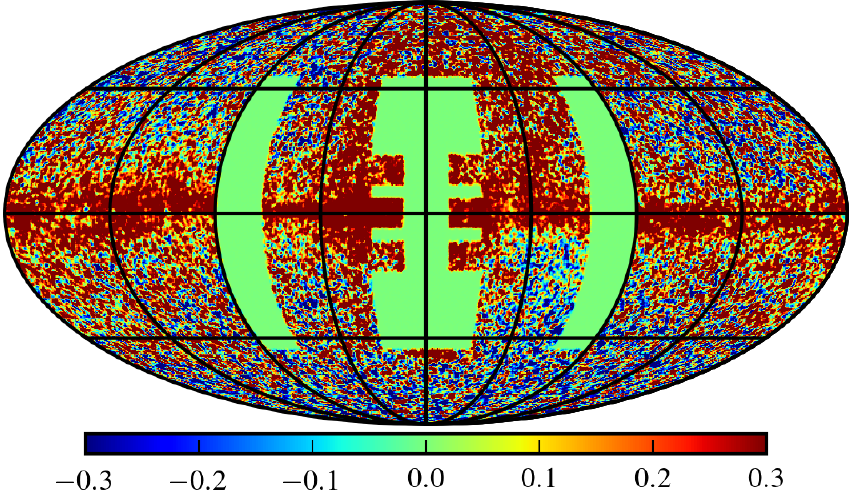}
}
\subfigure{
  \includegraphics{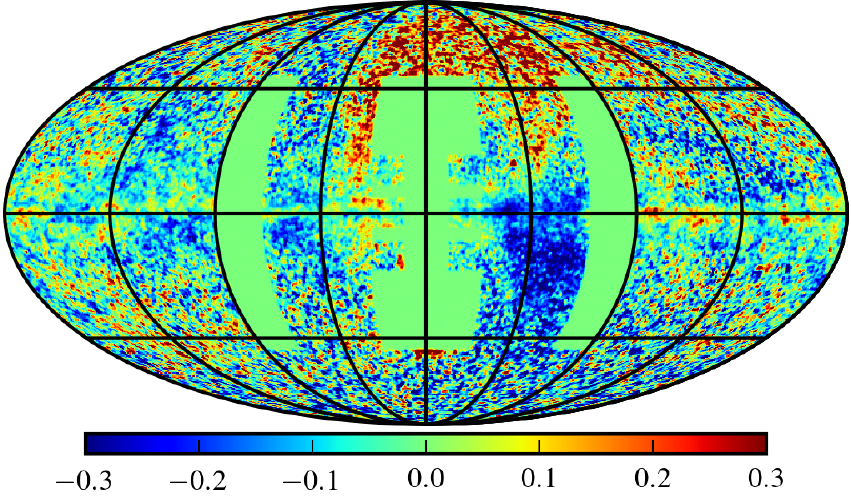}
  \includegraphics{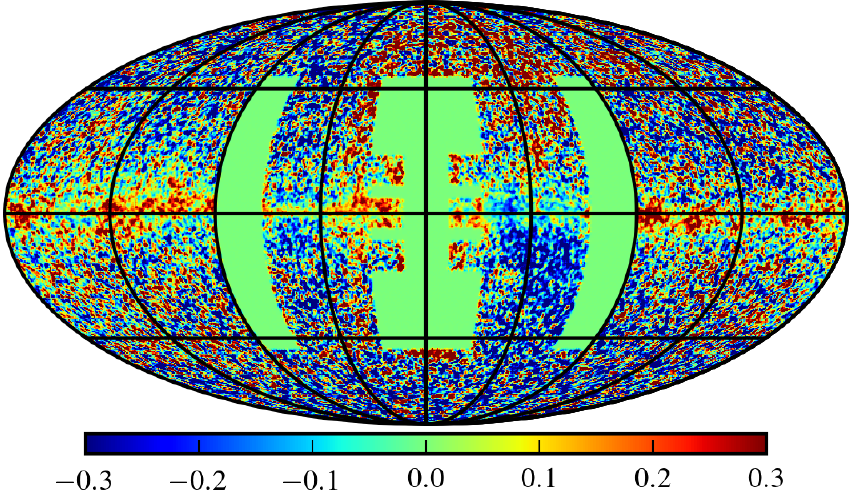}
}
\subfigure{
\includegraphics{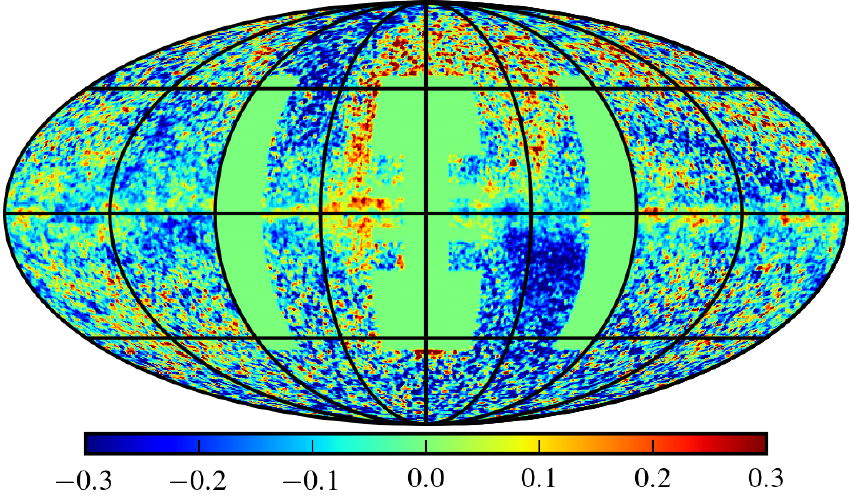}
\includegraphics{fig2d.pdf}
}
\caption{Residual fractional counts $(data - model)/model$ in the $1-3.16$~GeV (left) and $3.16-10$~GeV (right) energy ranges for the 
baseline Pulsars model (upper), 
the intensity-scaled Pulsars model (centre), 
and the index-scaled Pulsars model (lower) fitted following the procedure 
described in the text.
The isotropic component determined from the high-latitude fit is included
in the baseline model (upper panels) for the fractional residual calculation.
The baseline model does not include a model for the Loop-I SNR, resulting 
in the large positive residuals in the northern Galactic hemisphere.
The maps are calculated for a HEALPix order 8 pixelisation 
($\sim 0.23^\circ$ resolution) and smoothed with a 
1$^\circ$ FWHM Gaussian.
Regions not used for the IEM tuning procedure are masked.
The positive residual at mid-to-high latitudes interior to the solar circle
is due to mismatch between the data and the relatively simple Loop~I model.
The residuals close to the plane from this mismatch are lower and 
do not affect the analysis of the $15^\circ \times 15^\circ$ 
region about the GC.
\label{fig:residual_Base_Yusifov_Intensity_allsky_1to3.16GeV_and_3.16to10GeV}}
\end{figure*}

Two analysis approaches for studying the interstellar emission have been
used by the \fermilat\ Collaboration in previous works.
Templates tracing \gray{} emission processes 
were used to 
determine the \gray{} emissivity of the interstellar gas 
within several kpc of the 
Sun~\citep[e.g.,][]{2009ApJ...703.1249A,2010ApJ...710..133A,2011ApJ...726...81A}.
The \GP\ code was used in an extensive 
study of IEMs constrained
by local CR data and their correspondence with the 
\fermilat{} data \citep{2012ApJ...750....3A}.
There are merits to both approaches.
Fitting templates allows for fairly robust extraction of physical quantities, 
but is a method that is 
constrained by the assumption that interstellar medium (ISM) 
densities and other properties
(gas-to-dust ratio, \Xco-factor, etc.) and CR spectra remain constant 
throughout the template, and that a suitable template is available (e.g., 
the inverse Compton [IC] component of the interstellar emission must 
be obtained using modelling codes) -- see \citet{1988A&A...207....1S} and 
\citet{1996A&A...308L..21S}, and references therein.
The \GP\ code can be used to predict the diffuse \gray{} emission 
throughout the Galaxy, and is capable of
reproducing the observations at the $\sim 20$\% level.
But the predictions of the propagation model based approach 
are limited by the quality
of the inputs to the model calculations, which include the spatial 
distribution of CR sources and their injection spectra, and the spatial 
distribution of the interstellar gas density and the 
interstellar radiation field (ISRF) energy density.
In this paper a combination of these methods is used where the \GP\ code 
is employed to predict templates
for the interstellar emission that are fit to the \gray{} data to 
estimate the foreground and background emission toward the inner Galaxy.

The results of the study by \citet{2012ApJ...750....3A} are used for the 
baseline IEMs, which are further fit to the \fermilat\ data. 
As a reminder, the \citet{2012ApJ...750....3A} study 
used a grid of IEMs based on diffusion-reacceleration CR propagation 
models. 
The spatial distribution of CR sources, the \hi\ spin temperature, 
\hi\ column density corrections from dust emission, and the size of 
the CR confinement 
volume were the fixed parameters in the grid.
For each grid point the diffusion coefficient was obtained by adjusting it to 
reproduce the observed CR secondary/primary ratios iterating with a fit to 
the \gray{} data for the \Xco\ distribution for each CR source model. 
The \gray{} emission for each of the IEMs in the grid was then compared 
with the \fermilat\ data in the 200~MeV to 100~GeV energy range.
The models in the \citet{2012ApJ...750....3A} 
study agree at the $\sim 10-20$\% level with the LAT observations over the sky.

A major uncertainty affecting predictions of the interstellar 
emission toward the inner Galaxy is the spatial distribution of CR sources.
The \citet{2004A&A...422..545Y} pulsar distribution (``Pulsars'') 
and the distribution of OB-stars \citep[``OBstars'';][]{2000A&A...358..521B} 
encapsulate
this because they represent reasonable extremes for the Galactocentric radial 
dependence. 
Figure~\ref{fig:source_distribution} shows the Galactocentric radial 
distributions of these CR source models.
The Pulsars distribution is non-zero at the GC while the OBstars
distribution goes to zero near $\sim 2$~kpc.
The models\footnote{Specifically, the 
\model{Y}{6}{30}{150}{2} (Pulsars) and \model{O}{6}{30}{150}{2} (OBstars) 
models from \citet{2012ApJ...750....3A}.} 
assume an axisymmetric cylindrical geometry for the CR confinement 
volume with a halo height $z_h = 6$~kpc and maximum radial 
boundary $R_h = 30$~kpc.
This halo height is the closest in the IEM grid\footnote{Halo heights of 
4, 6, 8, and 10~kpc were used in the \citet{2012ApJ...750....3A} study.} 
to the halo height distribution mean ($\sim 5.5$~kpc) determined
by \citet{2011ApJ...729..106T};
the exact value of the halo height is not critical for the analysis.

For the IEM fitting procedure the \GP\ code is used 
to calculate all-sky \gray{} 
intensity maps from $1-100$~GeV for 10 logarithmically spaced energy bins 
per decade for the Pulsars and OBstars baseline models, which are normalised 
to local CR data, using the configuration
files for each available from the \GP\ website\footnote{http://galprop.stanford.edu/PaperIISuppMaterial/}.
The \GP\ code produces intensity maps in annuli that correspond to ranges
in Galactocentric radii; the total intensity map for a given 
\gray{} production process 
($\pi^0$-decay, IC, Bremsstrahlung) is the sum of all the 
annular intensity maps for that process, and the total predicted \gray{} 
sky from a \GP\ run is the sum of intensities from all processes.
Table~\ref{table:rings} lists the Galactocentric annuli and 
the corresponding longitude ranges for the full extent of each annulus, as well
as the `tangent' regions that are used in the fitting procedure for the
components interior to the solar circle 
\citep[see Appendix~B of][for a description of the generation of the \hi\ and 
CO gas annuli]{2012ApJ...750....3A}.

The annular intensity maps are used as templates 
together with an isotropic component and a model 
for \gray{} emission associated with Loop~I employing a two-component spatial 
template from \citet{2007ApJ...664..349W} with a power-law spectral model 
for each, and 
point sources from the 3FGL source catalog\footnote{This allows for 
discrimination between structured interstellar emission and point sources 
close to the Galactic plane when developing the fore-/background IEMs.}.
This combined model is fit to 
the \fermilat\ data excluding the $15^\circ \times 15^\circ$ region about
the GC using a maximum-likelihood method\footnote{The GaRDiAn code is used, 
which forward folds the model with the instrument response and PSF for 
the likelihood evaluation -- see Appendix~A of \citet{2012ApJ...750....3A}}, 
but with the point-source normalisations and spectral shapes held constant.
Because they make only a small contribution this does not significantly 
affect the determination of the IEM parameters.

Two IEMs for each of the Pulsars and OBstars models -- 4
in total -- are constructed.
The two variants for each model are termed ``intensity-scaled'' 
and ``index-scaled''.
The normalisation parameters for the templates are determined in a 
series of fits to the data, starting at high latitudes for the local 
components and then working from the outer Galaxy to the inner 
Galaxy, always fixing the already determined normalisation 
parameters in subsequent fits.
For the intensity-scaled variants only the normalisations of the
individual intensity maps are allowed to change.
For the index-scaled
variants the same fitting procedure is followed, but
additional degrees of freedom are allowed to the spectrum of the
gas-related interstellar emission when fitting to the annuli 
interior to the solar circle.
The details of the procedure for the intensity-scaled variants are
given in Appendix~\ref{appendix:IEM}.
The motivation for the index-scaled variants is described further below.

Figure~\ref{fig:residual_Base_Yusifov_Intensity_allsky_1to3.16GeV_and_3.16to10GeV} upper and centre panels show 
the fractional residuals, $(data - model)/model$, 
for $1-10$~GeV energies\footnote{The~$>10$~GeV residuals show similar 
characteristics to the~$3.16-10$~GeV energy band, but they are not 
shown here because of their relatively limited statistics.}    
for the baseline and intensity-scaled Pulsars model.
The isotropic component determined for the intensity-scaled
IEM has been included in the baseline model for the fractional calculation to
show the relative differences from the Galactic components of the IEMs.
The regions not used in the fitting procedure are explicitly masked in the
figure.
They are not used because of localised extended excesses that are most 
likely unrelated to the large-scale interstellar emission.
In particular, 
the band covering $70^\circ \leq l \leq 90^\circ$ includes the Cygnus region 
($l \sim 75-85^\circ$) around the Galactic plane;
the corresponding band for negative longitudes is a consequence of the 
axisymmetric nature of the model being used.
The range $90^\circ \leq l \leq 270^\circ$ is used to 
constrain the IC emission from annulus~5 so the data 
out of the plane from the $70^\circ \leq l \leq 90^\circ$ region 
is not required to constrain this component.
The $-20^\circ \leq l \leq 20^\circ, 10^\circ \leq |b| \leq 50^\circ$ region 
where the \fermi\ haze/bubbles 
have been detected is also excluded. 
Including these regions
in the fitting procedure would bias the normalisation
of the IEM components because models for these features are not 
included in this study.

Outside of the Galactic plane the fractional residuals are substantially 
reduced for the intensity-scaled IEM compared to the baseline.
This is due to the scaling of the $\pi^0$-decay interstellar emission for the
local annulus, and of the IC component generally.
The coefficient for the local gas (annulus~5) interstellar emission is adjusted
upward (see Table~\ref{table:coefficients} in Appendix~\ref{appendix:IEM} for a full list of IEM coefficients).
Meanwhile, the coefficients of the IC intensity for the local and other
annuli interior to the solar circle are increased compared to the baseline
IEM.
\citet{2012ApJ...750....3A} also found that better fits to the \gray{} data
were generally obtained by increasing the IC intensity for large regions
of the sky for the baseline IEMs.
While the \GP\ version used by \citet{2012ApJ...750....3A} only allowed
the calculation of all-sky IC intensity maps,
the decomposition of the IC intensity into Galactocentric annuli used here enables 
mismodelling of the IC emission due to uncertainties of the gradients in the
CR electron and ISRF distributions to be more accurately 
treated (Sec.~\ref{results:interstellar_emission}).

Along the Galactic plane the $\gtrsim 30$\% under-prediction
by the baseline model is reduced to $\lesssim \pm10$\% after this scaling, 
except for scattered regions.
The longitude range $l \sim -(15-70)^\circ$ is the largest such region 
where the intensity-scaled IEM over-predicts the data by $\sim 20-30$\% in the 
$1-3.16$~GeV band. 
Within $\sim 10^\circ$ of the mid-plane this may indicate that the
spectrum of the IEM related to the CR nuclei/gas interaction ($\pi^0$-decay)
in this region is too soft.
To account for this, additional
degrees of freedom to the spectrum\footnote{See Appendix~\ref{appendix:IEM}.}
of the $\pi^0$-decay interstellar emission are allowed,  
and the model is refitted for annuli 
interior to the solar circle following the same 
sequence of regions as for the intensity-scaled IEMs 
-- this is the `index-scaled' IEM variant.

Figure~\ref{fig:residual_Base_Yusifov_Intensity_allsky_1to3.16GeV_and_3.16to10GeV} lower panels show the fractional residuals for the index-scaled
variants of the Pulsars IEM.
The fractional difference 
for the $l \sim -(15-70)^\circ$ region in the $1-3.16$~GeV energy
range is reduced to $\lesssim \pm 10$\% with a slight increase in the
residual for the corresponding positive longitude range.
At mid-to-high latitudes the residual is reduced because the IC for the
annuli interior to the solar circle and Loop-I 
model are also refit. 

The intensity-scaled IEM gives a lower residual for positive longitudes
inside the solar circle around 1~GeV, while the index-scaled variant is closer
to the data at negative longitudes.
The converse appears at higher energies where instead the intensity-scaled IEM
gives lower residuals at negative longitudes than the index-scaled variant.
The underlying axisymmetric geometry for the IEMs is partly responsible 
for this: there is not enough freedom in the model parameters, even for the 
index-scaled variant, to account for differences due to any error in 
the assumption of
an average CR distribution for each Galactocentric annulus 
(the limitations of the IEM tuning procedure and resulting 
models are discussed further in Section~\ref{sec:limitations}).
Qualitatively, similar results for the scaled OBstars 
IEMs (not shown) are obtained.

It is not straightforward to identify a best IEM after fitting because the 
qualitative improvement for each over the corresponding 
baseline IEM is similar.
Consequently, all 4 (Pulsars/OBstars, intensity-/index-scaled) IEMs are used 
to estimate the 
fore-/background toward the $15^\circ \times 15^\circ$ region about the GC 
below.

\subsection{Modelling $15^\circ \times 15^\circ$ Region about the Galactic Centre}
\label{sec:15x15}

\subsubsection{Point-Source Candidates}
\label{sec:pointsource}

Point-source candidates (`seeds') are identified using the wavelet analysis 
algorithm {\it PGWave} \citep[][]{1997ApJ...483..350D,Ciprini:2007zz}, 
one of the source detection algorithms employed in the development of 
the \fermilat\ catalogs.
The method finds seeds subject to 
a user-specified signal-to-noise criterion
($3\sigma$ is used) based on the assumption of a locally constant background.
This step identifies true point sources, as well as 
structures in the interstellar emission that are indistinguishable from 
point sources due to the 
finite angular resolution and statistics of the \fermilat\ data, without 
dependence on the specifics of an IEM. 

Four energy intervals with 
spacing $\Delta \log_{10}$E~$=0.5$ covering $1-100$~GeV, i.e., 
$1-3.16$, $3.16-10$, $10-31.6$, and $31.6-100$~GeV are used.
{\it PGWave} is run for each energy interval and seeds that are 
above the signal/noise threshold are retained.
The seeds found for each energy interval are combined.
Seed locations within the 68\% containment
radius of the PSF for the highest energy interval ($\sim 0.1^\circ$) are 
considered duplicate.
Duplicate seeds are combined at the location determined from
the highest energy interval that exceeds the signal-to-noise criterion.
Over the energy bands there are 142 unique seeds.

{\it PGWave} does not provide spectral information for the point-source 
seeds. 
The spectra of the candidates are initially evaluated using
{\it PointLike}, a package for maximum-likelihood 
analysis of \fermilat\ data \citep{2010PhDT.......147K,2012ApJ...756....5L}.
{\it PointLike} also has the  
capability of optimising positions for seeds from the 
{\it PGWave}-determined list, but it requires an IEM. 
The Pulsar and OBstars intensity-scaled IEMs are used for this step.
The point-source parameters are allowed to vary for the {\it PointLike}
optimisation, while
the IEM and isotropic components of the background model are held constant.
This enables the optimisation of the positions for the point-source 
candidates as well as determine preliminary spectral parameters.

A PowerLaw (PL)\footnote{PL: $dN/dE = N_0 E^{-\alpha}$ with 
parameters $N_0$ and $\alpha$} is assigned as the spectral 
model to each seed and the spectral parameters are fit 
to make initial evaluations
for the fluxes, spectral indices, and a test statistic {\it 
(TS)} \citep{1996ApJ...461..396M}. 
Because no spectral information for the seeds are initially available 
the parameters for each are fit with 
all other candidates set to zero and held constant.
Using the trial spectra, the seeds are refit 10 at a time ordered by decreasing
{\it PointLike}-determined {\it TS}, with the flux normalisation 
of the other candidates  held constant. 
The combined list of candidates is refit 
using the information from the subsets to obtain a set of stable power-law 
spectral parameters.

Following the pass to determine initial PL spectral models, 
the individual candidates ranked by the {\it PointLike}-determined
{\it TS} in descending order are refit with two hypotheses for the 
spectral properties,
PL and LogParabola (LP)\footnote{LP: 
$dN/dE = N_0 (E/E_b)^{-\alpha - \beta \log (E/E_b)}$ 
with parameter tuple $N_0, \alpha,\beta$, and $E_b$}
\footnote{An exponential cut-off power law was not included in 
the spectral templates, 
as employed in the 3FGL, because of the limited energy range for the current 
study.}.
The following criteria are used to choose between the spectral models:  
if the {\it PointLike}-determined {\it TS} 
is $< 50$, a PL spectrum is assigned to the point-source seed.
If the {\it PointLike}-determined {\it TS} is $> 50$, the seed 
is fit with a PL and a LP 
spectrum, allowing also the spectral parameters for other candidates within 
$3^\circ$ to vary.
If {\it TS}(LP)~$>$~{\it TS}(PL)~+~9, a LP 
spectrum is assigned to the seed; a PL spectrum is assigned if this 
condition is not met.

The spectra of the combined list of candidates are 
refit and only candidates with {\it PointLike}-determined
{\it TS} $>9$ retained for the maximum-likelihood 
fit (Sec.~\ref{sec:maxlikelihood}).
The optimisation step reduces the number of candidates by $\approx60$\% 
from the initial {\it PGWave}-identified list.

For the only extended source that has been previously identified in the 
region, the W28 supernova remnant \citep{2010ApJ...718..348A}, the 
spatial template and spectral model employed for the 3FGL analysis were used.
The spectral parameters are refit during the maximum-likelihood procedure.

\begin{figure*}[ht]
\subfigure{
\includegraphics[trim = 0 98 10 98, scale=0.40]{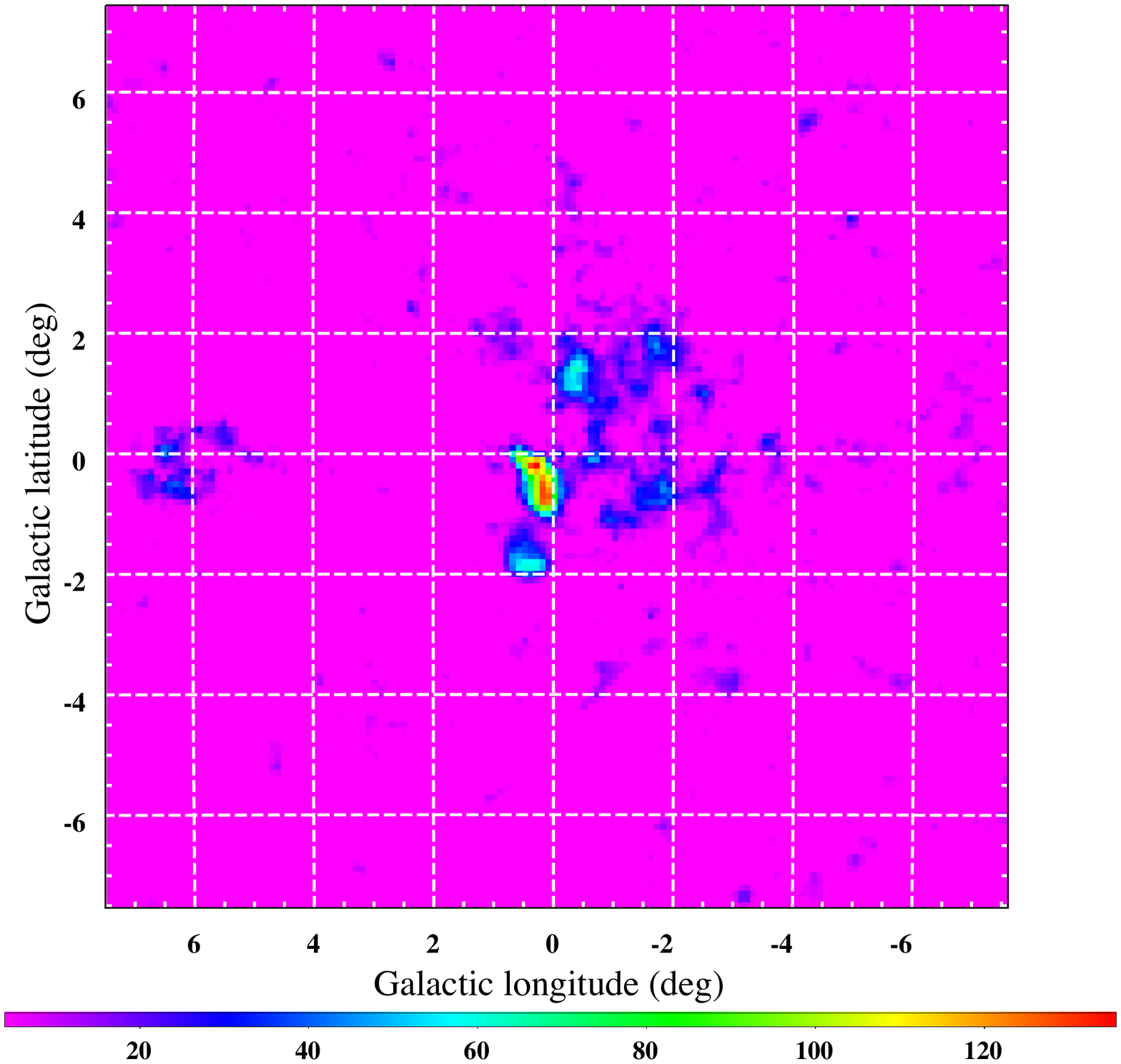}
\includegraphics[trim = 0 98 10 98, scale=0.40]{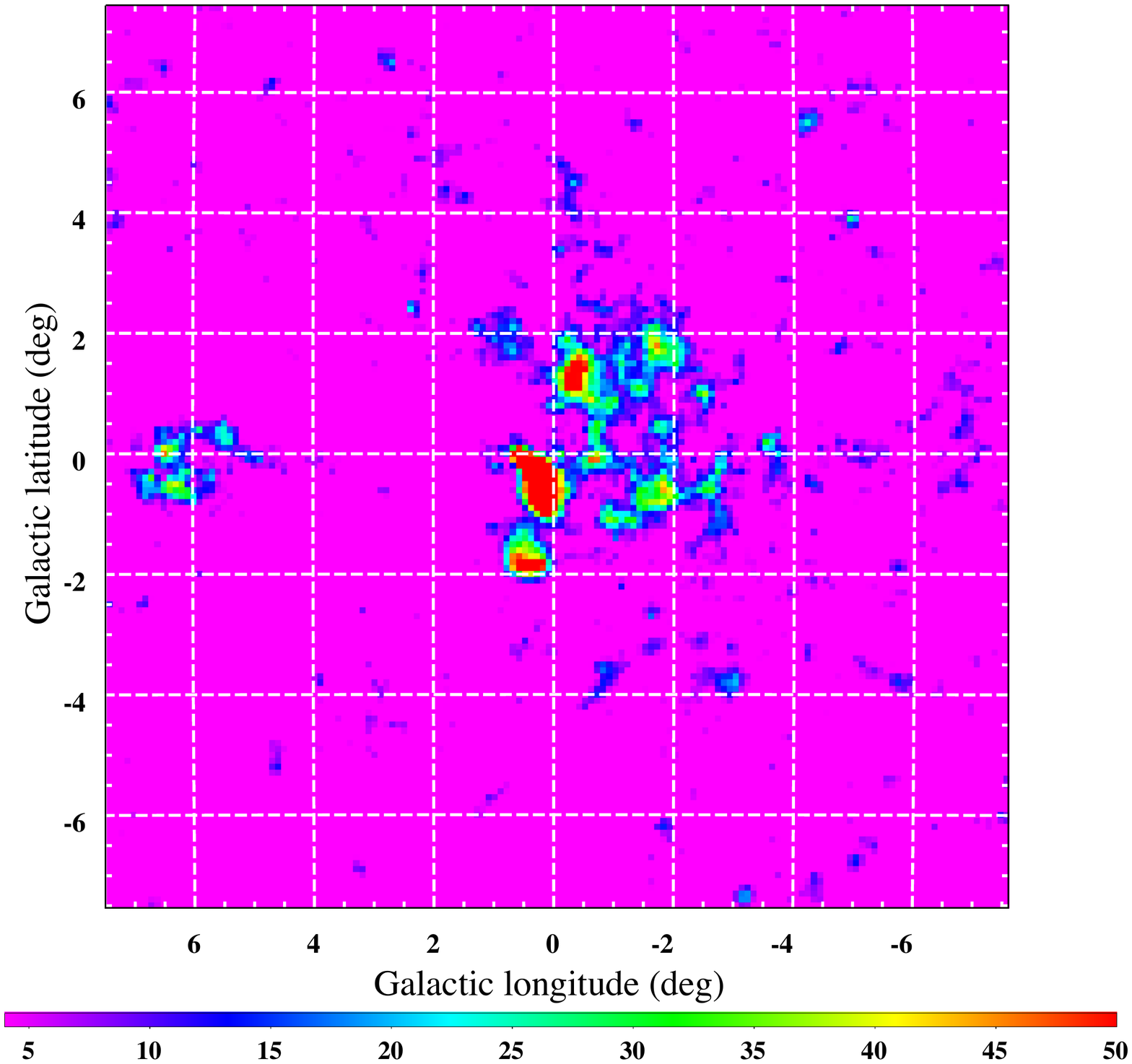}
}
\subfigure{
\includegraphics[trim = 0 98 10 98, scale=0.40]{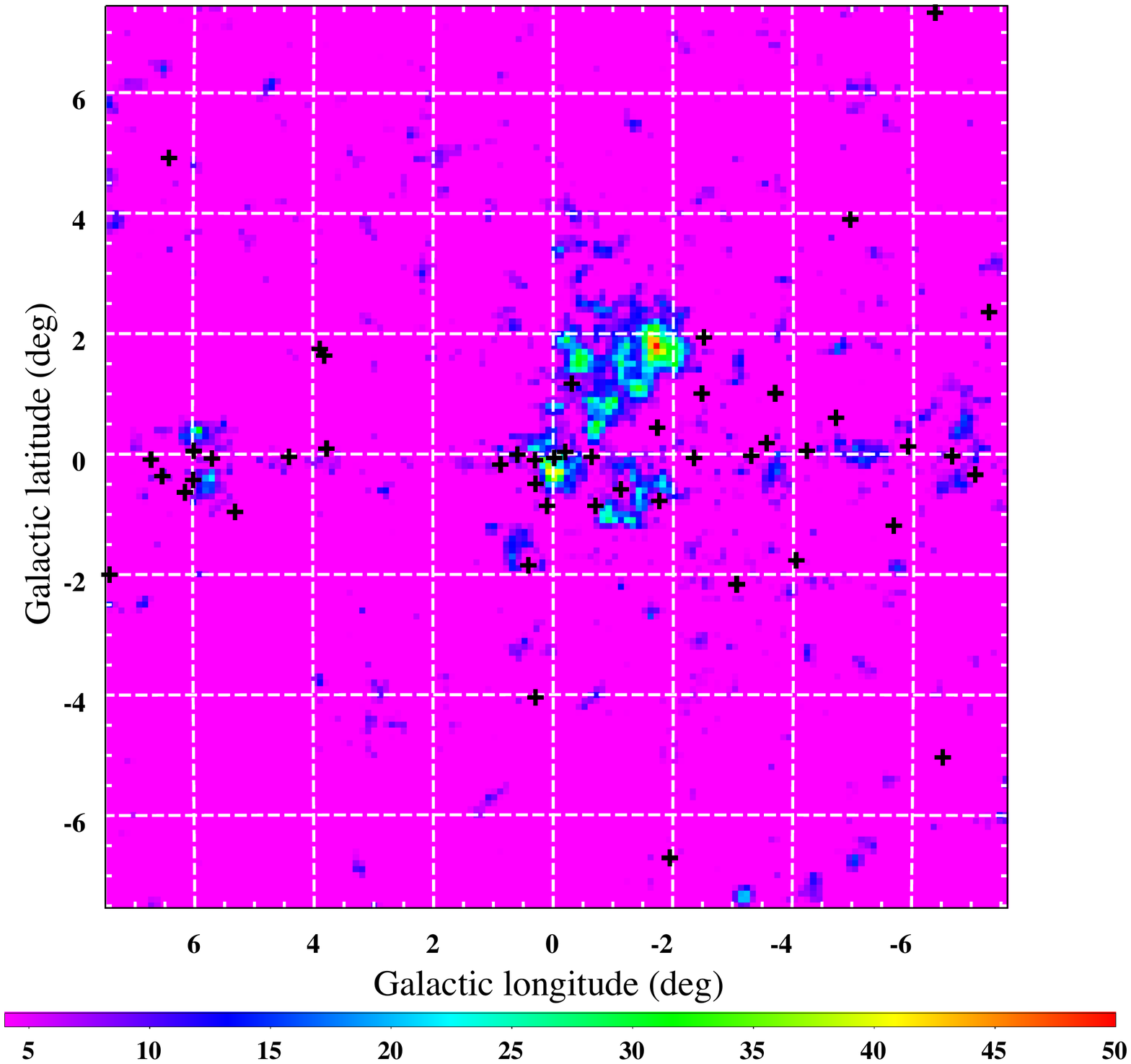}
\includegraphics[trim = 0 98 10 98, scale=0.40]{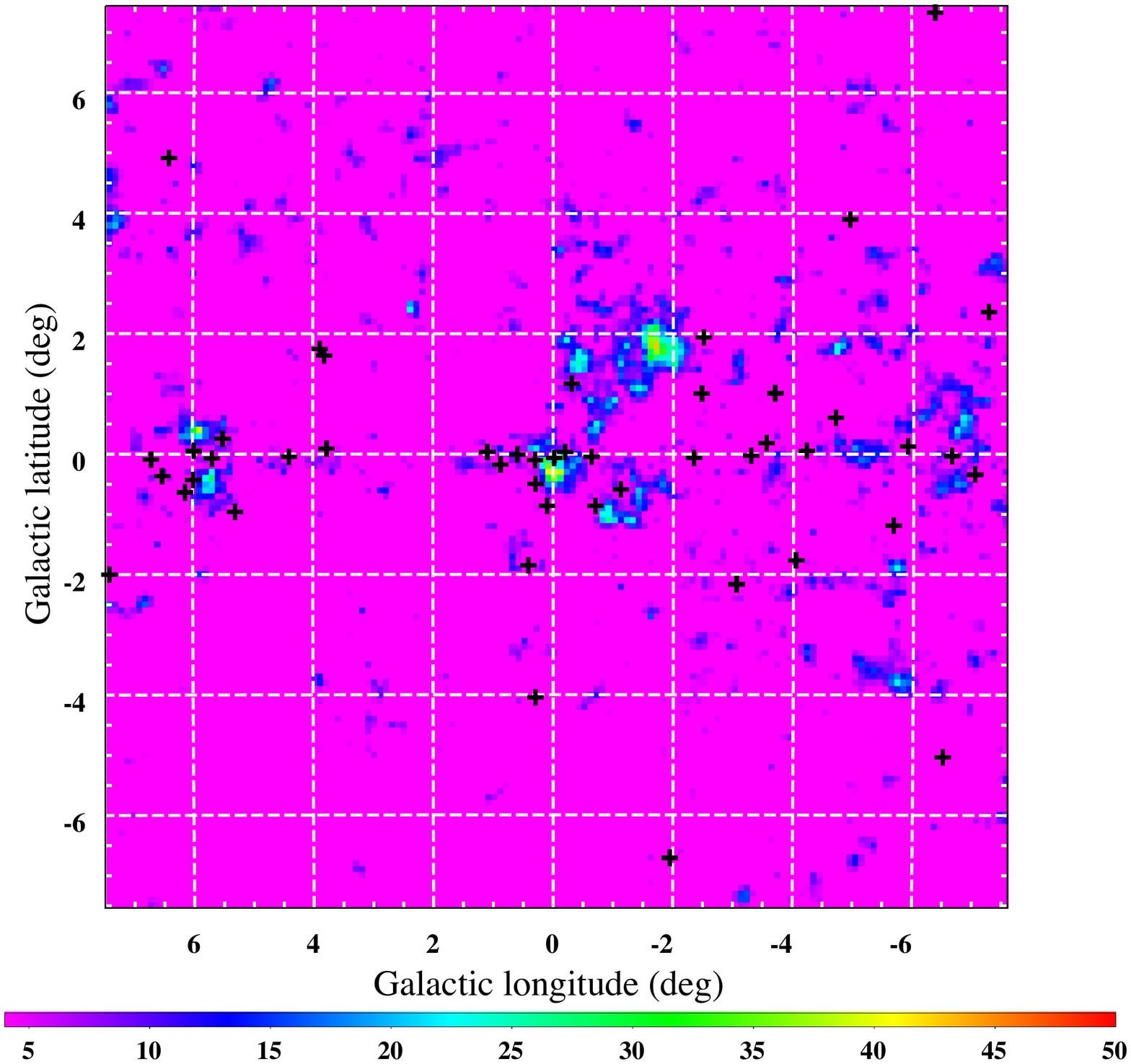}}
\caption{{\it TS} map corresponding to the maximum-likelihood 
result including the Pulsars intensity-scaled IEM after the first 
iteration for the $15^\circ \times 15^\circ$ region about the GC 
({\it top panels; left: full scale, right: TS$<$50}) and after the 
second iteration ({\it bottom, left panel}). 
The {\it TS} map for the only iteration of the analysis for the OBstars 
intensity-scaled model is also shown ({\it bottom, right panel}). 
The black crosses 
indicate the location of the $TS>$25 point sources.
 \label{fig:tsmaps}}
\end{figure*}

\subsubsection{Combined Interstellar Emission and Point Source Fit}
\label{sec:maxlikelihood}

Point-source candidates are combined with their {\it PointLike} trial 
spectra together with the IEMs described in 
Section~\ref{sec:interstellaremission}
in a second maximum-likelihood fit. 
A binned likelihood 
fit is performed using the {\it Fermi} ScienceTool {\it gtlike}.
The templates for the $\pi^0$-decay related \gray{} intensity from \hi\ and 
CO, and the IC emission, in annulus~1 are freely scaled in the 
fitting procedure.
The size of these templates is slightly larger than the 
$15^\circ \times 15^\circ$ region so the fit results are only strictly valid
within $\sim 1$~kpc of the GC rather than the formal $~1.5$~kpc extent of
annulus~1.
The annulus~1 templates are fixed in spatial distribution and spectra to the 
respective \GP\ predictions; for the index-scaled IEM variants allowing 
additional spectral freedom to the annulus~1 $\pi^0$-decay components 
was tried as well, but the fits were unstable.
The contributions of the IEM and the isotropic component, 
as determined by the procedure 
outlined in Section~\ref{sec:interstellaremission}, 
are held constant in the fit.
The scaling factors for the interstellar emission templates for annulus~1 are 
fit concurrently with the spectral parameters of the point-source seeds.
Because of the large number of point-source seeds, the fit is performed 
iteratively, starting from the largest {\it TS} candidates and progressively 
fitting the lower {\it TS} ones while 
the rest are fixed to their best fit values from the previous iteration.
The normalisation of the aforementioned innermost ring IEM intensities 
are free parameters in each iteration.

The results of the maximum-likelihood fit are values and confidence ranges 
for the coefficients of 
the \hi\ annulus~1, CO annulus~1, IC annulus~1, as well as the
{\it TS}, fluxes and spectra for the point sources.
All point sources with a maximum-likelihood determined 
$TS > 9$ are included in the model;
a $TS = 25$ threshold is used for a formal detection, 
corresponding to just over $4\sigma$ as for the 3FGL and other
\fermilat\ source catalogs.

\subsubsection{Residual Maps and Iteration}
\label{sec:residual_iteration}

A potential drawback of using the wavelet detection algorithm is that 
fainter point sources may be missed with a single iteration.
This is remedied in this analysis 
by iterating the point-source seed detection on
the residual maps following the maximum-likelihood fit and 
rerunning the analysis chain.
The iteration is made until there are no new significant excesses in 
the {\it TS} map of the region
that satisfy the signal-to-noise criterion adopted
in the {\it PGWave} detection step.
The {\it TS} map is determined by moving a putative point source with a PL 
with spectral index $-2$ using {\it PointLike} through a 
grid of locations 
in the region and by maximising the likelihood function at each location.  
The positions of peaks with {\it TS}$>9$ are added to the source model.

Figure~\ref{fig:tsmaps} (left panel) shows the 
{\it TS} map following the {\it gtlike} maximum-likelihood fit using these candidates 
in the first full iteration for the Pulsars intensity-scaled IEM.
Some significant excesses remain after the initial pass, in particular around 
the GC.
The {\it PGWave} and {\it PointLike} seed-finding and optimisation 
steps are repeated (Sec.~\ref{sec:pointsource}), 
finding 37 additional candidates with a {\it PointLike}-assigned {\it TS} 
$>9$ for the Pulsars intensity-scaled IEM.
The left bottom panel in Fig.~\ref{fig:tsmaps} shows the {\it TS} map 
after the second full iteration of the analysis 
of the $15^\circ \times 15^\circ$ region for this IEM. 
Some excesses with {\it TS} $> 25$ remain. 
However, these correspond to point-source candidates reported by
{\it PGWave} in the first iteration that were rejected 
because their {\it gtlike}-assigned {\it TS} $<9$, possibly due to
structures having angular extensions that are larger than the PSF core.

The procedure for the initial optimisation, binned 
likelihood fit, and iteration is also done using 
the OBstars intensity-scaled IEM.
This yields 85 point source candidates with {\it PointLike}-assigned 
{\it TS} $>9$.
The point-source candidates are combined with their {\it PointLike} trial 
spectra together with the OBstars intensity-scaled IEM
in a maximum-likelihood fit using {\it gtlike}, following the same procedure 
as outlined above.
The {\it TS} 
map after the {\it gtlike} maximum-likelihood fit for this model is shown in 
the bottom right panel of Fig.~\ref{fig:tsmaps}.
The results are very similar to the second iteration of the analysis 
using the Pulsars intensity-scaled model.
Consequently, no further iteration is made for the OBstars model.

For the Pulsars and OBstars index-scaled variants, the 
point-source seeds with {\it TS} $>9$ from the 
corresponding intensity-scaled IEM are used in the source model for 
the maximum-likelihood fit. 
This procedure yields {\it TS} maps 
that are very similar to the respective intensity-scaled counterparts.  
The results are summarised in Section~\ref{results}.

\section{Results}
\label{results}

\begin{figure*}[ht]
\subfigure{
\includegraphics[scale=0.32]{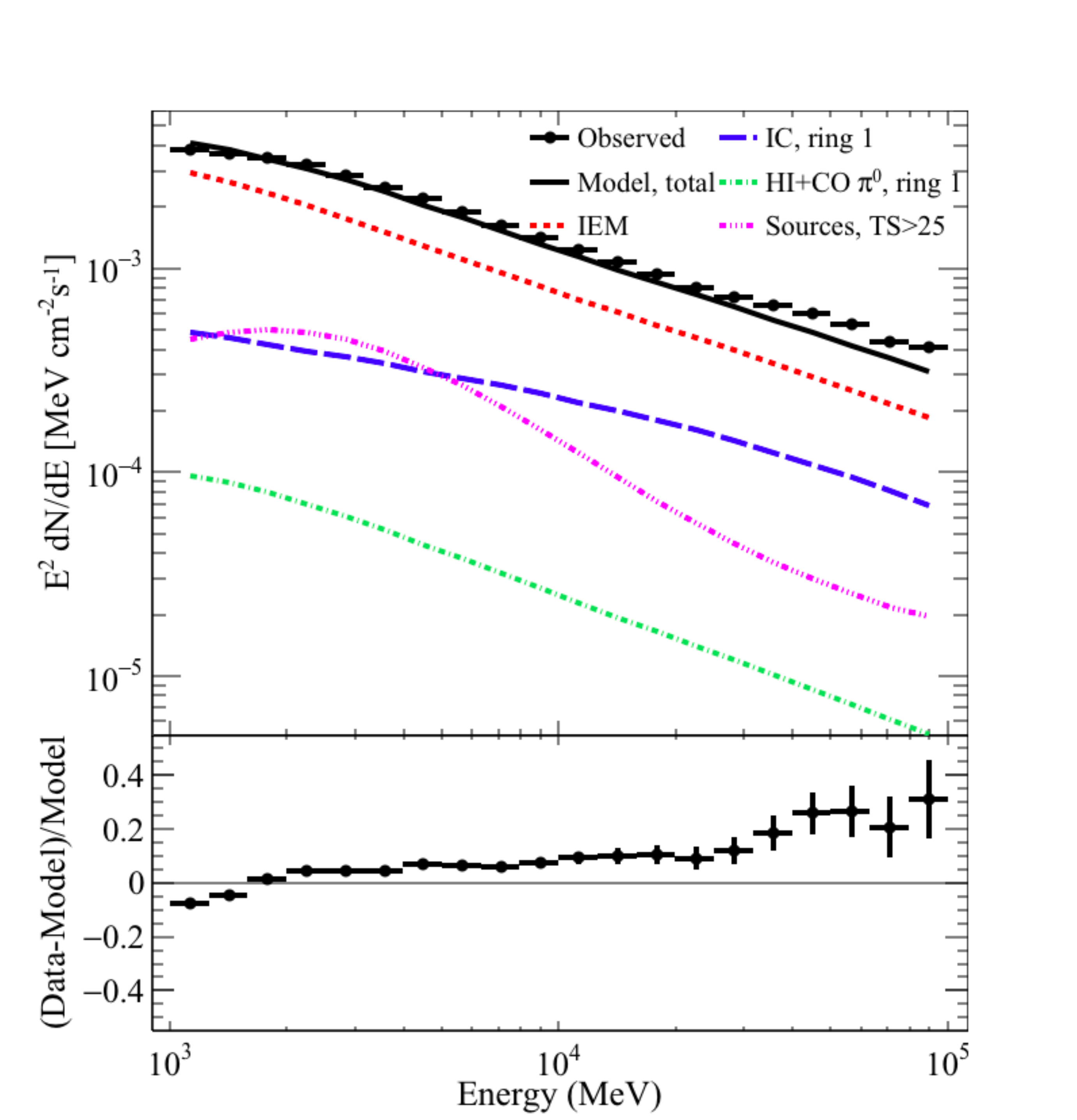}
\includegraphics[scale=0.32]{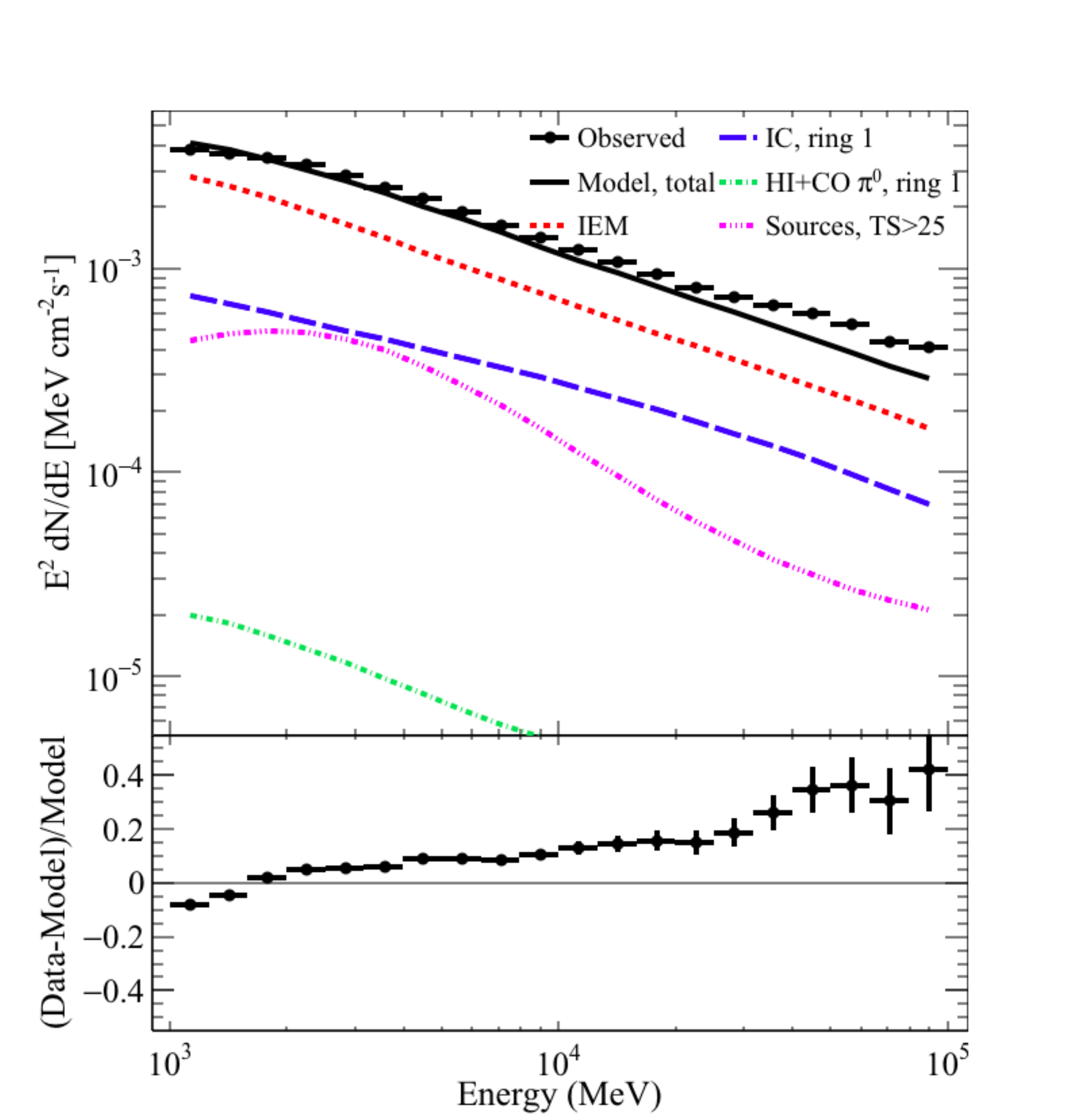}}
\subfigure{
\includegraphics[scale=0.32]{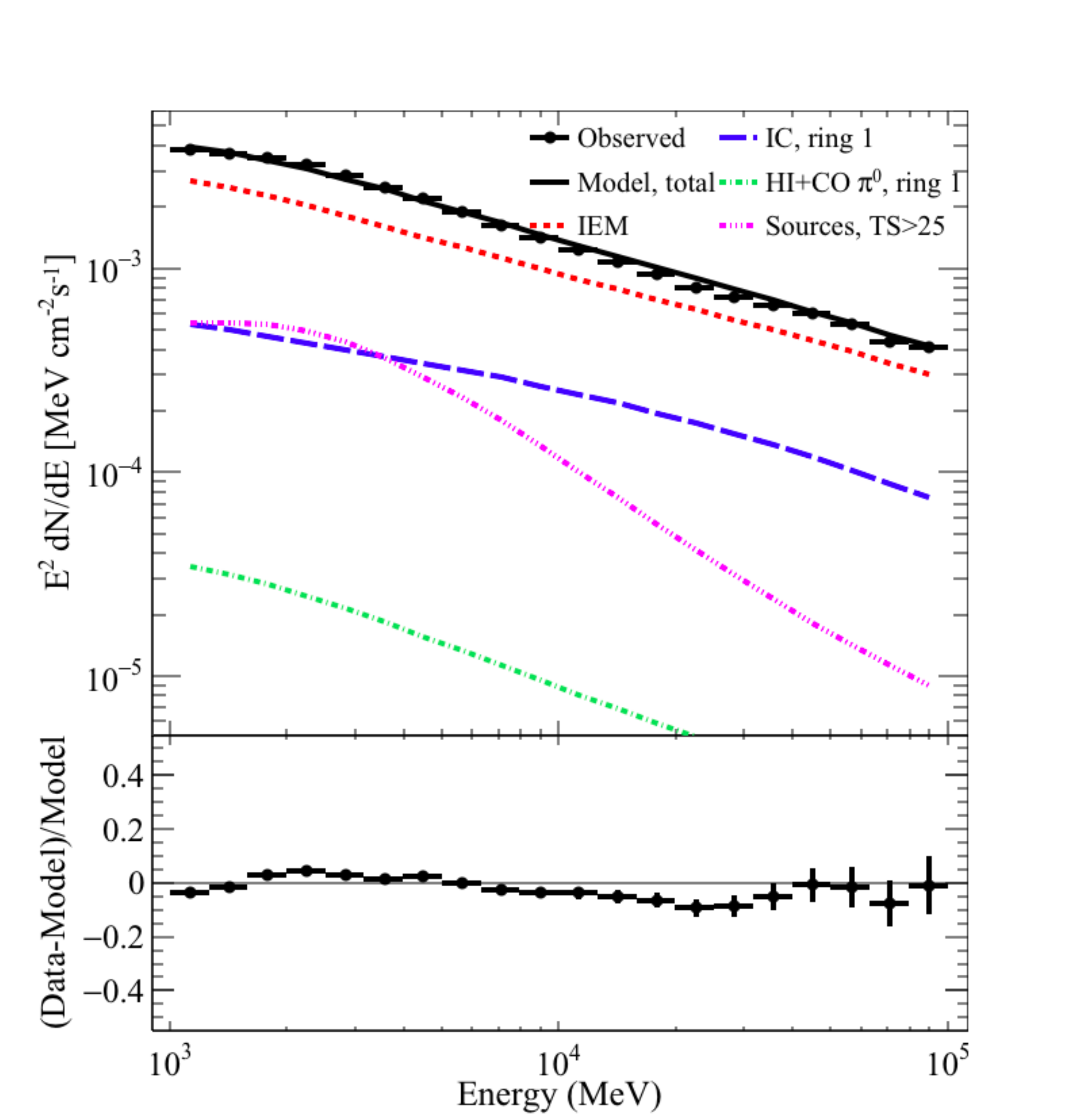}
\includegraphics[scale=0.32]{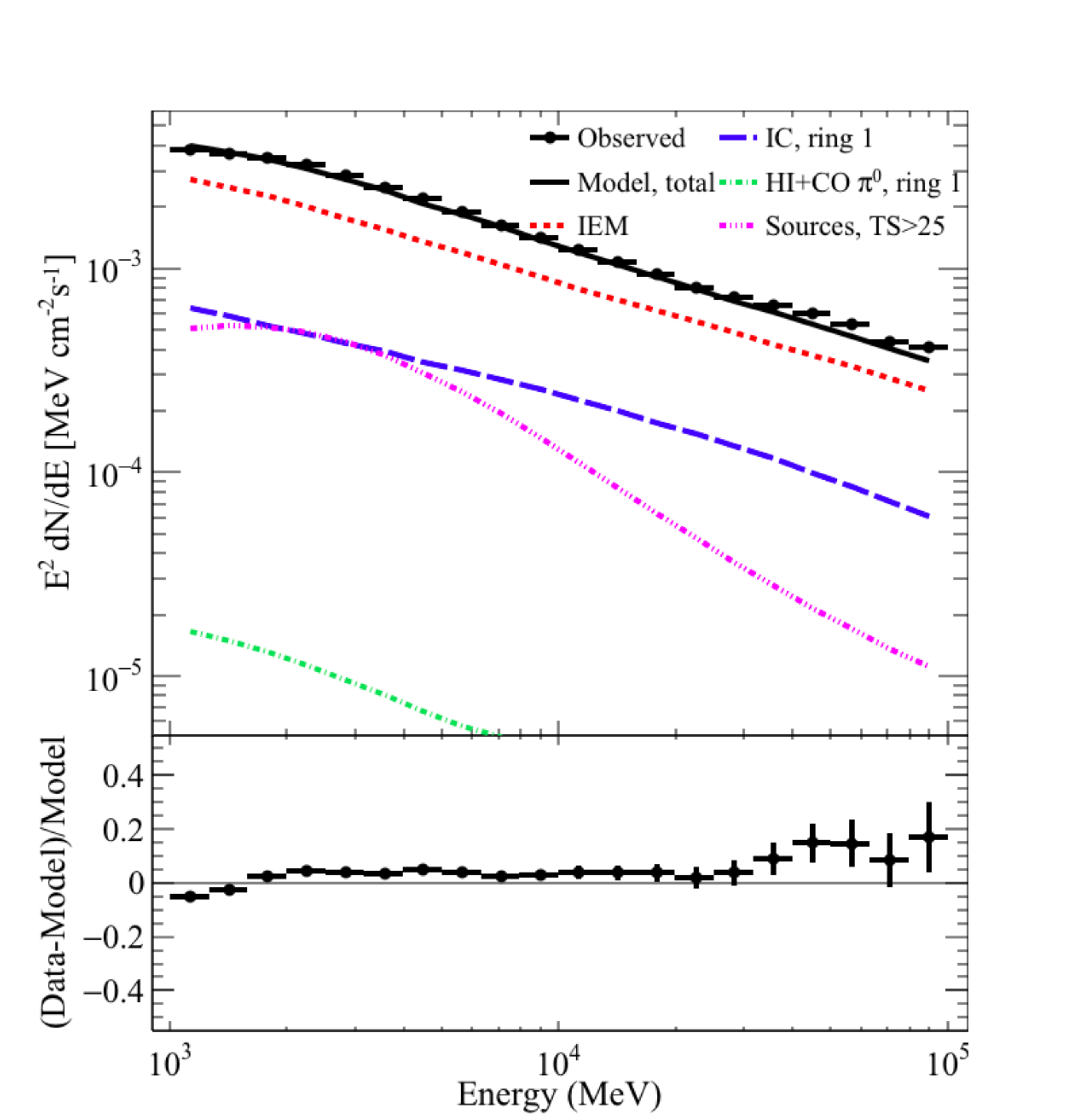}}
\caption{
Differential fluxes for the $15^\circ \times 15^\circ$ region about the GC 
for the four IEMs constrained as described in 
Section~\ref{sec:interstellaremission}.
Upper row shows the results for the intensity-scaled IEMs based on the 
Pulsars (left) and OBstars (right) source distributions.
Lower row shows the results for the index-scaled IEMs 
based on the Pulsars (left) and OBstars (right) source distributions.
Line styles: solid (total model), long-dash (IC, annulus 1), 
dot-dash (\hi\ and CO gas $\pi^0$-decay, annulus 1), 
dot-dot-dot-dash (point sources), dash (Galactic interstellar emission excluding annulus 1 for IC, \hi\ and CO gas $\pi^0$-decay). 
Solid circles: data.
\label{results:interstellar_emission_fluxes}}
\end{figure*}

\begin{figure*}[ht]
\subfigure{
\includegraphics[scale=0.4]{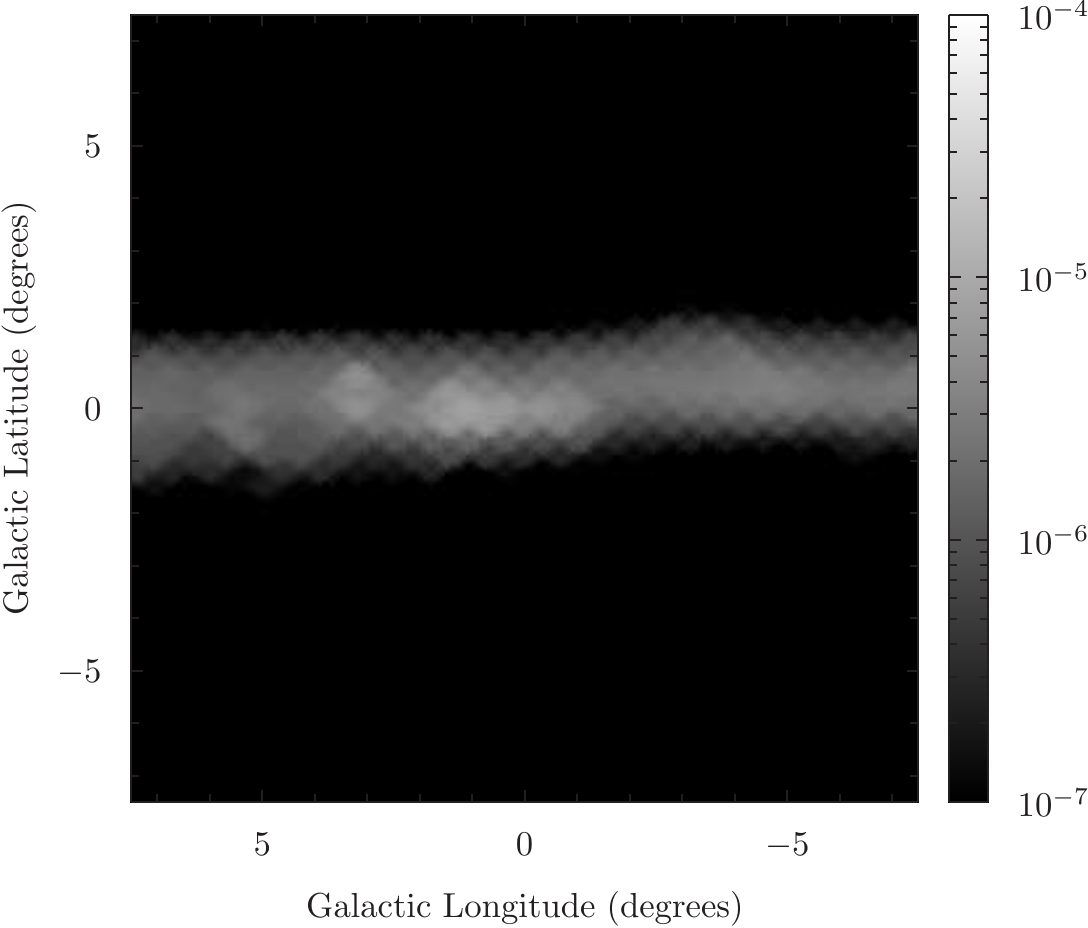}
\includegraphics[scale=0.4]{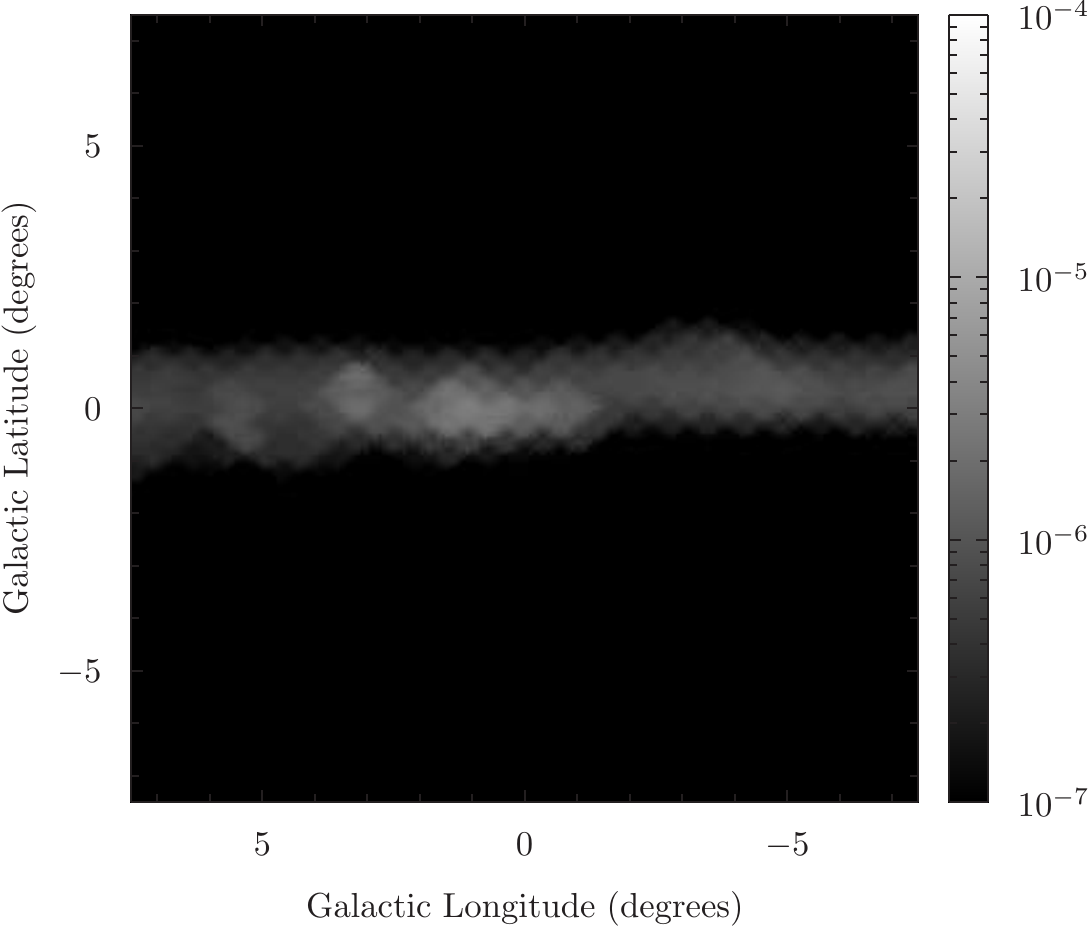}
\includegraphics[scale=0.4]{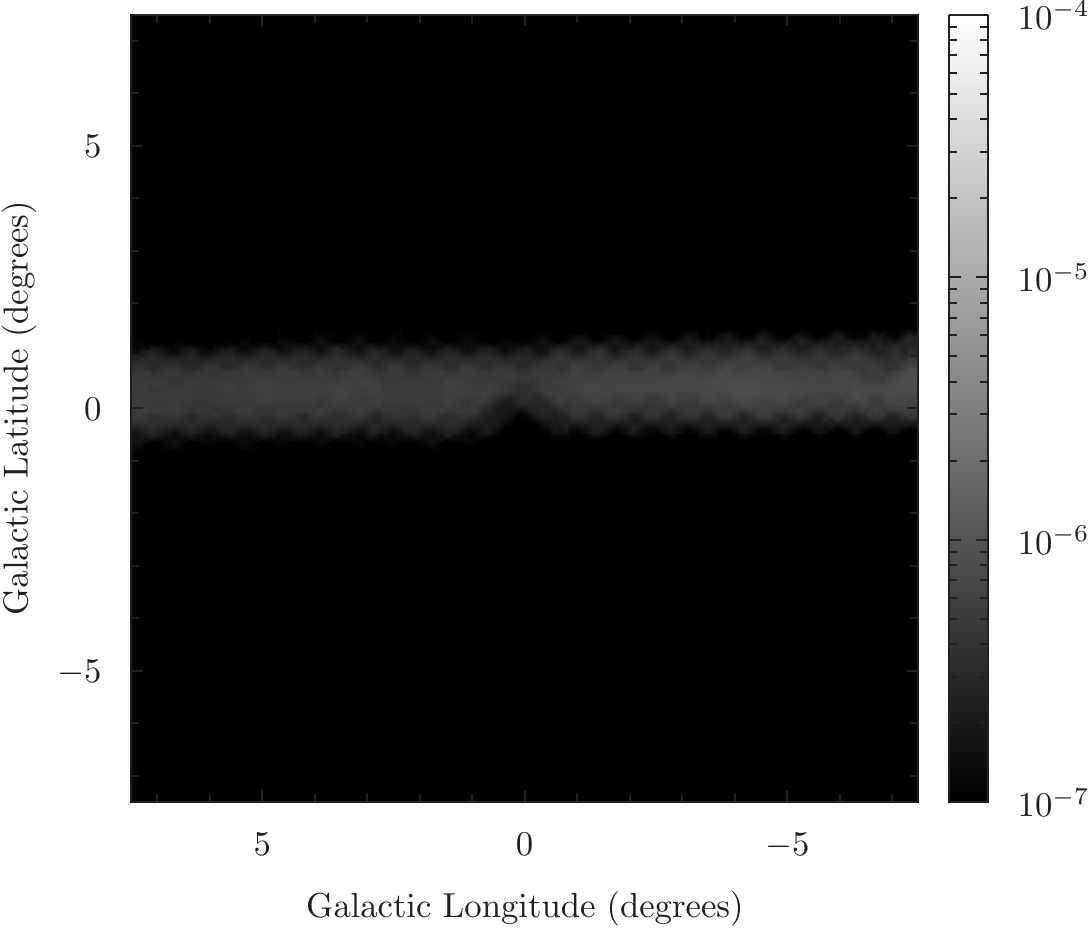}
\includegraphics[scale=0.4]{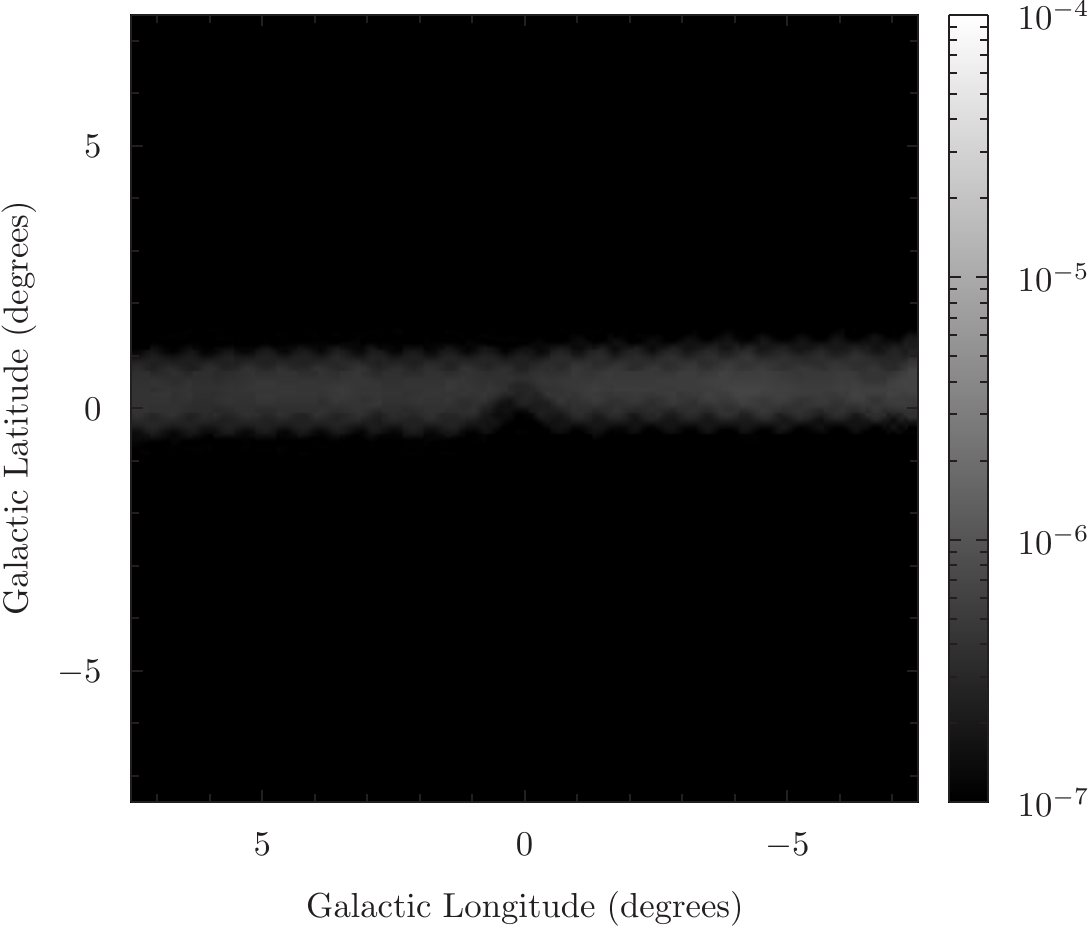}
}
\subfigure{
\includegraphics[scale=0.4]{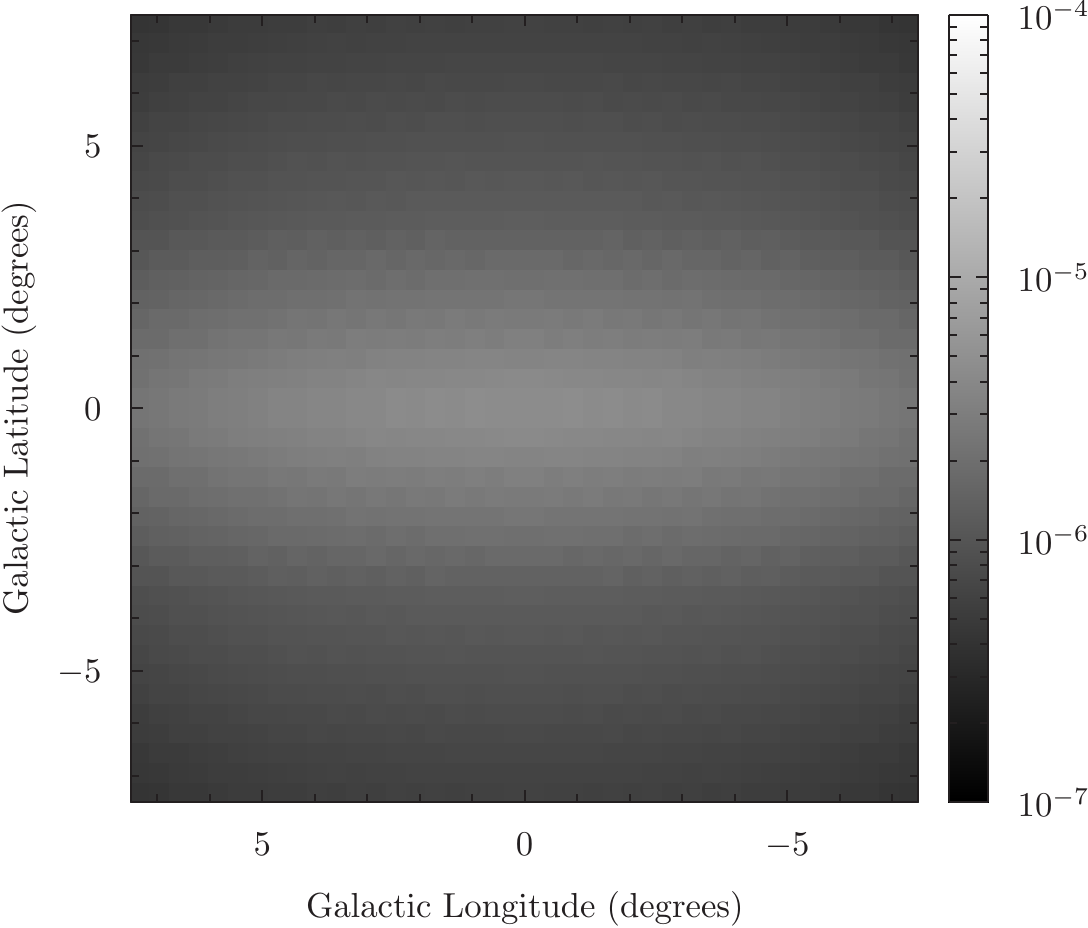}
\includegraphics[scale=0.4]{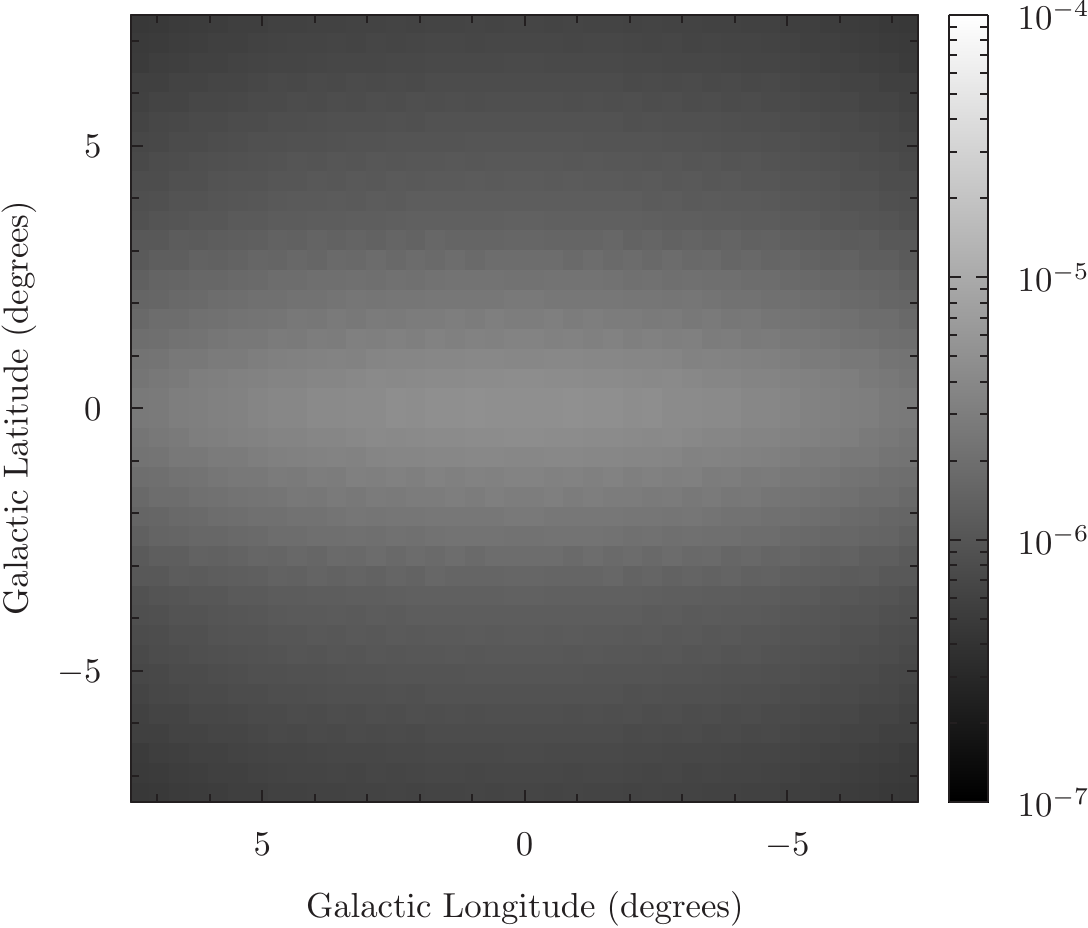}
\includegraphics[scale=0.4]{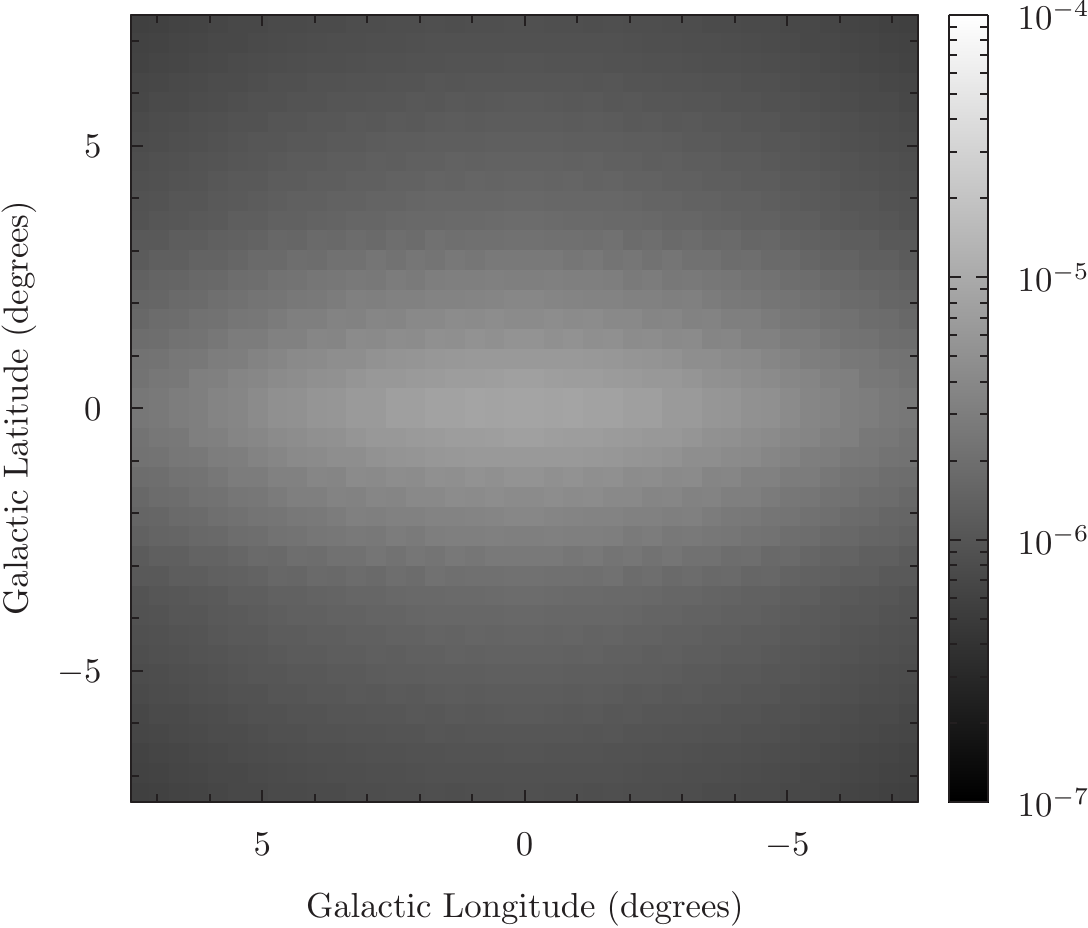}
\includegraphics[scale=0.4]{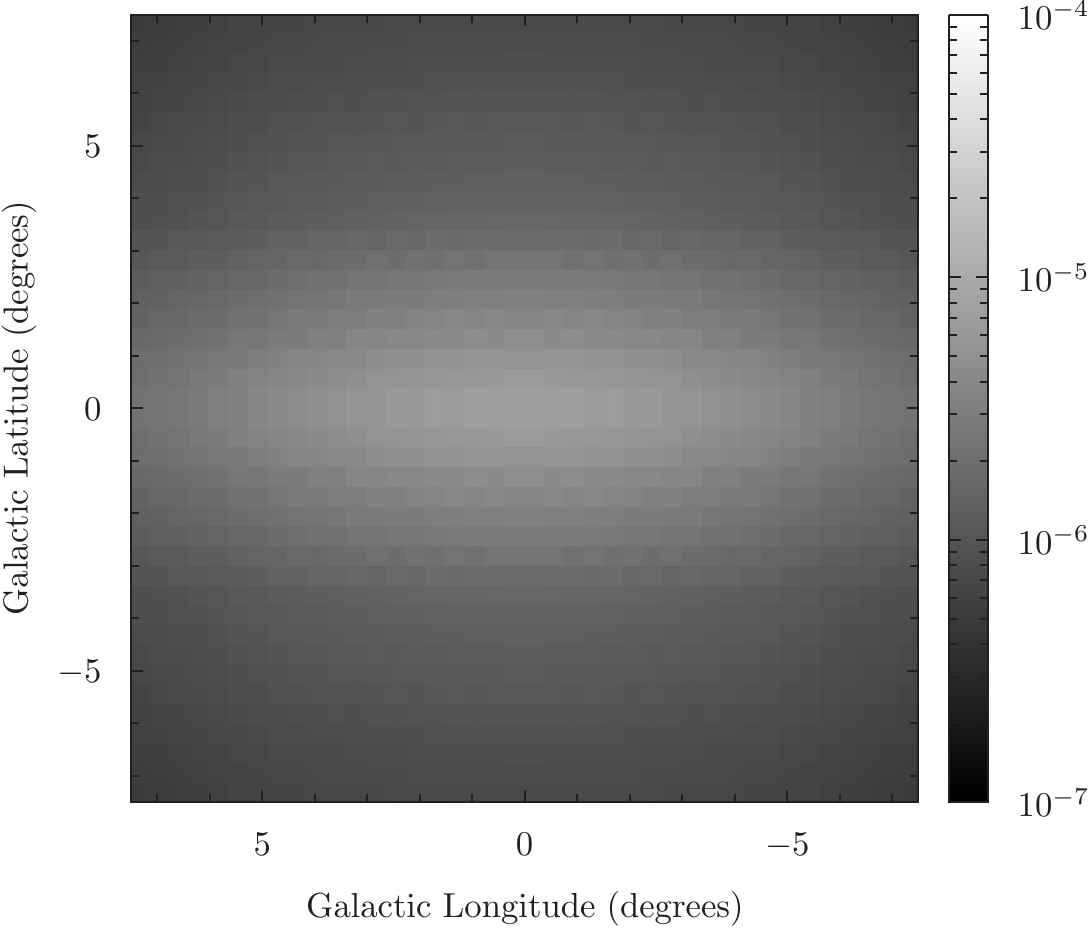}
}
\subfigure{
\includegraphics[scale=0.4]{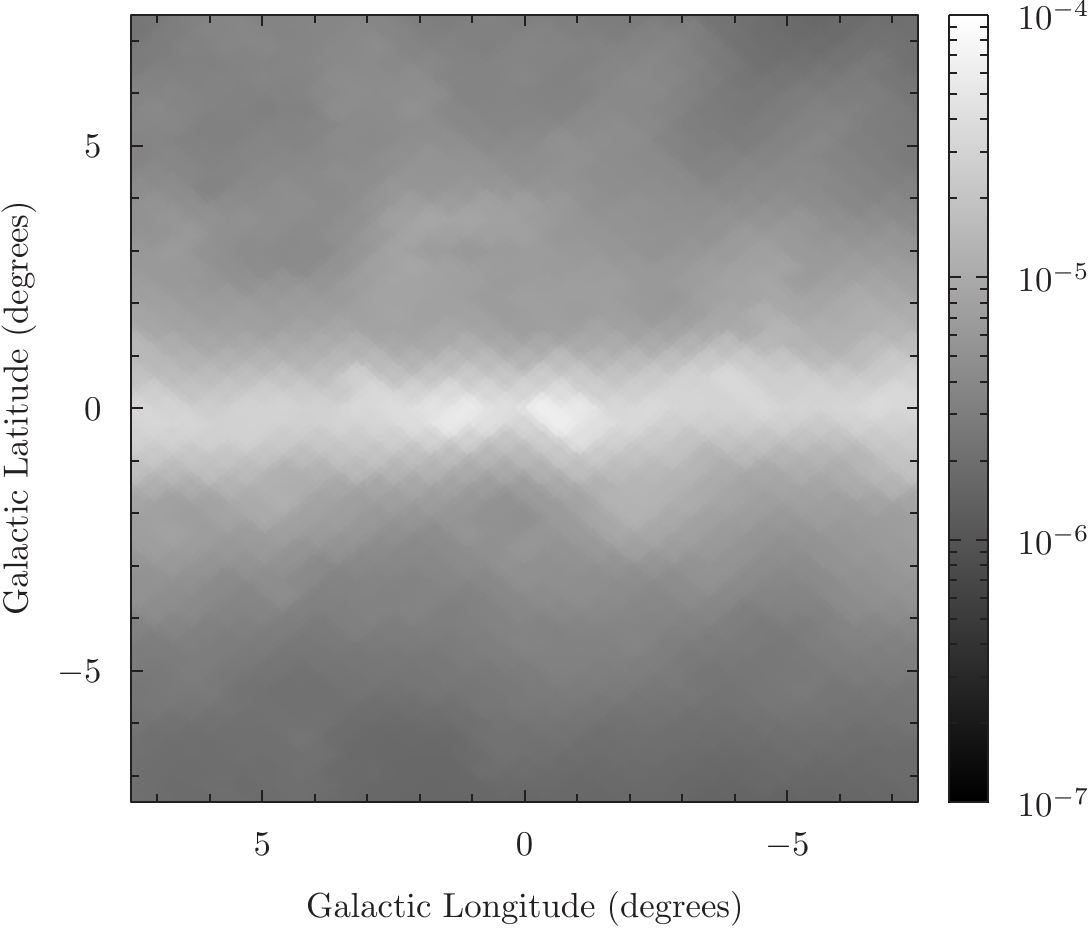}
\includegraphics[scale=0.4]{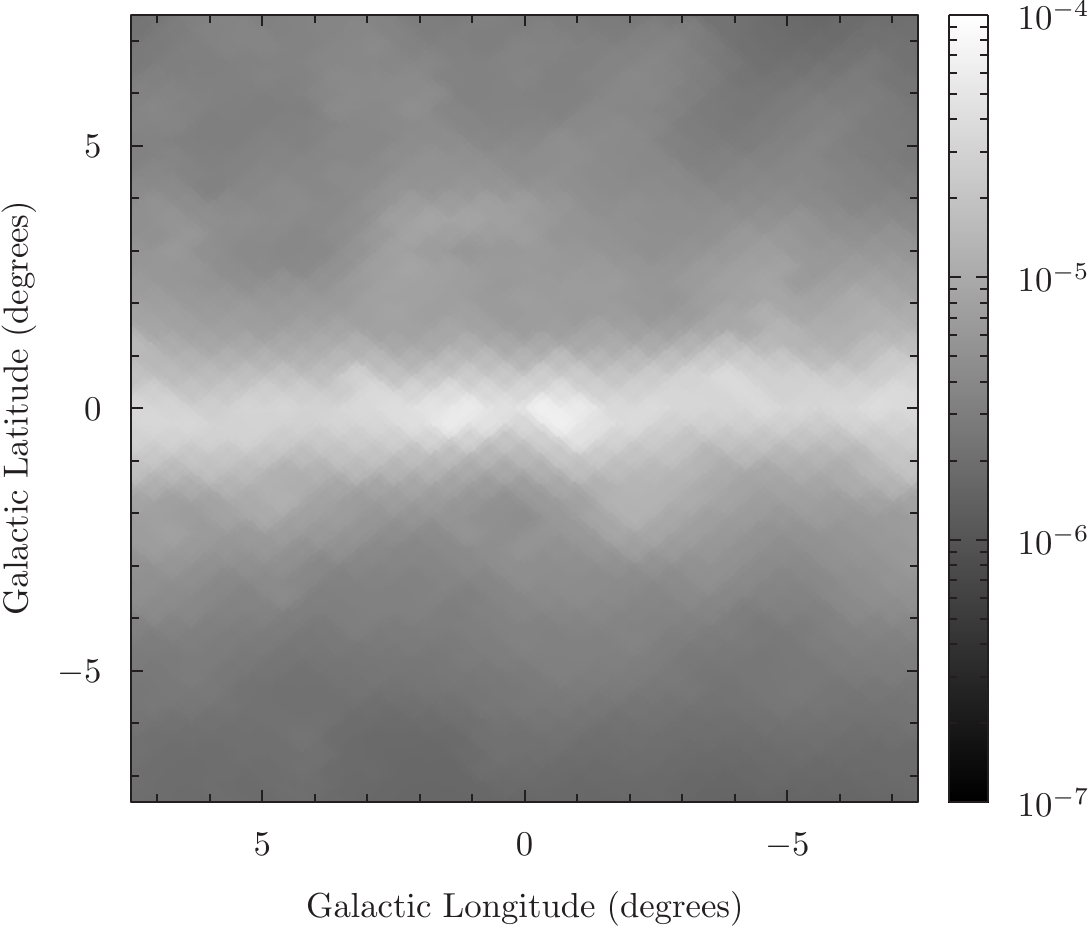}
\includegraphics[scale=0.4]{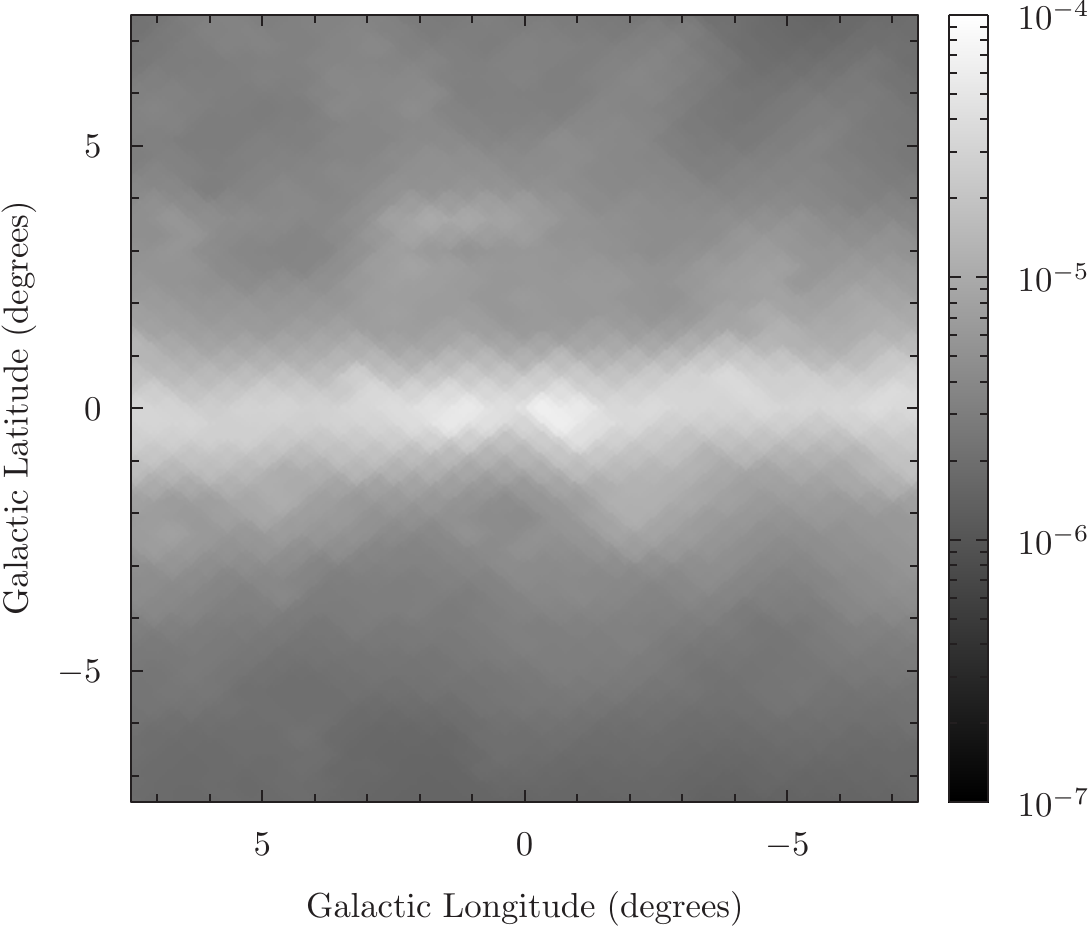}
\includegraphics[scale=0.4]{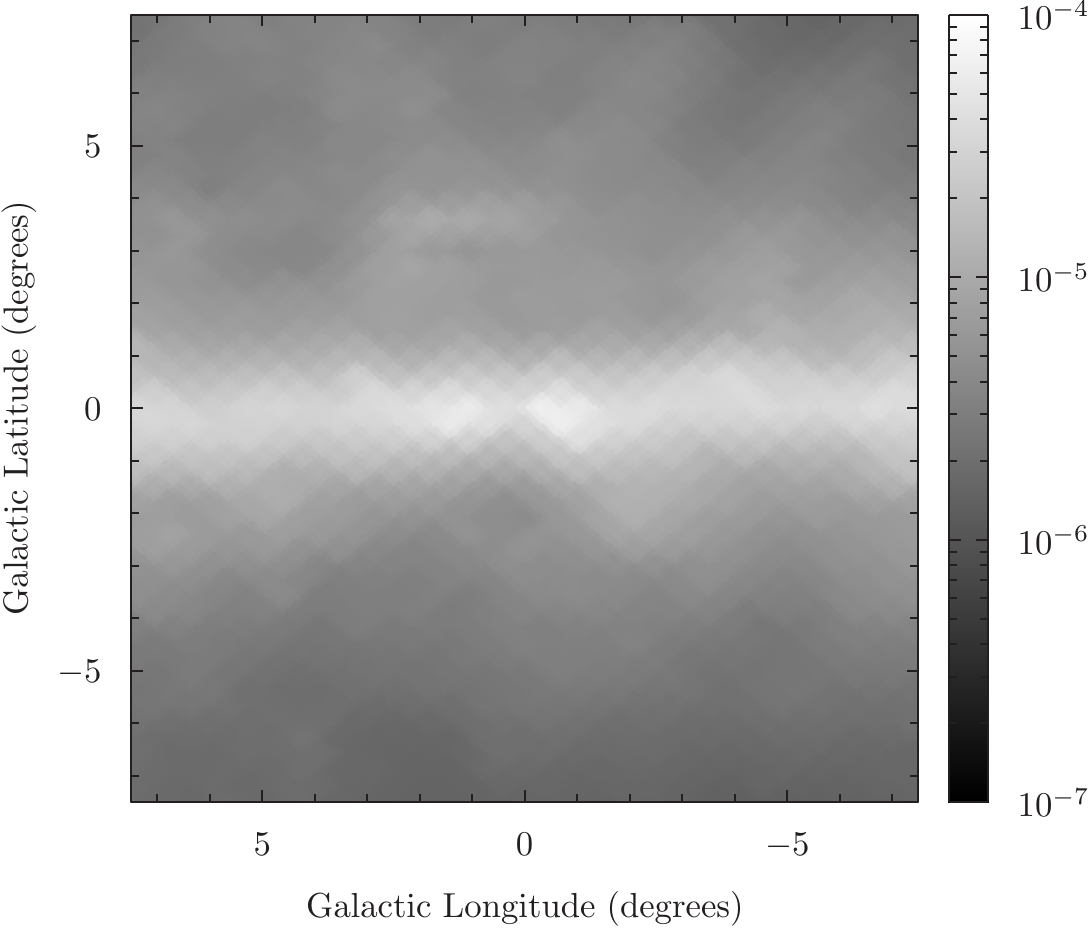}
}
\caption{
Interstellar emission model components $>1$~GeV for the fit results for the $15^\circ \times 15^\circ$ region about the GC and the fore-/background.
First column: Pulsars intensity-scaled; second column: Pulsars, index-scaled;
third column: OBstars, intensity-scaled; fourth column: OBstars, index-scaled.
First row: $\pi^0$-decay intensity for 
annulus~1 after the maximum-likelihood fit (Section~\ref{sec:maxlikelihood}); 
second row: IC intensity for annulus~1 after fitting; 
third row: total fore-/background interstellar emission.
Colour scale units: cm$^{-2}$ s$^{-1}$ sr$^{-1}$. 
\label{fig:results_ROI_components}}
\end{figure*}

\subsection{Interstellar Emission}
\label{results:interstellar_emission}

Figure~\ref{results:interstellar_emission_fluxes} shows the differential 
spectra of the individual components obtained for the 4~IEMs 
integrated over the $15^\circ \times 15^\circ$ region about the GC.
The figure separates the emission components in terms of the contributions
by $\pi^0$-decay and IC for annulus~1, the interstellar emission 
fore-/background,
and point sources over the region.
As expected, the fore-/background dominates for each IEM, which is
predominantly $\pi^0$-decay in origin.
IC scattering is the dominant interstellar emission component over the 
inner $\sim1$~kpc
This contrasts with the predictions of the baseline Pulsars and OBstars IEMs, 
in which the neutral gas $\pi^0$-decay interstellar emission 
components over the same region have similar fluxes.
Combined, the \GP-predicted 
\hi\ and CO-related $\pi^0$-decay annulus~1 templates are 
brighter by up to an order of magnitude than the IC emission for either model.
The fit for the $15^\circ \times 15^\circ$ region preferentially 
adjusts the annulus~1 IC component while suppressing the \hi\ and CO-related
$\pi^0$-decay templates for all IEMs.
The scaling factors for the annulus~1 IC templates are $\sim6-30$, with 
higher values for the OBstars IEM variants.
While the difference in the scaling factors between the IEMs is 
large, the final flux determined for the IC over annulus~1 over all 
four IEMs is within a factor $\sim 1.5$.

Figure~\ref{fig:results_ROI_components} shows the spatial distribution of the
individual components over the $15^\circ \times 15^\circ$ region about the GC 
for energies $>1$~GeV for each IEM.
As noted above, the fore-/background interstellar emission is clearly 
the brightest component.
Although the intensity scale does not directly show a strong variation within 
$\sim \pm 1^\circ$ of the mid-plane the fore-/background over the 
four IEMs varies by $\sim 30$\%, which is similar
to elsewhere in the plane.
Outside of the plane the brightness of individual features varies.
For example, around $l \sim 0-2^\circ, b \sim 3^\circ$ the Pulsars IEMs 
are dimmer.
Other low-intensity structures appear to change subtly for the different IEMs.
This is predominantly 
due to scaling of the $\pi^0$-decay emission from annuli $2-4$, which is discussed further below. 

The IC intensity is more peaked toward the GC for the OBstars 
model compared to the Pulsars IEM.
This is due to spatial distributions of CR sources employed 
for the respective IEMs, which result in different spatial distributions 
for the propagated CR electron intensities over the inner few kpcs.
The paucity of sources within a few kpc of the GC for the OBstars IEM gives
a constant intensity because the electrons must diffuse from larger 
Galactocentric radii.
On the other hand, the Pulsars source 
distribution peaks around a few kpc and is 
non-zero in toward the GC, giving a higher electron intensity with a gradient.
Combined with the spatial distribution over the same region 
of the ISRF intensity, which peaks at the GC, the results are 
IC templates that are broader (Pulsars) or more 
peaked (OBstars) for annulus~1.

Table~\ref{tab:results_fluxes_roi} gives the fluxes of the different 
components integrated over the 4 energy bins from $1-100$~GeV.
Trends that can explain the low level of emission associated with the 
annulus~1 $\pi^0$-decay component are not readily apparent.
Because the fore-/background is held constant for the maximum-likelihood fit
over the $15^\circ \times 15^\circ$ region (Sec.~\ref{sec:maxlikelihood})
there is no correlation matrix with the IEM that can be examined to determine
degeneracies for the structured interstellar emission component.

But some understanding of the effect of the fore-/background can be inferred
from examining the fluxes per annuli for each of the IEMs 
(see Table~\ref{table:fluxes} in Appendix~\ref{appendix:IEM}).
The IEM scaling procedure (Sec.~\ref{sec:interstellaremission}), 
gives similar contributions by IC emission 
for annuli $>1$ to the total flux over the $15^\circ \times 15^\circ$ region
for the Pulsars and OBstars IEMs.
Also, for each IEM the total emission from the scaled local and outer 
annulus (annuli 5~\&~6) across the $15^\circ \times 15^\circ$ region is 
very similar.

The major difference is the distribution of the 
$\pi^0$-decay flux over annuli $2-4$.
The standout feature is how the \hi-related $\pi^0$-decay
flux for annuli 3 and 4 (Table~\ref{table:fluxes}) is correlated
with the essentially complete suppression of the $\pi^0$-decay flux in 
annulus~1 for the OBstars index-scaled IEM.
Similar patterns appear with the other IEMs, but are less pronounced.
The scaling procedure results in combinations of the 
structured fore-/background emission that leave only a small amount of 
flux for the annulus~1 $\pi^0$-decay templates to be assigned by 
the likelihood maximisation (Sec.~\ref{sec:maxlikelihood}).

While the IC and combined point source fluxes are larger than the formal 
statistical uncertainties, and the variation across IEMs, this
is not true for the annulus~1 $\pi^0$-decay emission.
For this component the reliability of the fluxes 
obtained is uncertain: 
the intrinsic variation of the structured fore-/background across IEMs
is large in comparison, and 
there is also a potential correlation with some of the flux attributed
to the point sources, which is discussed further below.

\begin{deluxetable*}{llccccccccc}[ht]
\tablecolumns{11}
\tablewidth{0pt}
\tablecaption{Fluxes of the Components\tablenotemark{1} for $15^\circ \times 15^\circ$ region about GC.}
\tablehead{
\colhead{Interstellar} & 
\colhead{Energy Band} & 
\colhead{Annulus~1} &
\colhead{Annulus~1} &
\colhead{Point Sources} &
\colhead{} &
\colhead{Fore-/background} &
\colhead{} &
\colhead{Isotropic} &
\colhead{Model} &
\colhead{Data} \\
\colhead{Emission Model} &
\colhead{(GeV)} & 
\colhead{$\pi^0$-decay} &
\colhead{IC} &
\colhead{} &
\colhead{$\pi^0$-decay} &
\colhead{IC} &
\colhead{Brem} &
\colhead{} &
\colhead{Total} &
\colhead{}
}
\startdata
Pulsars & & & & & & & & & & \\
intensity-scaled
& $1.00-3.16$ & 6.1$\pm$1.1 & 32.5$\pm$0.6 & 36.3$\pm$1.2 & 135 & 23 & 24 & 2.3 & 259$\pm$3 & 251$\pm$13\tablenotemark{2}\\
& $3.16-10.00$ & 1.0$\pm$0.2 & 7.1$\pm$0.1 & 7.3$\pm$0.2 & 21 & 5.4 & 1.7 & 0.6 & 44.1$\pm$0.5 & 44$\pm$3\\
& $10.00-31.62$ & 0.13$\pm$0.02 & 1.41$\pm$0.03 & 0.81$\pm$0.04 & 2.9 & 1.2 & 0.14 & 0.17 & 6.7$\pm$0.1 & 6.8$\pm$0.7\\
& $31.62-100.00$ & 0.02\tablenotemark{3} & 0.24\tablenotemark{3} & 0.11$\pm$0.01 & 0.4 & 0.2 & 0.01 & 0.04 & 1.04$\pm$0.02 & 1.2$\pm$0.2\\
& & & & & & & & & & \\
Pulsars & & & & & & & & & & \\
index-scaled
& $1.00-3.16$ & 2.1$\pm$1.1 & 35.5$\pm$0.6 & 37.9$\pm$1.5 & 127 & 25 & & & 254$\pm$3 & \\
& $3.16-10.00$ & 0.3$\pm$0.2 & 7.8$\pm$0.1 & 6.6$\pm$0.2 & 25 & 6 & & & 48$\pm$0.5 & \\
& $10.00-31.62$ & 0.05$\pm$0.02 & 1.54$\pm$0.03 & 0.62$\pm$0.03 & 4.2 & 1.3 & & & 8.0$\pm$0.1 & \\
& $31.62-100.00$ & 0.01\tablenotemark{3} & 0.3\tablenotemark{3} & 0.07$\pm$0.01 & 0.75 & 0.23 & & & 1.37$\pm$0.01 & \\
OBstars & & & & & & & & & & \\
intensity-scaled
& $1.00-3.16$ & 1.3$\pm$0.5 & 47.0$\pm$0.6 & 35.7$\pm$1.2 & 128 & 23 & 21 & 2.6 & 259$\pm$2 & \\
& $3.16-10.00$ & 0.2$\pm$0.1 & 9.1$\pm$0.1 & 7.3$\pm$0.2 & 19 & 5.1 & 1.4 & 0.7 & 43.3$\pm$0.4 & \\
& $10.00-31.62$ & 0.02$\pm$0.01 & 1.62$\pm$0.02 & 0.8$\pm$0.1 & 2.6 & 1.1 & 0.12 & 0.16 & 6.4$\pm$0.1 & \\
& $31.62-100.00$ & --\tablenotemark{3} & 0.25\tablenotemark{3} & 0.11$\pm$0.01 & 0.4 & 0.2 & 0.01 & 0.03 & 0.97$\pm$0.02 & \\
& & & & & & & & & & \\
OBstars & & & & & & & & & & \\
index-scaled
& $1.00-3.16$ & 1.0$\pm$0.5 & 40.9$\pm$0.6 & 38.3$\pm$1.3 & 135 & 19 & & & 257$\pm$2 & \\
& $3.16-10.00$ & 0.14$\pm$0.07 & 7.9$\pm$0.1 & 6.8$\pm$0.2 & 24 & 4.3 &  & & 45.6$\pm$0.4 & \\
& $10.00-31.62$ & 0.02$\pm$0.01 & 1.41$\pm$0.02 & 0.69$\pm$0.04 & 3.9 & 0.9 &  & & 7.2$\pm$0.1 & \\
& $31.62-100.00$ & --\tablenotemark{3} & 0.22\tablenotemark{3} & 0.08$\pm$0.01 & 0.6 & 0.2 & & & 1.15$\pm$0.01 & 
\enddata
\tablenotetext{1}{Units: $10^{-8}$ ph~cm$^{-2}$~s$^{-1}$.}
\tablenotetext{2}{The errors are dominated by systematic uncertainties from the effective area, see \citet{2012ApJS..203....4A} for details.}
\tablenotetext{3}{Flux and/or statistical uncertainty below $10^{-10}$ ph~cm$^{-2}$~s$^{-1}$.}
\label{tab:results_fluxes_roi}
\end{deluxetable*}

\subsection{Point Sources}
\label{results:point_sources}
\begin{figure}[ht]
\includegraphics[scale=0.8]{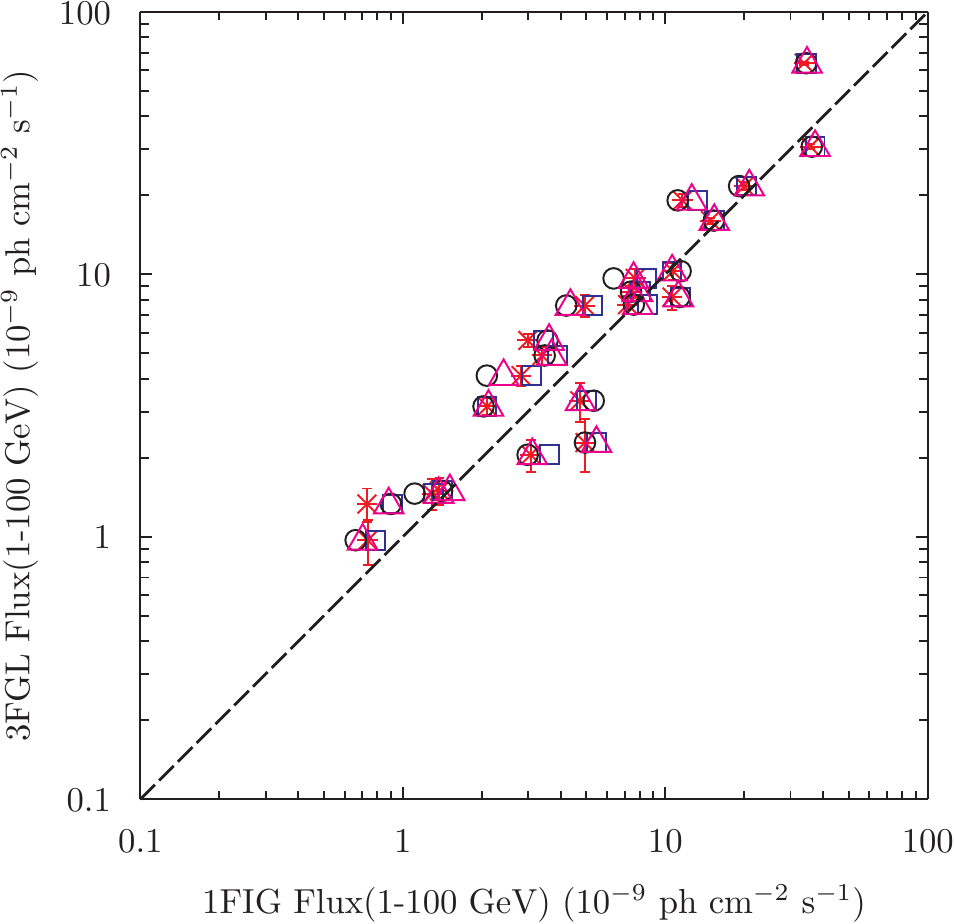}
\caption{
Flux of 1FIG sources associated with 3FGL sources. 
Symbols: crosses, Pulsars intensity-scaled; open squares, Pulsars index-scaled; 
open circles, OBstars intensity-scaled; open triangles, OBstars index-scaled.
The dashed line is a guide for the eye.
\label{fig:1fig_3fgl_associations}}
\end{figure}

Table~\ref{tab:Sources_3fgl} summarises the 
properties of the 48 point sources ordered by increasing right ascension
over the $15^\circ \times 15^\circ$ region with $TS \geq 25$ for the 
Pulsars intensity-scaled IEM, which was the model 
used for the point-source positions and localisation uncertainties.
The list is termed the first {\it Fermi} Inner Galaxy point source 
catalog (1FIG).
27 of the 1FIG sources have a 95\% containment localisation error ellipse 
that intersects the 95\% containment radius of a 3FGL point source.
These associations are also given in the table. 
The correlation plot (Fig.~\ref{fig:1fig_3fgl_associations}) shows that 
the fluxes of the 1FIG sources compared with their associations in the 3FGL are
in good agreement.

Fourteen sources in the 3FGL have a multi-wavelength association. 
For example, the
3FGL source J1701.2-3006 is associated with the globular cluster NGC~6266. 
For the given $TS \geq 25$ detection threshold used for 1FIG source detection
10 counterparts in the 3FGL are obtained that have a multi-wavelength 
association.
The W28 supernova remnant (3FGL~J1801.3-2326e) is included in the model 
of the region as an extended source (see Sec.~\ref{sec:pointsource}), while
the 3FGL sources J1716.6-2812 (NGC~6316), J1746.3-2851c (PWN~G0.13-0.11), and 
J1750.2-3704 (Terzan~5) are missing from 1FIG.
The latter missing counterparts are discussed in 
Sec.~\ref{sec:point_source_discussion}, together with possible 
multi-wavelength associations for other 1FIG sources.

\begin{deluxetable*}{lccccccl}[ht]
\tablecolumns{8}
\tablewidth{0pt}
\tablecaption{Point Sources Detected with $TS > 25$ for Pulsars intensity-scaled IEM for the $15^\circ \times 15^\circ$ region about the GC. 
\label{tab:Sources_3fgl}}
\tablehead{
\colhead{Name} & 
\colhead{$l$} &
\colhead{$b$} &
\colhead{$\Delta\theta$} &
\colhead{$TS$} & 
\colhead{$F_{1-100\,{\rm GeV}}$\tablenotemark{1}} &
\colhead{Type\tablenotemark{2}} &
\colhead{{\it Fermi} Catalog} \\
\colhead{1FIG} & 
\colhead{degrees} &
\colhead{degrees} &
\colhead{degrees} &
\colhead{} & 
\colhead{$10^{-9}$ ph cm$^{-2}$ s$^{-1}$} &
\colhead{} & 
\colhead{Association}}
\startdata
J1701.1-3004 & $353.60$ & $7.34$ & $0.03$ & 165 & 2.61/2.62/2.54/2.65 & LP & 3FGL J1701.2-3006\\
J1717.5-3342 & $352.74$ & $2.36$ & $0.03$ & 113 & 3.13/3.58/3.73/3.78 & LP & 3FGL J1717.8-3342\\
J1718.0-3056 & $355.05$ & $3.90$ & $0.04$ & 26 & 0.74/0.93/0.92/0.90 & PL & 3FGL J1718.1-3056\\
J1728.6-3433 & $353.36$ & $-0.03$ & $0.04$ & 42 & 3.65/4.38/3.70/3.85 & PL & \\
J1729.1-3502 & $352.98$ & $-0.34$ & $0.05$ & 87 & 3.76/4.10/4.01/3.79 & LP & \\
J1730.2-3351 & $354.10$ & $0.13$ & $0.08$ & 34 & 3.41/4.22/3.43/3.58 & PL & \\
J1731.3-3235 & $355.30$ & $0.60$ & $0.05$ & 80 & 3.97/5.01/4.19/4.89 & LP & \\
J1731.6-3001 & $357.49$ & $1.94$ & $0.03$ & 120 & 3.13/3.68/3.04/3.18 & LP & 3FGL J1731.8-3001\\
J1732.3-3131 & $356.31$ & $1.01$ & $0.01$ & 3339 & 36.00/37.44/36.32/37.46 & LP & 3FGL J1732.5-3130\\
J1734.6-3228 & $355.78$ & $0.06$ & $0.07$ & 30 & 2.01/2.53/1.72/1.54 & PL & \\
J1735.4-3030 & $357.52$ & $1.00$ & $0.05$ & 40 & 2.82/3.40/2.54/2.95 & PL & \\
J1736.1-3150 & $356.45$ & $0.18$ & $0.09$ & 35 & 1.74/1.64/1.53/1.40 & PL & \\
J1736.1-3422 & $354.34$ & $-1.19$ & $0.05$ & 47 & 2.34/2.64/2.68/2.74 & PL & \\
J1737.4-3144 & $356.71$ & $-0.03$ & $0.06$ & 67 & 3.52/3.82/2.80/2.88 & LP & \\
J1739.4-3010 & $358.27$ & $0.44$ & $0.09$ & 39 & 3.61/4.64/2.44/3.17 & PL & \\
J1740.1-3057 & $357.66$ & $-0.06$ & $0.04$ & 43 & 3.11/3.48/2.10/2.86 & PL & \\
J1740.2-2834 & $359.69$ & $1.17$ & $0.11$ & 76 & 3.96/4.73/2.81/3.53 & LP & 3FGL J1740.5-2843\\
J1741.5-2538 & $2.37$ & $2.44$ & $0.04$ & 25 & 0.53/0.61/0.51/0.00 & PL & 3FGL J1741.9-2539\\
J1741.5-2054 & $6.41$ & $4.92$ & $0.02$ & 1679 & 15.16/15.72/15.60/15.67 & LP & 3FGL J1741.9-2054\\
J1742.5-3318 & $355.96$ & $-1.77$ & $0.07$ & 104 & 3.42/3.92/3.51/3.73 & LP & 3FGL J1742.6-3321\\
J1744.2-2930 & $359.36$ & $-0.05$ & $0.05$ & 106 & 7.51/9.06/6.68/7.86 & LP & \\
J1744.3-3051 & $358.23$ & $-0.78$ & $0.15$ & 31 & 2.36/2.33/1.73/1.85 & PL & 3FGL J1744.7-3043\\
J1745.0-2905 & $359.80$ & $0.03$ & $0.04$ & 270 & 16.14/15.98/14.52/17.38 & LP & \\
J1745.1-3012 & $358.87$ & $-0.59$ & $0.05$ & 123 & 7.79/8.57/6.42/7.76 & LP & 3FGL J1745.1-3011\\
J1745.5-2859 & $359.98$ & $-0.07$ & $0.01$ & 3063 & 56.82/57.61/56.57/56.99 & LP & \\
J1746.4-2843 & $0.30$ & $-0.10$ & $0.04$ & 330 & 19.00/18.53/18.90/18.93 & LP & \\
J1746.5-3240 & $356.95$ & $-2.16$ & $0.03$ & 358 & 7.53/8.15/7.51/7.94 & LP & 3FGL J1746.8-3240\\
J1747.0-2826 & $0.59$ & $-0.01$ & $0.04$ & 169 & 11.24/13.11/11.01/12.45 & LP & 3FGL J1747.0-2828\\
J1747.2-2959 & $359.29$ & $-0.86$ & $0.02$ & 879 & 20.38/20.61/19.30/21.12 & LP & 3FGL J1747.2-2958\\
J1747.6-2442 & $3.88$ & $1.75$ & $0.04$ & 36 & 0.87/1.38/0.89/1.02 & LP & \\
J1748.1-2449 & $3.81$ & $1.64$ & $0.03$ & 446 & 9.07/9.29/9.36/9.56 & LP & 3FGL J1748.0-2447\\
J1748.2-2856 & $0.29$ & $-0.50$ & $0.17$ & 90 & 6.75/7.36/6.89/7.30 & LP & 3FGL J1747.7-2904\\
J1748.2-2816 & $0.88$ & $-0.18$ & $0.02$ & 377 & 11.20/11.97/11.86/11.78 & LP & 3FGL J1748.3-2815c\\
J1749.1-2917 & $0.10$ & $-0.86$ & $0.12$ & 92 & 4.97/5.30/4.21/4.43 & LP & 3FGL J1749.2-2911\\
J1750.2-3705 & $353.50$ & $-5.04$ & $0.06$ & 49 & 1.41/1.45/1.45/1.55 & PL & 3FGL J1750.2-3704\\
J1753.5-2931 & $0.41$ & $-1.85$ & $0.10$ & 73 & 2.87/3.15/2.14/2.47 & LP & 3FGL J1754.0-2930\\
J1753.6-2539 & $3.77$ & $0.09$ & $0.02$ & 276 & 7.21/8.67/7.68/8.02 & LP & 3FGL J1754.0-2538\\
J1755.5-2511 & $4.39$ & $-0.04$ & $0.05$ & 59 & 3.46/4.34/3.56/3.84 & LP & \\
J1758.5-2405 & $5.68$ & $-0.07$ & $0.04$ & 95 & 4.89/5.52/5.06/5.58 & LP & 3FGL J1758.8-2402\\
J1759.0-2345 & $5.98$ & $0.05$ & $0.03$ & 115 & 4.69/5.04/5.33/4.75 & LP & 3FGL J1758.8-2346\\
J1800.5-2359 & $5.99$ & $-0.43$ & $0.03$ & 276 & 10.92/10.89/11.71/10.88 & LP & 3FGL J1800.8-2402\\
J1801.1-2313 & $6.69$ & $-0.09$ & $0.03$ & 137 & 9.25/8.34/9.22/8.84 & LP & \\
J1801.2-2451 & $5.29$ & $-0.96$ & $0.07$ & 47 & 3.01/3.70/3.75/4.34 & PL & \\
J1801.4-2330 & $6.51$ & $-0.36$ & $0.02$ & 234 & 14.26/13.28/13.65/12.17 & LP & \\
J1801.6-2358 & $6.13$ & $-0.64$ & $0.04$ & 29 & 2.40/2.86/2.62/3.18 & PL & \\
J1802.2-3043 & $0.29$ & $-4.05$ & $0.05$ & 32 & 0.75/0.80/0.67/0.71 & PL & 3FGL J1802.4-3043\\
J1808.2-3358 & $358.05$ & $-6.72$ & $0.07$ & 51 & 1.33/1.35/1.15/1.41 & PL & 3FGL J1808.3-3357\\
J1809.5-2332 & $7.39$ & $-2.00$ & $0.01$ & 7791 & 64.82/66.11/65.87/66.49 & LP & 3FGL J1809.8-2332
\enddata
\tablenotetext{1}{The localisations and $TS$ are for the Pulsars intensity-scaled IEM, but the fluxes for each IEM are also supplied ordered as Pulsars intensity/index-scaled, OBstars intensity/index-scaled, respectively.}
\tablenotetext{2}{Table~\ref{table:source_properties} in Appendix~\ref{appendix:spectral_parameters} lists the corresponding spectral parameters for each IEM.}
\end{deluxetable*}

\begin{figure*}[ht]
\subfigure{
\includegraphics[scale=0.8]{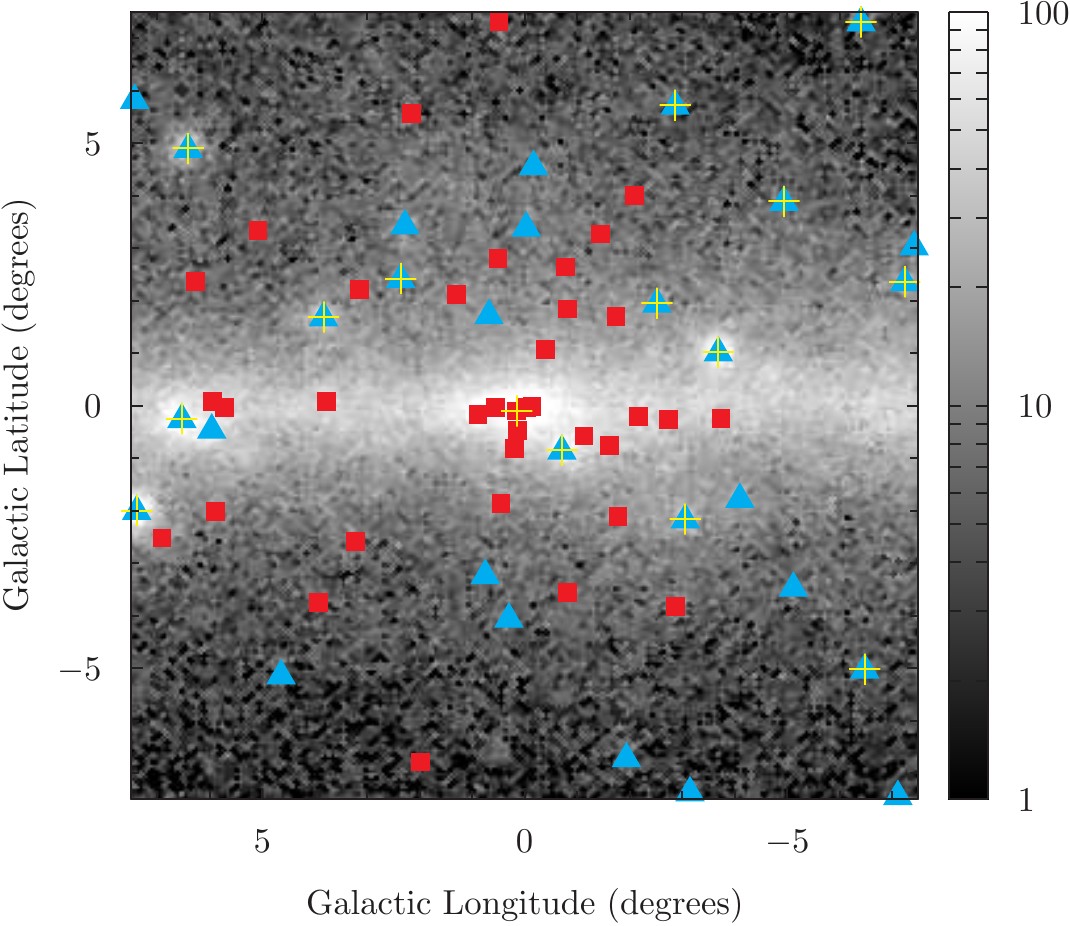}
\includegraphics[scale=0.8]{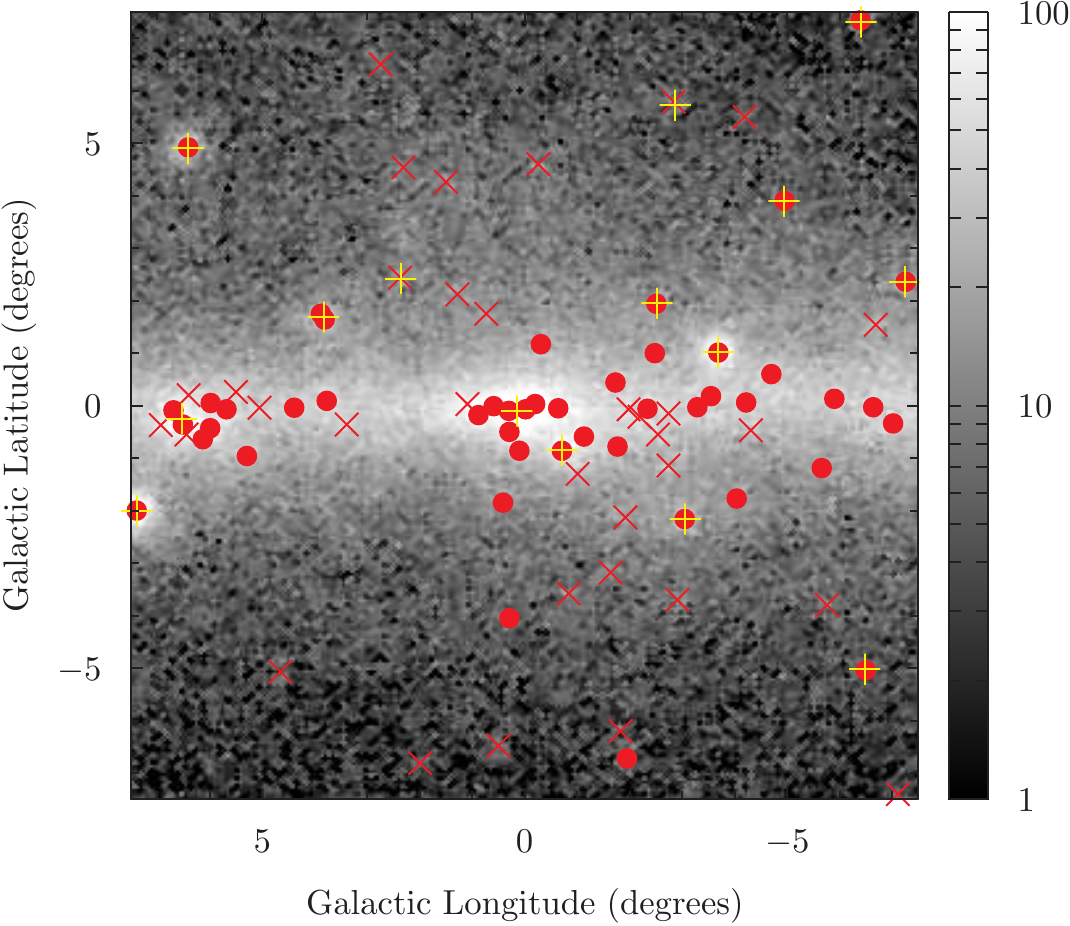}
}
\caption{
Point sources for 3FGL (left panel) and 1FIG (right panel, for Pulsars 
intensity-scaled IEM) overlaid on 
the total counts for the $15^\circ \times 15^\circ$ region about the GC.
Left panel symbol key: filled squares, `flagged' 3FGL sources; 
filled triangles, other 3FGL sources; upright crosses, 3FGL sources with a 
multi-wavelength association. 
Right panel symbol key: filled circles, 1FIG sources with $TS \geq 25$; 
angled crosses, 1FIG source candidates with $TS < 25$; upright crosses, 
as in left panel.
Colour scale is in counts per $0.05^2$ degree pixel.
\label{fig:3fgl_1fig_overlay}}
\end{figure*}

\begin{figure*}[ht]
\includegraphics[scale=0.8]{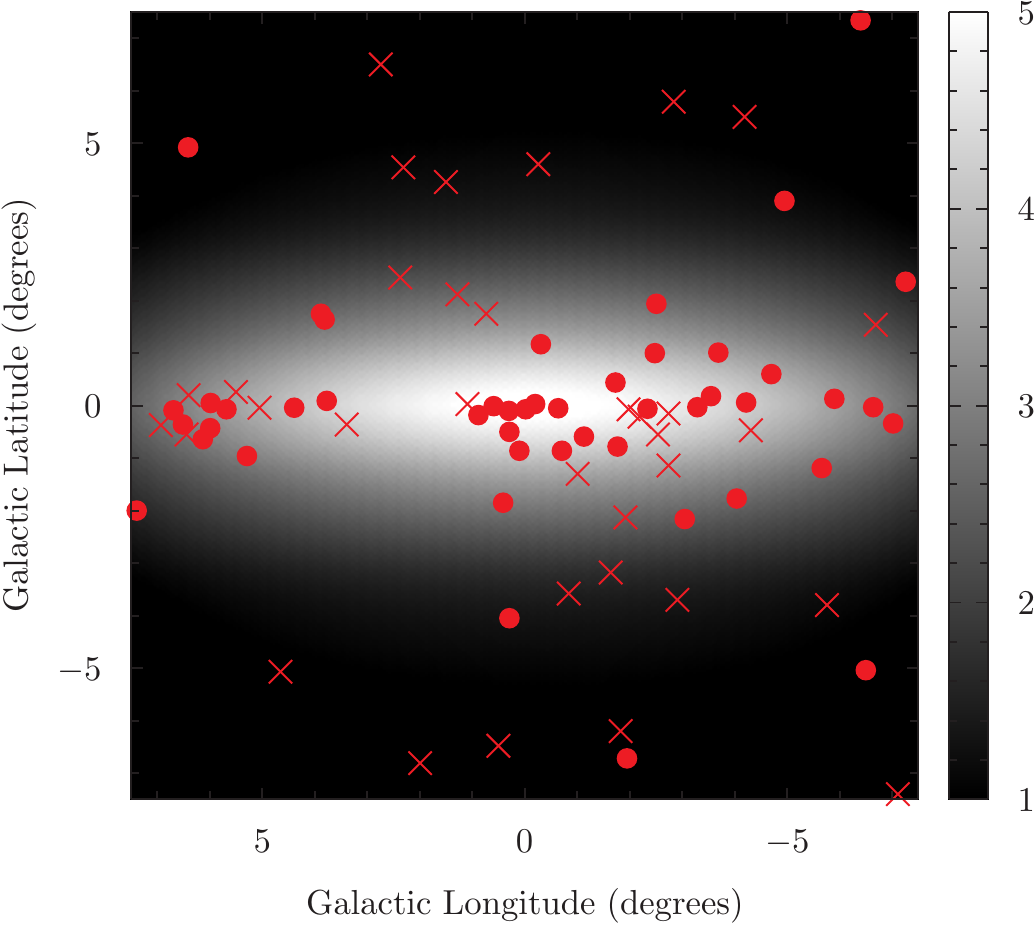}
\includegraphics[scale=0.8]{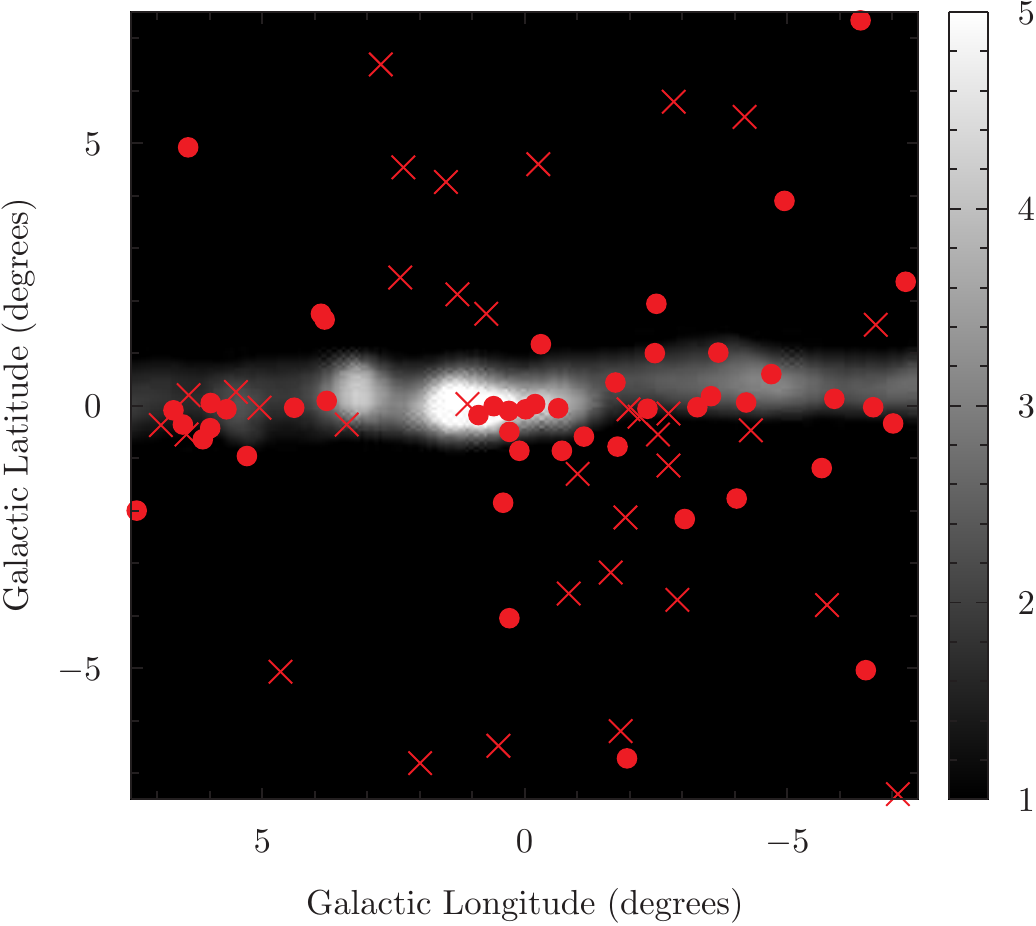}
\caption{
Point sources from 1FIG overlaid on the fitted 
components (Section~\ref{sec:maxlikelihood}) of the interstellar
emission from annulus~1 for the Pulsars intensity-scaled IEM. 
Left panel shows the overlay on the annulus~1 IC while the right panel 
shows the overlay on the annulus~1 $\pi^0$-decay components from atomic and 
molecular gas.
Symbol key: filled circles, 1FIG sources with 
$TS > 25$; angled crosses, 1FIG sources candidates with $TS < 25$.
Colour scale is in counts per $0.05^2$ degree pixel.
\label{fig:1fig_overlay}}
\end{figure*}

\begin{figure}[ht]
\includegraphics[scale=0.8]{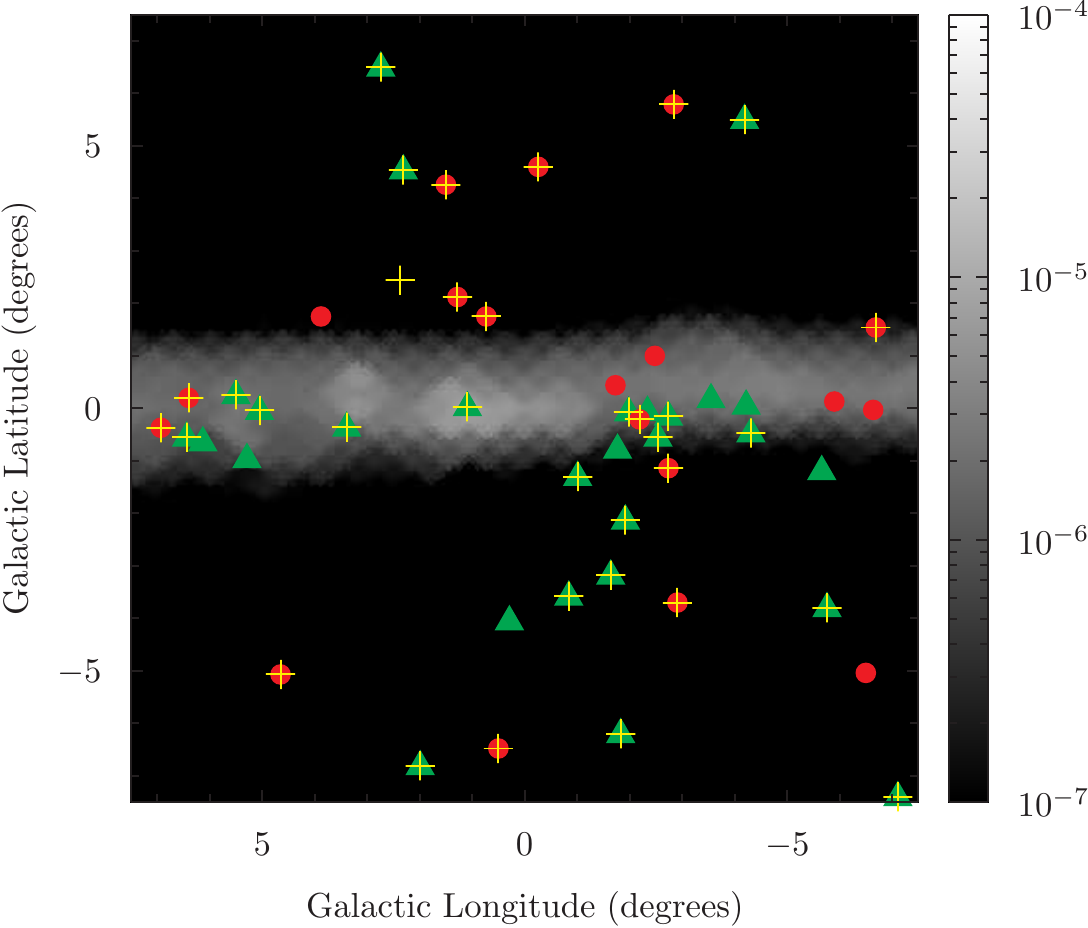}
\caption{
Sources and source candidates with {\it TS}~$<50$ overlaid on the fitted  
$\pi^0$-decay emission from annulus~1 for the Pulsars intensity-scaled IEM. 
Symbol key: filled circles, sources and source candidates 
with power-law indices $>2.5$; 
filled triangles, sources and source candidates 
with power-law indices $\le 2.5$; crosses, source candidates 
with {\it TS}~$<25$. 
The highest flux $>1$~GeV for a point-source shown 
is $3.7\times 10^{-9}$~ph~cm$^{-2}$~s$^{-1}$.
Colour scale units: cm$^{-2}$ s$^{-1}$ sr$^{-1}$. 
\label{fig:1fig_pl_overlay}}
\end{figure}

Figure~\ref{fig:3fgl_1fig_overlay} shows the point sources from the 1FIG and
3FGL overlaid on the total photon counts for the $15^\circ \times 15^\circ$
region about the GC.
The 3FGL sources are separated
according to whether they have an analysis flag set in the 3FGL catalog: 
flagged sources indicate their properties depend on the IEM or other details
of the analysis in the region.
The density of flagged 3FGL sources is higher out of the Galactic plane than 
that of the 1FIG sources, even if the $TS < 25$ source candidates are included.
This can be partly attributed to differences in the treatment of the IC 
emission, and  its interplay with the gas components, for the IEMs employed for the respective analyses.
The 3FGL IEM uses an 
all-sky IC map based on a \GP\ calculation, with its 
spatial distribution taken as a fixed template and the spectral parameters
adjusted to improve the correspondence with the data \citep{3fgl_iem_paper}.
The decomposition of the IC intensity map 
into Galactocentric annulus templates employed here for the first time 
introduces 
additional degrees of freedom that can account for Galactocentric radial 
gradients in the IC emissivities. 
This allows more flexibility to fit for a spatial distribution of IC 
emission that is not correctly represented by the baseline \GP\ calculations.

The 3FGL source density is higher out of the Galactic plane compared to the 
1FIG, while the reverse is the case closer to the plane.
The 1FIG sources in the plane do 
cluster in approximately the same regions as the high-density
clusters for the 3FGL: near the W28 supernova remnant and the GC.
Outside these regions the density of sources is higher than the 3FGL.
That many of these additional sources appear to trace 
features in at least one of the annulus~1 templates is suggestive that they 
may be misattributed interstellar emission.
This can be seen in Fig.~\ref{fig:1fig_overlay} where the 
1FIG sources are overlaid 
on the fitted components of the interstellar emission for
annulus~1 for the Pulsars intensity-scaled IEM.
By way of example, many of the $TS > 25$ sources appear to trace the edge 
of the fitted neutral gas $\pi^0$-decay template.
Because many of these sources lack a multi-wavelength association 
it is not straightforward to determine whether they are true point sources.

The combined flux of 1FIG 
point sources and point source candidates 
across the $15^\circ \times 15^\circ$ region for the Pulsars 
intensity-scaled IEM $> 1$~GeV is 
$44.6\pm1.4 \times 10^{-8}$ ph cm$^{-2}$ s$^{-1}$.
Of this total, only 20\% is due to point sources with a 
multi-wavelength association in the 3FGL, while 
10\% of the total is due to $TS < 25$ source candidates.
Focussing on the 
region $-7.5^\circ \leq l \leq -0.5^\circ, -1.5^\circ \leq b \leq 1.5^\circ$
(the region with the highest density of 1FIG sources 
without 3FGL counterparts and where they 
appear to trace the edge of the $\pi^0$-decay
template), 
the combined flux from the $TS > 25$ 1FIG sources is $\sim 20$\% of the total
point-source flux over the $15^\circ \times 15^\circ$ region.
The total flux for the annulus~1 $\pi^0$-decay along the entire plane 
from the fit is about the same as that from these sources alone.

Although the absolute values differ 
similar relative contributions to the total point 
source flux determined for each of the other IEMs are obtained.
However, for the Pulsars index-scaled, and both OBstars IEM variants, the 
annulus~1 $\pi^0$-decay template is even less intense than for the Pulsars 
intensity-scaled IEM.
Over all IEMs, the 17 1FIG sources with $TS > 100$ have a 
variation in the combined flux that is $\lesssim 1$\%.
For sources without a 3FGL multi-wavelength association 
and with 1FIG {\it TS} in the range $25 \leq TS < 100$ (24 sources) the 
combined flux over this region is more strongly dependent 
on the IEM: $6.7-8.3\times10^{-8}$ ph cm$^{-2}$ s$^{-1}$.
The larger of these values is 
comparable to the variation to the total fore-/background for the 
IEMs over this region (Table~\ref{tab:results_fluxes_roi}).

It is probable that there is some misattribution of interstellar emission 
to low-flux (e.g., less than $\sim$ 
few $\times 10^{-9}$ ph cm$^{-2}$ s$^{-1}$ $> 1$~GeV) point sources.
The low-flux sources are all relatively 
low-significance sources and modelled using power-law spectra 
(Section~\ref{sec:pointsource}).
The distribution of their spectral indices over the $15^\circ \times 15^\circ$ 
region may provide some information: softer spectral indices (e.g., $\gtrsim 2.5$ in spectral index) can indicate that the
low-flux sources are more likely associated with the structured/gas-related
interstellar emission, while harder indices can indicate a more ``IC-like'' 
distribution.
Figure~\ref{fig:1fig_pl_overlay} shows all point sources and candidates 
with a $TS < 50$ overlaid on the fitted $\pi^0$-decay annulus~1 
template for the Pulsars intensity-scaled IEM.
The point sources are coded according to the spectral indices: circles 
show those with indices $>2.5$, while triangles show those with 
indices $\leq 2.5$.
There is no clear trend of softer spectrum point sources tracing the structured
emission, nor one where the harder spectrum point sources have a high density
out of the plane. 
It is difficult to identify the exact fraction of the emission, or to 
what component (gas-related, IC), the low-flux point sources could 
be ascribed if they are indeed due to mismodelling of the interstellar emission
over the region.

\subsection{Residuals}
\label{results:residuals}

\begin{figure*}[htb]
\subfigure{
\includegraphics[scale=0.6]{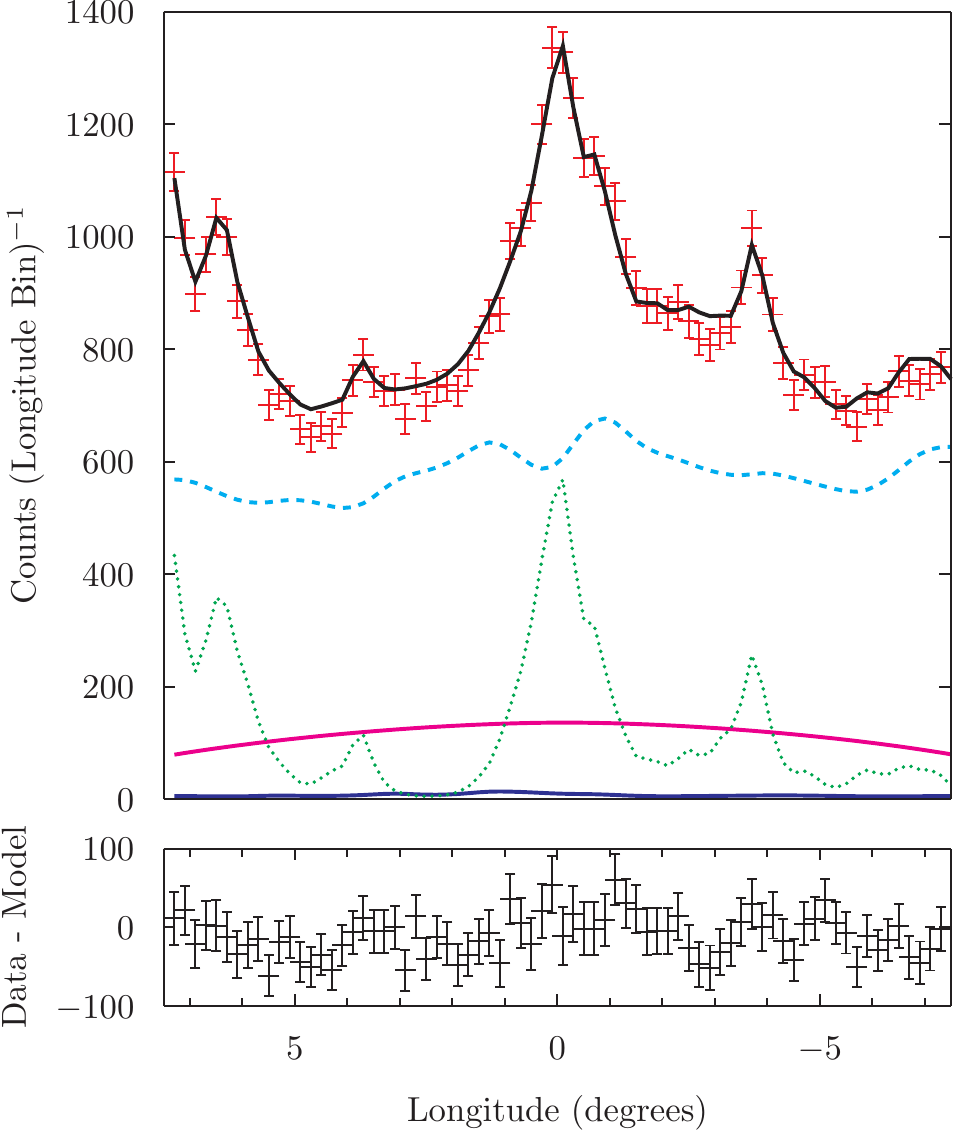}
\includegraphics[scale=0.6]{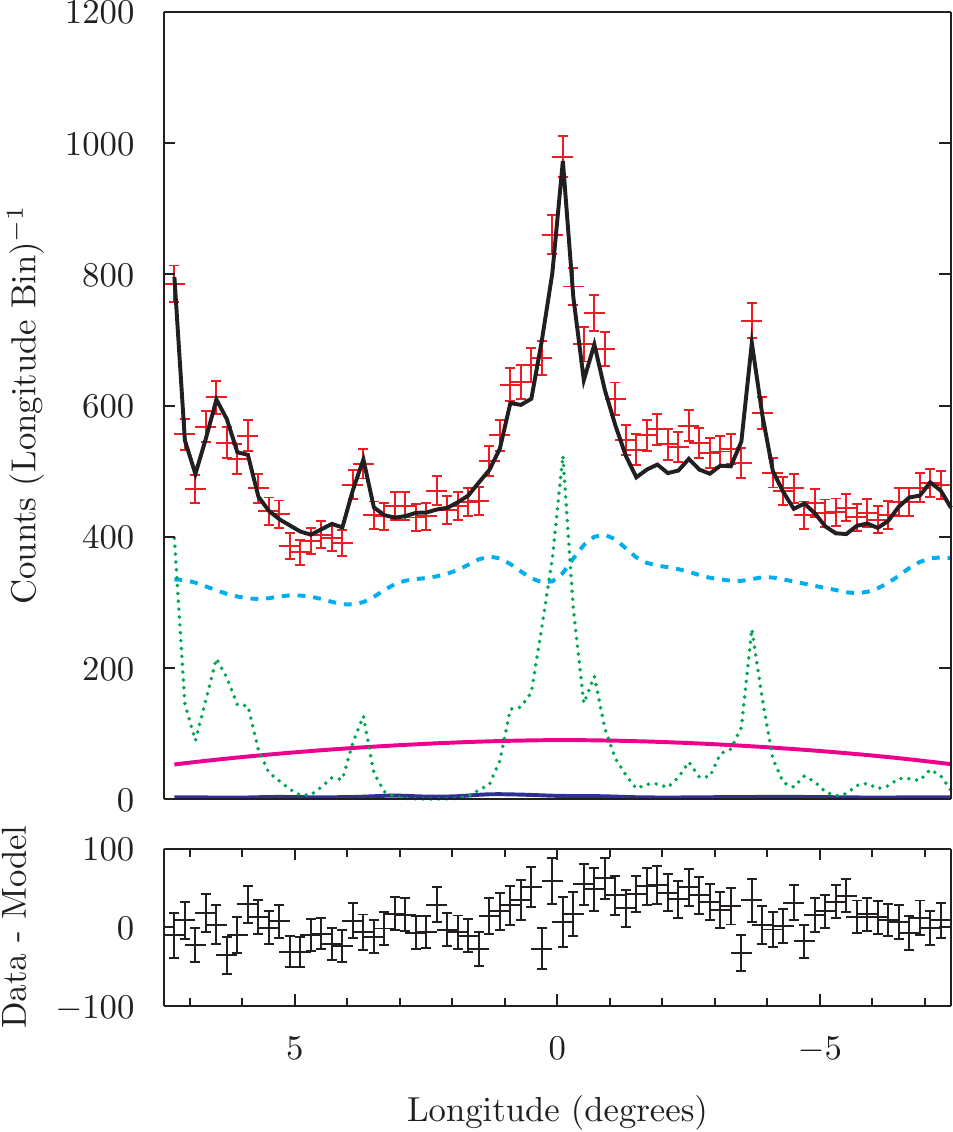}
\includegraphics[scale=0.6]{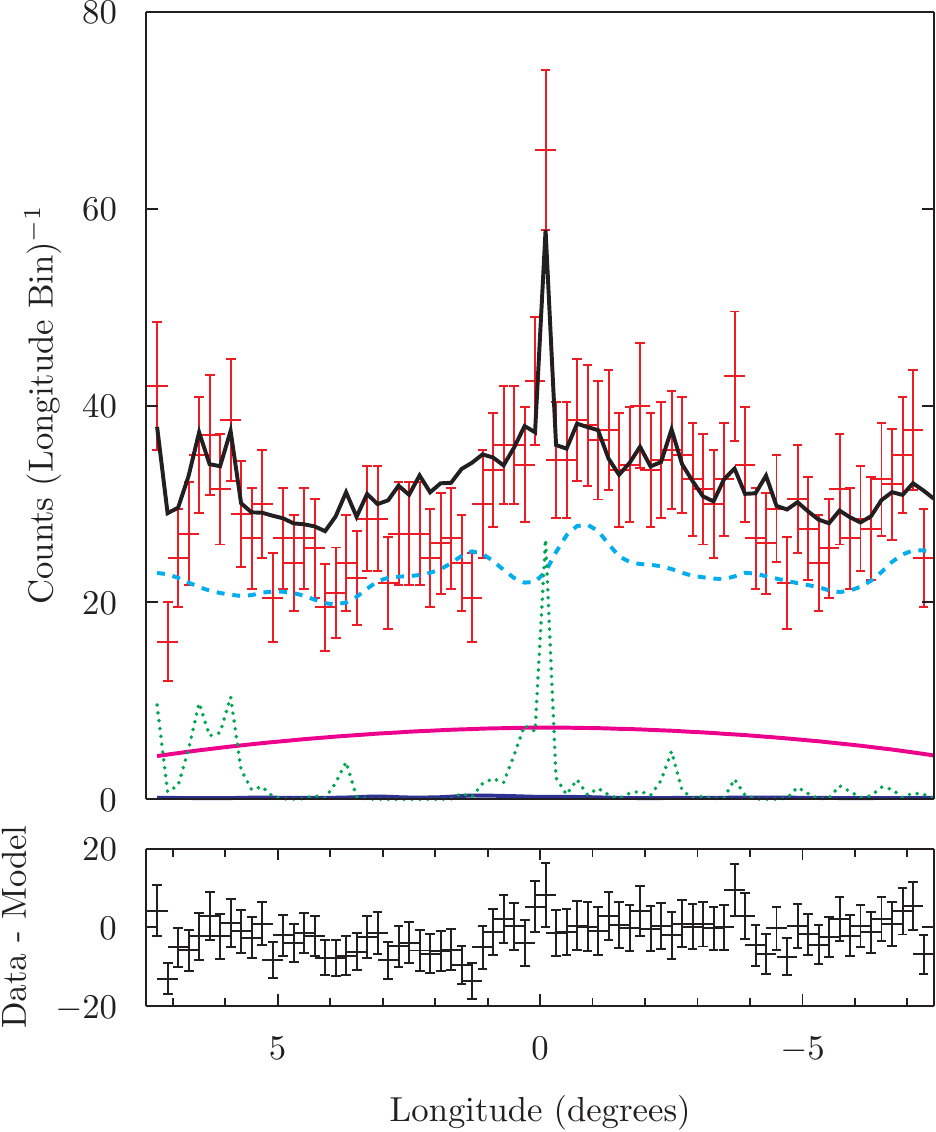}
}
\subfigure{
\includegraphics[scale=0.6]{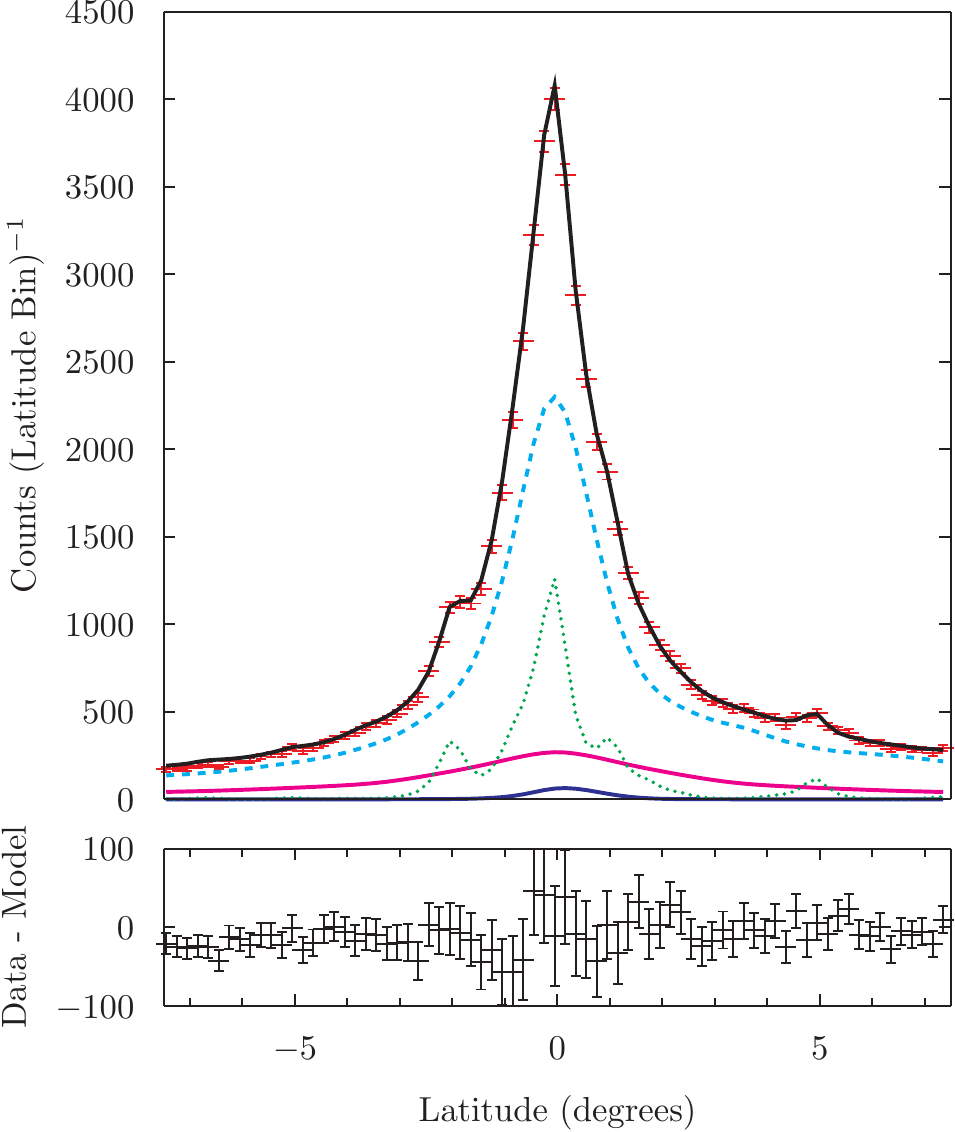}
\includegraphics[scale=0.6]{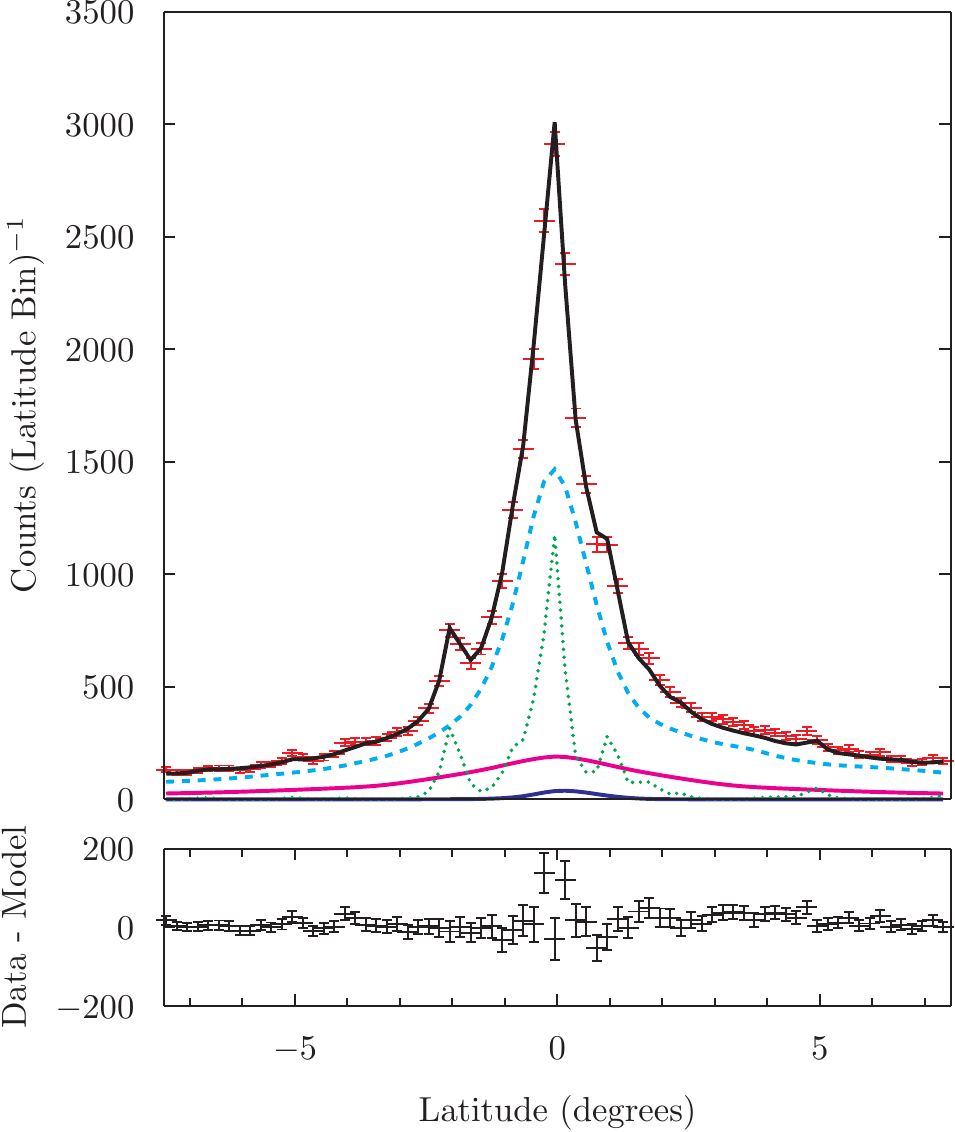}
\includegraphics[scale=0.6]{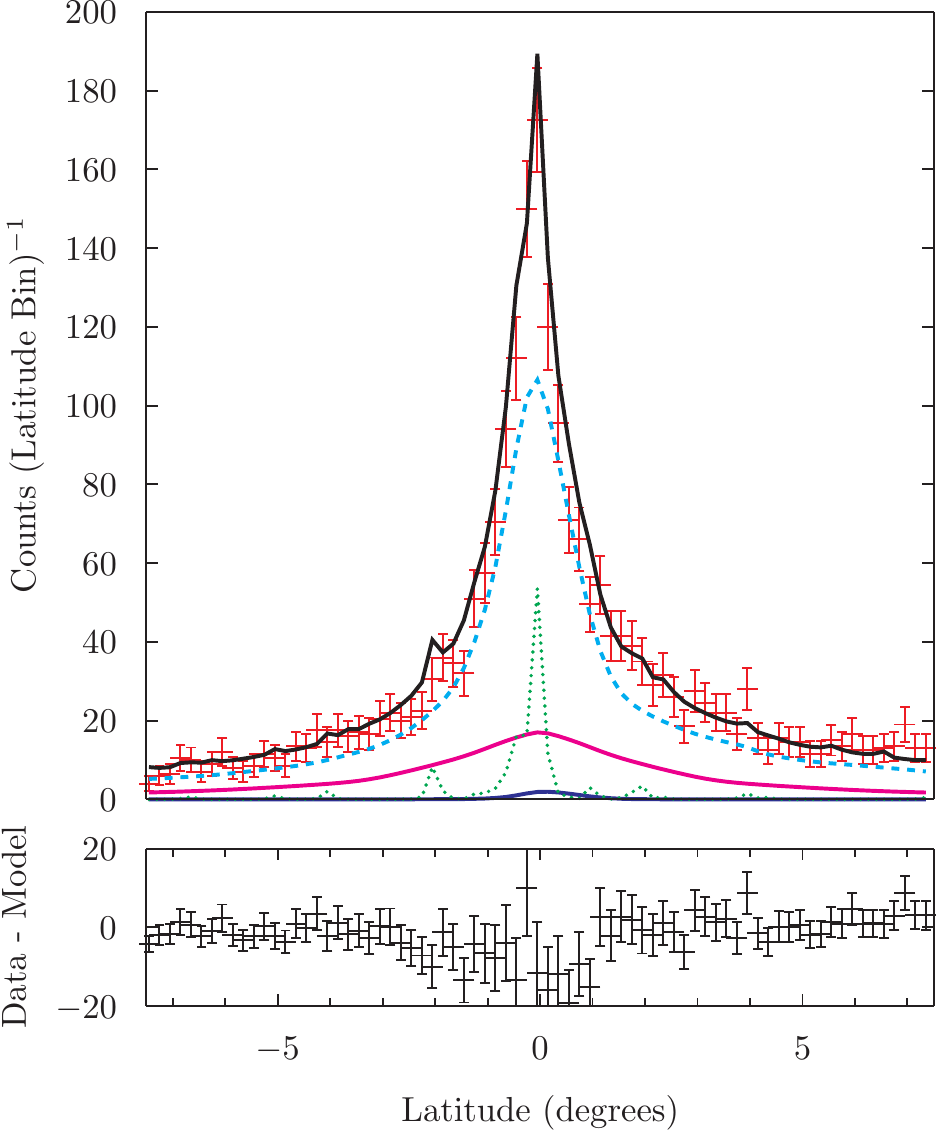}
}
\caption{
\label{fig:res_profiles}
Longitude (upper) and latitude (lower) profiles for $1-1.6$ (left), $1.6-10$ (middle), and $10-100$~GeV energies (left), respectively, of the residual 
counts $(data-model)$ for the Pulsars 
index-scaled IEM after fitting for interstellar emission and point sources 
across the $15^\circ \times 15^\circ$ region.
Line styles: black/solid, total model; cyan/dashed, fore/background interstellar
emission; green/dotted, point sources; magenta/solid, IC from annulus~1; 
blue/solid, $\pi^0$-decay from annulus~1.
Point styles: red, data; black, residual counts.
The lower sub-panel for each profile gives the residual counts after the model
has been subtracted from the data.
The error bars are statistical.
Profiles for the residuals counts for other IEMs display similar features with the major difference being the number of counts.}
\end{figure*}

\begin{figure*}[htb]
\subfigure{
\includegraphics[scale=0.18]{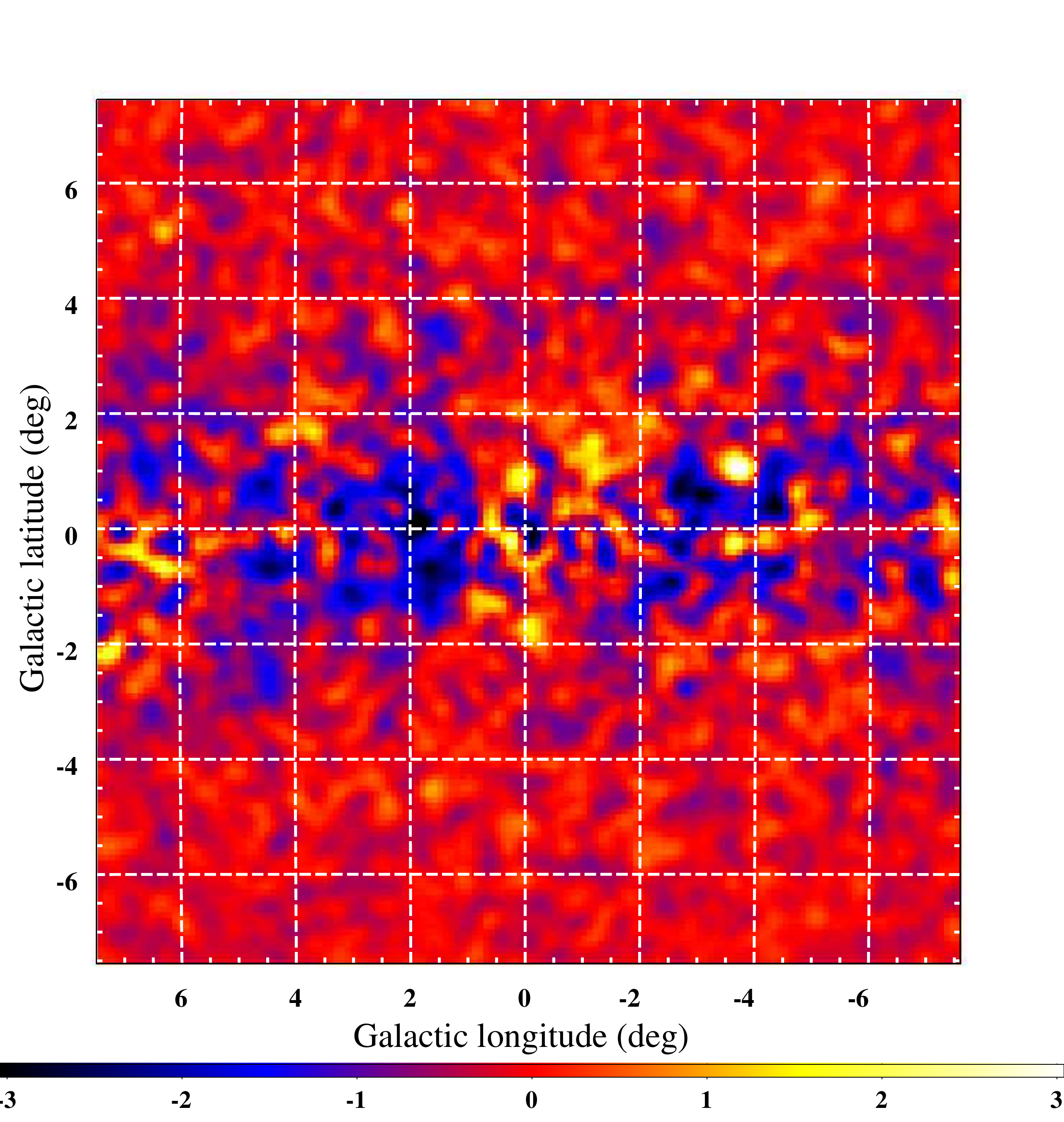}
\includegraphics[scale=0.18]{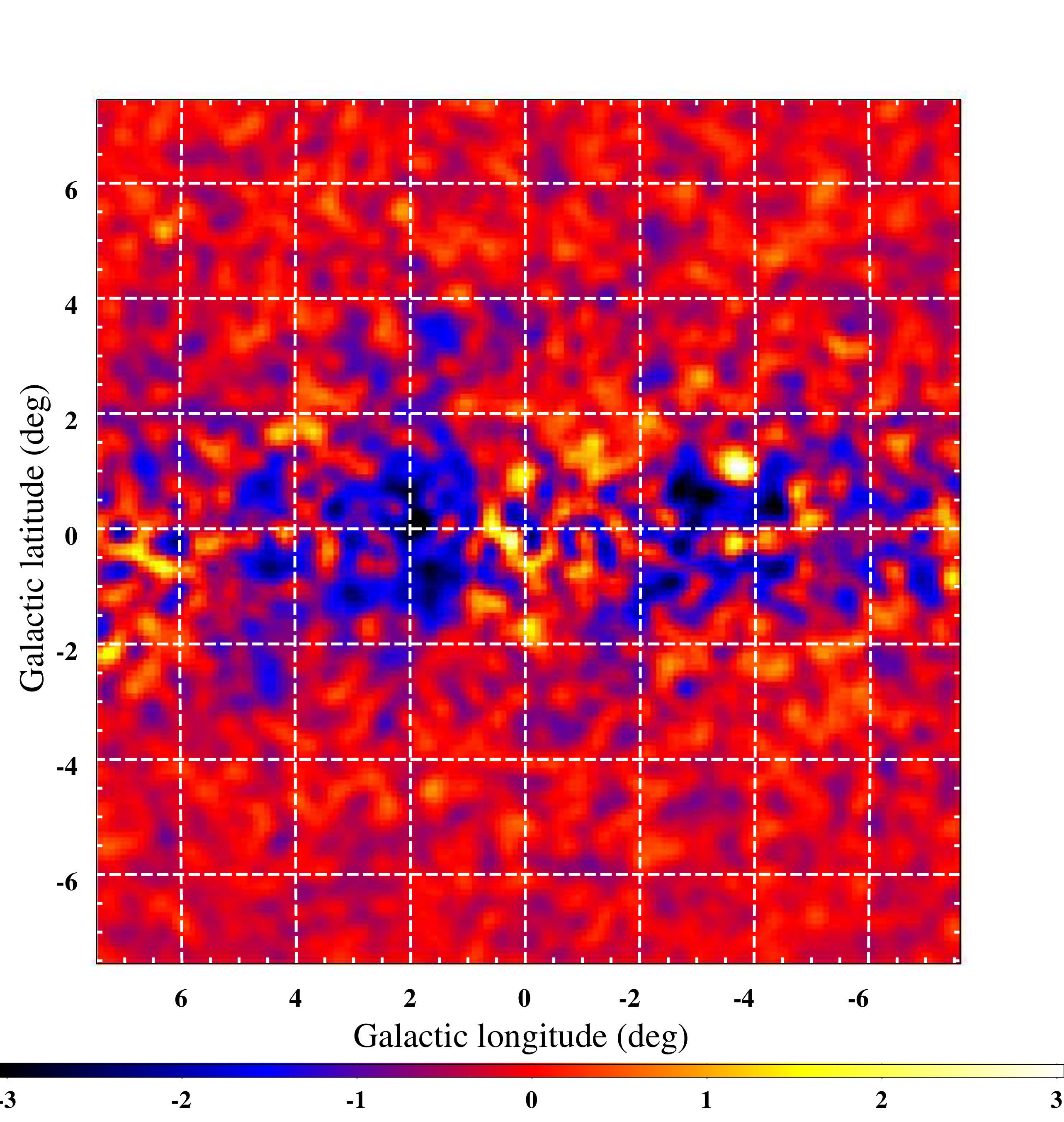}
\includegraphics[scale=0.18]{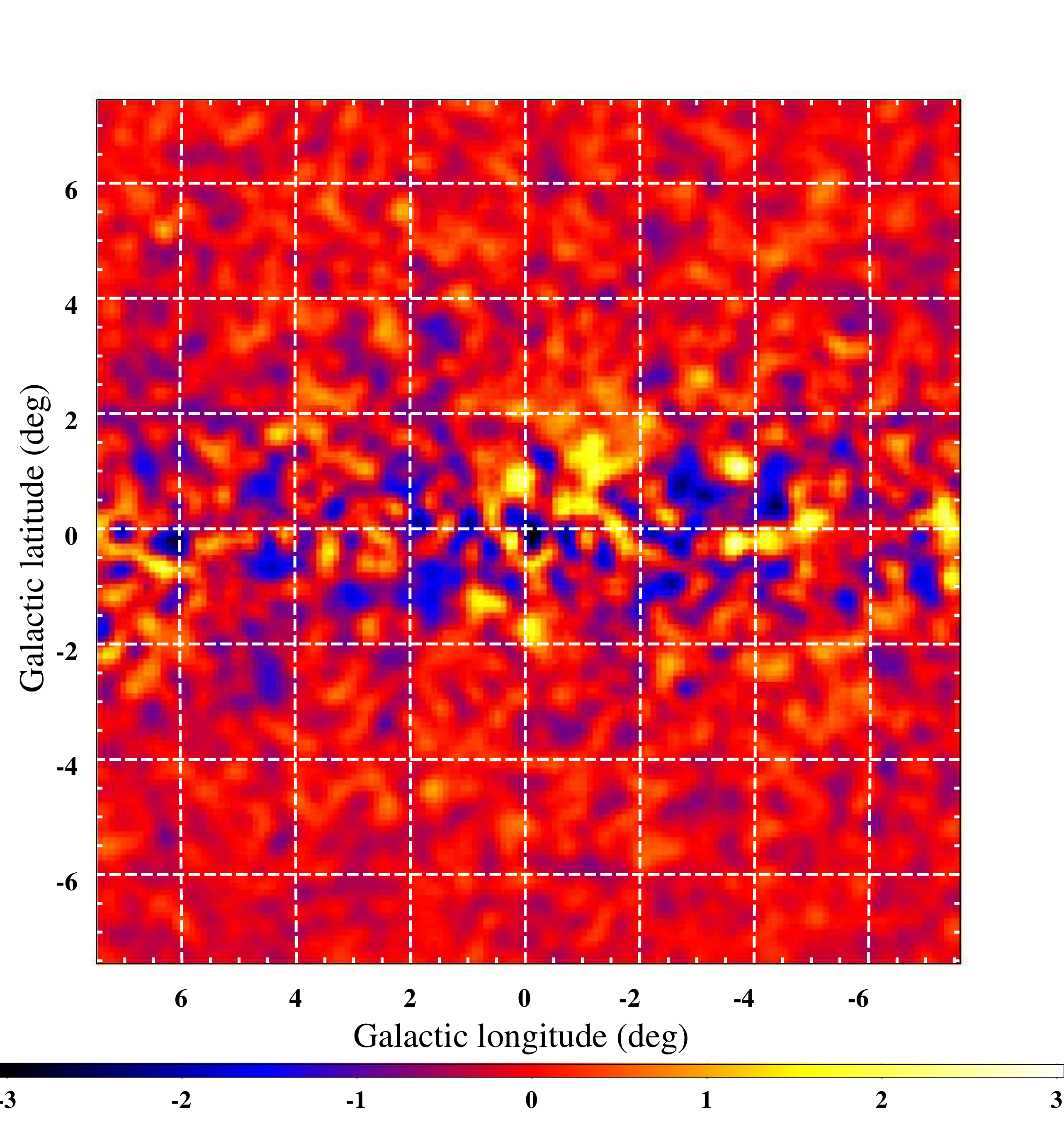}
\includegraphics[scale=0.18]{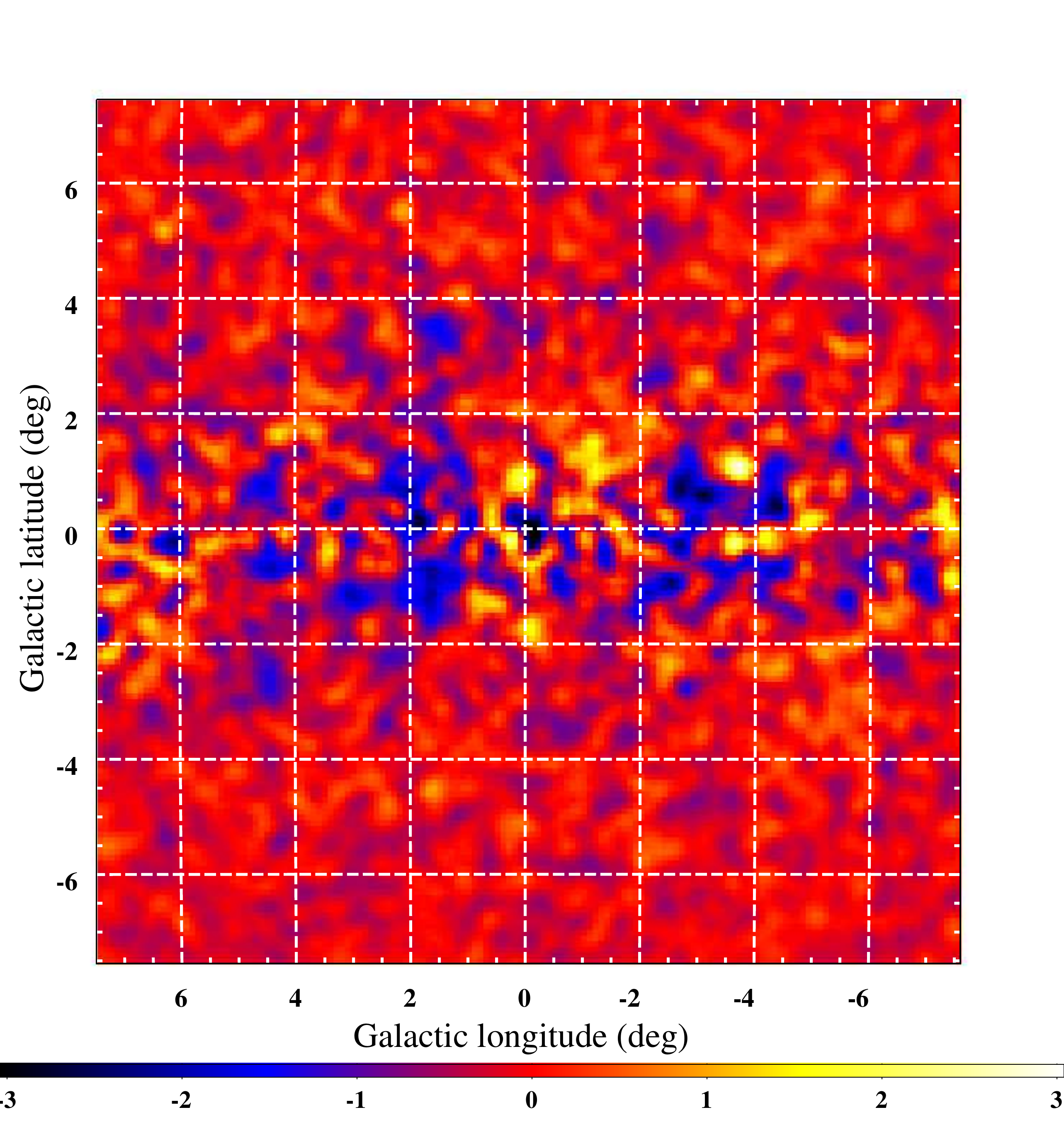}
}
\subfigure{
\includegraphics[scale=0.18]{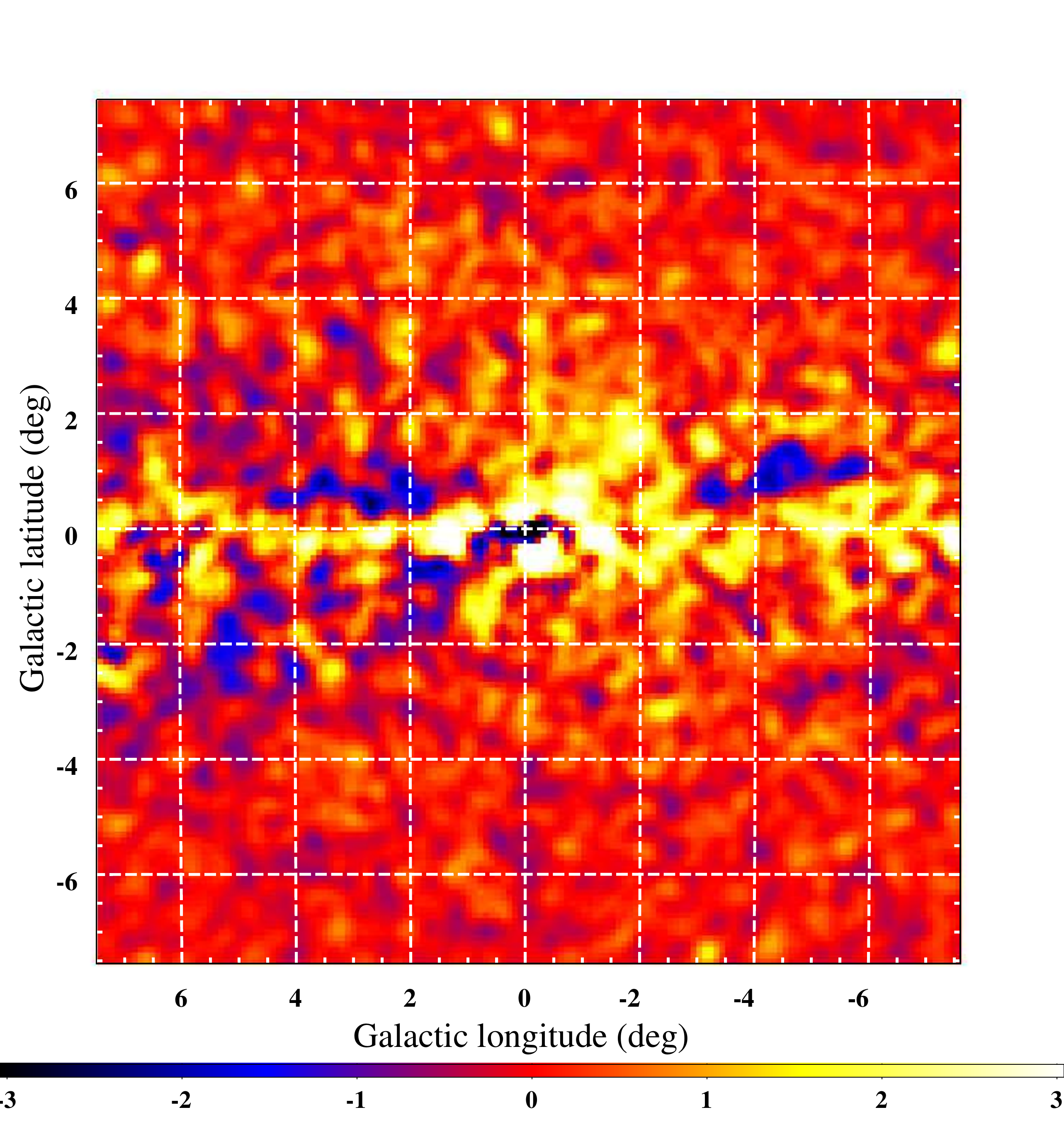}
\includegraphics[scale=0.18]{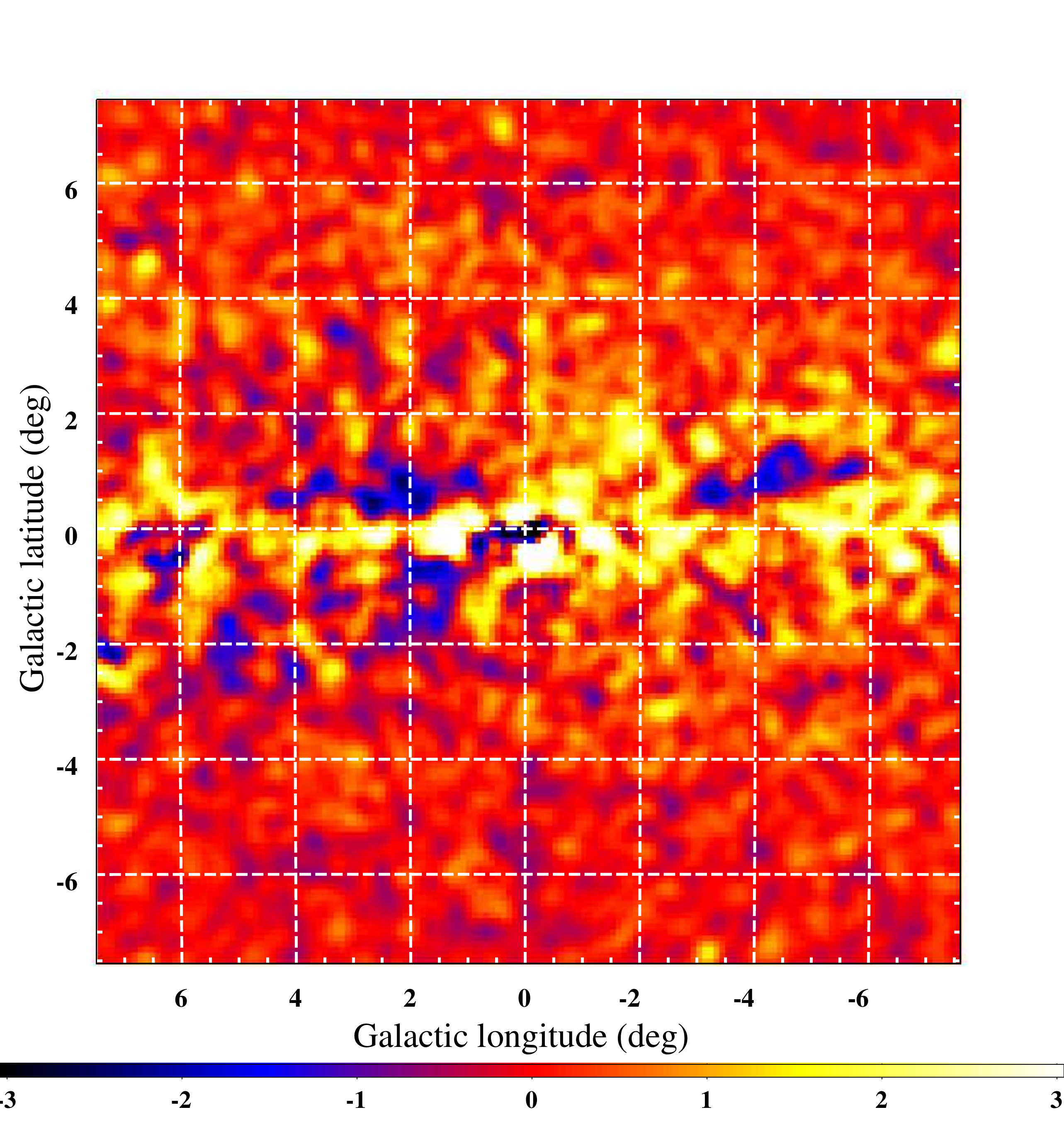}
\includegraphics[scale=0.18]{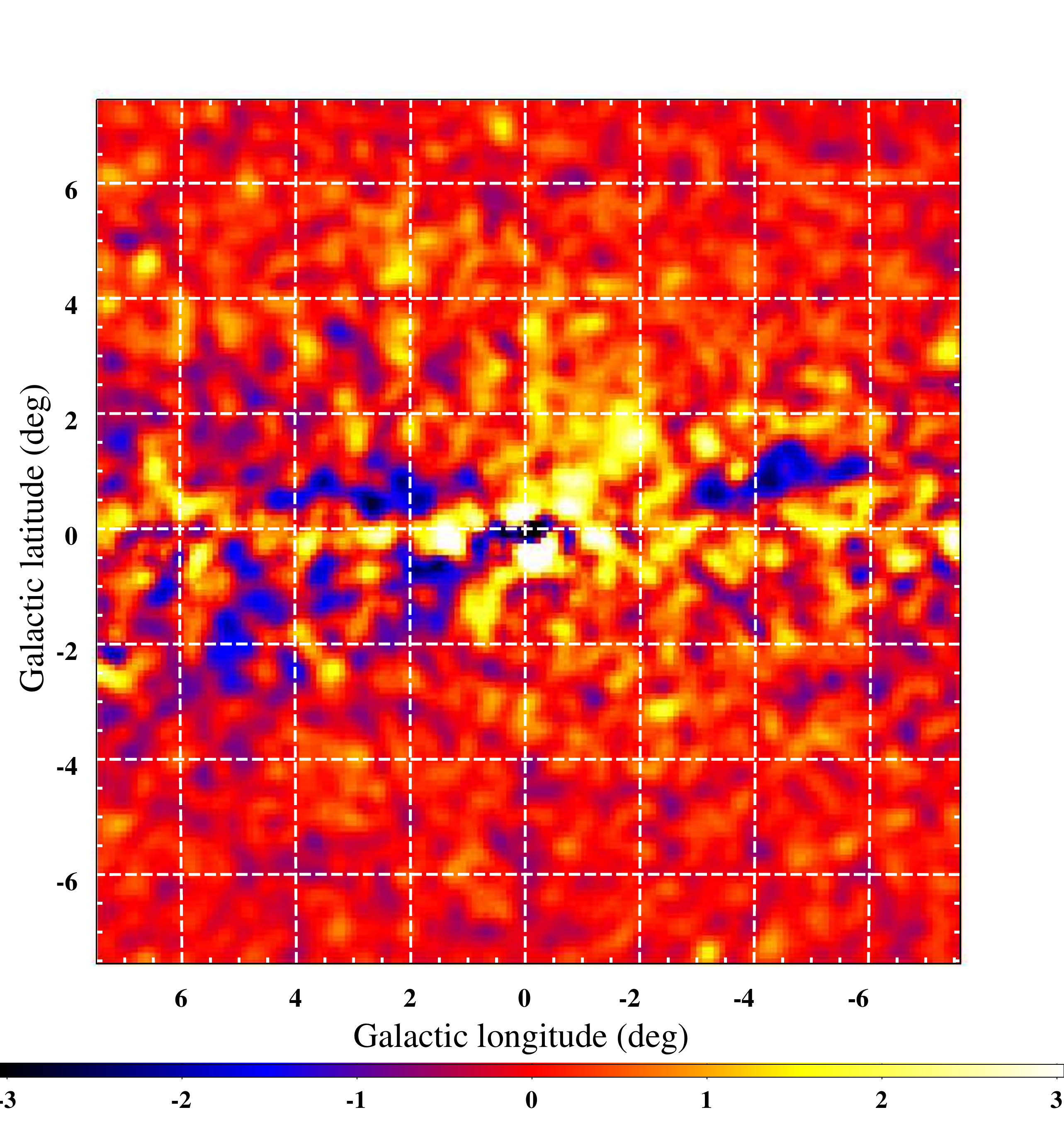}
\includegraphics[scale=0.18]{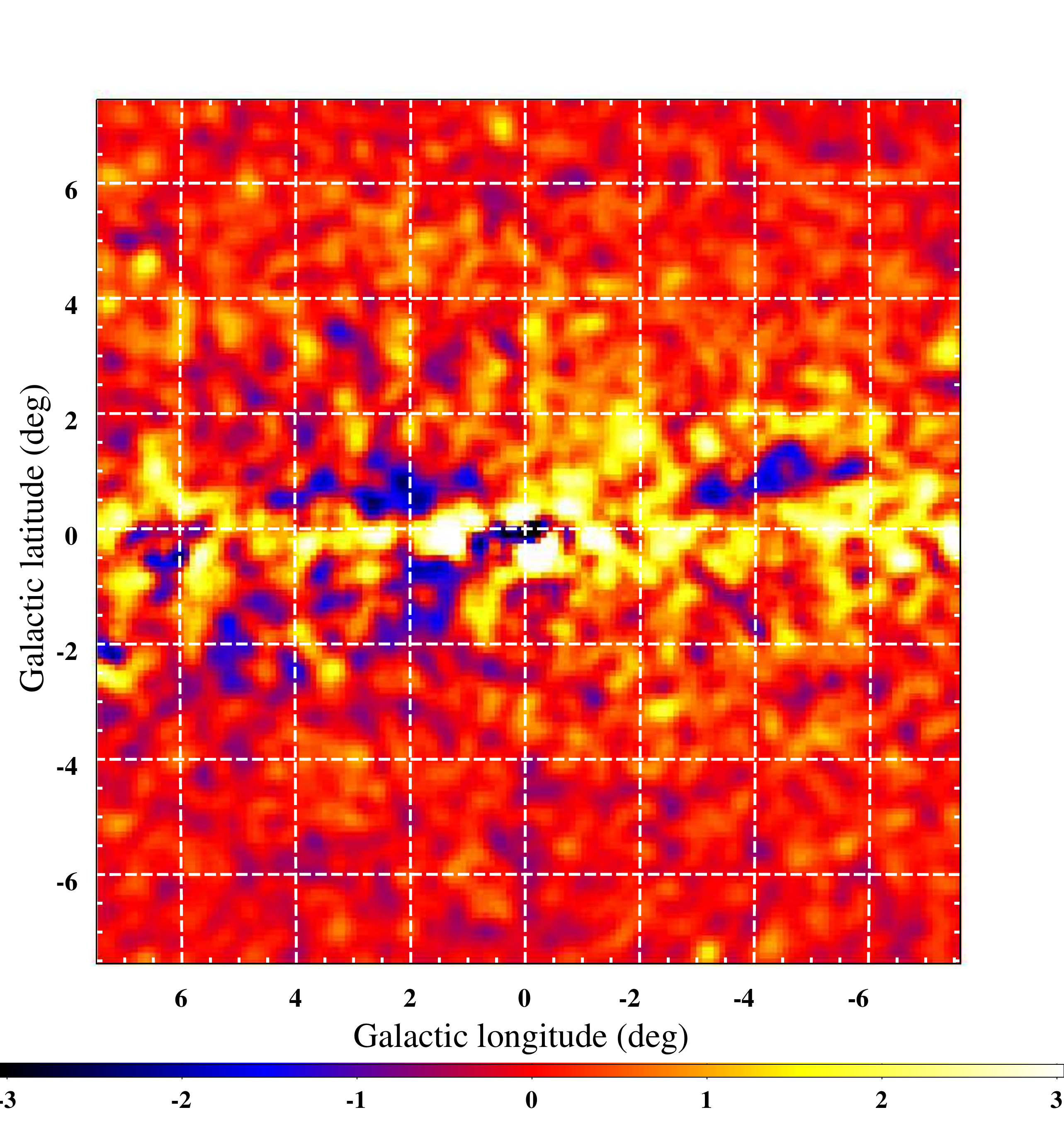}
}
\subfigure{
\includegraphics[scale=0.18]{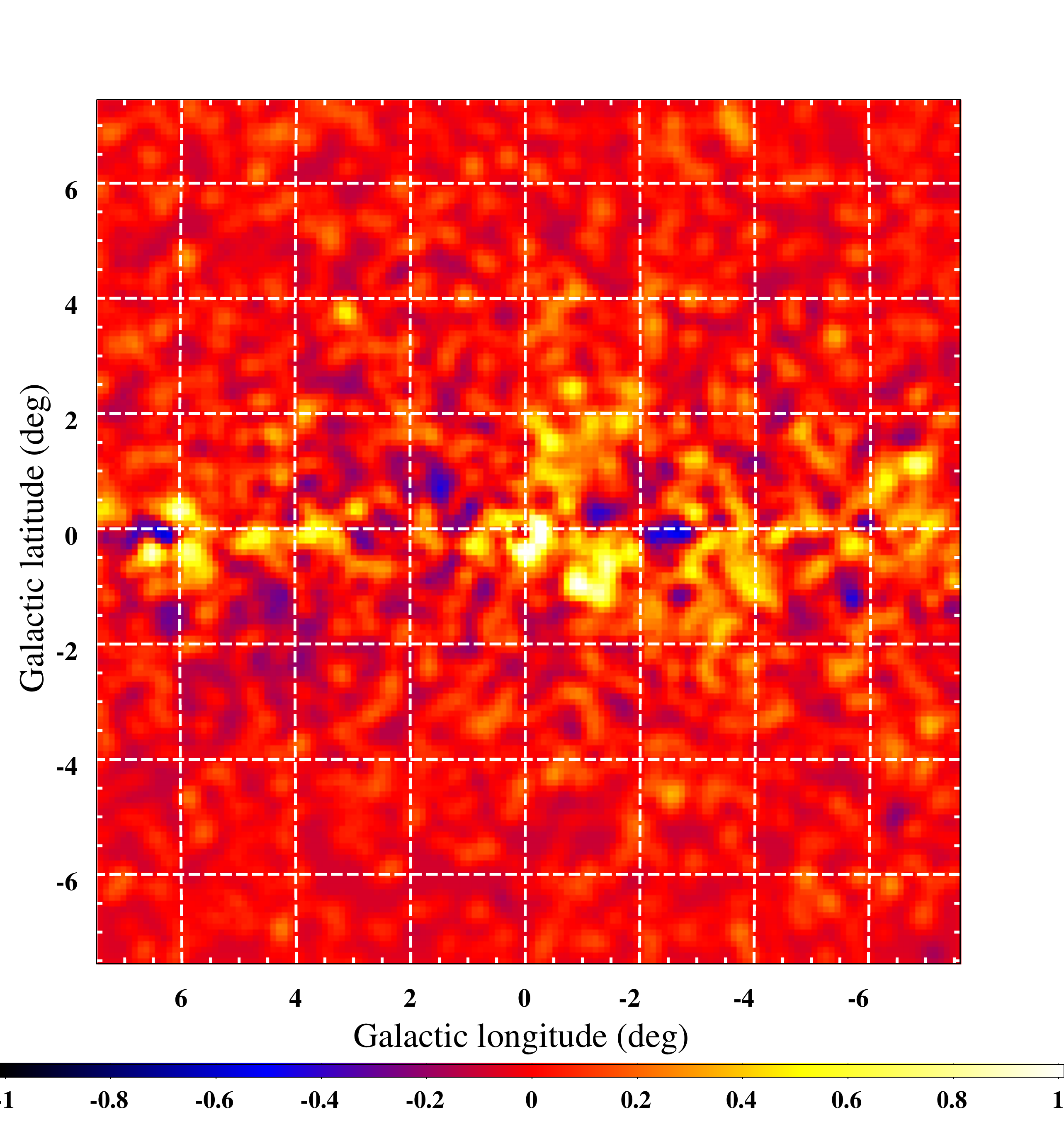}
\includegraphics[scale=0.18]{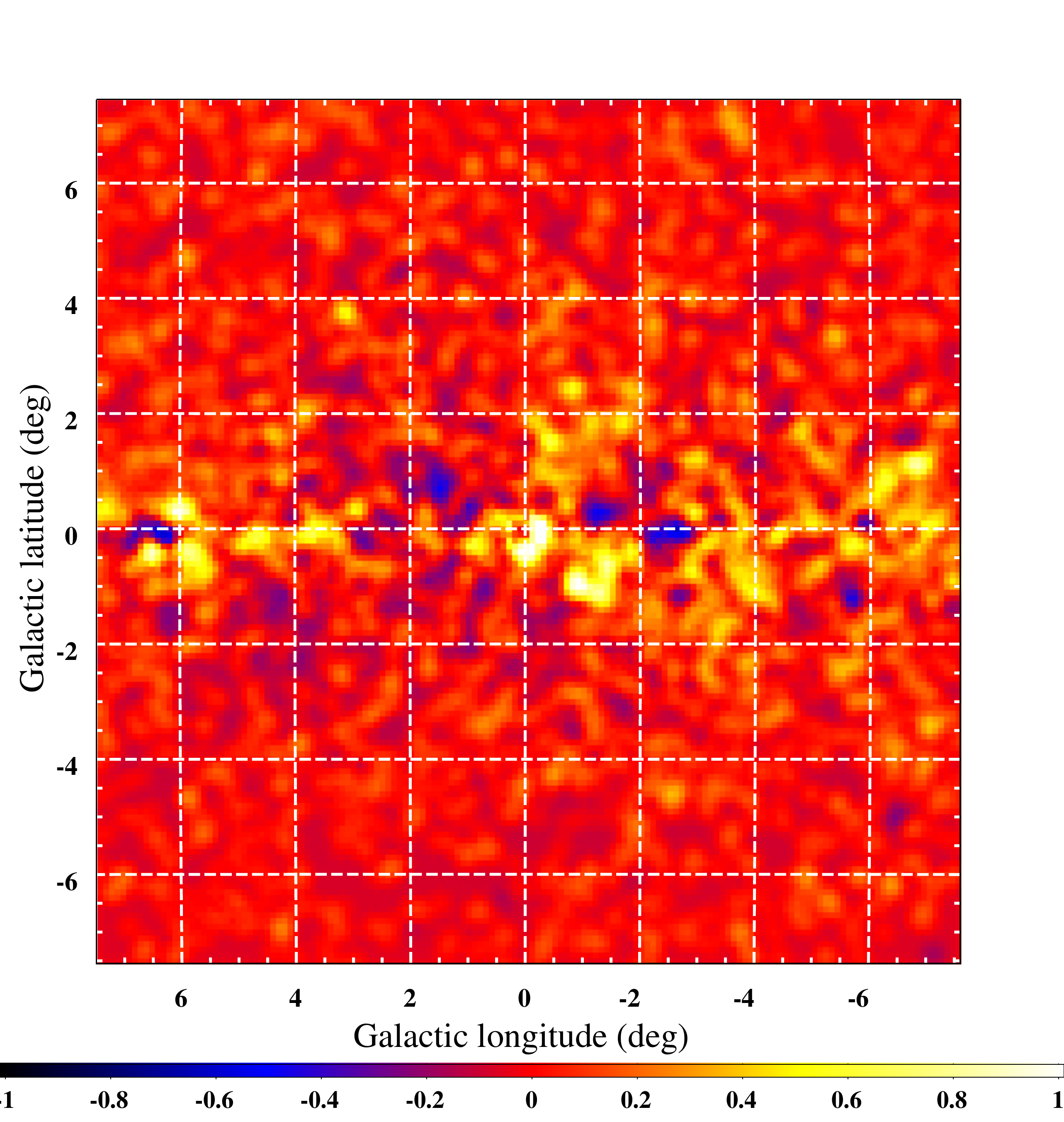}
\includegraphics[scale=0.18]{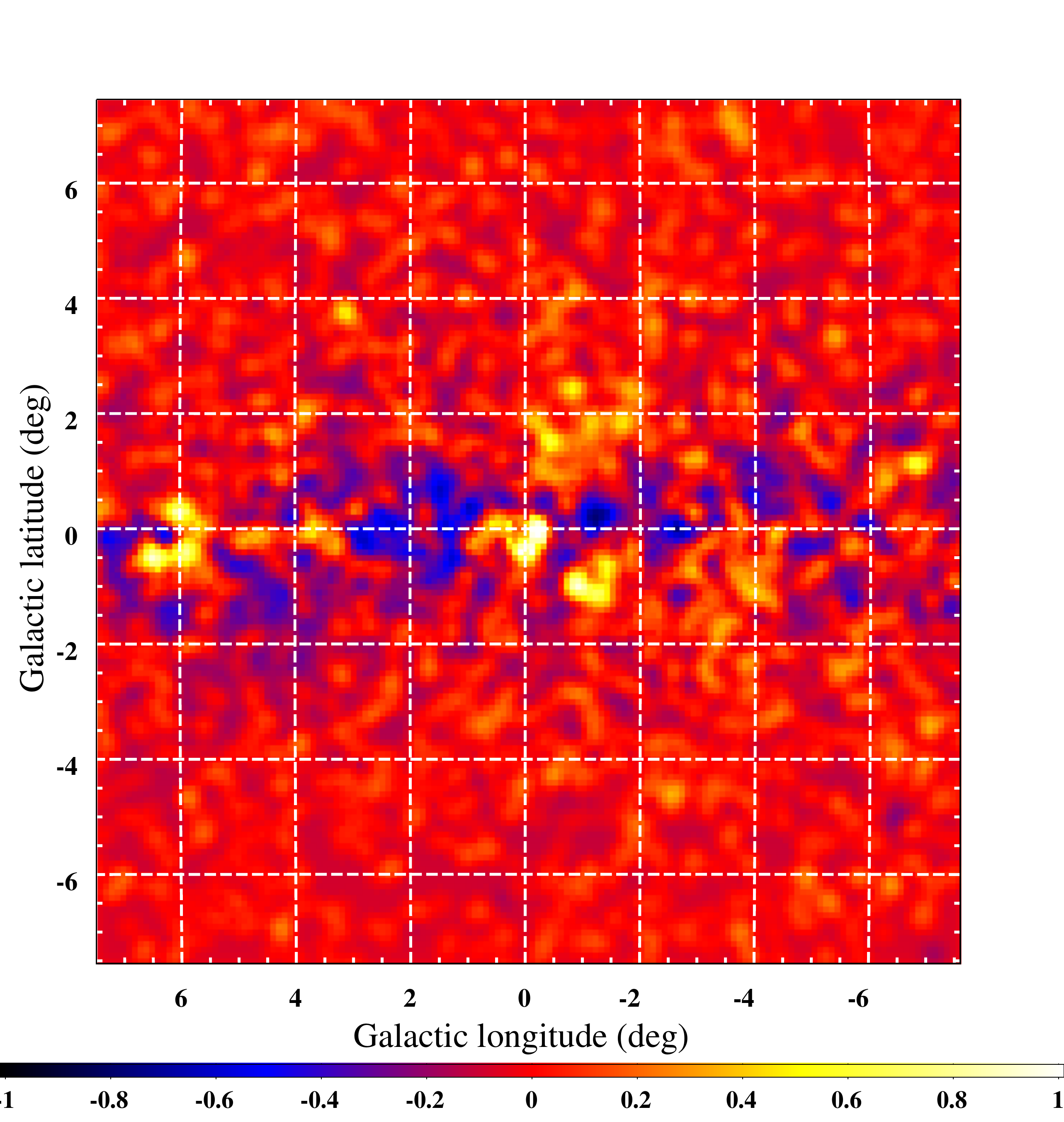}
\includegraphics[scale=0.18]{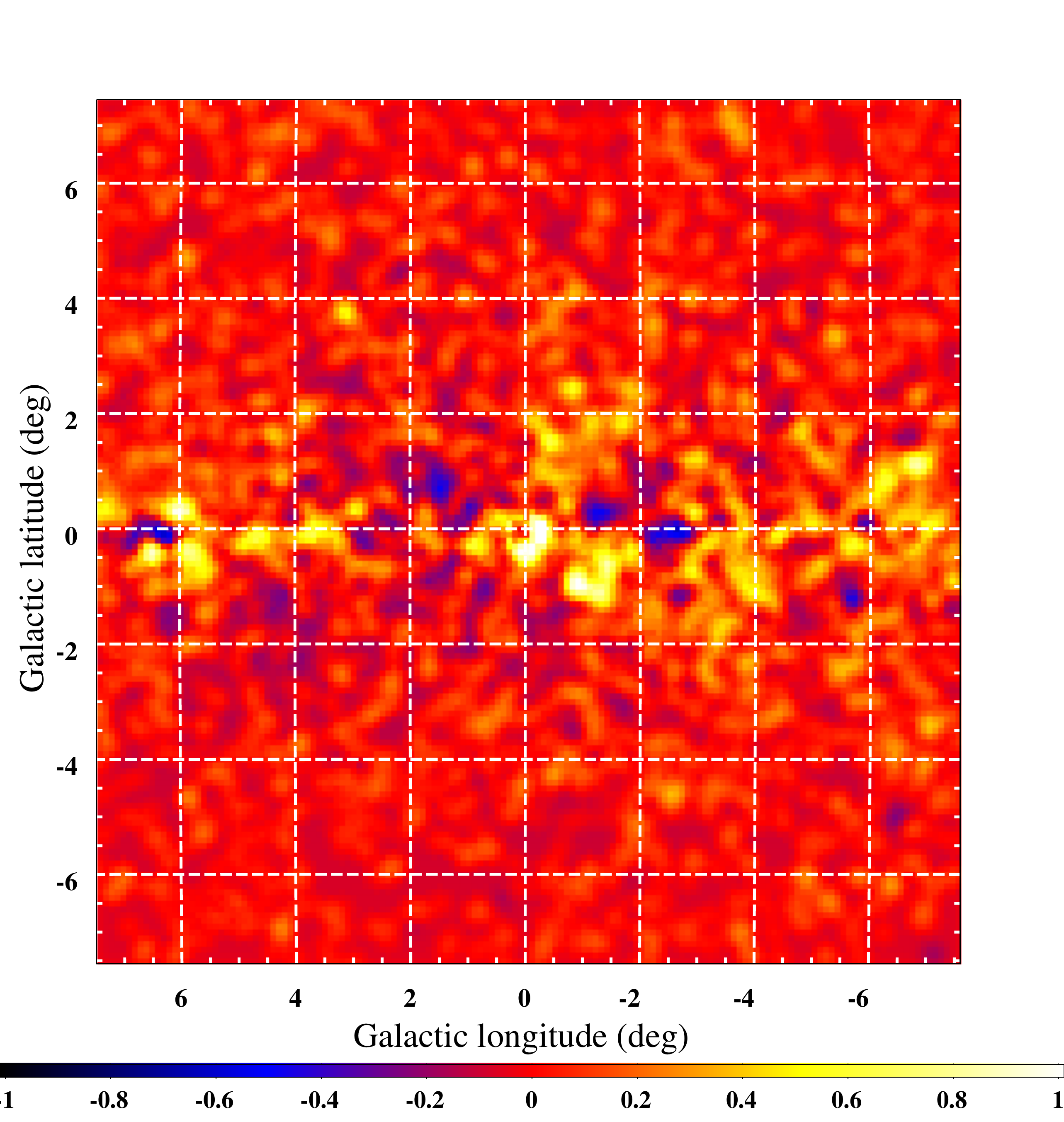}
}
\caption{Residual counts for the $15^\circ \times 15^\circ$ region about
the GC for the Pulsars and OBstars 
IEMs for energy ranges $1-1.6$~GeV (upper row), $1.6-10$~GeV (middle row), and 
$>10$~GeV (bottom row).
The two leftmost columns show the residual counts for the intensity-scaled variant
for the Pulsars and  OBstars, respectively.
The two rightmost columns show the 
residual counts for the index-scaled variant
for the Pulsars and OBstars, respectively.
The colour scale is in counts/0.1 deg$^2$ pixel.
\label{fig:residual_IEM_ROI}}
\end{figure*}

\begin{figure*}[htb]
\subfigure{
\includegraphics[scale=0.32]{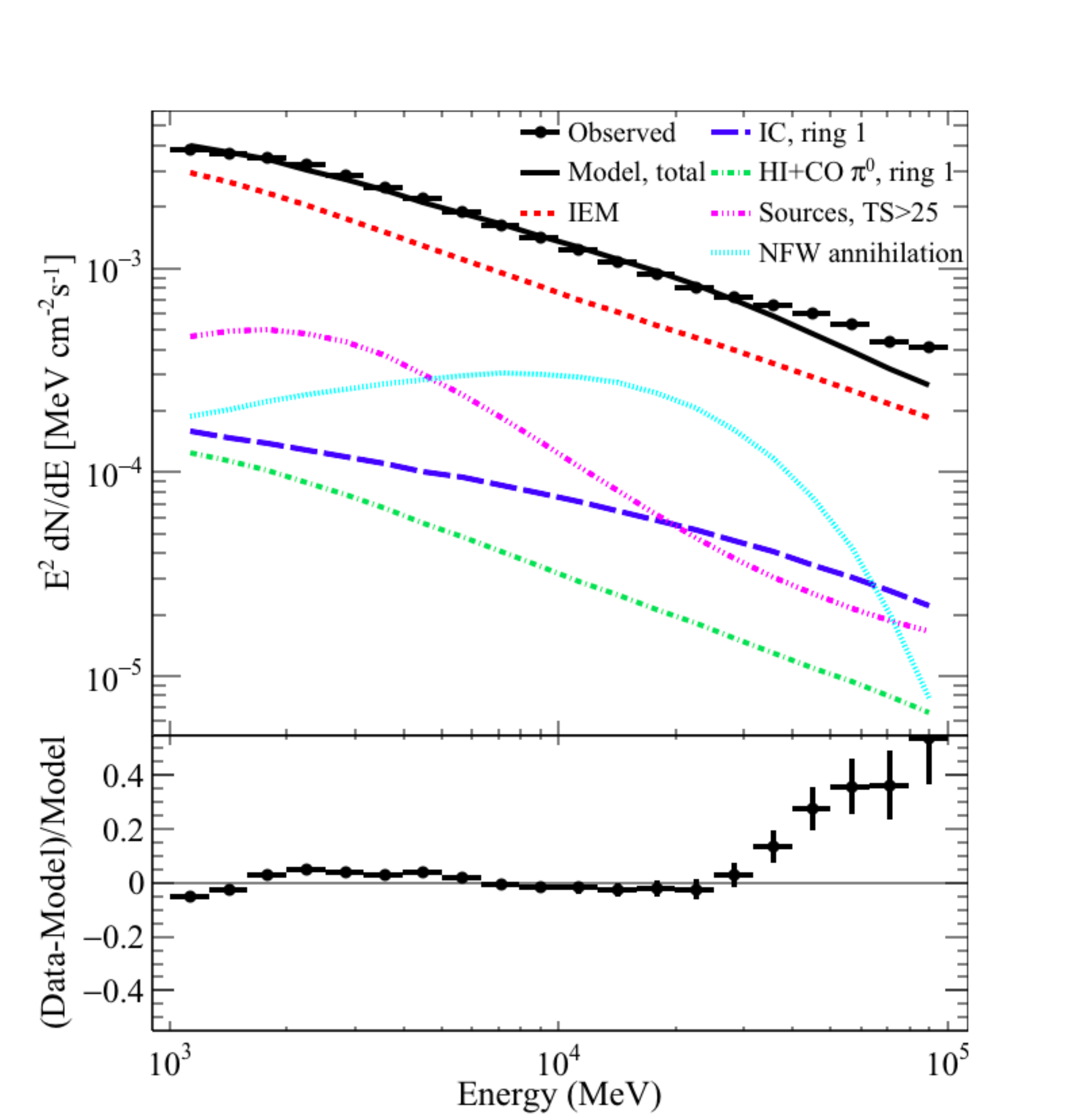}
\includegraphics[scale=0.32]{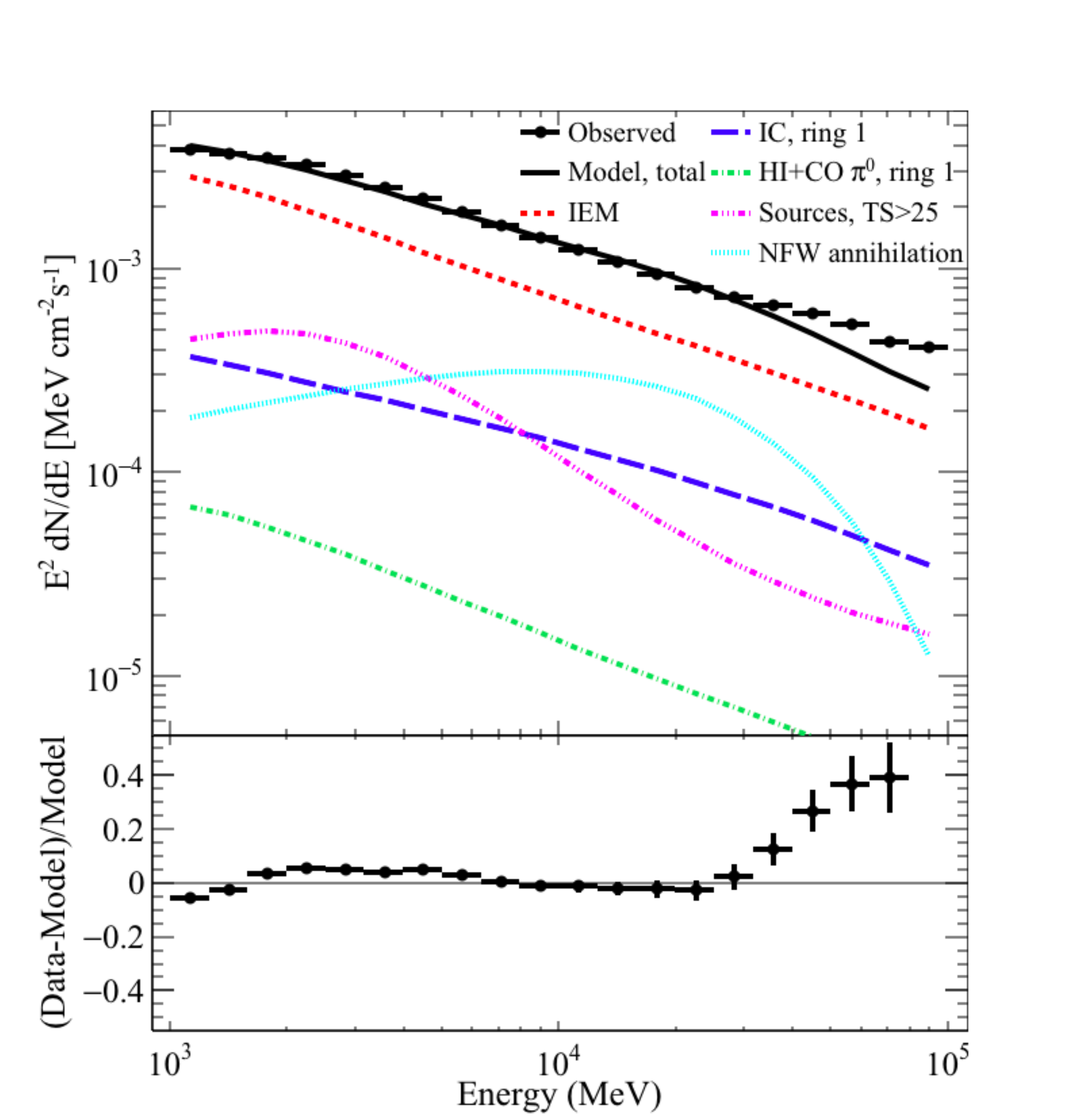}}
\subfigure{
\includegraphics[scale=0.32]{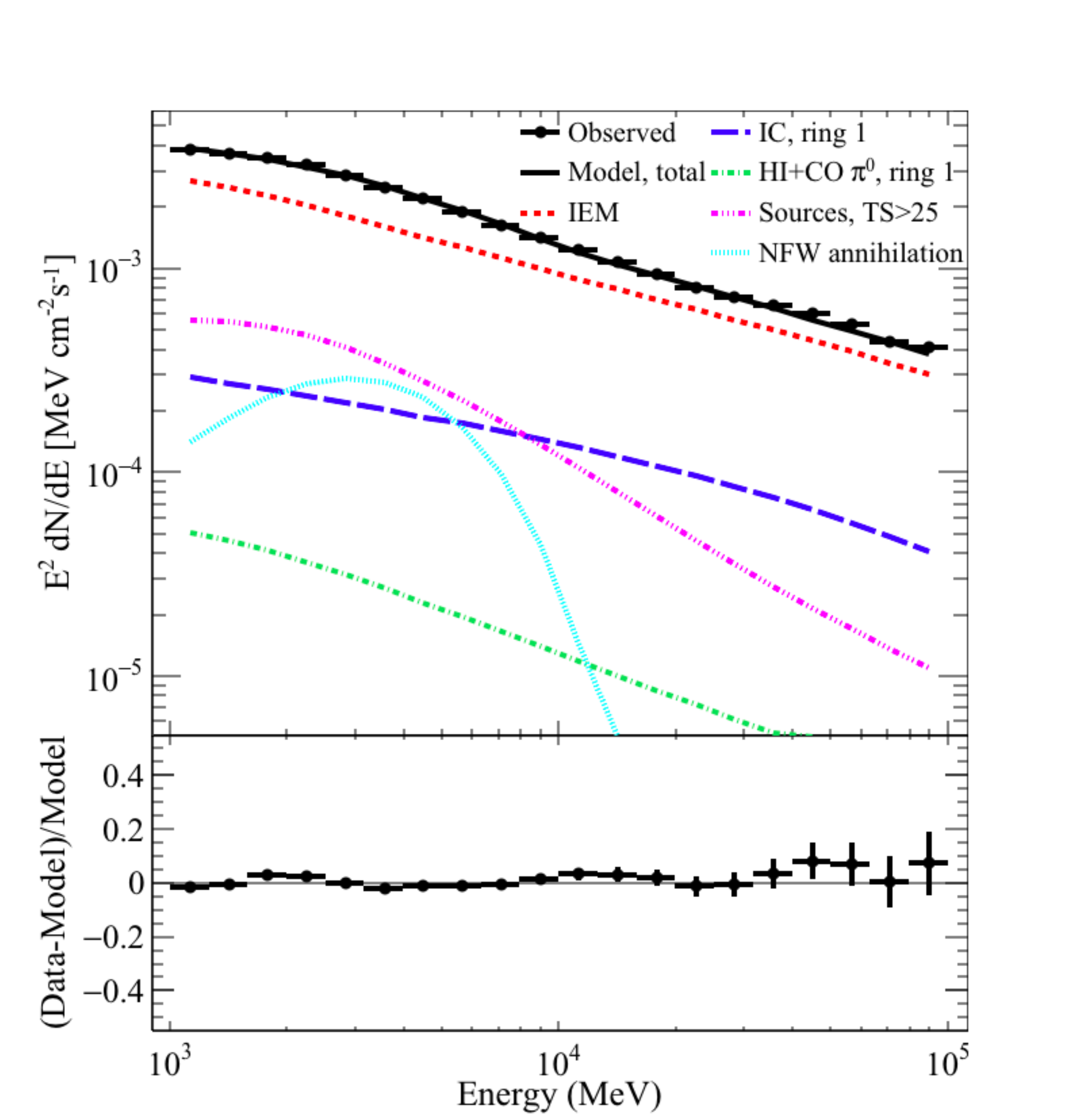}
\includegraphics[scale=0.32]{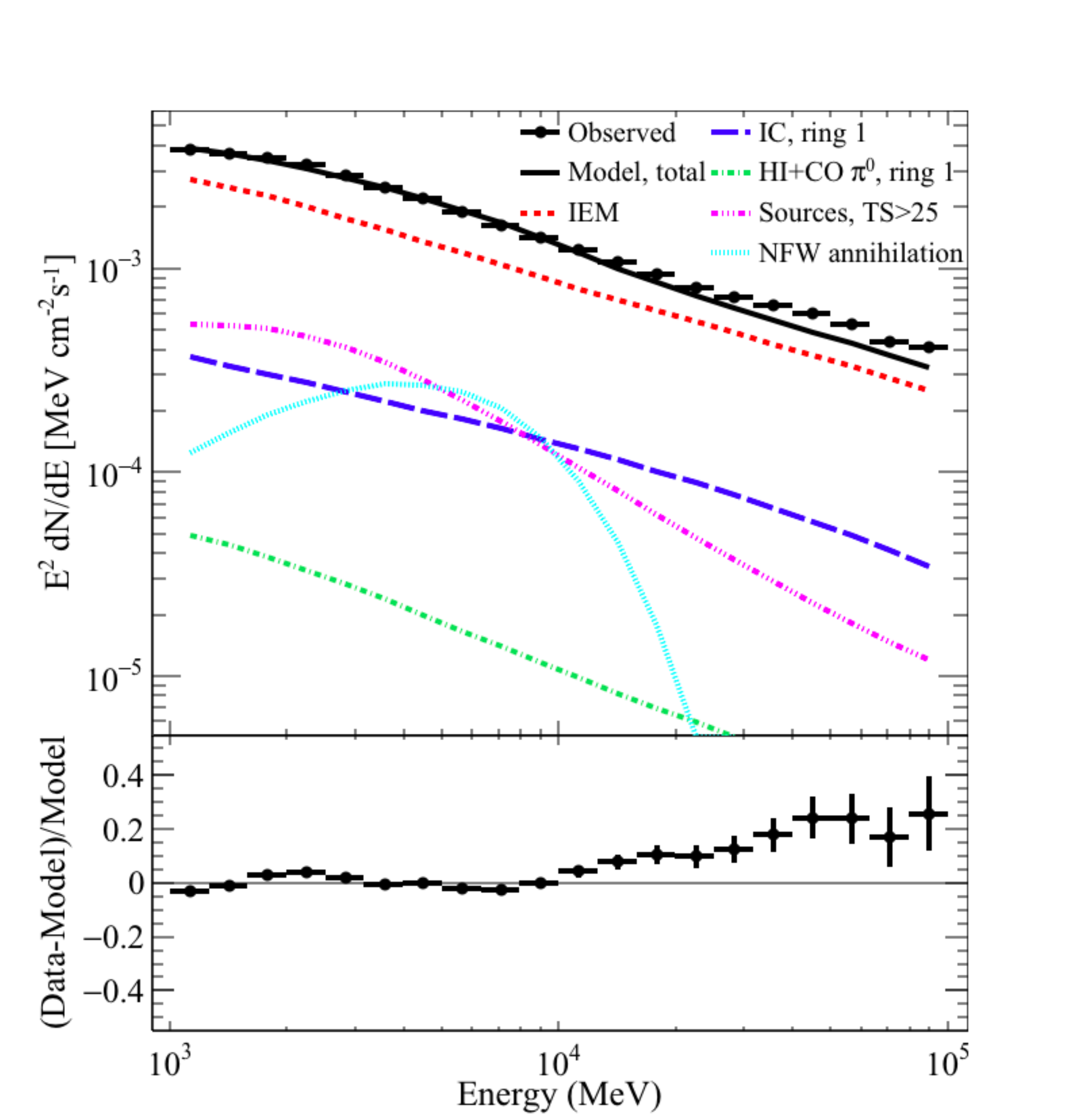}}
\caption{
Differential fluxes for the $15^\circ \times 15^\circ$ region about the GC 
for the four IEMs constrained as described in 
Sec.~\ref{sec:interstellaremission}  using an NFW profile centred on the GC 
as an additional spatial template for the maximum-likelihood fit with spectrum
modeled with a exponential cut-off power law function.
Upper row shows the intensity-scaled for the Pulsars (left) and 
OBstars (right) IEMs.
Lower row shows the index-scaled for the Pulsars (left) and 
OBstars (right) IEMs. Line styles: solid (total model), long-dash (IC, annulus 1), 
dot-dash (\hi\ and CO gas $\pi^0$-decay, annulus 1), dot-dot-dot-dash (point sources), dash (Galactic interstellar emission excluding annulus 1 for IC, \hi\ and CO gas $\pi^0$-decay), dot (new component).  Solid circles: data.
\label{results:nfw_fluxes}}
\end{figure*}

Figure~\ref{results:interstellar_emission_fluxes} shows the fractional residuals
below the differential flux spectrum for each IEM  
integrated over the  $15^\circ \times 15^\circ$ region as a function 
of energy. 
Some trends are evident:
each model over-predicts the data
below $\sim2$~GeV and under-predicts above $\sim2$~GeV, except for
the Pulsars index-scaled IEM that over-predicts the data $\gtrsim 5$~GeV.

\begin{deluxetable*}{lccccc}[ht]
\tablecolumns{11}
\tablewidth{0pt}
\tablecaption{Residual Component Fit Parameters.\label{tab:res_spectra}}
\tablehead{
\colhead{IEM Model} & 
\colhead{Spatial Template} & 
\colhead{Spectral Index} &
\colhead{Cut-off Energy (MeV)} &
\colhead{$\Delta \log L$} &
\colhead{$\log L$} 
}
\startdata
\\Pulsars & & & & & \\
intensity-scaled & NFW annihilation	 & -1.5 $\pm$ 0.1   & 16360 $\pm$ 2945 &   282 & -83027 \\  
& NFW-c\tablenotemark{1} annihilation	 & -1.4 $\pm$ 0.1   & 13120 $\pm$ 2075 &   272 & -83037 \\  
& NFW decay				 & 2.5 $\pm$ 0.4    & 1012 $\pm$ 130 &     99 & -83210 \\   	
& $1^\circ$\tablenotemark{2} 		 & -1.6 $\pm$ 0.1   & 20210 $\pm$ 7451 &   118 & -83191 \\  
& $2^\circ$\tablenotemark{2} 		 & -0.4 $\pm$ 0.5   & 3804 $\pm$ 1706 &    157 & -83152 \\  
& $5^\circ$\tablenotemark{2} 		 & -0.1 $\pm$ 0.7   & 2999 $\pm$ 1470 &    154 & -83155 \\  
& $10^\circ$\tablenotemark{2} 		 & 3.3 $\pm$ 0.5    & 819 $\pm$ 100 &      85 & -83224 \\   	
& Unresolved sources 			 & 0.7 $\pm$ 1.0    & 2313 $\pm$ 1350 &    84 & -83225 \\   	
& & & & & \\				                                                           
Pulsars & & & & & \\			                                                           
index-scaled & NFW annihilation		 & 0.2 $\pm$ 0.3    & 1346 $\pm$ 177 &     165 & -82757 \\  	
& NFW-c annihilation     		 & 0.5 $\pm$ 0.1    & 1132 $\pm$ 30 &      166 & -82755 \\  		
& NFW decay				 & 4.1 $\pm$ 0.4    & 500 $\pm$ 4 &	   67 & -82854 \\     	
& $1^\circ$ 				 & 0.0 $\pm$ 1.1   & 1241 $\pm$ 641 &     76 & -82846 \\   
& $2^\circ$ 				 & 2.0 $\pm$ 0.2    & 693 $\pm$ 31 &       100 & -82822 \\  	
& $5^\circ$ 				 & 2.3 $\pm$ 2.0    & 684 $\pm$ 356 &      101 & -82820 \\  	
& $10^\circ$ 				 & 4.1 $\pm$ 0.5    & 500 $\pm$ 8 &	   59 & -82862 \\     	
& Unresolved sources 			 & 3.3 $\pm$ 0.3    & 500 $\pm$ 2 &	   36 & -82885 \\     	
& & & & & \\				                                                           
OBstars & & & & & \\			                                                           
intensity-scaled & NFW	annihilation	 & -1.5 $\pm$ 0.1   & 18100 $\pm$ 2939 &   298 & -83163 \\  
& NFW-c annihilation	   		 & -1.3 $\pm$ 0.1   & 12610 $\pm$ 2062 &   236 & -83225 \\  
& NFW decay				 & -0.9 $\pm$ 0.3   & 10540 $\pm$ 6265 &   159 & -83302 \\  
& $1^\circ$ 				 & 0.3 $\pm$ 1.7    & 2348 $\pm$ 2426 &    23 & -83438 \\   	
& $2^\circ$ 				 & 0.6 $\pm$ 1.7    & 2251 $\pm$ 2076 &    80 & -83381 \\   	
& $5^\circ$ 				 & -1.2 $\pm$ 0.2   & 12680 $\pm$ 3860 &   213 & -83248 \\  
& $10^\circ$ 				 & 0.2 $\pm$ 0.1    & 3001 $\pm$ 207 &     144 & -83317 \\  	
& Unresolved sources 			 & 0.1 $\pm$ 0.6    & 3513 $\pm$ 1543 &    74 & -83387 \\   	
& & & & & \\				                                                           
OBstars & & & & & \\ 			                                                           
index-scaled & NFW annihilation		 & -0.5 $\pm$ 0.5   & 2682 $\pm$ 912 &     165 & -82819 \\  
& NFW-c  annihilation     		 & -0.4 $\pm$ 0.4   & 2528 $\pm$ 696 &     148 & -82836 \\  	
& NFW decay				 & 3.5 $\pm$ 0.5    & 664 $\pm$ 74 &       102 & -82882 \\  	
& $1^\circ$ 				 & 1.1 $\pm$ 0.2    & 1057 $\pm$ 68 &      42 & -82942 \\   	
& $2^\circ$ 				 & 3.4 $\pm$ 0.7    & 644 $\pm$ 102 &      58 & -82926 \\   	
& $5^\circ$ 				 & 1.9 $\pm$ 2.4    & 962 $\pm$ 695 &      118 & -82866 \\  	
& $10^\circ$ 				 & 3.8 $\pm$ 0.5    & 625 $\pm$ 69 &       96 & -82888 \\
 & Unresolved sources 			 & 4.7 $\pm$ 0.7    & 500 $\pm$ 7 &        28 & -82956 \\    
\enddata
\tablenotetext{1}{NFW-contracted profile with index $\gamma = 1.2$.}
\tablenotetext{2}{Two-dimensional Gaussian with corresponding half-width, half-maximum.}
\end{deluxetable*}

Figure~\ref{fig:res_profiles} shows the longitude and latitude profiles for 
the energy ranges\footnote{Each band
approximately covering the energy intervals where the under/over-predictions 
in the fractional residuals are more prominent. Note that the profiles are 
essentially the same even if, e.g., a smaller latitude band is used to construct
the longitude profiles because the majority of counts are concentrated near 
the plane.} 
$1-1.6$, $1.6-10$, and $>10$~GeV for the Pulsars index-scaled model 
shown in Fig.~\ref{results:interstellar_emission_fluxes}, which has 
the lowest fractional residual across the $1-100$~GeV energy range.
(The features are mostly the same for the other IEMs with the major 
difference their magnitude in terms of counts, hence 
these profiles are not shown because of their similarity.)

The lower sub-panel for each figure gives the residual counts ($data-model$).
While there is considerable statistical noise, 
the total residual counts may be distributed 
asymmetrically in longitude about the GC below 10~GeV.
However, quantifying such an asymmetry using a purely data-driven 
method, e.g., by forming 
the ratio $A = (f_+ - f_-)/(f_+ + f_-)$ where $f_+$ and $f_-$ are the counts
for some equally sized regions about some symmetry line, is not useful 
here because the residuals are of mixed sign.

Figure~\ref{fig:residual_IEM_ROI} shows in greater detail 
the spatial distributions of the residual for each of the IEMs and in the 
three energy bands.
Common features across IEMs are present: 
the model is too bright compared to the data mostly along the Galactic plane 
for the lowest energy band 
($1-1.6$~GeV) and this behaviour is more pronounced for the intensity-scaled 
IEMs, while the models under-predict the data around the GC in 
the $1.6-10$~GeV energy band.

Note that the ecliptic crosses the $15^\circ \times 15^\circ$ region 
and therefore the Sun and Moon contribute to the observed 
emission~\cite{2011ApJ...734..116A}.
The emissions from these objects are not included 
in the fore-/background models employed in this analysis.
But above 1~GeV it is small relative to the observed residual emission
\footnote{The solar \gray{} flux $>1$~GeV within $5^\circ$ of the Sun 
track 
on the sky is $\sim 2\times10^{-8}$ ph cm$^{-2}$ s$^{-1}$ \citep{2011ApJ...734..116A}, while the Lunar \gray{} flux $>1$~GeV is 
$\sim2\times 10^{-9}$ ph cm$^{-2}$ s$^{-1}$ \citep{2012ApJ...758..140A}. 
The fraction of the data taking period spent in the $15^\circ \times 15^\circ$
region by either object is $\sim 5$\% of the total, and their emission is 
distributed about the ecliptic.
For the Sun this corresponds to $\sim 50-100$ counts $>1$~GeV, which is 
$\lesssim 0.1$\% of the total counts. 
The Lunar contribution is lower.}.

Although the spatial distribution of the residuals is not suggestive of a
contribution by the \fermi\ bubbles, it is also possible that there is 
some emission from them over the $15^\circ \times 15^\circ$ region. 
Without a spatial template for the \fermi\ bubbles over the region their 
contribution is tested using a model with an isotropic spatial distribution 
across the 
$15^\circ \times 15^\circ$ region with intensity and spectrum as determined 
from analyses at higher latitudes \citep[e.g.,][]{2014ApJ...793...64A}.
The $data-model$ agreement only marginally improves if this contribution
is included.

The model over-prediction at the lowest energies is 
primarily correlated 
with the Galactic plane, which could be due to mismodelling of the 
gas component of the IEMs. 
Some of the positive residual in the few GeV range could be due to an
extended component that is more concentrated toward the GC compared to the 
IEM components.
But, the profiles shown in Fig.~\ref{fig:res_profiles} and the 
spatial distributions shown in Fig.~\ref{fig:residual_IEM_ROI} represent
the situation if the fit is made only for interstellar emission about the GC 
and point sources.
As a consequence it is difficult to establish properties for an additional
component not presently included in the model for the region.
A spatial and spectral model needs to be assumed and fit to 
the data together with the interstellar emission and point sources.

A set of templates for the spatial 
distribution of the additional component is selected with each fit together 
with the interstellar emission components and point sources using the 
maximum-likelihood procedure described above.
Because the excess emission in the few GeV range is distributed around the 
GC, templates that peak there are considered. 
A set of two-dimensional Gaussians with 
varying HWHM ($1^\circ$, $2^\circ$, $5^\circ$, $10^\circ$) are used. 
While these spatial distributions do not have an obvious physical 
interpretation, they can be used to gauge the 
radial extent of the positive residuals. 
Spatial templates to model
the predicted distribution for \gray{s} produced by 
dark matter (DM) particles annihilating or decaying in the Galaxy are included.  
The inner region of the Galaxy is predicted to be the brightest 
site for a DM signal in \gray{s} and could 
be well within the sensitivity of \fermilat.
To model the DM density distribution, the  
Navarro, Frenk, and White (NFW)~\citep{1997ApJ...490..493N} profile is employed~\footnote{The following parametrization is employed: \\
$\rho(r)=\rho_0 \left( \frac{r}{R_s}\right)^{-\gamma}\left( 1+\frac{r}{R_s} \right) ^{\gamma-3}$, 
where $\gamma$ is the slope of the DM distribution in the innermost region
and its value is discussed in the text. As for the other parameters, in this work
$R_s$~=~20~kpc and $\rho_0$ corresponds to a local DM 
density $\rho_{\sun}=0.3$~GeV/cm$^3$.} 
with different choices for the slope
of the profile in the innermost region, $\gamma=1, 1.2$.     
The NFW profile is predicted by simulations of cold DM, 
while the more peaked distribution with $\gamma=1.2$ (NFW-c) is
motivated by earlier work~\citep[e.g.,][]{2011PhLB..697..412H,2012PhRvD..86h3511A} 
and could  arise when baryonic effects are included in simulations. 
The square of the NFW profile is used as a template for DM annihilation, 
hereafter
referred to as simply the ``NFW profile''.
The possibility that an unresolved population of \gray{} 
point sources such as pulsars distributed along the Galactic plane
is contributing to the observed emission is also considered. 
Predictions of the \gray{} emission from unresolved pulsars
exist and they span a  range of possibilities \citep[e.g.,][]{2007ApJ...671..713S,2010JCAP...01..005F}. 
Here, the spatial distribution of an unresolved pulsar population
is modelled using the distribution of the CO gas in annulus~1, because this 
is a likely tracer for regions of high-mass star formation, and  
smooth it with a $2^\circ$ Gaussian to account for the scale height of the
associated pulsar population.
 
For each of the spatial templates listed above, the spectrum is modelled 
with an exponential cut-off power law. 
This form has some flexibility to model a pulsar or a DM annihilation 
spectrum without supposing specific scenarios.
For each of the spatial templates listed above and for each of the IEMs, 
a maximum-likelihood 
fit is made in the $15^\circ \times 15^\circ$ region as described in 
Section~\ref{sec:maxlikelihood}.

\begin{figure}[t]
\includegraphics[trim = 90 50 0 0,scale=0.30]{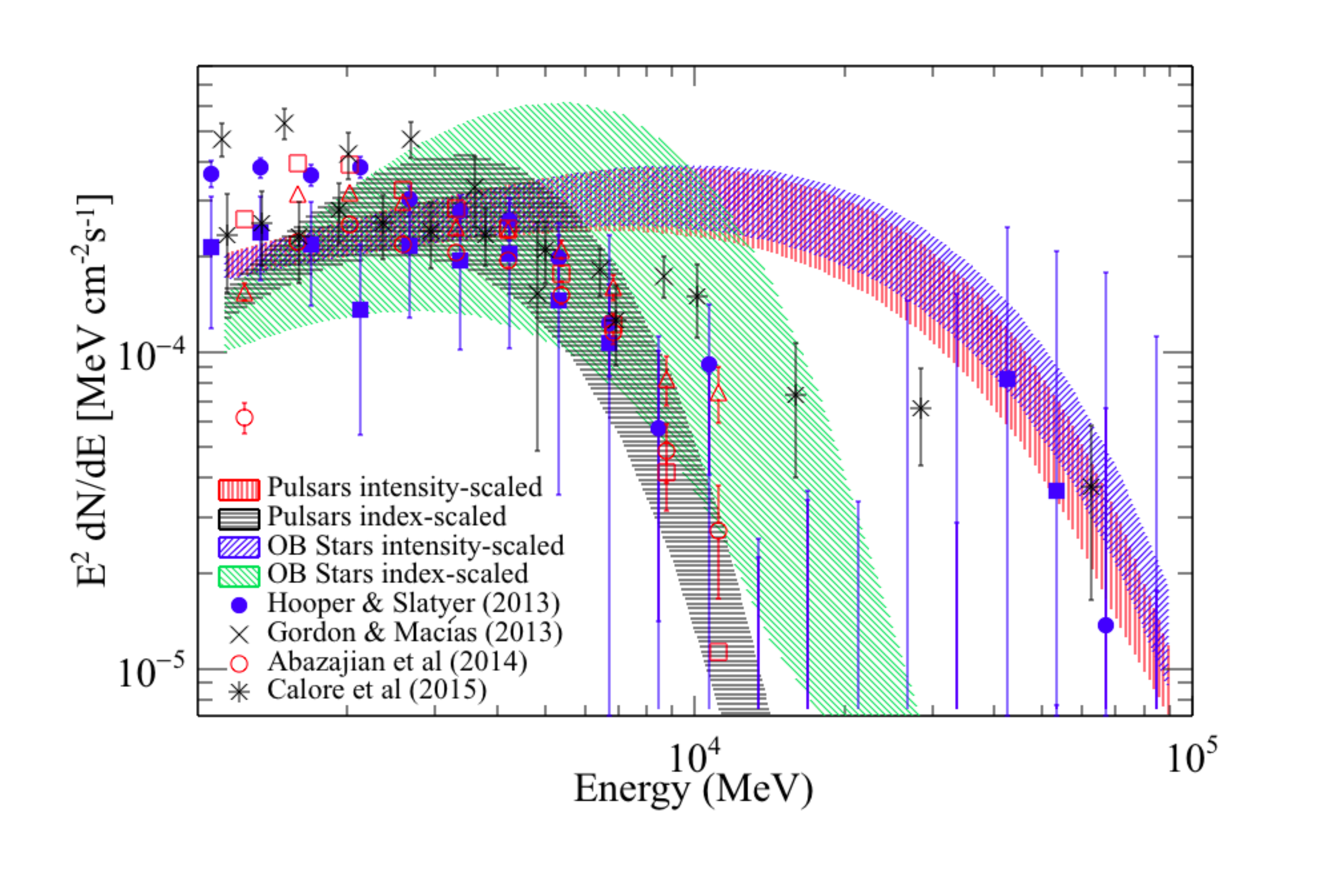}
\caption{
Differential fluxes for the $15^\circ \times 15^\circ$ region 
about the GC of the NFW component with spectrum
modelled with an exponential cut-off power law. 
The envelopes include the fit uncertainties for the normalisation and 
spectral index.
Hatch styles: Pulsars, intensity-scaled (red, vertical); 
Pulsars, index-scaled (black, horizontal); 
OBstars, intensity-scaled (blue, diagonal-right); 
OBstars, index-scaled (green, diagonal-left).
Results from selected other works are overlaid.
{\it Filled symbols}: ~\cite{2013PDU.....2..118H}, different symbols bracket the results obtained when different regions of the sky are considered in the fit; 
{\it Angled crosses}: ~\cite{2013PhRvD..88h3521G}; {\it Open symbols}: ~\cite{2014PhRvD..90b3526A}, front-converting events shown with triangles, front- and back-converting events shown with squares and circles, depending on the modelling of the fore-/background. {\it Stars }: ~\cite{2015PhRvD..91f3003C}.
Note: the overlaid results are rescaled to the DM content over the $15^\circ \times 15^\circ$ region for an NFW profile with index $\gamma$=1.
\label{fig:plec_overlay}}
\end{figure}

The improvement in likelihood as well as the resulting 
best-fit parameters for the spectrum of the additional 
component are summarised in Table~\ref{tab:res_spectra}\footnote{500 MeV 
is the lowest value of the energy cutoff allowed in the fit.}.
All templates yield statistically significant improvements 
compared to the model without the additional component.
The largest improvements are observed for the NFW annihilation templates, 
whereas the unresolved source component yields the smallest improvements.

The new component spectra present harder spectral 
indices and lower energy cutoffs for the index-scaled IEMs  
compared to the intensity-scaled variants. 
This is consistent with the index-scaled models having overall better 
agreement with the data at higher energy, and therefore attributing the 
positive residual found for the intensity-scaled IEMs $\gtrsim 10$ GeV 
to gas related emission rather than to the new component.
Within the same IEM, the spectrum for the more peaked templates 
(NFW and NFW-c for DM annihilation, and the $1^\circ$ gaussian) 
present softer indices and higher energy cutoffs. 
The NFW decay and the $10^\circ$ gaussian (the more extended templates)
perform similarly to each other for most IEMs.

Among the gaussian templates, the $2^\circ$ and $5^\circ$ gaussians perform 
better for the Pulsar IEMs, while the $5^\circ$ and $10^\circ$ gaussians
for the OB stars IEMs. This result is an indication that the gaussian templates 
might be compensating for mismodelling of the IC contribution, 
whose morphology differs for the OB stars and Pulsars IEMs.

By including the NFW profile component the agreement with the data has an 
overall 
improvement for all the models up to $\sim 30$~GeV, as 
shown in Fig.~\ref{results:nfw_fluxes}, with the Pulsars index-scaled
variant yielding the best agreement over the full energy range.
However, a broad range for the best-fit parameters of the 
spectral model is found. 
The variation is not easily ascribed
to a covariance with only a single component of the model that is fitted
over the $15^\circ \times 15^\circ$ region.
For example, the annulus~1 IC and \hi-related $\pi^0$-decay normalisations 
adjust in the fit to compensate for the additional template.
But the spectral parameters of the residual template are not solely determined
by the fit with the interstellar emission components and point sources over
the inner region about the GC; 
the fore-/background interstellar emission has an effect as well.

The intensity-scaled IEMs yield similar spectral parameters for the 
NFW template, but the results for the index-scaled IEMs have a 
stronger variation.
This can be seen in Fig.~\ref{fig:plec_overlay}, which shows the 
flux spectral envelopes 
from including the uncertainties on the normalisation and spectral index
obtained for the NFW template for the 4~IEMs.
The index-scaled IEMs have the distinction of harder spectra for the 
$\pi^0$-decay interstellar emission for annuli $2-4$ 
(Table~\ref{table:coefficients}), but also modified IC contributions for 
annuli~2 and~3 compared to their intensity-scaled counterparts.
The majority of the $\pi^0$-decay fore-/background interstellar emission
is due to annulus~4 and, as already noted in Sec.~\ref{results:interstellar_emission}, even small variations in 
the structured fore-/background interstellar emission can have a 
follow-on effect on the spatial distribution of the residual emission over 
the $15^\circ \times 15^\circ$ region. 
It is difficult to test how small variations in the $\pi^0$-decay 
fore-/background from this annulus affect the residual model parameters 
because the annulus~4 fit parameters are 
determined at an intermediate step in the fitting.
But the comparison between the results for the Pulsars and OBstars index-scaled
IEMs show that the different spectral parameters obtained for the 
structured interstellar emission fore-/background can alter the final 
fitted values for all components over the $15^\circ \times 15^\circ$ 
region\footnote{The Pulsars index-scaled IEM has the same 
spectral parameters across all annuli interior to the solar circle for the 
separate \hi- and CO-components and the lowest cut-off energy for the residual
template, while the OBstars index-scaled IEM has 
the annulus~2 and~3 components set to the \GP\ predictions because they did not 
converge in the IEM fitting.
Whether annuli~2 and~3 have a 
significant effect on the residual spectral parameters for the Pulsars 
index-scaled IEM was tested 
by also setting them to the \GP\ predictions and refitting
for the annulus~1 interstellar emission, point sources, and residual model 
parameters.
The normalisation and cut-off energy of the residual model did not appreciably
change, indicating that the majority of any effect related to the 
structured fore-/background from the index-scaled IEMs is likely from annulus~4.}.

With the interstellar emission fore-/background held constant for each IEM, 
the interplay between the centrally peaked positive residual template 
and the interstellar emission components is not surprising.
Because the IC component is maximally peaked toward the GC for all IEMs 
an additional template that is also peaked there will also be attributed 
some flux when fit.
Over all IEMs the effect of including the NFW model for the residual 
results in an IC annulus~1 contribution that is up to three times smaller 
and \hi\ annulus~1 contribution that is up to three times larger.

\begin{figure*}[t]
\subfigure{
\includegraphics[trim = 0 60 0 0, scale=0.265]{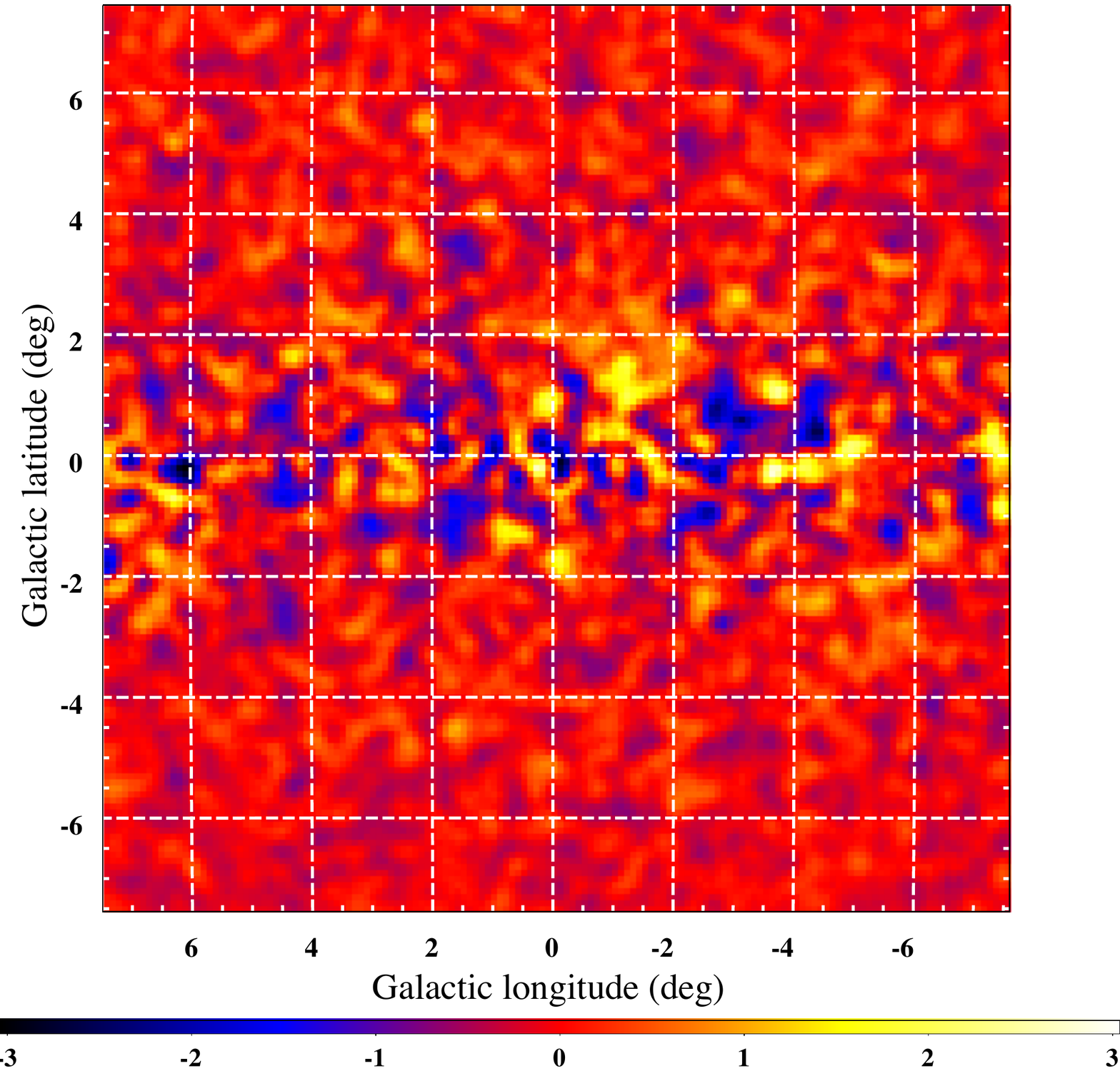}
\includegraphics[trim = 0 60 0 0, scale=0.265]{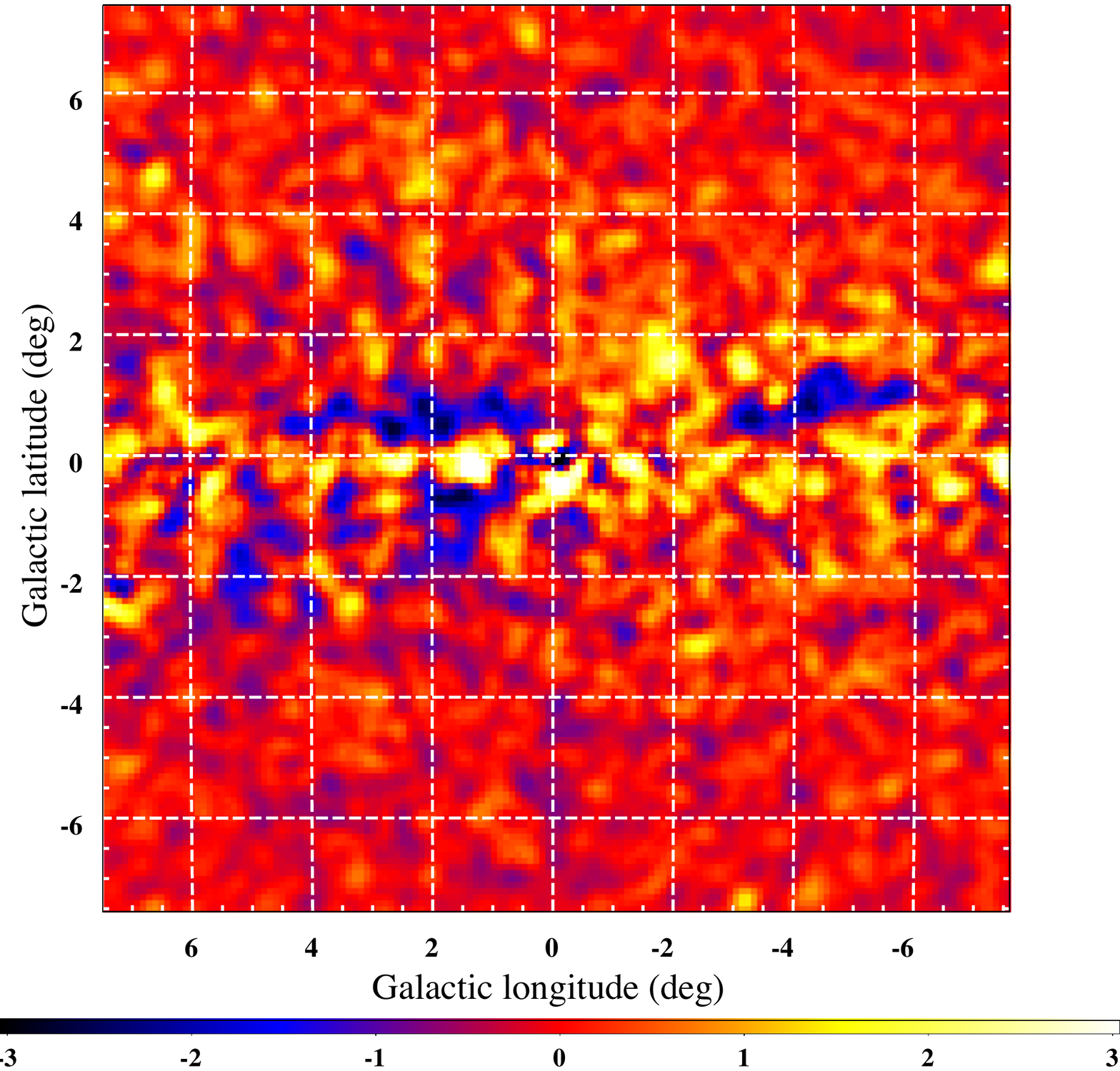}
\includegraphics[trim = 0 60 0 0, scale=0.265]{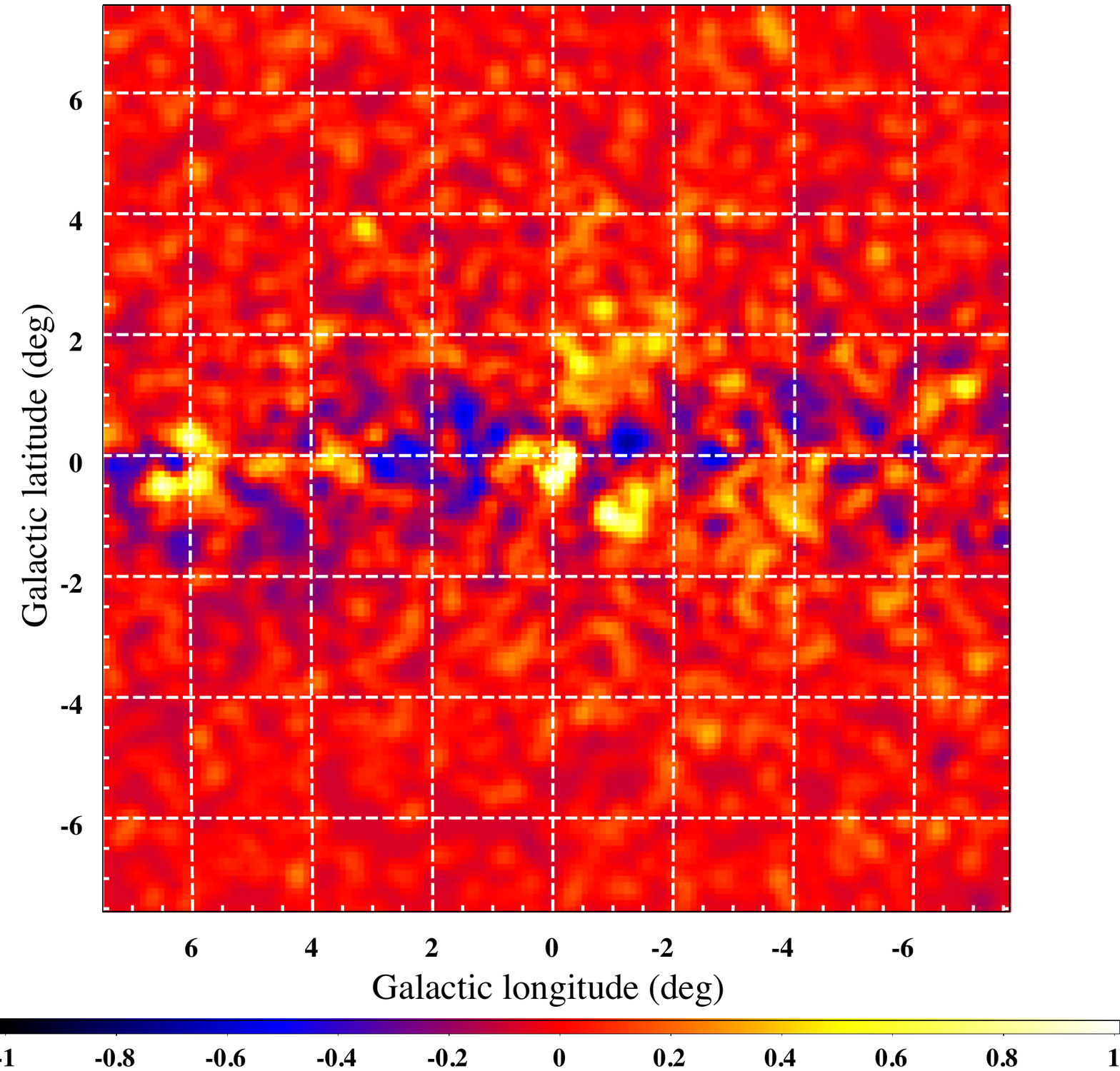}
}
\caption{Residual counts for the $15^\circ \times 15^\circ$ region about
the GC for the Pulsars index-scaled IEM together with the NFW profile 
template for energy ranges $1-1.6$~GeV (left), $1.6-10$~GeV (middle), and 
$>10$~GeV (right).
The colour scale is in counts/0.1 deg$^2$ pixel.
\label{fig:residual_NFW}}
\end{figure*}

Note that even if a centrally peaked template is included as a 
model for the positive residual, it does not account for all of the emission.
This can be seen in Fig.~\ref{fig:residual_NFW}, which shows the residual
counts for the NFW template and IEM with the best spectral residuals 
(Pulsars index-scaled).
Qualitatively, the remainder does not appear distributed symmetrically about
the GC below 10~GeV, and still has extended positive residuals even at higher
energies along and about the plane.

\section{Discussion}
\label{sec:discussion}
\subsection{Interstellar Emission}
\label{sec:interstellar_emission_discussion}

\begin{figure*}[ht]
\subfigure{
\includegraphics[scale=0.9]{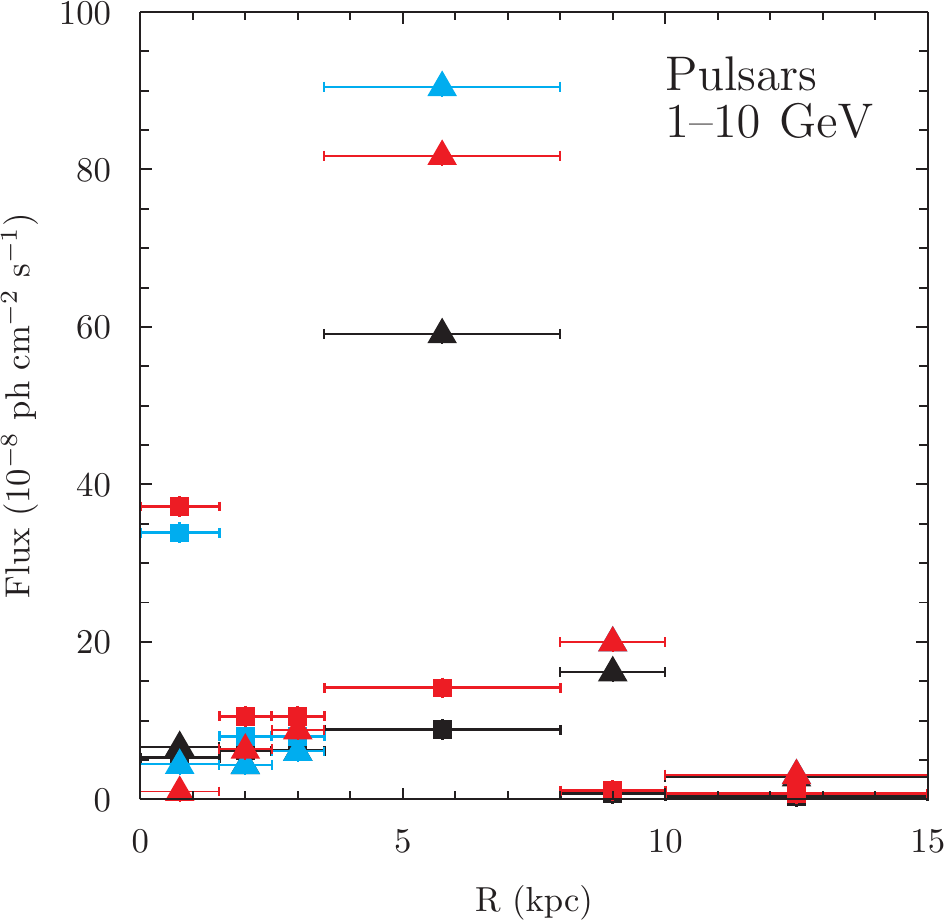}
\includegraphics[scale=0.9]{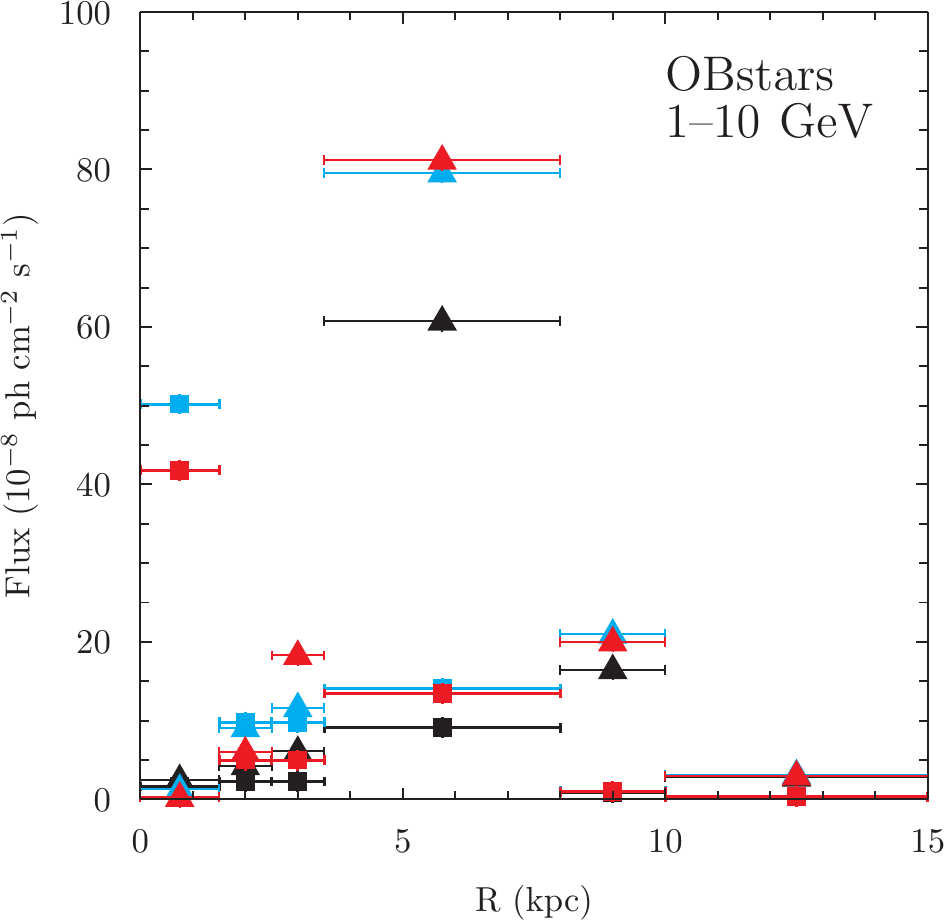}
}\\
\subfigure{
\includegraphics[scale=0.9]{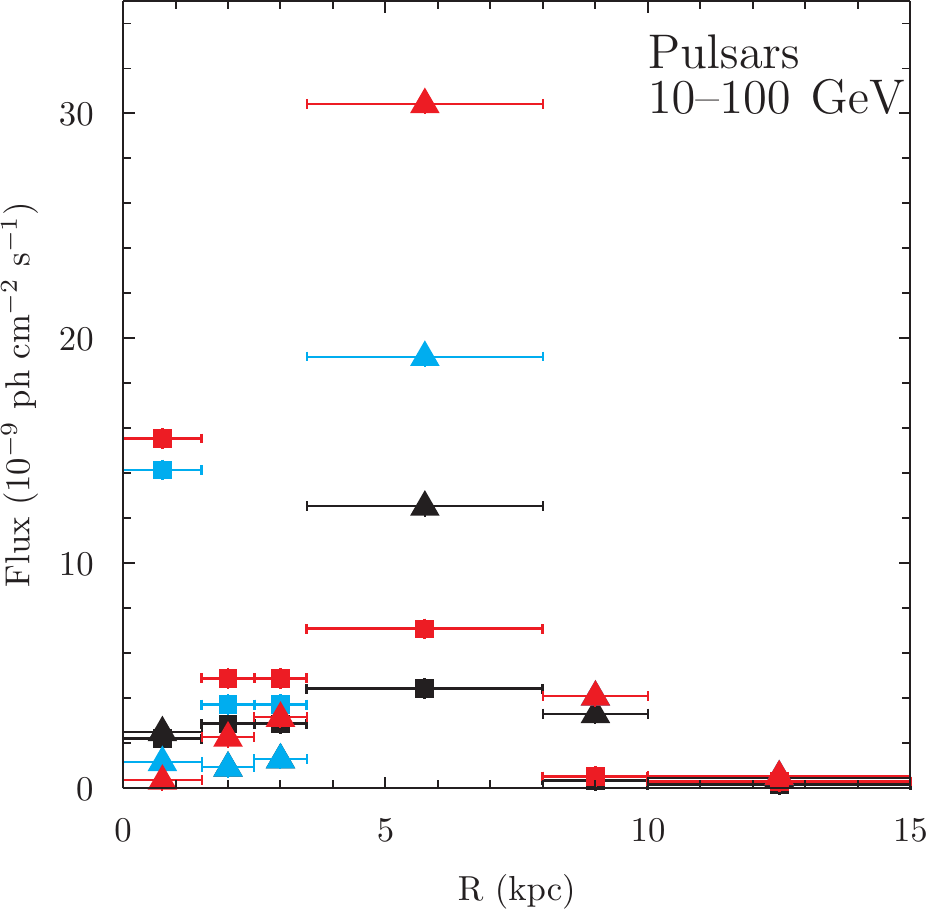}
\includegraphics[scale=0.9]{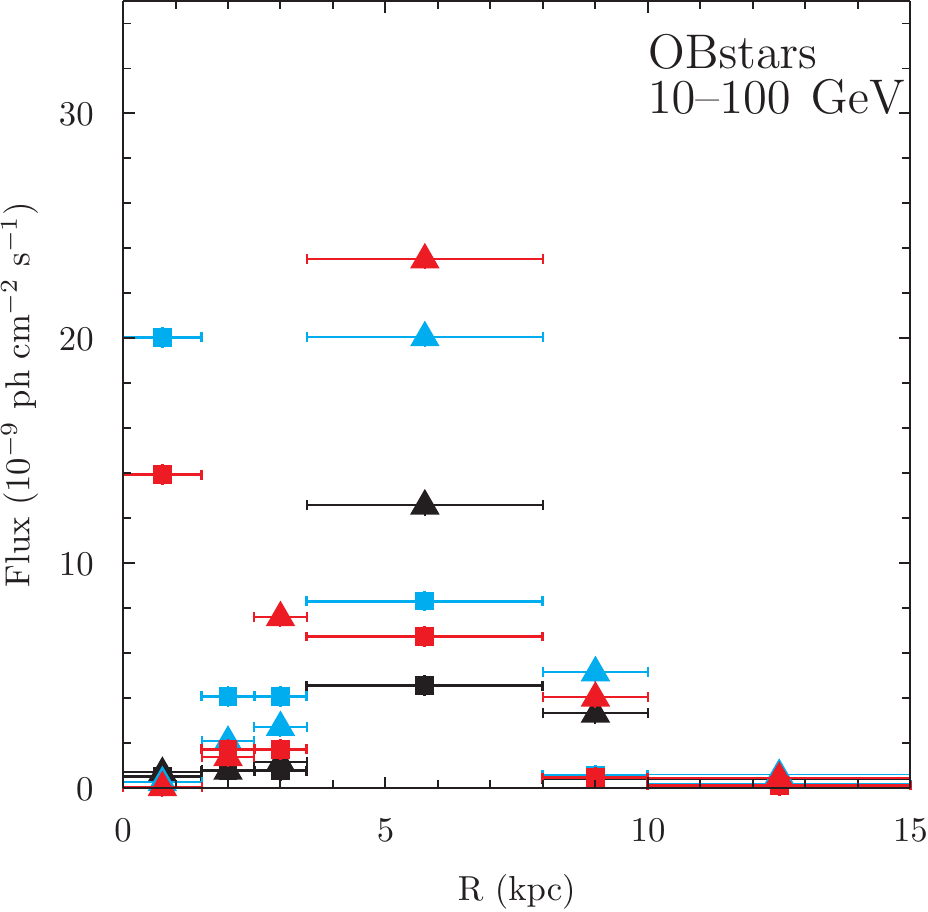}
}
\caption{Flux per ring for the $15^\circ \times 15^\circ$ region about the 
GC for the Pulsars (left) and OBstars (right) IEM variants for the $1-10$ (top) and $10-100$~GeV (bottom) 
energy ranges.
Point colours: black, \GP-predicted; cyan, intensity-scaled; red, index-scaled.
Point types: filled square, IC; filled triangle, $\pi^0$-decay.
Note some symbols are obscured for annuli at larger radii because the 
intensity-/index-scaled variants are the same outside the solar circle.
\label{fig:fluxes}}
\end{figure*}

This study is the first using the \fermilat\ data 
that has made a separation between the large-scale 
interstellar emission of the Galaxy and that from the 
inner $\sim 1$~kpc about the GC.
The IC emission from annulus~1 is found to dominate the interstellar emission
from the innermost region, and represents the majority of the IC 
brightness from this component along and through the line-of-sight toward 
the GC.
The contribution by the IC from annulus~1 
to the total flux depends on the IEM and whether the residual
is fitted (Sec.~\ref{results:residuals}).
For the latter case the IC from annulus~1 is still up-scaled compared to the
\GP\ predictions, but by a 
factor $\sim 2$ lower than if fitted solely for the interstellar emission 
components and point sources. 
The remainder is distributed across 
the \hi-related $\pi^0$-decay annulus~1 component and the template used
to fit the residual centred on the GC.
For either case (residual template used/not-used), the fitted {\it fluxes} 
attributed to the IC annulus~1 component across all IEMs are 
within a factor $\sim 2$ -- the flux and its range
is the important quantity, instead of the individual (model-dependent) 
scaling factors.

The Pulsars intensity-scaled IEM with the residual template gives the 
minimal `enhanced' flux for IC annulus~1.
The average 
CR electron intensity $\gtrsim 5$~GeV in the Galactic plane is estimated 
for this model within $\sim 1$~kpc of the GC 
as $\sim 2.8\pm0.1\times10^{-4}$ cm$^{-2}$ s$^{-1}$ sr$^{-1}$, 
where the uncertainty is statistical only.
This energy range is used because its lower bound corresponds to the 
CR electron energies producing $\sim 1$~GeV IC \gray{s}.
This is $\sim$ a factor of two higher than the local total CR electron
density for this same energy range for the Pulsars baseline model.
On the other hand, the OBstars intensity-scaled IEM fitted without the 
residual component gives the maximal `enhanced' flux for IC annulus~1.
The average CR electron intensity $\gtrsim 5$~GeV in the Galactic plane 
within $\sim 1$~kpc of the GC for this IEM is 
$\sim 9.4\pm0.1\times10^{-4}$ cm$^{-2}$ s$^{-1}$ sr$^{-1}$.

Measurements of the interstellar emission at hard X-ray energies to MeV 
\gray{s} by INTEGRAL/SPI \citep{2011ApJ...739...29B} show that the majority
is due to IC scattering by $\sim$GeV energy CR electrons off the infrared
component of the ISRF\footnote{The majority of the IC \gray{s} in the 
energy range of this study are produced by scattering off the optical component
of the ISRF.}.
The \GP\ calculations, which follow the same ``conventional'' model 
normalisation condition to local CR measurements as used in this paper, made to 
interpret the SPI measurements indicate
that IEMs with at least factor of 2 higher CR densities toward the inner 
Galaxy are a plausible explanation for the data.
Another possible explanation is a higher intensity for the radiation
field energy density in the 
inner Galaxy than used in the standard ISRF model of \citet{Porter2008}; 
these possibilities are not tested here because
they require detailed investigations that are beyond the scope of the current
work.
The higher CR electron densities obtained from this analysis are plausible
given the same electrons are IC scattering different components of the ISRF 
to produce the interstellar emission $\gtrsim 1$~GeV and 
at SPI energies.

The purpose for fitting the baseline IEMs to the data was to obtain 
estimates for the interstellar emission fore-/background.
However, the fit results for the individual rings for each IEM 
potentially give some information on the large-scale distribution of CRs 
througout the Galaxy. 
Tables~\ref{table:coefficients} and~\ref{table:fluxes} in 
Appendix~A.1 give the fit coefficients and fluxes for the scaled 
IEMs, while Fig.\ref{fig:fluxes} shows the 
integrated fluxes for the 1--10 (top) and 10--100~GeV (bottom) 
energy ranges, respectively, 
over the $15^\circ \times 15^\circ$ region 
for the \GP-predicted and scaled version of each IEM for the Pulsars (left) and OBstars (right) source distributions.

The fitting procedure generally increases the intensity of each annulus relative
to the nominal model.
The coefficients for the intensity-scaled Pulsars and OBstars IEMs are 
mostly higher than the \GP\ predictions toward the inner Galaxy 
(annuli~$2-3$).
Those for the OBstars IEM are higher than the Pulsars, which reflects 
the fact that the spatial distribution for the CR sources in this model 
cuts off within $\sim 2$~kpc of the GC.
The cut-off in the OBstars source spatial distribution produces 
a {\it predicted} CR intensity that is lower 
compared to the Pulsars IEM over this region.
The fitting procedure adjusts the OBstars predictions upward more than the 
Pulsars to compensate.
This indicates that a Pulsars-style spatial source distribution is 
closer to the real spatial distribution of sources within $\sim 2$~kpc of
the GC.
But, even the Pulsars spatial source distribution is scaled up by the fit 
over this region, indicating that even more `peaked' source models, or some 
modification to the propagation model, is required to 
describe the distribution of CRs toward the inner Galaxy.
Meanwhile, there is more similarity in the scaling coefficients 
for annuli~$4-6$.
This reflects that the CR source distributions and propagation 
conditions for both IEMs are not 
significantly different in their Galactocentric radial distributions in
these annuli.

This spectral parameters for the annuli interior to the solar circle for
the index-scaled variants give results that are strongly dependent on the
IEM being fit.
For the Pulsars IEM the spectrum of the CR nuclei/gas interstellar emission 
is consistently harder across annuli~$2-4$ for both CO and \hi\ components than
the intensity-scaled IEMs.
For the OBstars IEM only the \hi\ component has a hardening to the spectrum
across annuli~$2-4$.
For this IEM the fits for annuli $2-3$ were unstable when fitting both
CO and \hi\ components.
Because the size of the regions are small, the low flux of the annuli $2-3$ 
components in comparison to those that are 
already-determined from fitting to the outer longitude ranges
means that the data are insufficiently constraining.
However, a convergent fit is obtained if the CO-related $\pi^0$-decay
templates is set to the \GP\ prediction.
The motivation for allowing the additional freedom to fit the spectrum for
the gas-related interstellar emission interior to the solar circle 
is solely to improve the fit residuals.
But, the harder index for the \hi\ and CO component 
when fitting the Pulsars IEM 
can be an indication that the assumption of a uniform CR source spectrum 
across the Galaxy is insufficient, or that the diffusive propagation
of CRs is non-uniform.

Generally, the fitting results can be interpreted as a reconfirmation 
that the CR gradient in the Galaxy is flatter than 
expected based on current knowledge of the Galactocentric radial 
distribution of CR sources, which has been known 
since the SAS-2 \citep{1977ApJ...217..843S}, 
COS-B \citep{1986A&A...154...25B,1988A&A...207....1S}, 
and EGRET \citep{1997ApJ...481..205H,2001ApJ...555...12D} all-sky surveys.
The explanation is not clear. 
\citet{1993A&A...267..372B} suggested that the radial distribution of 
CR sources derived from observations may be biased and their real 
distribution is flatter or the diffusion parameters derived from the local 
CR measurements are not the same throughout the Galaxy.
Solutions to this issue in terms of CR propagation phenomenology 
have been proposed:  
CR-driven Galactic winds and 
anisotropic diffusion \citep{2002A&A...385..216B}, or 
non-uniform diffusion coefficient 
that increases with Galactocentric radius and the distance from the 
Galactic plane \citep{2007APh....27..411S}.

The current analysis has focussed on finding IEMs to estimate the 
fore-/background toward the inner Galaxy.
The broader implications of our
scaled IEMs for the large-scale distribution of CRs in the Galaxy are 
deferred to future work.

\subsection{Point Sources}
\label{sec:point_source_discussion}

\begin{figure}[ht]
\includegraphics[scale=0.85]{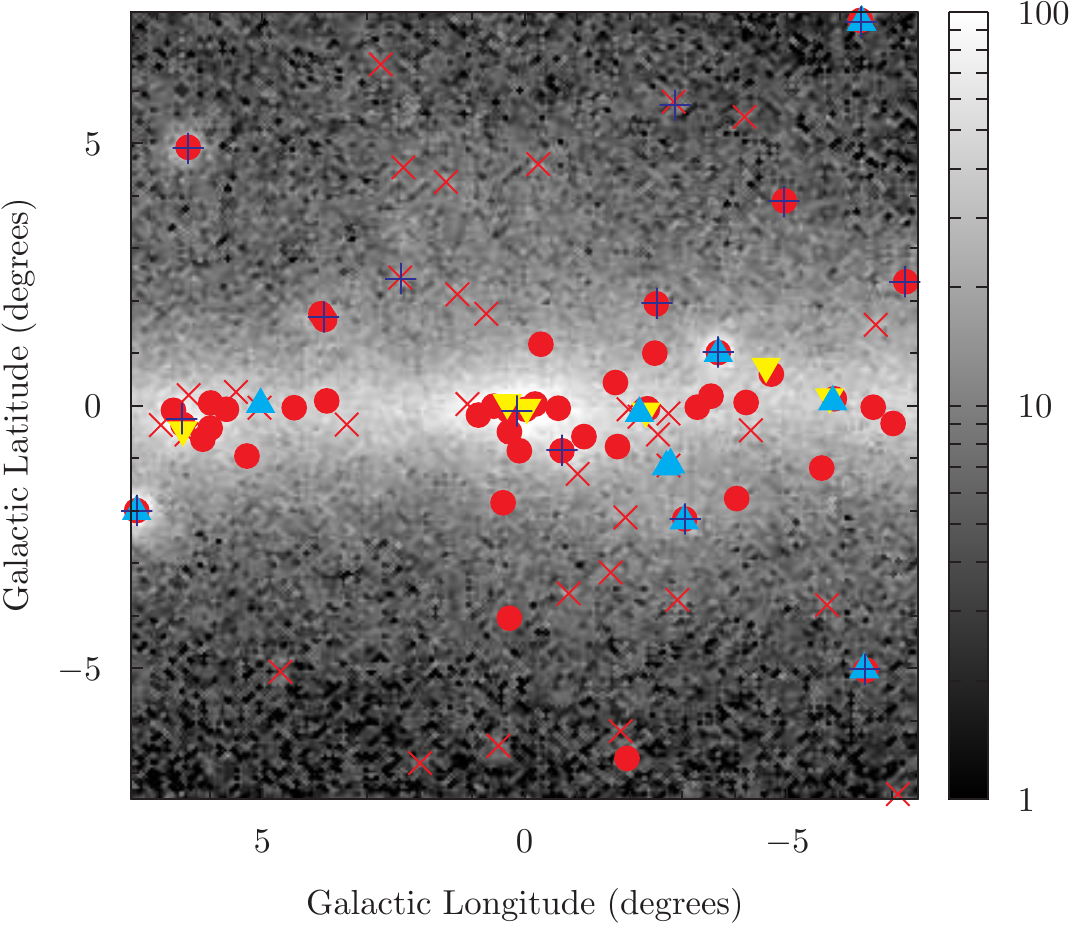}
\caption{
1FIG sources and source candidates and \fermilat\ data used in this paper with pulsars from the ATNF catalog \citep{2005AJ....129.1993M} and Green's SNR 
catalog \citep{2014BASI...42...47G} overlaid. 
Symbol key: filled circles, 1FIG sources; angled crosses, {\it TS} $<25$ source candidates; upright crosses, 3FGL multi-wavelength 
associated sources; inverted triangles, SNRs from Green's catalog with 
spatial overlap with a 1FIG source within its 95\% error ellipse; upright 
triangles, pulsars from the ATNF catalog within $3\times$ the 
1FIG source error ellipse.
Colour scale units: counts per $0.05^2$ deg pixel. 
\label{fig:1fig_pulsars_snrs_overlay}}
\end{figure}

Figure~\ref{fig:1fig_pulsars_snrs_overlay} shows the 1FIG sources and source 
candidates overlaid on the \fermilat\ data used in this paper, and 3FGL 
multi-wavelength associated sources, together with SNRs from Green's SNR 
catalog\footnote{http://www.mrao.cam.ac.uk/surveys/snrs/} \citep{2014BASI...42...47G}
and pulsars from the ATNF catalog\footnote{http://www.atnf.csiro.au/people/pulsar/psrcat/} \citep{2005AJ....129.1993M}, respectively, that are within 
95\% of the 1FIG source/source candidate error ellipse.
The 3FGL sources that have likely counterparts at other wavelengths that 
are listed in the catalog {\it not} detected
in the 1FIG are either due to a too 
low {\it TS} (3FGL J1716.6-2812 -- NGC~6316), or are more than the 
95\% containment radius of the error ellipse from a potential 1FIG 
counterpart (3FGL~J1750.2-3704 -- Terzan~5 and 
3FGL~J1746.3-2851c -- PWN G0.13-0.11). 
 
There are 14 1FIG sources and source candidates with overlaps with  
the above mentioned SNR and pulsar catalogs.
Multiple overlaps occur across and within the catalogs, e.g., 
SNR~354.1+00.1 and PSR~J1701-3006A,D,E overlap with 1FIG~J1701.1-3004.

The 1FIG source J1801.4-2330 overlaps with SNR~006.5-00.4, which has been 
detected in the first LAT catalog of SNRs \citep{SNRcatref}.
1FIG~J1740.1-3057 overlaps with SNR~357.7-00.1 (MSH~17-39), and has 
been detected previously in \fermilat\ data \citep{2013ApJ...774...36C}.
The source 1FIG~J1745.5-2859 
overlaps with SNR~000.0+00.0 (Sgr~A~East), but this is 
in a strongly confused region and other counterparts may be possible.
The 3 other SNRs (SNR~354.1+00.1, SNR~355.4+00.7, and SNR~000.3+00.0 
corresponding to 1FIG~J1730.2-3351, 1FIG~J1731.3-3235, and 1FIG~J1746.4-2843, 
respectively) are new detections in high-energy \gray{s} at \fermi\ 
energies.
Follow-on studies are required to better characterise their spatial and 
spectral properties.

The comparison with the ATNF catalog yields 9~1FIG sources overlapping 
with known pulsars.
The 1FIG source J1750.2-3705 is the counterpart of the globular 
cluster NGC~6441, which
has been detected in high-energy \gray{s} \citep{2011ApJ...729...90T}.
Four of the remaining 8 overlap with nearby pulsars ($\lesssim 0.2$~kpc) and 
are listed in the LAT Second Catalog of Gamma-ray Pulsars\footnote{http://fermi.gsfc.nasa.gov/ssc/data/access/lat/2nd\_PSR\_catalog/} \citep{2013ApJS..208...17A}.
The remaining 4 sources have been identified previously and searches for 
pulsed emission have been made but with no detections \citep{2013ApJS..208...17A}.

Obviously the comparison made here between the 1FIG sources and \gray{} 
source classes is not exhaustive.
However, more than two thirds of the 1FIG sources do not have 
associations with sources in known classes of \gray{} emitters.
The unassociated 1FIG sources tend to be close to the Galactic plane.
It remains a significant 
possiblity that a majority of the point sources found 
over the $15^\circ \times 15^\circ$ region can be attributed to 
mis-identified interstellar emission, as already discussed in 
Sec.~\ref{results:point_sources}.

\subsection{Residuals}
\label{sec:residuals_discussion}

A number of studies of the residual emission toward the 
inner Galaxy have been performed, as described in Sec.~\ref{intro}.
Figure~\ref{fig:plec_overlay} compares the results from this 
analysis with selected results from the literature.
The comparison is useful because a similar spectral model to other 
authors is used when fitting the residual emission associated with the centrally
peaked spatial templates.
The IEMs that are used in this paper 
are scaled to the data outside of the $15^\circ \times 15^\circ$ region to 
reduce the discrepancies with the data, particularly along the Galactic 
plane, by the a-priori GALPROP-generated IEMs from \citet{2012ApJ...750....3A}. 
The developments also made as part of this work have allowed
additional degrees of freedom to be included for the scaling of the IEMs --  
treating the IC template as the sum of individual templates with the 
same Galactocentric radial boundaries as the $\pi^0$-decay templates, and 
allowing for spectral variations in the $\pi^0$-decay templates from those 
predicted interior to the solar circle -- that go beyond the modelling 
of the interstellar emission employed by other 
analyses \citep[e.g.,][]{2015JCAP...03..038C}.
In addition, a catalogue of point sources is derived for 
each IEM that is used in the analysis of the inner 
region about the GC.
The prescriptive method of determining the fore-/background 
interstellar emission, together with the self-consistent treatment of the
point sourcess\footnote{Note: this includes the sub-threshold 
point source candidates and those that satisfy the $TS > 25$ criterion for ``detection'' used for the 1FIG.} and interstellar emission 
for the inner $\sim$kpc about the 
GC allows the least biased estimate to-date to be made of the positive residual 
emission about the GC.
This work finds that for individual IEMs the spectral parameters for a 
spatial template that peaks at the GC, such as the NFW profile, can be 
relatively tightly constrained.
However, over all IEMs considered in this work the variation of, for example, 
the cut-off energy for an exponential power-law spectral model is much wider
than that for any individual model.

\begin{figure*}[htb]
\subfigure{
\includegraphics[scale=0.32]{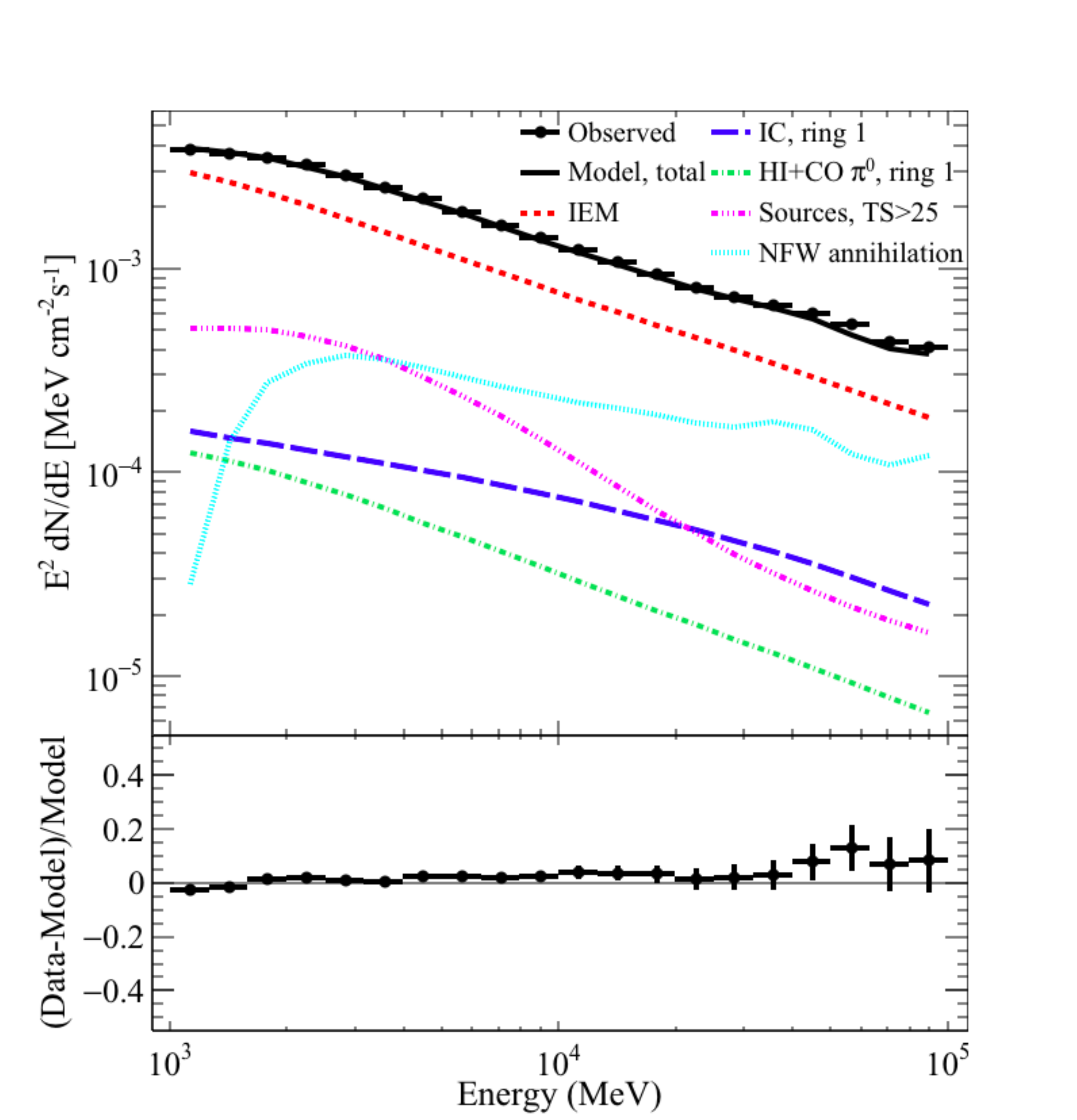}
\includegraphics[scale=0.32]{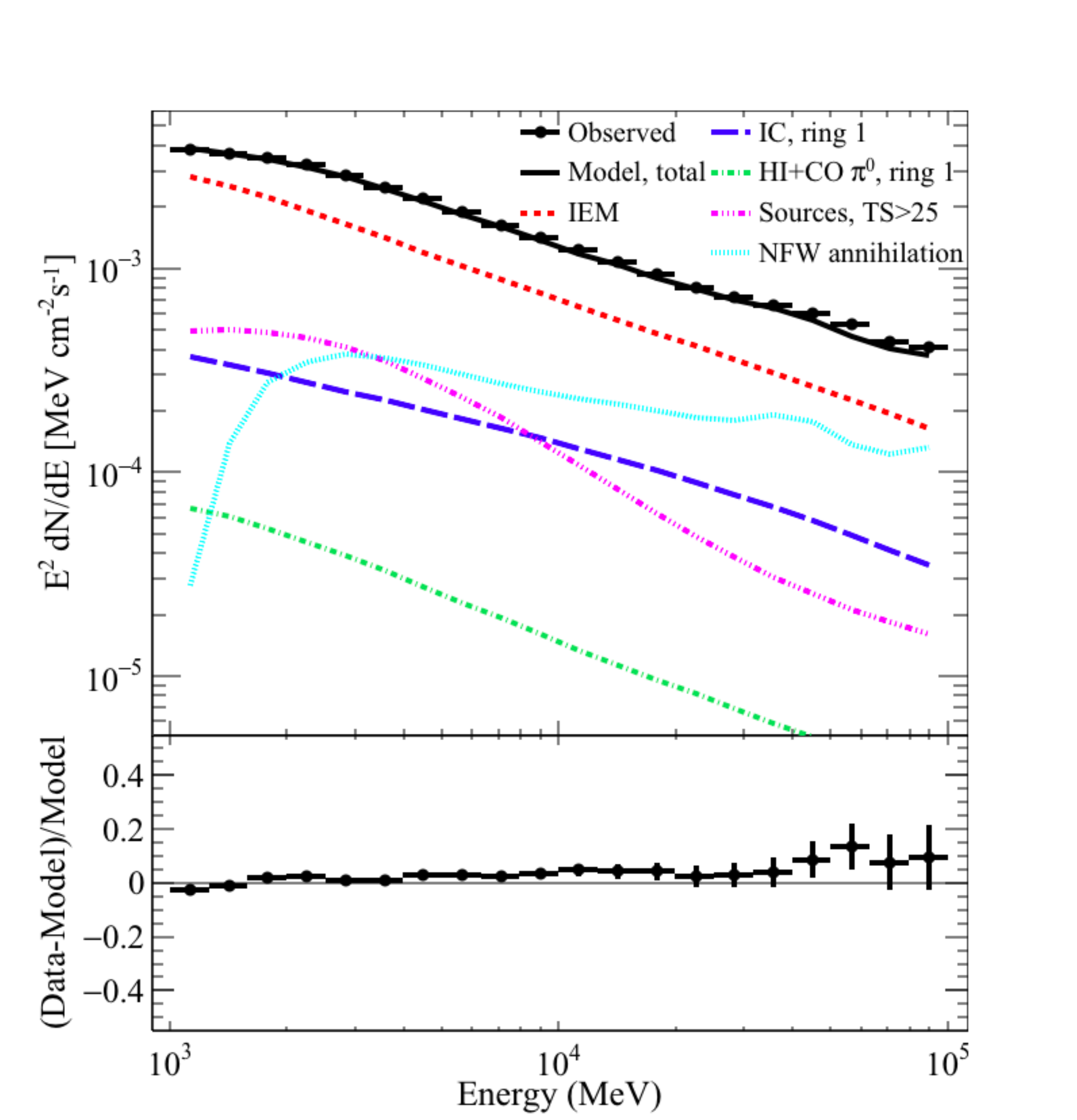}}
\subfigure{
\includegraphics[scale=0.32]{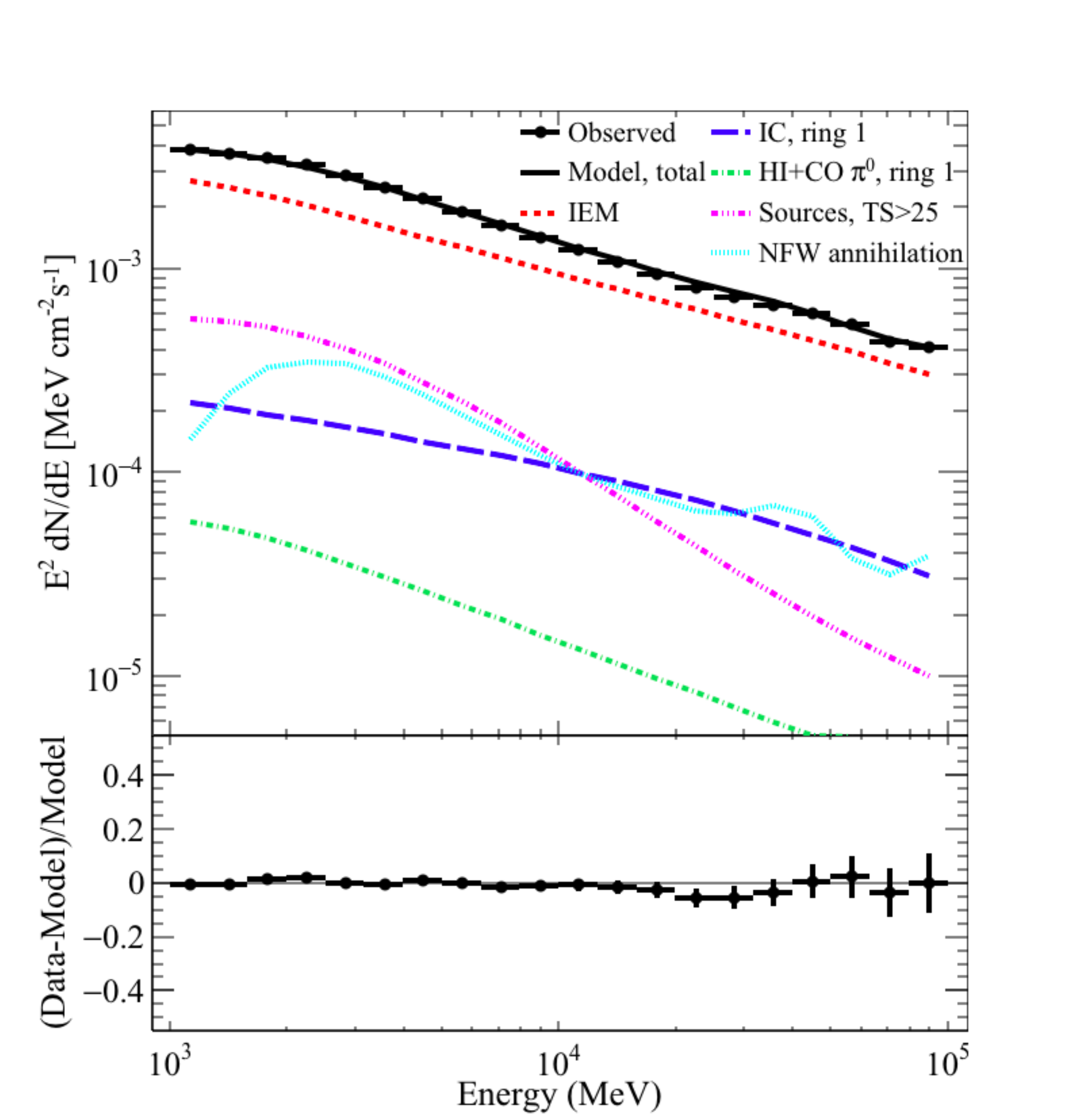}
\includegraphics[scale=0.32]{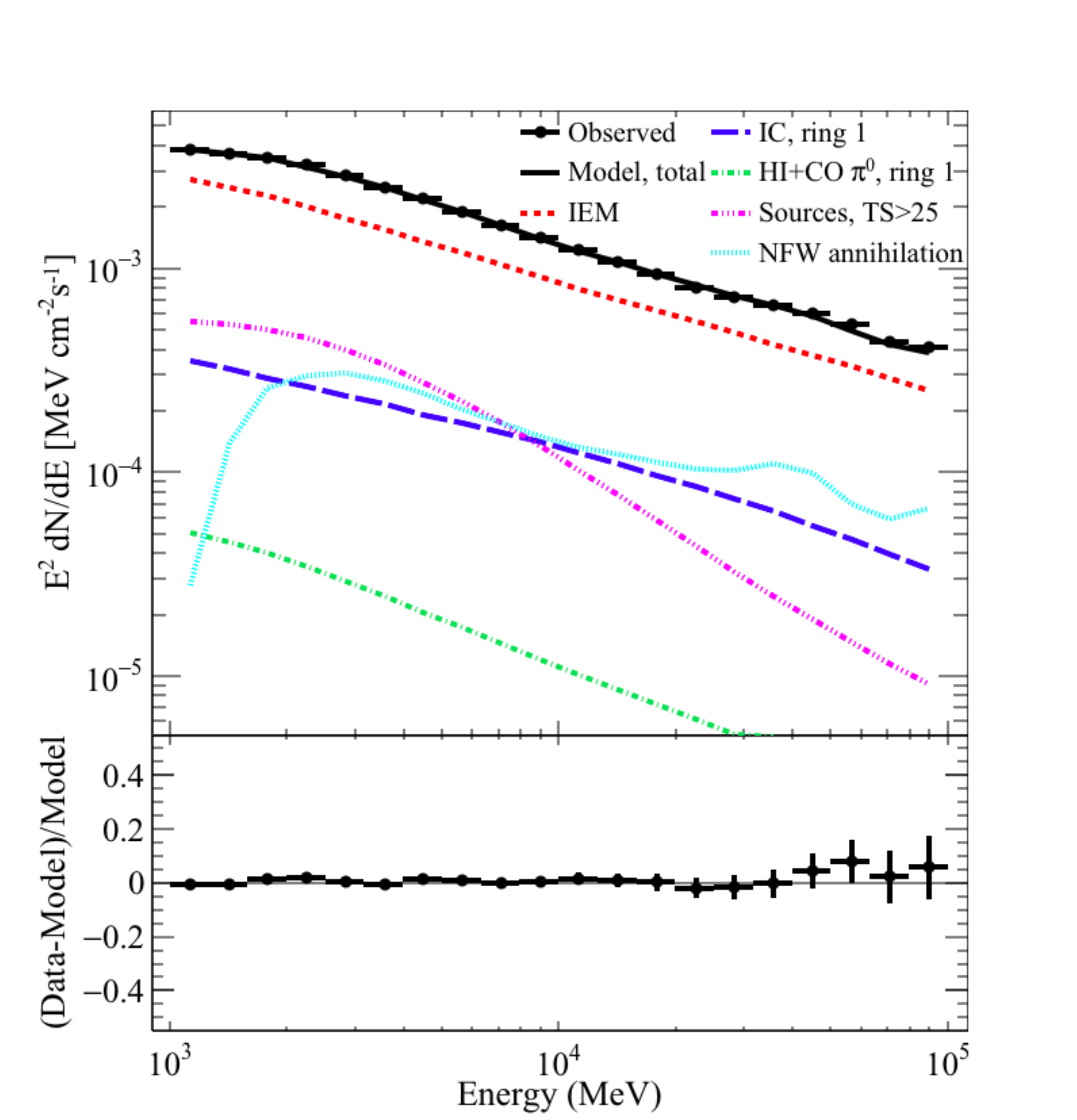}}
\caption{
Same as in Figure~\ref{results:nfw_fluxes}, but  with the spectrum of the NFW profile 
modeled with a power-law per energy band over the $1-100$~GeV range.
\label{results:mbpl_fluxes}}
\end{figure*}

\begin{figure}[ht]
\includegraphics[trim = 90 50 0 0,scale=0.30]{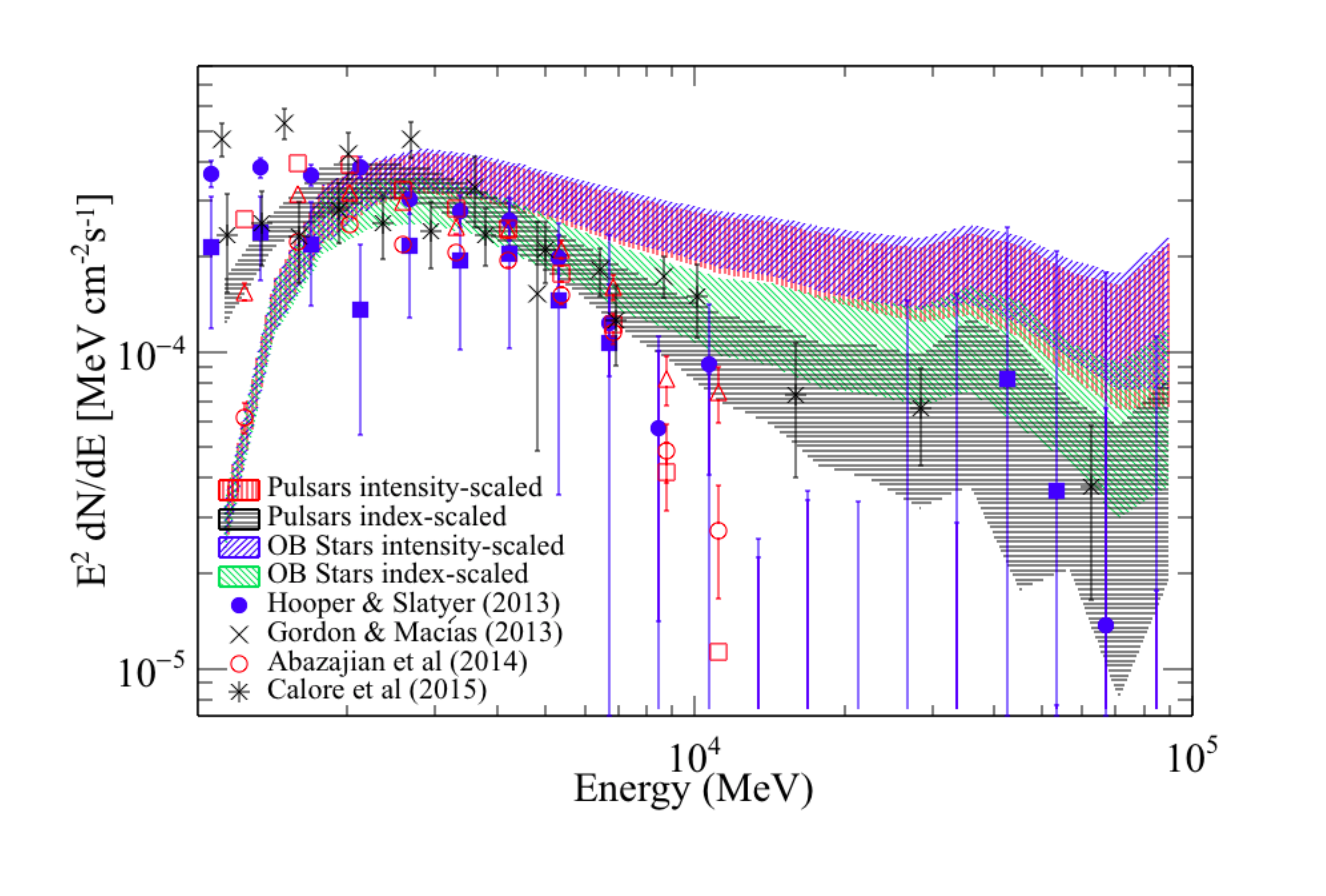}
\caption{
Same as in Figure~\ref{fig:plec_overlay}, but  with the spectrum of the NFW profile 
modeled with a power-law per energy band over the $1-100$~GeV range. The envelopes include the fit uncertainties for the normalisation and spectral indices.
\label{fig:mbpl_overlay}}
\end{figure}

Although the spectral residuals are generally 
improved by an additional template, discrepancies remain that are 
more pronounced for the intensity-scaled variants of the IEMs.
It should be emphasised that despite this observation the intensity-scaled 
IEMs cannot be 
excluded on the basis of fits made to the $15^\circ \times 15^\circ$ region about
the GC.
All four of the IEMs are tuned to data outside this relatively 
small region, providing similar improvements to the all-sky residuals, and 
hence are equivalent 
representations of the fore-/background toward and through the GC. 
Because of the limitations 
in modelling the interstellar emission, the higher energy cutoff spectra for 
the NFW profile component with the intensity-scaled IEMs cannot be ruled out.
With the limited freedom for the interstellar emission components 
(only the normalisation for the IC, \hi\ and CO-related $\pi^0$-decay 
intensity maps are allowed to vary -- 3 parameters) and in the specification
of the spectral model for the positive residual (normalisation, spectral 
index, and cut-off energy -- 3 parameters) the spread in the positive
residual template parameters is considerable (see above.) 
If more freedom is allowed for the spectrum of the positive residual then 
the spectral residuals for all four IEMs over the $15^\circ \times 15^\circ$ region 
can be very small.
Figures~\ref{results:mbpl_fluxes} and~\ref{fig:mbpl_overlay} show the 
results for all IEMs if more degrees of freedom are allowed to model the 
spectrum of the NFW profile\footnote{The spectral model is a 
power-law function per energy bin, with 10 bins
equally spaced in logarithmic energy over the $1-100$~GeV energy range. 
This  model is defined by a normalisation and spectral index in each bin, 
for a total of 11 parameters.} (note that the feature at $\sim$40~GeV 
 is not  significant when the fit uncertainties
are considered, as shown in Figure~\ref{fig:mbpl_overlay}.)
For this choice of spectral model indeed the residuals are very good for 
all IEMs.
It is therefore premature, because of the variations in the IEMs and their limitations,
to favor a specific IEM among those we considered 
and to attribute the high energy residual to a particular origin.

Although a large formal 
statistical significance may be indicated 
for the detection of a new component, note 
that fitting a centrally peaked profile does not account for all of
the positive residual over the $15^\circ \times 15^\circ$ region.
Ascribing a singular origin to such a residual component is premature given 
the limited constraints on the other emission components over the 
$15^\circ \times 15^\circ$ region.
A complete assessment of the uncertainties (see Sec.~\ref{sec:limitations}) 
is required to understand the nature of its spatial and spectral parameters.
The current work demonstrates that even in the optimistic 
scenario where the presence of a DM component in the data might be established 
based on the spatial distribution of the associated \gray{} emission, 
important information on the DM particle such as its mass and 
annihilation spectrum is strongly dependent on the IEM.
This was first demonstrated by~\cite{2015JCAP...05..011A} using preliminary 
results based on this work.

\subsection{Limitations of the Analysis}
\label{sec:limitations}

The IEMs used in this analysis are cylindrically axisymmetric averaging 
the CR densities and other details azimuthally about the GC.
Some shortcomings of these models were 
noted earlier (Sec.~\ref{sec:interstellaremission}), 
particularly when fitting for the large-scale interstellar emission 
interior to the solar circle\footnote{Toward the outer Galaxy the axisymmetric
models are likely a very good approximation because the long propagation times
for CRs to diffuse from regions with higher source density (inside the solar
circle) to the outer Galaxy mean that any spatial granularity 
is effectively washed out.}. 
The density of CR sources and other ISM components 
is highest inside the solar circle and 
is most likely radially and azimuthally dependent, e.g., being associated with 
spiral arms or the Galactic bar/bulge.
However, in the absence of detailed three-dimensional models for the 
interstellar 
gas, radiation field, and CR sources the axisymmetric models are the 
only viable method for estimating the fore-/background toward and through 
the line-of-sight to the GC.

The IEM fitting interior to the solar circle 
uses the tangent ranges for positive and negative longitudes to obtain 
parameters for the annuli~$2-4$ (Table~\ref{table:coefficients}).
To examine the effect of the azimuthal averaging, 
fits to the tangent ranges were made for positive and negative longitudes 
to gauge the difference in the parameters for the IEMs 
obtained when considering each separately.
The scaling factors for annulus~4 obtained when fitting negative and 
positive longitude ranges were statistically consistent
\footnote{The average statistical uncertainty for the normalisation
of each interstellar emission component per annulus is $\sim10$\%, except for 
annuli~2 and 3; see Appendix~\ref{appendix:IEM}.} 
with those found when fitting both ranges combined.
For annuli~2 and~3 the fits to the positive and negative
tangent longitude ranges result in scaling parameters that differ by 
factors up to $\sim 2$ from each other, which is well beyond the statistical
uncertainty; the average value obtained by fitting 
both tangent ranges together is approximately in-between for the 
intensity-scaled IEMs over annuli~2 and~3.
For the index-scaled IEMs the spectral parameters are harder or softer 
than the average when using the positive/negative tangent ranges individually 
for annuli~$2-4$.
However, there is no clear trend and the over/under-prediction is not
confined to a particular energy interval.

The uncertainty for the IEM fore-/background flux 
toward the GC due to the 
azimuthally averaged IEMs is difficult to quantify precisely.
A minimal estimate can be made from the statistical uncertainty for 
the annulus~4 $\pi^0$-decay flux for each IEM, because the
fit results for the combined tangent ranges are within these uncertainties 
when fitted to the positive and negative ranges individually.
Above 1~GeV this is $\sim 4 \times 10^{-8}$ ph~cm$^{-2}$~s$^{-1}$ for the 
$15^\circ \times 15^\circ$ region about the GC across all IEMs.
This is comparable to the fitted flux from annulus~1 $\pi^0$-decay or
the $TS < 25$ point sources over the same region.

Any analysis employing the Galactocentric annulus
decomposition for the gas column densities is subject to the loss of 
kinematic resolution for sight lines within $l \sim \pm12^\circ$ 
of the GC/anti-GC.
Appendix~B of \citet{2012ApJ...750....3A} details the transformation of 
\hi\ and CO gas-survey data into the column density distributions over 
Galactocentric annuli used in this analysis, and employed
by many others.
The assumptions made in the transformation for the site lines over the 
$15^\circ \times 15^\circ$ region about the GC have an impact on 
the interstellar emission and point sources in the 
maximum-likelihood fitting and consequently the spatial distribution of 
residuals.
Approximations made interpolating the gas column density across the 
$l \pm 10^\circ$ range can result in an incorrect gas density distribution 
along the line-of-sight.
Spurious point sources in the analysis and structure in residuals 
can result from this because a higher/lower CR intensity compared to 
where the gas should be placed is used in creating the 
interstellar emission templates. 
The scaling procedure for the IEM then adjusts the individual annuli 
potentially producing low-level artifacts due to a combination of 
the effects described above.

To obtain an estimate of the uncertainties associated with misplacement of 
the gas new maps of the column density per annuli are created. 
10\% of the \hi\ gas column density is randomly displaced over the annuli
and recombined with the $\pi^0$-decay emissivity
\footnote{The contribution by CO-related $\pi^0$-decay emission is the same 
as that obtained from the scaling procedure.} in each annulus to create
modified intensity maps for this process, which are summed to produce 
new fore-/background intensity maps.
The 68\% fractional change per pixel from 100 such realisations for each 
IEM is compared with the fore-/background resulting from the scaling 
procedure (Sec.~\ref{sec:interstellaremission}).
Depending on the IEM and energy range,  variations from 1\% to 15\% in the 
intensity per pixel for the fore-/background from the structured interstellar
emission across the $15^\circ \times 15^\circ$ region are obtained, with the 
largest for OBstars index-scaled and smallest for the Pulsar 
intensity-scaled IEM, respectively.
Because of the somewhat arbitrary choice of the precise fraction of 
\hi\ column density\footnote{Similar modifications of the CO column density 
distribution are not explored because the detailed knowledge to make a truly 
informed estimate is not available.} that is redistributed over the 
annuli these variations are illustrative rather than providing 
a true `systematic uncertainty' associated with 
the gas misplacement.
Note that the uncertainty is maximised toward the GC because it is furthest
away from the gas column density interpolation base points 
at $l \sim \pm12^\circ$.

\section{Summary}
\label{sec:summary}

The analysis described in this paper 
employs specialised IEMs that are fit to the 
\gray{} data without reference to the $15^\circ \times 15^\circ$ region about 
the GC.
Finding point-source seeds for the same region using a method that does 
not rely on detailed IEMs, the source-seeds and IEMs 
are combined in a maximum-likelihood fit to determine the 
interstellar emission across the inner $\sim 1$~kpc about the GC and 
point sources over the region.
The overwhelming majority of \gray{} emission from the 
$15^\circ \times 15^\circ$ region is due to interstellar emission and point
sources.
To summarise the results for these aspects of the analysis:

\begin{itemize}

\item{The interstellar emission over the $15^\circ \times 15^\circ$ region 
is $\sim 85$\% of the total.
For the case of fitting only `standard' interstellar emission 
processes and point sources the fore-/background is $\sim 80$\% with the 
remaining $\sim 20$\% mainly due to IC from the inner region.
The contribution by the $\pi^0$-decay process over the inner region is 
much less than the IC, with the relative contributions by the 
\hi- and CO-related emission suppressed compared to the \GP\ predictions.

With this scenario 
there are residual counts that are distributed with some 
general peak around the GC.
If a model for the positive residual with a spatial distribution
that peaks near the GC is simultaneously fit with interstellar 
emission and point sources, the IC flux is reduced 
by a factor $\sim 3$ and the \hi-related $\pi^0$-decay is increased by a 
factor $\sim 3$.
Even with the additional parameters introduced by a model for the
positive residual the IC flux is considerably enhanced compare to
those predicted by the baseline IEMs, and remains the dominant 
interstellar emission component over the inner region about the GC.
This indicates that 
the CR electron and/or ISRF intensities in the region are higher 
than those in the baseline IEMs.
}

\item{The total flux of point sources over the $15^\circ \times 15^\circ$ 
region is $15$\% of the total flux over the four IEMs, and this is stable 
whether or not a model for the residual counts is included in the fitting 
procedure. 
Only $\sim 20$\% of this is attributed to point sources with a 3FGL 
counterpart with a multi-wavelength association.
Approximately 60\% of the 1FIG sources have a 3FGL counterpart, with a good 
correlation between the 1FIG and 3FGL fluxes.
However, the spatial density of 1FIG sources is more
closely distributed near the Galactic plane than the 3FGL.

The 1FIG contains 11 out of the 14 3FGL sources with a multi-wavelength 
association but the parameters of the spectral model for each vary according to 
the IEM that is employed. 
The three sources with a multi-wavelength association in the 3FGL
not included in the 1FIG are either due to a {\it TS} below the detection 
threshold, or because they are more than the 
95\% containment radius of the error ellipse from a potential 1FIG 
counterpart.
In addition, sources listed in other \fermilat\ derived catalogs 
of SNRs and pulsars \citep[]{2013ApJS..208...17A,SNRcatref}, and other 
individual analyses are found.
Spatial overlaps for 3~1FIG sources with SNRs listed in 
Green's catalog \citep{2014BASI...42...47G} are obtained.
These are previously undetected, but further characterisation of their 
spatial and spectral properties in \gray{s} awaits more detailed follow-up
analyses.
}

\end{itemize}

A critical aspect of this analysis is the determination of point source 
localisations, fluxes, and spectral properties for each IEM.
Over the $15^\circ \times 15^\circ$ region spatial distribution of the source
density differs between the 1FIG and 3FGL. 
The IEMs constructed for the present analysis employ similar gas maps to the
3FGL IEM, but use a different fitting methodology.
The IEMs are not optimised to flatten residual features across
the $15^\circ \times 15^\circ$ region, and allow for more freedom to fitting
the IC intensity distribution.

Many of the 1FIG point sources lie close to the Galactic plane.
Although there is no definitive tracing of individual interstellar emission
components by these point sources, there is a possibility that a fraction of 
them are misattributed interstellar emission.
The spatial distribution of point sources and point-source candidates 
are essential pieces of information for understanding the contribution
of unresolved source populations across the region.
Better quantification of the misattributed fraction is necessary to 
determine the correct spatial distribution of point sources over the region.
However, this is beyond the scope of the current analysis.

The separation of the fore-/background interstellar emission employed 
in this work is not without its limitations.
But, the major ones that should be 
investigated for an improved analysis of the high-energy \gray{} emission from 
this region are described.

The residual flux in the $15^\circ \times 15^\circ$ region 
only becomes significant
with respect to the interstellar emission components and point sources 
for energies $\gtrsim2$~GeV.
If only interstellar emission and point sources are fit to the data 
the residual emission is weakly asymmetric about the GC, but the statistical
noise is large.
This may be suggestive of an excess in the data 
that is not symmetric with respect to the GC. 
However, the extended over-subtraction
and the paucity of point sources in the region around $l \sim$ $2^\circ$ are
indicative of mismodelling of the interstellar emission in that region, possibly
due to inadequacies in the treatment of interstellar gas along and through 
the line-of-sight toward the region.
Because of this uncertainty, it cannot presently be established if this 
feature is caused by an asymmetric excess in the data 
due to something other than standard astrophysical production mechanisms.
 
If a model for the positive residual that uses a spatial template that is 
centrally peaked toward the GC with an exponentially cut-off power law spectrum
is fit together with the interstellar emission and point sources, 
the spectral parameters are tightly constrained.
However, this analysis shows that the range of spectral parameters
using such a model for the positive residual 
is much wider when considering multiple IEMs.
Flat spectral residuals across the whole $1-100$~GeV energy range are only
obtained for an IEM with a significant modification to the spectra of the 
structured component of the interstellar emission fore-/background, compared
to a \GP\ model normalised to local CR spectra.
Otherwise, the spectral residuals are flat only up to $\sim 30$~GeV.

The analysis described in this paper 
contrasts with other works examining the \gray{} emission observed 
by the \fermilat\ toward the GC because multiple specialised 
IEMs are developed 
to estimate the fore-/background without reference to the data in 
the region of interest about the GC.
The self-consistent determination of point sources and point-source 
candidates using these specialised IEMs
is another element of the analysis that has previously not been employed.
After subtraction of interstellar emission and point sources, 
an extended residual is present. 
It can be fit with a centrally peaked profile with a specified spectral model,
but not all of the positive residual is accounted for by such a model.
Because of the uncertain nature of the properties of the 
positive residual due to the IEM and point source determination,
a precise physical interpretation of its origin is premature.

\acknowledgements

The \fermilat\ Collaboration acknowledges generous ongoing support from a 
number of agencies and institutes that have supported both the development 
and the operation of the LAT as well as scientific data analysis.  
These include the National Aeronautics and Space Administration and the 
Department of Energy in the United States, the Commissariat \`a l'Energie 
Atomique and the Centre National de la Recherche Scientifique / Institut 
National de Physique Nucl\'eaire et de Physique des Particules in France, the 
Agenzia Spaziale Italiana and the Istituto Nazionale di Fisica Nucleare in 
Italy, the Ministry of Education, Culture, Sports, Science and 
Technology (MEXT), High Energy Accelerator Research Organization (KEK) and 
Japan Aerospace Exploration Agency (JAXA) in Japan, and the K.~A.~Wallenberg 
Foundation, the Swedish Research Council and the Swedish National Space 
Board in Sweden.

Additional support for science analysis during the operations phase is 
gratefully acknowledged from the Istituto Nazionale di Astrofisica in 
Italy and the Centre National d'\'Etudes Spatiales in France.

GALPROP development is partially funded via NASA grants NNX09AC15G, 
NNX10AE78G, and NNX13AC47G.

Some of the results in this paper have been derived using the 
HEALPix~\citep{2005ApJ...622..759G} package.

\bibliographystyle{apj}
\bibliography{ms_1_gc}

\begin{thebibliography}{67}
\expandafter\ifx\csname natexlab\endcsname\relax\def\natexlab#1{#1}\fi

\bibitem[{{Abazajian} {et~al.}(2014){Abazajian}, {Canac}, {Horiuchi}, \&
  {Kaplinghat}}]{2014PhRvD..90b3526A}
{Abazajian}, K.~N., {Canac}, N., {Horiuchi}, S., \& {Kaplinghat}, M. 2014,
  \prd, 90, 023526

\bibitem[{{Abazajian} \& {Kaplinghat}(2012)}]{2012PhRvD..86h3511A}
{Abazajian}, K.~N., \& {Kaplinghat}, M. 2012, \prd, 86, 083511

\bibitem[{{Abdo} {et~al.}(2009{\natexlab{a}}){Abdo}, {Ackermann}, {Ajello},
  {Atwood}, {Axelsson}, {Baldini}, {Ballet}, {Barbiellini}, {Bastieri},
  {Baughman}, {Bechtol}, {Bellazzini}, {Berenji}, {Bloom}, {Bonamente},
  {Borgland}, {Bregeon}, {Brez}, {Brigida}, {Bruel}, {Burnett}, {Caliandro},
  {Cameron}, {Caraveo}, {Carlson}, {Casandjian}, {Cecchi}, {{\c C}elik},
  {Chekhtman}, {Cheung}, {Ciprini}, {Claus}, {Cohen-Tanugi}, {Conrad},
  {Cutini}, {Dermer}, {de Angelis}, {de Palma}, {Digel}, {Silva}, {Drell},
  {Dubois}, {Dumora}, {Farnier}, {Favuzzi}, {Fegan}, {Focke}, {Frailis},
  {Fukazawa}, {Funk}, {Fusco}, {Gargano}, {Gasparrini}, {Gehrels}, {Germani},
  {Giebels}, {Giglietto}, {Giordano}, {Glanzman}, {Godfrey}, {Grenier},
  {Grondin}, {Grove}, {Guillemot}, {Guiriec}, {Hanabata}, {Harding},
  {Hayashida}, {Hays}, {Hughes}, {J{\'o}hannesson}, {Johnson}, {Johnson},
  {Johnson}, {Kamae}, {Katagiri}, {Kawai}, {Kerr}, {Kn{\"o}dlseder}, {Kocian},
  {Kuehn}, {Kuss}, {Lande}, {Latronico}, {Lemoine-Goumard}, {Longo}, {Loparco},
  {Lott}, {Lovellette}, {Lubrano}, {Makeev}, {Mazziotta}, {McEnery}, {Meurer},
  {Michelson}, {Mitthumsiri}, {Mizuno}, {Moiseev}, {Monte}, {Monzani},
  {Morselli}, {Moskalenko}, {Murgia}, {Nolan}, {Norris}, {Nuss}, {Ohsugi},
  {Okumura}, {Omodei}, {Orlando}, {Ormes}, {Ozaki}, {Paneque}, {Panetta},
  {Parent}, {Pepe}, {Pesce-Rollins}, {Piron}, {Pohl}, {Porter}, {Rain{\`o}},
  {Rando}, {Razzano}, {Reimer}, {Reimer}, {Reposeur}, {Ritz}, {Rochester},
  {Rodriguez}, {Ryde}, {Sadrozinski}, {Sanchez}, {Sander}, {Saz Parkinson},
  {Schalk}, {Sellerholm}, {Sgr{\`o}}, {Smith}, {Smith}, {Spandre}, {Spinelli},
  {Starck}, {Stecker}, {Strickman}, {Strong}, {Suson}, {Tajima}, {Takahashi},
  {Takahashi}, {Tanaka}, {Thayer}, {Thayer}, {Thompson}, {Tibaldo}, {Torres},
  {Tosti}, {Tramacere}, {Uchiyama}, {Usher}, {Vasileiou}, {Vilchez}, {Vitale},
  {Waite}, {Wang}, {Winer}, {Wood}, {Ylinen}, \&
  {Ziegler}}]{2009ApJ...703.1249A}
{Abdo}, A.~A., {et~al.} 2009{\natexlab{a}}, \apj, 703, 1249

\bibitem[{{Abdo} {et~al.}(2009{\natexlab{b}}){Abdo}, {Ackermann}, {Ajello},
  {Ampe}, {Anderson}, {Atwood}, {Axelsson}, {Bagagli}, {Baldini}, {Ballet}, \&
  et~al.}]{2009APh....32..193A}
---. 2009{\natexlab{b}}, Astroparticle Physics, 32, 193

\bibitem[{{Abdo} {et~al.}(2010{\natexlab{a}}){Abdo}, {Ackermann}, {Ajello},
  {Allafort}, {Antolini}, {Atwood}, {Axelsson}, {Baldini}, {Ballet},
  {Barbiellini}, \& et~al.}]{2010ApJS..188..405A}
---. 2010{\natexlab{a}}, \apjs, 188, 405

\bibitem[{{Abdo} {et~al.}(2010{\natexlab{b}}){Abdo}, {Ackermann}, {Ajello},
  {Allafort}, {Baldini}, {Ballet}, {Barbiellini}, {Bastieri}, {Bechtol},
  {Bellazzini}, {Berenji}, {Blandford}, {Bloom}, {Bonamente}, {Borgland},
  {Bouvier}, {Brandt}, {Bregeon}, {Brigida}, {Bruel}, {Buehler}, {Buson},
  {Caliandro}, {Cameron}, {Caraveo}, {Carrigan}, {Casandjian}, {Cecchi}, {{\c
  C}elik}, {Chekhtman}, {Chiang}, {Ciprini}, {Claus}, {Cohen-Tanugi}, {Conrad},
  {Dermer}, {de Palma}, {Silva}, {Drell}, {Dubois}, {Dumora}, {Farnier},
  {Favuzzi}, {Fegan}, {Fukazawa}, {Fukui}, {Funk}, {Fusco}, {Gargano},
  {Gehrels}, {Germani}, {Giglietto}, {Giordano}, {Glanzman}, {Godfrey},
  {Grenier}, {Grove}, {Guiriec}, {Hadasch}, {Hanabata}, {Harding}, {Hays},
  {Horan}, {Hughes}, {J{\'o}hannesson}, {Johnson}, {Johnson}, {Kamae},
  {Katagiri}, {Kataoka}, {Kn{\"o}dlseder}, {Kuss}, {Lande}, {Latronico}, {Lee},
  {Lemoine-Goumard}, {Llena Garde}, {Longo}, {Loparco}, {Lovellette},
  {Lubrano}, {Makeev}, {Mazziotta}, {Michelson}, {Mitthumsiri}, {Mizuno},
  {Moiseev}, {Monte}, {Monzani}, {Morselli}, {Moskalenko}, {Murgia},
  {Nakamori}, {Nolan}, {Norris}, {Nuss}, {Ohno}, {Ohsugi}, {Omodei}, {Orlando},
  {Ormes}, {Ozaki}, {Panetta}, {Parent}, {Pelassa}, {Pepe}, {Pesce-Rollins},
  {Piron}, {Porter}, {Rain{\`o}}, {Rando}, {Razzano}, {Reimer}, {Reimer},
  {Reposeur}, {Rodriguez}, {Roth}, {Sadrozinski}, {Sander}, {Saz Parkinson},
  {Sgr{\`o}}, {Siskind}, {Smith}, {Smith}, {Spandre}, {Spinelli}, {Strickman},
  {Suson}, {Tajima}, {Takahashi}, {Takahashi}, {Tanaka}, {Thayer}, {Thayer},
  {Thompson}, {Tibaldo}, {Tibolla}, {Torres}, {Tosti}, {Uchiyama}, {Uehara},
  {Usher}, {Vasileiou}, {Vilchez}, {Vitale}, {Waite}, {Wang}, {Winer}, {Wood},
  {Yamamoto}, {Yamazaki}, {Yang}, {Ylinen}, \& {Ziegler}}]{2010ApJ...718..348A}
---. 2010{\natexlab{b}}, \apj, 718, 348

\bibitem[{{Abdo} {et~al.}(2010{\natexlab{c}}){Abdo}, {Ackermann}, {Ajello},
  {Baldini}, {Ballet}, {Barbiellini}, {Bastieri}, {Baughman}, {Bechtol},
  {Bellazzini}, {Berenji}, {Bloom}, {Bonamente}, {Borgland}, {Bregeon}, {Brez},
  {Brigida}, {Bruel}, {Burnett}, {Buson}, {Caliandro}, {Cameron}, {Caraveo},
  {Casandjian}, {Cecchi}, {{\c C}elik}, {Chekhtman}, {Cheung}, {Chiang},
  {Ciprini}, {Claus}, {Cohen-Tanugi}, {Cominsky}, {Conrad}, {Dermer}, {de
  Palma}, {Digel}, {Silva}, {Drell}, {Dubois}, {Dumora}, {Farnier}, {Favuzzi},
  {Fegan}, {Focke}, {Fortin}, {Frailis}, {Fukazawa}, {Funk}, {Fusco},
  {Gargano}, {Gehrels}, {Germani}, {Giavitto}, {Giebels}, {Giglietto},
  {Giordano}, {Glanzman}, {Godfrey}, {Grenier}, {Grondin}, {Grove},
  {Guillemot}, {Guiriec}, {Harding}, {Hayashida}, {Horan}, {Hughes}, {Jackson},
  {J{\'o}hannesson}, {Johnson}, {Johnson}, {Kamae}, {Katagiri}, {Kataoka},
  {Kawai}, {Kerr}, {Kn{\"o}dlseder}, {Kuss}, {Lande}, {Latronico},
  {Lemoine-Goumard}, {Longo}, {Loparco}, {Lott}, {Lovellette}, {Lubrano},
  {Makeev}, {Mazziotta}, {McEnery}, {Meurer}, {Michelson}, {Mitthumsiri},
  {Mizuno}, {Monte}, {Monzani}, {Morselli}, {Moskalenko}, {Murgia}, {Nolan},
  {Norris}, {Nuss}, {Ohsugi}, {Okumura}, {Omodei}, {Orlando}, {Ormes},
  {Paneque}, {Pelassa}, {Pepe}, {Pesce-Rollins}, {Piron}, {Porter},
  {Rain{\`o}}, {Rando}, {Razzano}, {Reimer}, {Reimer}, {Reposeur}, {Rodriguez},
  {Ryde}, {Sadrozinski}, {Sanchez}, {Sander}, {Saz Parkinson}, {Sgr{\`o}},
  {Siskind}, {Smith}, {Spandre}, {Spinelli}, {Starck}, {Strickman}, {Strong},
  {Suson}, {Takahashi}, {Tanaka}, {Thayer}, {Thayer}, {Thompson}, {Tibaldo},
  {Torres}, {Tosti}, {Tramacere}, {Uchiyama}, {Usher}, {Vasileiou}, {Vilchez},
  {Vitale}, {Waite}, {Wang}, {Winer}, {Wood}, {Ylinen}, {Ziegler}, \&
  {Fermi/LAT Collaboration}}]{2010ApJ...710..133A}
---. 2010{\natexlab{c}}, \apj, 710, 133

\bibitem[{{Abdo} {et~al.}(2011){Abdo}, {Ackermann}, {Ajello}, {Baldini},
  {Ballet}, {Barbiellini}, {Bastieri}, {Bechtol}, {Bellazzini}, {Berenji},
  {Bonamente}, {Borgland}, {Bouvier}, {Bregeon}, {Brez}, {Brigida}, {Bruel},
  {Buehler}, {Buson}, {Caliandro}, {Cameron}, {Caraveo}, {Casandjian},
  {Cecchi}, {Charles}, {Chekhtman}, {Chiang}, {Ciprini}, {Claus},
  {Cohen-Tanugi}, {Conrad}, {Cutini}, {de Angelis}, {de Palma}, {Dermer},
  {Digel}, {Silva}, {Drell}, {Dubois}, {Favuzzi}, {Fegan}, {Focke}, {Fortin},
  {Frailis}, {Funk}, {Fusco}, {Gargano}, {Gasparrini}, {Gehrels}, {Germani},
  {Giglietto}, {Giordano}, {Giroletti}, {Glanzman}, {Godfrey}, {Grenier},
  {Grillo}, {Guiriec}, {Hadasch}, {Hays}, {Hughes}, {Iafrate},
  {J{\'o}hannesson}, {Johnson}, {Johnson}, {Kamae}, {Katagiri}, {Kataoka},
  {Kn{\"o}dlseder}, {Kuss}, {Lande}, {Latronico}, {Lee}, {Lionetto}, {Longo},
  {Loparco}, {Lott}, {Lovellette}, {Lubrano}, {Makeev}, {Mazziotta}, {McEnery},
  {Mehault}, {Michelson}, {Mitthumsiri}, {Mizuno}, {Moiseev}, {Monte},
  {Monzani}, {Morselli}, {Moskalenko}, {Murgia}, {Nakamori}, {Naumann-Godo},
  {Nolan}, {Norris}, {Nuss}, {Ohsugi}, {Okumura}, {Omodei}, {Orlando}, {Ormes},
  {Ozaki}, {Paneque}, {Pelassa}, {Pesce-Rollins}, {Pierbattista}, {Piron},
  {Porter}, {Rain{\`o}}, {Rando}, {Razzano}, {Reimer}, {Reimer}, {Reposeur},
  {Ritz}, {Sadrozinski}, {Schalk}, {Sgr{\`o}}, {Share}, {Siskind}, {Smith},
  {Spandre}, {Spinelli}, {Strickman}, {Strong}, {Takahashi}, {Tanaka},
  {Thayer}, {Thayer}, {Thompson}, {Tibaldo}, {Torres}, {Tosti}, {Tramacere},
  {Troja}, {Uchiyama}, {Usher}, {Vandenbroucke}, {Vasileiou}, {Vianello},
  {Vilchez}, {Vitale}, {Vladimirov}, {Waite}, {Wang}, {Winer}, {Wood}, {Yang},
  \& {Ziegler}}]{2011ApJ...734..116A}
---. 2011, \apj, 734, 116

\bibitem[{{Abdo} {et~al.}(2012){Abdo}, {Ackermann}, {Ajello}, {Atwoo},
  {Baldini}, {Ballet}, {Barbiellini}, {Bastieri}, {Bechtol}, {Bellazzini},
  {Berenji}, {Blandford}, {Bonamente}, {Borgland}, {Bottacini}, {Bouvier},
  {Bregeon}, {Brigida}, {Bruel}, {Buehler}, {Buson}, {Caliandro}, {Cameron},
  {Caraveo}, {Casandjian}, {Cecchi}, {Charles}, {Chekhtman}, {Chiang},
  {Ciprini}, {Claus}, {Cohen-Tanugi}, {Conrad}, {Cutini}, {D'Ammando}, {de
  Angelis}, {de Palma}, {Dermer}, {Digel}, {Silva}, {Drell}, {Drlica-Wagner},
  {Dubois}, {Favuzzi}, {Fegan}, {Focke}, {Fortin}, {Fukazawa}, {Funk}, {Fusco},
  {Gargano}, {Gehrels}, {Germani}, {Giglietto}, {Giommi}, {Giordano},
  {Giroletti}, {Glanzman}, {Godfrey}, {Gomez-Vargas}, {Grenier}, {Grove},
  {Guiriec}, {Hadasch}, {Hays}, {Hill}, {Horan}, {Hou}, {Hughes}, {Iafrate},
  {Jackson}, {J{\'o}hannesson}, {Johnson}, {Kamae}, {Katagiri}, {Kataoka},
  {Kn{\"o}dlseder}, {Kuss}, {Lande}, {Larsson}, {Latronico}, {Lemoine-Goumard},
  {Longo}, {Loparco}, {Lott}, {Lovellette}, {Lubrano}, {Mazziotta}, {McEnery},
  {Mehault}, {Michelson}, {Mitthumsiri}, {Mizuno}, {Moiseev}, {Monte},
  {Monzani}, {Morselli}, {Moskalenko}, {Murgia}, {Naumann-Godo}, {Nolan},
  {Norris}, {Nuss}, {Ohno}, {Ohsugi}, {Okumura}, {Omodei}, {Orienti},
  {Orlando}, {Ormes}, {Ozaki}, {Paneque}, {Panetta}, {Parent}, {Pesce-Rollins},
  {Pierbattista}, {Piron}, {Pivato}, {Poon}, {Porter}, {Prokhorov},
  {Rain{\`o}}, {Rando}, {Razzano}, {Razzaque}, {Reimer}, {Reimer}, {Reposeur},
  {Rochester}, {Roth}, {Sadrozinski}, {Sanchez}, {Sbarra}, {Schalk},
  {Sgr{\`o}}, {Share}, {Siskind}, {Spandre}, {Spinelli}, {Stawarz},
  {Takahashi}, {Tanaka}, {Thayer}, {Thayer}, {Thompson}, {Tibaldo},
  {Tinivella}, {Torres}, {Tosti}, {Troja}, {Uchiyama}, {Usher},
  {Vandenbroucke}, {Vasileiou}, {Vianello}, {Vitale}, {Waite}, {Wang}, {Winer},
  {Wood}, {Wood}, {Yang}, \& {Zimmer}}]{2012ApJ...758..140A}
---. 2012, \apj, 758, 140

\bibitem[{{Abdo} {et~al.}(2013){Abdo}, {Ajello}, {Allafort}, {Baldini},
  {Ballet}, {Barbiellini}, {Baring}, {Bastieri}, {Belfiore}, {Bellazzini}, \&
  et~al.}]{2013ApJS..208...17A}
---. 2013, \apjs, 208, 17

\bibitem[{{Acero} {et~al.}(2015{\natexlab{a}})}]{3fgl_iem_paper}
{Acero}, F., {et~al.} 2015{\natexlab{a}}, Submitted

\bibitem[{{Acero} {et~al.}(2015{\natexlab{b}}){Acero}, {Ackermann}, {Ajello},
  {Albert}, {Atwood}, {Axelsson}, {Baldini}, {Ballet}, {Barbiellini},
  {Bastieri}, {Belfiore}, {Bellazzini}, {Bissaldi}, {Blandford}, {Bloom},
  {Bogart}, {Bonino}, {Bottacini}, {Bregeon}, {Britto}, {Bruel}, {Buehler},
  {Burnett}, {Buson}, {Caliandro}, {Cameron}, {Caputo}, {Caragiulo}, {Caraveo},
  {Casandjian}, {Cavazzuti}, {Charles}, {Chaves}, {Chekhtman}, {Cheung},
  {Chiang}, {Chiaro}, {Ciprini}, {Claus}, {Cohen-Tanugi}, {Cominsky}, {Conrad},
  {Cutini}, {D'Ammando}, {de Angelis}, {DeKlotz}, {de Palma}, {Desiante},
  {Digel}, {Di Venere}, {Drell}, {Dubois}, {Dumora}, {Favuzzi}, {Fegan},
  {Ferrara}, {Finke}, {Franckowiak}, {Fukazawa}, {Funk}, {Fusco}, {Gargano},
  {Gasparrini}, {Giebels}, {Giglietto}, {Giommi}, {Giordano}, {Giroletti},
  {Glanzman}, {Godfrey}, {Grenier}, {Grondin}, {Grove}, {Guillemot}, {Guiriec},
  {Hadasch}, {Harding}, {Hays}, {Hewitt}, {Hill}, {Horan}, {Iafrate}, {Jogler},
  {J{\'o}hannesson}, {Johnson}, {Johnson}, {Johnson}, {Johnson}, {Kamae},
  {Kataoka}, {Katsuta}, {Kuss}, {La Mura}, {Landriu}, {Larsson}, {Latronico},
  {Lemoine-Goumard}, {Li}, {Li}, {Longo}, {Loparco}, {Lott}, {Lovellette},
  {Lubrano}, {Madejski}, {Massaro}, {Mayer}, {Mazziotta}, {McEnery},
  {Michelson}, {Mirabal}, {Mizuno}, {Moiseev}, {Mongelli}, {Monzani},
  {Morselli}, {Moskalenko}, {Murgia}, {Nuss}, {Ohno}, {Ohsugi}, {Omodei},
  {Orienti}, {Orlando}, {Ormes}, {Paneque}, {Panetta}, {Perkins},
  {Pesce-Rollins}, {Piron}, {Pivato}, {Porter}, {Racusin}, {Rando}, {Razzano},
  {Razzaque}, {Reimer}, {Reimer}, {Reposeur}, {Rochester}, {Romani},
  {Salvetti}, {S{\'a}nchez-Conde}, {Saz Parkinson}, {Schulz}, {Siskind},
  {Smith}, {Spada}, {Spandre}, {Spinelli}, {Stephens}, {Strong}, {Suson},
  {Takahashi}, {Takahashi}, {Tanaka}, {Thayer}, {Thayer}, {Thompson},
  {Tibaldo}, {Tibolla}, {Torres}, {Torresi}, {Tosti}, {Troja}, {Van Klaveren},
  {Vianello}, {Winer}, {Wood}, {Wood}, \& {Zimmer}}]{2015ApJS..218...23A}
---. 2015{\natexlab{b}}, \apjs, 218, 23

\bibitem[{{Ackermann} {et~al.}(2011){Ackermann}, {Ajello}, {Baldini}, {Ballet},
  {Barbiellini}, {Bastieri}, {Bechtol}, {Bellazzini}, {Berenji}, {Bloom},
  {Bonamente}, {Borgland}, {Brandt}, {Bregeon}, {Brez}, {Brigida}, {Bruel},
  {Buehler}, {Buson}, {Caliandro}, {Cameron}, {Caraveo}, {Casandjian},
  {Cecchi}, {Charles}, {Chekhtman}, {Chiang}, {Ciprini}, {Claus},
  {Cohen-Tanugi}, {Conrad}, {Dermer}, {de Palma}, {Digel}, {Drell}, {Dubois},
  {Favuzzi}, {Ferrara}, {Focke}, {Fukazawa}, {Funk}, {Fusco}, {Gargano},
  {Germani}, {Giglietto}, {Giordano}, {Giroletti}, {Glanzman}, {Godfrey},
  {Grenier}, {Guiriec}, {Hadasch}, {Hanabata}, {Harding}, {Hayashi},
  {Hayashida}, {Hughes}, {Itoh}, {J{\'o}hannesson}, {Johnson}, {Johnson},
  {Kamae}, {Katagiri}, {Kataoka}, {Kn{\"o}dlseder}, {Kuss}, {Lande},
  {Latronico}, {Lee}, {Llena Garde}, {Longo}, {Loparco}, {Lovellette},
  {Lubrano}, {Makeev}, {Martin}, {Mazziotta}, {McEnery}, {Mehault},
  {Michelson}, {Mizuno}, {Monte}, {Monzani}, {Morselli}, {Moskalenko},
  {Murgia}, {Naumann-Godo}, {Nishino}, {Nolan}, {Norris}, {Nuss}, {Ohsugi},
  {Okumura}, {Omodei}, {Orlando}, {Ormes}, {Ozaki}, {Parent}, {Pelassa},
  {Pepe}, {Pesce-Rollins}, {Piron}, {Porter}, {Rain{\`o}}, {Rando}, {Razzano},
  {Reimer}, {Reimer}, {Ripken}, {Sada}, {Sadrozinski}, {Sgr{\`o}}, {Siskind},
  {Spandre}, {Spinelli}, {Strickman}, {Strong}, {Suson}, {Takahashi},
  {Takahashi}, {Tanaka}, {Thayer}, {Thompson}, {Tibaldo}, {Torres},
  {Tramacere}, {Uchiyama}, {Uehara}, {Usher}, {Vandenbroucke}, {Vasileiou},
  {Vilchez}, {Vitale}, {Vladimirov}, {Waite}, {Wang}, {Wood}, {Yang}, \&
  {Ziegler}}]{2011ApJ...726...81A}
{Ackermann}, M., {et~al.} 2011, \apj, 726, 81

\bibitem[{{Ackermann} {et~al.}(2012{\natexlab{a}}){Ackermann}, {Ajello},
  {Atwood}, {Baldini}, {Ballet}, {Barbiellini}, {Bastieri}, {Bechtol},
  {Bellazzini}, {Berenji}, {Blandford}, {Bloom}, {Bonamente}, {Borgland},
  {Brandt}, {Bregeon}, {Brigida}, {Bruel}, {Buehler}, {Buson}, {Caliandro},
  {Cameron}, {Caraveo}, {Cavazzuti}, {Cecchi}, {Charles}, {Chekhtman},
  {Chiang}, {Ciprini}, {Claus}, {Cohen-Tanugi}, {Conrad}, {Cutini}, {de
  Angelis}, {de Palma}, {Dermer}, {Digel}, {Silva}, {Drell}, {Drlica-Wagner},
  {Falletti}, {Favuzzi}, {Fegan}, {Ferrara}, {Focke}, {Fortin}, {Fukazawa},
  {Funk}, {Fusco}, {Gaggero}, {Gargano}, {Germani}, {Giglietto}, {Giordano},
  {Giroletti}, {Glanzman}, {Godfrey}, {Grove}, {Guiriec}, {Gustafsson},
  {Hadasch}, {Hanabata}, {Harding}, {Hayashida}, {Hays}, {Horan}, {Hou},
  {Hughes}, {J{\'o}hannesson}, {Johnson}, {Johnson}, {Kamae}, {Katagiri},
  {Kataoka}, {Kn{\"o}dlseder}, {Kuss}, {Lande}, {Latronico}, {Lee},
  {Lemoine-Goumard}, {Longo}, {Loparco}, {Lott}, {Lovellette}, {Lubrano},
  {Mazziotta}, {McEnery}, {Michelson}, {Mitthumsiri}, {Mizuno}, {Monte},
  {Monzani}, {Morselli}, {Moskalenko}, {Murgia}, {Naumann-Godo}, {Norris},
  {Nuss}, {Ohsugi}, {Okumura}, {Omodei}, {Orlando}, {Ormes}, {Paneque},
  {Panetta}, {Parent}, {Pesce-Rollins}, {Pierbattista}, {Piron}, {Pivato},
  {Porter}, {Rain{\`o}}, {Rando}, {Razzano}, {Razzaque}, {Reimer}, {Reimer},
  {Sadrozinski}, {Sgr{\`o}}, {Siskind}, {Spandre}, {Spinelli}, {Strong},
  {Suson}, {Takahashi}, {Tanaka}, {Thayer}, {Thayer}, {Thompson}, {Tibaldo},
  {Tinivella}, {Torres}, {Tosti}, {Troja}, {Usher}, {Vandenbroucke},
  {Vasileiou}, {Vianello}, {Vitale}, {Waite}, {Wang}, {Winer}, {Wood}, {Wood},
  {Yang}, {Ziegler}, \& {Zimmer}}]{2012ApJ...750....3A}
---. 2012{\natexlab{a}}, \apj, 750, 3

\bibitem[{{Ackermann} {et~al.}(2012{\natexlab{b}}){Ackermann}, {Ajello},
  {Albert}, {Allafort}, {Atwood}, {Axelsson}, {Baldini}, {Ballet},
  {Barbiellini}, {Bastieri}, {Bechtol}, {Bellazzini}, {Bissaldi}, {Blandford},
  {Bloom}, {Bogart}, {Bonamente}, {Borgland}, {Bottacini}, {Bouvier}, {Brandt},
  {Bregeon}, {Brigida}, {Bruel}, {Buehler}, {Burnett}, {Buson}, {Caliandro},
  {Cameron}, {Caraveo}, {Casandjian}, {Cavazzuti}, {Cecchi}, {{\c C}elik},
  {Charles}, {Chaves}, {Chekhtman}, {Cheung}, {Chiang}, {Ciprini}, {Claus},
  {Cohen-Tanugi}, {Conrad}, {Corbet}, {Cutini}, {D'Ammando}, {Davis}, {de
  Angelis}, {DeKlotz}, {de Palma}, {Dermer}, {Digel}, {Silva}, {Drell},
  {Drlica-Wagner}, {Dubois}, {Favuzzi}, {Fegan}, {Ferrara}, {Focke}, {Fortin},
  {Fukazawa}, {Funk}, {Fusco}, {Gargano}, {Gasparrini}, {Gehrels}, {Giebels},
  {Giglietto}, {Giordano}, {Giroletti}, {Glanzman}, {Godfrey}, {Grenier},
  {Grove}, {Guiriec}, {Hadasch}, {Hayashida}, {Hays}, {Horan}, {Hou}, {Hughes},
  {Jackson}, {Jogler}, {J{\'o}hannesson}, {Johnson}, {Johnson}, {Johnson},
  {Kamae}, {Katagiri}, {Kataoka}, {Kerr}, {Kn{\"o}dlseder}, {Kuss}, {Lande},
  {Larsson}, {Latronico}, {Lavalley}, {Lemoine-Goumard}, {Longo}, {Loparco},
  {Lott}, {Lovellette}, {Lubrano}, {Mazziotta}, {McConville}, {McEnery},
  {Mehault}, {Michelson}, {Mitthumsiri}, {Mizuno}, {Moiseev}, {Monte},
  {Monzani}, {Morselli}, {Moskalenko}, {Murgia}, {Naumann-Godo}, {Nemmen},
  {Nishino}, {Norris}, {Nuss}, {Ohno}, {Ohsugi}, {Okumura}, {Omodei},
  {Orienti}, {Orlando}, {Ormes}, {Paneque}, {Panetta}, {Perkins},
  {Pesce-Rollins}, {Pierbattista}, {Piron}, {Pivato}, {Porter}, {Racusin},
  {Rain{\`o}}, {Rando}, {Razzano}, {Razzaque}, {Reimer}, {Reimer}, {Reposeur},
  {Reyes}, {Ritz}, {Rochester}, {Romoli}, {Roth}, {Sadrozinski}, {Sanchez},
  {Saz Parkinson}, {Sbarra}, {Scargle}, {Sgr{\`o}}, {Siegal-Gaskins},
  {Siskind}, {Spandre}, {Spinelli}, {Stephens}, {Suson}, {Tajima}, {Takahashi},
  {Tanaka}, {Thayer}, {Thayer}, {Thompson}, {Tibaldo}, {Tinivella}, {Tosti},
  {Troja}, {Usher}, {Vandenbroucke}, {Van Klaveren}, {Vasileiou}, {Vianello},
  {Vitale}, {Waite}, {Wallace}, {Winer}, {Wood}, {Wood}, {Wood}, {Yang}, \&
  {Zimmer}}]{2012ApJS..203....4A}
---. 2012{\natexlab{b}}, \apjs, 203, 4

\bibitem[{{Ackermann} {et~al.}(2014){Ackermann}, {Albert}, {Atwood}, {Baldini},
  {Ballet}, {Barbiellini}, {Bastieri}, {Bellazzini}, {Bissaldi}, {Blandford},
  {Bloom}, {Bottacini}, {Brandt}, {Bregeon}, {Bruel}, {Buehler}, {Buson},
  {Caliandro}, {Cameron}, {Caragiulo}, {Caraveo}, {Cavazzuti}, {Cecchi},
  {Charles}, {Chekhtman}, {Chiang}, {Chiaro}, {Ciprini}, {Claus},
  {Cohen-Tanugi}, {Conrad}, {Cutini}, {D'Ammando}, {de Angelis}, {de Palma},
  {Dermer}, {Digel}, {Di Venere}, {Silva}, {Drell}, {Favuzzi}, {Ferrara},
  {Focke}, {Franckowiak}, {Fukazawa}, {Funk}, {Fusco}, {Gargano}, {Gasparrini},
  {Germani}, {Giglietto}, {Giordano}, {Giroletti}, {Godfrey}, {Gomez-Vargas},
  {Grenier}, {Guiriec}, {Hadasch}, {Harding}, {Hays}, {Hewitt}, {Hou},
  {Jogler}, {J{\'o}hannesson}, {Johnson}, {Johnson}, {Kamae}, {Kataoka},
  {Kn{\"o}dlseder}, {Kocevski}, {Kuss}, {Larsson}, {Latronico}, {Longo},
  {Loparco}, {Lovellette}, {Lubrano}, {Malyshev}, {Manfreda}, {Massaro},
  {Mayer}, {Mazziotta}, {McEnery}, {Michelson}, {Mitthumsiri}, {Mizuno},
  {Monzani}, {Morselli}, {Moskalenko}, {Murgia}, {Nemmen}, {Nuss}, {Ohsugi},
  {Omodei}, {Orienti}, {Orlando}, {Ormes}, {Paneque}, {Panetta}, {Perkins},
  {Pesce-Rollins}, {Petrosian}, {Piron}, {Pivato}, {Rain{\`o}}, {Rando},
  {Razzano}, {Razzaque}, {Reimer}, {Reimer}, {S{\'a}nchez-Conde}, {Schaal},
  {Schulz}, {Sgr{\`o}}, {Siskind}, {Spandre}, {Spinelli}, {Stawarz}, {Strong},
  {Suson}, {Tahara}, {Takahashi}, {Thayer}, {Tibaldo}, {Tinivella}, {Torres},
  {Tosti}, {Troja}, {Uchiyama}, {Vianello}, {Werner}, {Winer}, {Wood}, {Wood},
  \& {Zaharijas}}]{2014ApJ...793...64A}
---. 2014, \apj, 793, 64

\bibitem[{{Ackermann} {et~al.}(2015)}]{SNRcatref}
---. 2015, In preparation

\bibitem[{{Agrawal} {et~al.}(2015){Agrawal}, {Batell}, {Fox}, \&
  {Harnik}}]{2015JCAP...05..011A}
{Agrawal}, P., {Batell}, B., {Fox}, P.~J., \& {Harnik}, R. 2015, \jcap, 5, 11

\bibitem[{{Aharonian} {et~al.}(2006{\natexlab{a}}){Aharonian}, {Akhperjanian},
  {Bazer-Bachi}, {Beilicke}, {Benbow}, {Berge}, {Bernl{\"o}hr}, {Boisson},
  {Bolz}, {Borrel}, {Braun}, {Breitling}, {Brown}, {Chadwick}, {Chounet},
  {Cornils}, {Costamante}, {Degrange}, {Dickinson}, {Djannati-Ata{\"i}},
  {Drury}, {Dubus}, {Emmanoulopoulos}, {Espigat}, {Feinstein}, {Fontaine},
  {Fuchs}, {Funk}, {Gallant}, {Giebels}, {Gillessen}, {Glicenstein}, {Goret},
  {Hadjichristidis}, {Hauser}, {Hauser}, {Heinzelmann}, {Henri}, {Hermann},
  {Hinton}, {Hofmann}, {Holleran}, {Horns}, {Jacholkowska}, {de Jager},
  {Kh{\'e}lifi}, {Klages}, {Komin}, {Konopelko}, {Latham}, {Le Gallou},
  {Lemi{\`e}re}, {Lemoine-Goumard}, {Leroy}, {Lohse}, {Marcowith}, {Martin},
  {Martineau-Huynh}, {Masterson}, {McComb}, {de Naurois}, {Nolan}, {Noutsos},
  {Orford}, {Osborne}, {Ouchrif}, {Panter}, {Pelletier}, {Pita},
  {P{\"u}hlhofer}, {Punch}, {Raubenheimer}, {Raue}, {Raux}, {Rayner}, {Reimer},
  {Reimer}, {Ripken}, {Rob}, {Rolland}, {Rowell}, {Sahakian}, {Saug{\'e}},
  {Schlenker}, {Schlickeiser}, {Schuster}, {Schwanke}, {Siewert}, {Sol},
  {Spangler}, {Steenkamp}, {Stegmann}, {Tavernet}, {Terrier}, {Th{\'e}oret},
  {Tluczykont}, {van Eldik}, {Vasileiadis}, {Venter}, {Vincent}, {V{\"o}lk}, \&
  {Wagner}}]{2006Natur.439..695A}
{Aharonian}, F., {et~al.} 2006{\natexlab{a}}, \nat, 439, 695

\bibitem[{{Aharonian} {et~al.}(2006{\natexlab{b}}){Aharonian}, {Akhperjanian},
  {Bazer-Bachi}, {Beilicke}, {Benbow}, {Berge}, {Bernl{\"o}hr}, {Boisson},
  {Bolz}, {Borrel}, {Braun}, {Breitling}, {Brown}, {Chadwick}, {Chounet},
  {Cornils}, {Costamante}, {Degrange}, {Dickinson}, {Djannati-Ata{\"i}},
  {Drury}, {Dubus}, {Emmanoulopoulos}, {Espigat}, {Feinstein}, {Fontaine},
  {Fuchs}, {Funk}, {Gallant}, {Giebels}, {Gillessen}, {Glicenstein}, {Goret},
  {Hadjichristidis}, {Hauser}, {Heinzelmann}, {Henri}, {Hermann}, {Hinton},
  {Hofmann}, {Holleran}, {Horns}, {Jacholkowska}, {de Jager}, {Kh{\'e}lifi},
  {Komin}, {Konopelko}, {Latham}, {Le Gallou}, {Lemi{\`e}re},
  {Lemoine-Goumard}, {Leroy}, {Lohse}, {Martin}, {Martineau-Huynh},
  {Marcowith}, {Masterson}, {McComb}, {de Naurois}, {Nolan}, {Noutsos},
  {Orford}, {Osborne}, {Ouchrif}, {Panter}, {Pelletier}, {Pita},
  {P{\"u}hlhofer}, {Punch}, {Raubenheimer}, {Raue}, {Raux}, {Rayner}, {Reimer},
  {Reimer}, {Ripken}, {Rob}, {Rolland}, {Rowell}, {Sahakian}, {Saug{\'e}},
  {Schlenker}, {Schlickeiser}, {Schuster}, {Schwanke}, {Siewert}, {Sol},
  {Spangler}, {Steenkamp}, {Stegmann}, {Tavernet}, {Terrier}, {Th{\'e}oret},
  {Tluczykont}, {Vasileiadis}, {Venter}, {Vincent}, {V{\"o}lk}, \&
  {Wagner}}]{2006ApJ...636..777A}
---. 2006{\natexlab{b}}, \apj, 636, 777

\bibitem[{Atwood {et~al.}(2009)}]{Atwood:2009}
Atwood, W.~B., {et~al.} 2009, \apj, 697, 1071

\bibitem[{{Bergstr{\"o}m}(2000)}]{2000RPPh...63..793B}
{Bergstr{\"o}m}, L. 2000, Reports on Progress in Physics, 63, 793

\bibitem[{{Bloemen} {et~al.}(1993){Bloemen}, {Dogiel}, {Dorman}, \&
  {Ptuskin}}]{1993A&A...267..372B}
{Bloemen}, J.~B.~G.~M., {Dogiel}, V.~A., {Dorman}, V.~L., \& {Ptuskin}, V.~S.
  1993, \aap, 267, 372

\bibitem[{{Bloemen} {et~al.}(1986){Bloemen}, {Strong}, {Mayer-Hasselwander},
  {Blitz}, {Cohen}, {Dame}, {Grabelsky}, {Thaddeus}, {Hermsen}, \&
  {Lebrun}}]{1986A&A...154...25B}
{Bloemen}, J.~B.~G.~M., {et~al.} 1986, \aap, 154, 25

\bibitem[{{Bouchet} {et~al.}(2011){Bouchet}, {Strong}, {Porter}, {Moskalenko},
  {Jourdain}, \& {Roques}}]{2011ApJ...739...29B}
{Bouchet}, L., {et~al.} 2011, \apj, 739, 29

\bibitem[{{Breitschwerdt} {et~al.}(2002){Breitschwerdt}, {Dogiel}, \&
  {V{\"o}lk}}]{2002A&A...385..216B}
{Breitschwerdt}, D., {Dogiel}, V.~A., \& {V{\"o}lk}, H.~J. 2002, \aap, 385, 216

\bibitem[{{Bronfman} {et~al.}(2000){Bronfman}, {Casassus}, {May}, \&
  {Nyman}}]{2000A&A...358..521B}
{Bronfman}, L., {Casassus}, S., {May}, J., \& {Nyman}, L.-{\AA}. 2000, \aap,
  358, 521

\bibitem[{{Calore} {et~al.}(2015{\natexlab{a}}){Calore}, {Cholis}, {McCabe}, \&
  {Weniger}}]{2015PhRvD..91f3003C}
{Calore}, F., {Cholis}, I., {McCabe}, C., \& {Weniger}, C. 2015{\natexlab{a}},
  \prd, 91, 063003

\bibitem[{{Calore} {et~al.}(2015{\natexlab{b}}){Calore}, {Cholis}, \&
  {Weniger}}]{2015JCAP...03..038C}
{Calore}, F., {Cholis}, I., \& {Weniger}, C. 2015{\natexlab{b}}, \jcap, 3, 38

\bibitem[{{Castro} {et~al.}(2013){Castro}, {Slane}, {Carlton}, \&
  {Figueroa-Feliciano}}]{2013ApJ...774...36C}
{Castro}, D., {Slane}, P., {Carlton}, A., \& {Figueroa-Feliciano}, E. 2013,
  \apj, 774, 36

\bibitem[{{Ciprini} {et~al.}(2007){Ciprini}, {Tosti}, {Marcucci}, {Cecchi},
  {Discepoli}, {Bonamente}, {Germani}, {Impiombato}, {Lubrano}, \&
  {Pepe}}]{Ciprini:2007zz}
{Ciprini}, S., {et~al.} 2007, in American Institute of Physics Conference
  Series, Vol. 921, The First GLAST Symposium, ed. S.~{Ritz}, P.~{Michelson},
  \& C.~A. {Meegan}, 546--547

\bibitem[{{Damiani} {et~al.}(1997){Damiani}, {Maggio}, {Micela}, \&
  {Sciortino}}]{1997ApJ...483..350D}
{Damiani}, F., {Maggio}, A., {Micela}, G., \& {Sciortino}, S. 1997, \apj, 483,
  350

\bibitem[{{Daylan} {et~al.}(2014){Daylan}, {Finkbeiner}, {Hooper}, {Linden},
  {Portillo}, {Rodd}, \& {Slatyer}}]{2014arXiv1402.6703D}
{Daylan}, T., {et~al.} 2014, ArXiv e-prints 1402.6703

\bibitem[{{Digel} {et~al.}(2001){Digel}, {Grenier}, {Hunter}, {Dame}, \&
  {Thaddeus}}]{2001ApJ...555...12D}
{Digel}, S.~W., {Grenier}, I.~A., {Hunter}, S.~D., {Dame}, T.~M., \&
  {Thaddeus}, P. 2001, \apj, 555, 12

\bibitem[{{Dobler} {et~al.}(2010){Dobler}, {Finkbeiner}, {Cholis}, {Slatyer},
  \& {Weiner}}]{2010ApJ...717..825D}
{Dobler}, G., {Finkbeiner}, D.~P., {Cholis}, I., {Slatyer}, T., \& {Weiner}, N.
  2010, \apj, 717, 825

\bibitem[{{Faucher-Gigu{\`e}re} \& {Loeb}(2010)}]{2010JCAP...01..005F}
{Faucher-Gigu{\`e}re}, C.-A., \& {Loeb}, A. 2010, \jcap, 1, 5

\bibitem[{{Feng}(2010)}]{2010ARA&A..48..495F}
{Feng}, J.~L. 2010, \araa, 48, 495

\bibitem[{{Genzel} {et~al.}(2010){Genzel}, {Eisenhauer}, \&
  {Gillessen}}]{2010RvMP...82.3121G}
{Genzel}, R., {Eisenhauer}, F., \& {Gillessen}, S. 2010, Reviews of Modern
  Physics, 82, 3121

\bibitem[{{Goodenough} \& {Hooper}(2009)}]{2009arXiv0910.2998G}
{Goodenough}, L., \& {Hooper}, D. 2009, ArXiv e-prints 0910.2998

\bibitem[{{Gordon} \& {Mac{\'{\i}}as}(2013)}]{2013PhRvD..88h3521G}
{Gordon}, C., \& {Mac{\'{\i}}as}, O. 2013, \prd, 88, 083521

\bibitem[{{G{\'o}rski} {et~al.}(2005){G{\'o}rski}, {Hivon}, {Banday},
  {Wandelt}, {Hansen}, {Reinecke}, \& {Bartelmann}}]{2005ApJ...622..759G}
{G{\'o}rski}, K.~M., {et~al.} 2005, \apj, 622, 759

\bibitem[{{Green}(2014)}]{2014BASI...42...47G}
{Green}, D.~A. 2014, Bulletin of the Astronomical Society of India, 42, 47

\bibitem[{{Hooper} \& {Goodenough}(2011)}]{2011PhLB..697..412H}
{Hooper}, D., \& {Goodenough}, L. 2011, Physics Letters B, 697, 412

\bibitem[{{Hooper} \& {Slatyer}(2013)}]{2013PDU.....2..118H}
{Hooper}, D., \& {Slatyer}, T.~R. 2013, Physics of the Dark Universe, 2, 118

\bibitem[{{Huang} {et~al.}(2013){Huang}, {Urbano}, \&
  {Xue}}]{2013arXiv1307.6862H}
{Huang}, W.-C., {Urbano}, A., \& {Xue}, W. 2013, ArXiv e-prints 1307.6862

\bibitem[{{Hunter} {et~al.}(1997){Hunter}, {Bertsch}, {Catelli}, {Dame},
  {Digel}, {Dingus}, {Esposito}, {Fichtel}, {Hartman}, {Kanbach}, {Kniffen},
  {Lin}, {Mayer-Hasselwander}, {Michelson}, {von Montigny}, {Mukherjee},
  {Nolan}, {Schneid}, {Sreekumar}, {Thaddeus}, \&
  {Thompson}}]{1997ApJ...481..205H}
{Hunter}, S.~D., {et~al.} 1997, \apj, 481, 205

\bibitem[{{Jungman} {et~al.}(1996){Jungman}, {Kamionkowski}, \&
  {Griest}}]{1996PhR...267..195J}
{Jungman}, G., {Kamionkowski}, M., \& {Griest}, K. 1996, \physrep, 267, 195

\bibitem[{{Kerr}(2010)}]{2010PhDT.......147K}
{Kerr}, M. 2010, PhD thesis, University of Washington

\bibitem[{{Lande} {et~al.}(2012){Lande}, {Ackermann}, {Allafort}, {Ballet},
  {Bechtol}, {Burnett}, {Cohen-Tanugi}, {Drlica-Wagner}, {Funk}, {Giordano},
  {Grondin}, {Kerr}, \& {Lemoine-Goumard}}]{2012ApJ...756....5L}
{Lande}, J., {et~al.} 2012, \apj, 756, 5

\bibitem[{{Manchester} {et~al.}(2005){Manchester}, {Hobbs}, {Teoh}, \&
  {Hobbs}}]{2005AJ....129.1993M}
{Manchester}, R.~N., {Hobbs}, G.~B., {Teoh}, A., \& {Hobbs}, M. 2005, \aj, 129,
  1993

\bibitem[{{Mattox} {et~al.}(1996){Mattox}, {Bertsch}, {Chiang}, {Dingus},
  {Digel}, {Esposito}, {Fierro}, {Hartman}, {Hunter}, {Kanbach}, {Kniffen},
  {Lin}, {Macomb}, {Mayer-Hasselwander}, {Michelson}, {von Montigny},
  {Mukherjee}, {Nolan}, {Ramanamurthy}, {Schneid}, {Sreekumar}, {Thompson}, \&
  {Willis}}]{1996ApJ...461..396M}
{Mattox}, J.~R., {et~al.} 1996, \apj, 461, 396

\bibitem[{{Mayer-Hasselwander} {et~al.}(1998){Mayer-Hasselwander}, {Bertsch},
  {Dingus}, {Eckart}, {Esposito}, {Genzel}, {Hartman}, {Hunter}, {Kanbach},
  {Kniffen}, {Lin}, {Michelson}, {Muecke}, {von Montigny}, {Mukherjee},
  {Nolan}, {Pohl}, {Reimer}, {Schneid}, {Sreekumar}, \&
  {Thompson}}]{EGRETGCref}
{Mayer-Hasselwander}, H.~A., {et~al.} 1998, \aap, 335, 161

\bibitem[{{Moskalenko} \& {Strong}(1998)}]{1998ApJ...493..694M}
{Moskalenko}, I.~V., \& {Strong}, A.~W. 1998, \apj, 493, 694

\bibitem[{{Navarro} {et~al.}(1997){Navarro}, {Frenk}, \&
  {White}}]{1997ApJ...490..493N}
{Navarro}, J.~F., {Frenk}, C.~S., \& {White}, S.~D.~M. 1997, \apj, 490, 493

\bibitem[{{Nolan} {et~al.}(2012){Nolan}, {Abdo}, {Ackermann}, {Ajello},
  {Allafort}, {Antolini}, {Atwood}, {Axelsson}, {Baldini}, {Ballet}, \&
  et~al.}]{2FGLref}
{Nolan}, P.~L., {et~al.} 2012, \apjs, 199, 31

\bibitem[{{Porter} {et~al.}(2008){Porter}, {Moskalenko}, {Strong}, {Orlando},
  \& {Bouchet}}]{Porter2008}
{Porter}, T.~A., {Moskalenko}, I.~V., {Strong}, A.~W., {Orlando}, E., \&
  {Bouchet}, L. 2008, \apj, 682, 400

\bibitem[{{Shibata} {et~al.}(2007){Shibata}, {Honda}, \&
  {Watanabe}}]{2007APh....27..411S}
{Shibata}, T., {Honda}, N., \& {Watanabe}, J. 2007, Astroparticle Physics, 27,
  411

\bibitem[{{Stecker} \& {Jones}(1977)}]{1977ApJ...217..843S}
{Stecker}, F.~W., \& {Jones}, F.~C. 1977, \apj, 217, 843

\bibitem[{{Story} {et~al.}(2007){Story}, {Gonthier}, \&
  {Harding}}]{2007ApJ...671..713S}
{Story}, S.~A., {Gonthier}, P.~L., \& {Harding}, A.~K. 2007, \apj, 671, 713

\bibitem[{{Strong} \& {Mattox}(1996)}]{1996A&A...308L..21S}
{Strong}, A.~W., \& {Mattox}, J.~R. 1996, \aap, 308, L21

\bibitem[{{Strong} {et~al.}(1988){Strong}, {Bloemen}, {Dame}, {Grenier},
  {Hermsen}, {Lebrun}, {Nyman}, {Pollock}, \& {Thaddeus}}]{1988A&A...207....1S}
{Strong}, A.~W., {et~al.} 1988, \aap, 207, 1

\bibitem[{{Su} {et~al.}(2010){Su}, {Slatyer}, \&
  {Finkbeiner}}]{2010ApJ...724.1044S}
{Su}, M., {Slatyer}, T.~R., \& {Finkbeiner}, D.~P. 2010, \apj, 724, 1044

\bibitem[{{Tam} {et~al.}(2011){Tam}, {Kong}, {Hui}, {Cheng}, {Li}, \&
  {Lu}}]{2011ApJ...729...90T}
{Tam}, P.~H.~T., {et~al.} 2011, \apj, 729, 90

\bibitem[{{Trotta} {et~al.}(2011){Trotta}, {J{\'o}hannesson}, {Moskalenko},
  {Porter}, {Ruiz de Austri}, \& {Strong}}]{2011ApJ...729..106T}
{Trotta}, R., {et~al.} 2011, \apj, 729, 106

\bibitem[{{Vladimirov} {et~al.}(2012){Vladimirov}, {J{\'o}hannesson},
  {Moskalenko}, \& {Porter}}]{2012ApJ...752...68V}
{Vladimirov}, A.~E., {J{\'o}hannesson}, G., {Moskalenko}, I.~V., \& {Porter},
  T.~A. 2012, \apj, 752, 68

\bibitem[{{Wolleben}(2007)}]{2007ApJ...664..349W}
{Wolleben}, M. 2007, \apj, 664, 349

\bibitem[{{Yusifov} \& {K{\"u}{\c c}{\"u}k}(2004)}]{2004A&A...422..545Y}
{Yusifov}, I., \& {K{\"u}{\c c}{\"u}k}, I. 2004, \aap, 422, 545

\end{thebibliography}

\begin{appendix}

\section{Interstellar Emission Model Scaling Procedure}
\label{appendix:IEM}

The sky is sub-divided into regions where individual
annuli for the different emission processed calculated by the \GP\ code 
have high signal-to-noise contributions.
For the structured interstellar emission, the 
contributions by the Bremsstrahlung and $\pi^0$-decay 
from CR nuclei interacting with ionised hydrogen in the ISM, and the 
Bremsstrahlung contribution from the neutral gas, are held
constant throughout the scaling procedure at their respective 
\GP\ predictions.
These are sub-dominant compared to those of CR nuclei interacting
with the neutral gas and it was found in trial fits that the scaling procedure
typically set these contributions to zero, which is unrealistic.

\begin{figure*}[htb]
\subfigure{
\includegraphics[scale=0.65]{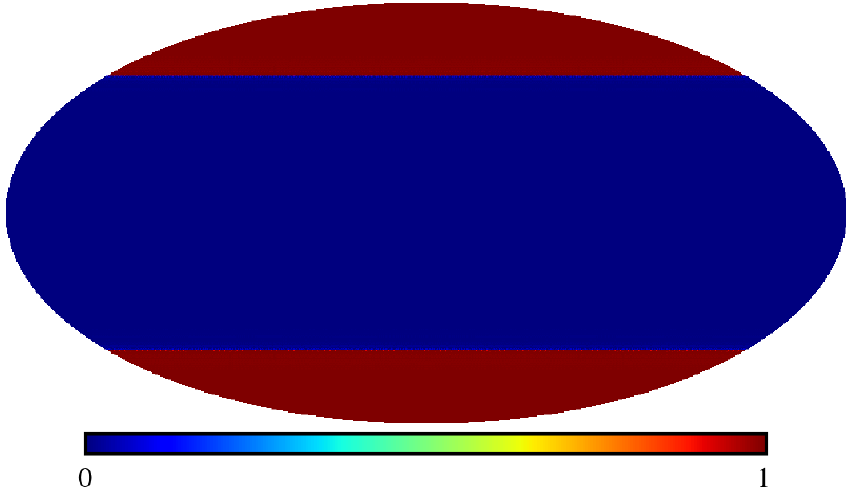}
\includegraphics[scale=0.65]{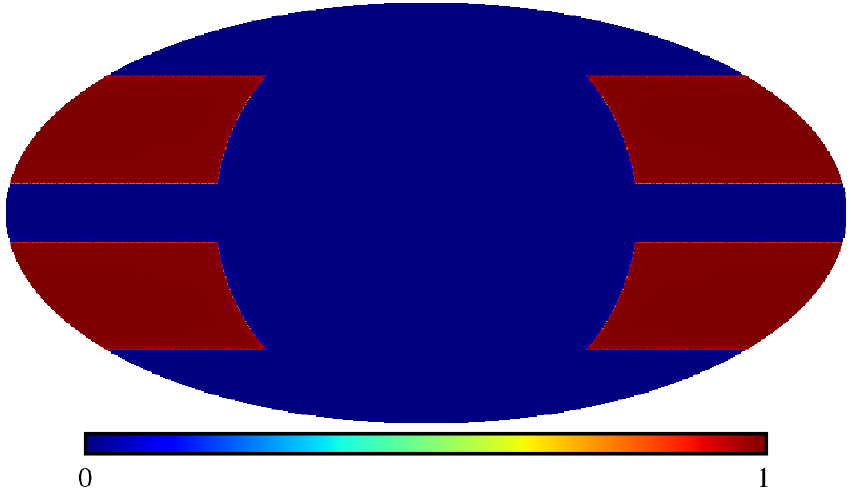}
\includegraphics[scale=0.65]{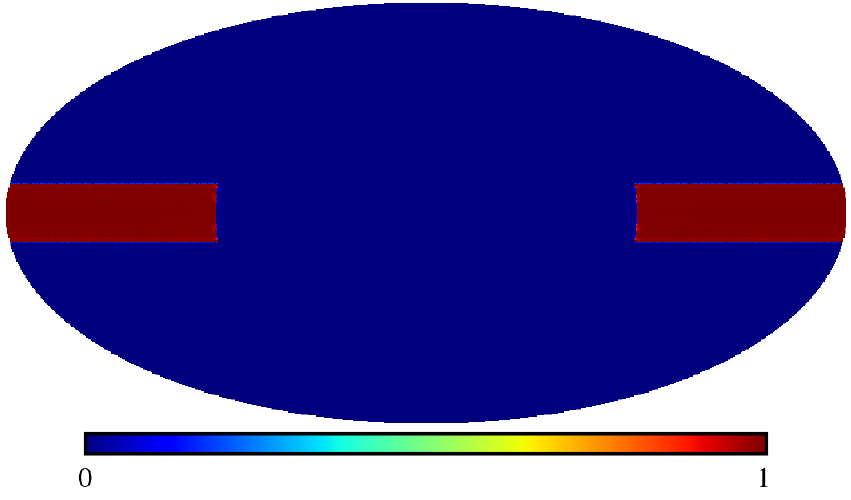}
}
\subfigure{
\includegraphics[scale=0.65]{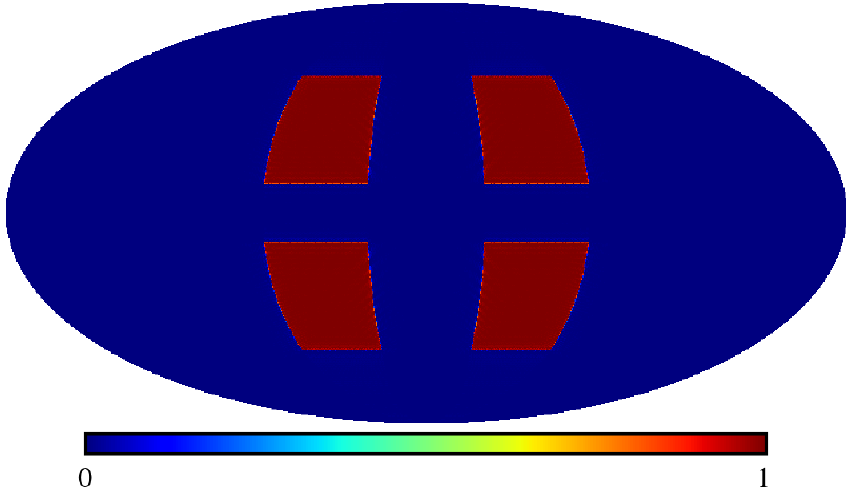}
\includegraphics[scale=0.65]{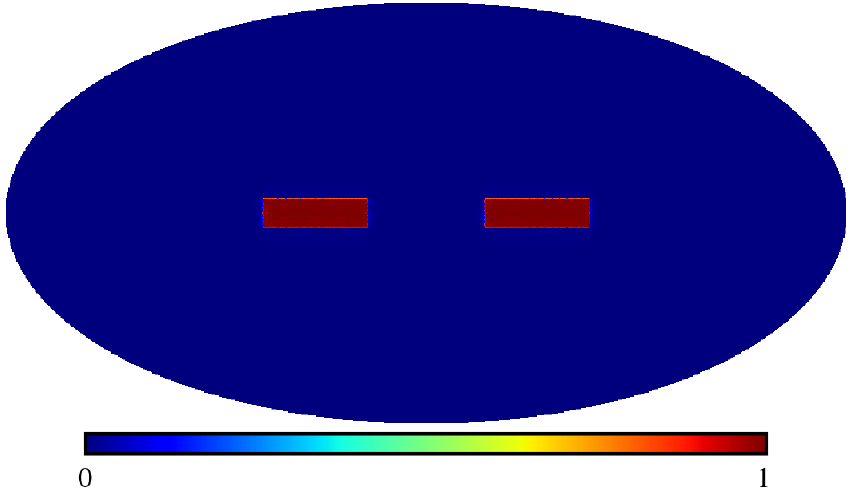}
\includegraphics[scale=0.65]{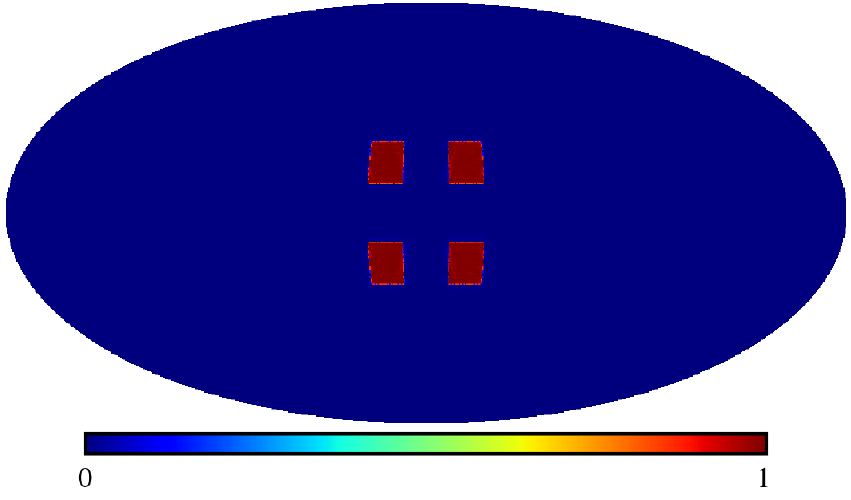}
}
\subfigure{
\includegraphics[scale=0.65]{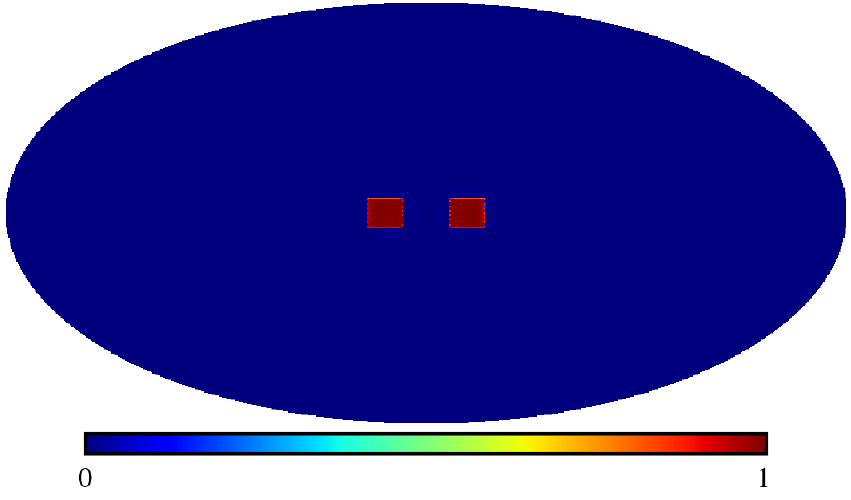}
}
\caption{Sky regions used for fitting the IEMs. 
The fitting sequence follows left-to-right, top-to-bottom where a non-zero 
value 
shows the active region used for each step of the procedure.
See the text for a description of the steps.
\label{fig:region_subdivision}}
\end{figure*}

Figure~\ref{fig:region_subdivision} shows the decomposition of the sky that 
is used for the IEM tuning.
For each of baseline IEMs described in Section~\ref{sec:interstellaremission} 
the coefficients for the per-annulus intensity maps are fit following the 
sequence described below.

The isotropic component for each IEM is determined first by fitting 
it to the data for $|b| \geq 50^\circ$ 
with one free parameter per energy bin for its 
intensity.
The coefficient of the local (annulus 5) \hi-related $\pi^0$-decay is allowed to 
vary in the fit to account for possible high-latitude structure.
The other components of the baseline IEM are held constant.
The isotropic component includes residual-charged particle 
background and astrophysical 
signals that are isotropic or near-to isotropically distributed at high
Galactic latitudes.
The purpose of this step of the procedure is to determine the 
level of emission in the data that is structureless, regardless of its origin.
The intensity of the isotropic component is held constant for the rest of 
IEM fitting procedure.

The intensities of the IC, \hi- and CO-associated $\pi^0$-decay
for the local and outer annuli (5 \& 6) are obtained by fitting in latitude
bands decreasing from $|b| = 50^\circ$ to the plane 
in the outer Galaxy ($90^\circ \leq l \leq 270^\circ$).
The latitude bands are chosen so that the components being fit dominate
the emission according to the baseline IEM.
Once fit the intensities of the components are held constant for the remainder
of the procedure.

The local annulus IC and \hi-related $\pi^0$-decay intensities are obtained
from $20^\circ \leq |b| \leq 50^\circ$.
The intensity for the local \hi-related $\pi^0$-decay intensity obtained
from the high-latitude region for determining the isotropic component is
used as a seed value for the mid-latitude region fit.
Then the local CO-associated $\pi^0$-decay and outer annulus IC intensities 
are obtained from fitting the region $5^\circ \leq |b| \leq 20^\circ$.
The outer annulus (annulus~6) \hi- and CO-associated $\pi^0$-decay intensities are 
determined from fitting to $0^\circ \leq |b| \leq 5^\circ$. 
This step of the procedure also determines the interstellar emission from
the longitude range $-70^\circ \leq l \leq 70^\circ$ by the local and 
outer annuli beyond the GC.

The contributions by IC, \hi- and
CO-associated $\pi^0$-decay for annuli $2-4$ 
are determined by fitting to the `tangent' longitude ranges for each annulus, as 
given in Table~\ref{table:rings}.
The intensities of the templates for the different emission processes 
are fit by decreasing annulus number.
Concentrically fitting the tangent ranges inward, 
under the axisymmetric assumption used here, enables the 
remaining fore-/background interstellar emission to the inner 
$\sim 1$~kpc to be estimated without including the data from 
the $15^\circ \times 15^\circ$ region about the GC.

The intensities of the IC for annulus~4, the Loop-I model, and 
the \hi-related $\pi^0$-decay  
are obtained from a fit in the latitude range $|b| \geq 20^\circ$.
The IC is held constant and the intensities for the \hi- and CO-related 
$\pi^0$-decay annulus~4 component and Loop-I model (using the intensities for the
Loop-I and \hi\ $\pi^0$-decay from
the higher latitude fit as seed values) are found by fitting to the 
data for the latitude range $|b| \leq 10^\circ$.
Because the IC is smoothly varying and the individual longitude ranges 
for annuli 2 and 3 are small, these are combined as a single annulus 
to determine the IC intensity for the longitude range $10^\circ \leq |l| 
\leq 25^\circ$.
The IC intensity is fit using 
the latitude band $10^\circ \leq |b| \leq 20^\circ$.
Higher latitudes are not employed 
because of the presence of the \fermi\ haze/``bubbles'' 
\citep{2010ApJ...717..825D,2010ApJ...724.1044S}, 
which are not modelled by the \GP\ code.
The \hi- and CO-related $\pi^0$-decay intensities for annuli 2 and 3 are 
obtained by fitting over the respective tangent ranges for $|b| \leq 5^\circ$ 
with the IC intensity obtained using the combined `annulus' held constant.
This latitude band is smaller than for annulus~4 because the bulk of the gas 
column density for annuli 2 and 3 is within $10^\circ$ of the mid-plane.

Table~\ref{table:coefficients} lists the coefficients obtained from 
the scaling procedure for both intensity-scaled and index-scaled IEMs.
These coefficients are applied to the \GP\ predictions for each of
the Pulsars and OBstars IEMs.
For entries that have a single number this is the scaling factor applied to
the whole intensity map output by \GP\ .
For entries with a tuple entry (e.g., $\pi^0$-decay for annuli $2-4$) the first
number is the scaling factor for the intensity map, the second and third 
numbers are the changes in spectral index of the intensity map above/below
the break energy.
For the break function that is used for the index-scaled IEMs 
the following form is used:

\begin{equation}
f(E) = f_0 E^{\gamma_1} (0.5 + 0.5(E/E_{\rm break})^{(\gamma_2 - \gamma_1)/\beta})^\beta
\label{eq:breakfn}
\end{equation}

\noindent
where $E_{\rm break} = 2$~GeV is the break energy, $\beta = 0.2$ is the smoothing
parameter with the value chosen to 
mitigate sensitivity to the precise value of the break energy, 
and $\gamma_2 - \gamma_1$ are the change in the spectral index obtained 
from the fit above/below $E_{\rm break}$.

\begin{deluxetable}{llccccc}[tb]
\tablecolumns{7}
\tablecaption{Scaling coefficients with respect to the baseline IEM. 
\label{table:coefficients}}
\tablehead{
\colhead{Model} &
\colhead{Process} &  
\colhead{Annulus~2} &
\colhead{Annulus~3} &
\colhead{Annulus~4} &
\colhead{Annulus~5} &
\colhead{Annulus~6}
}
\startdata
Pulsars & & & & & &  \\
intensity-scaled & IC & 1.3 & 1.3 & 1.6 & 1.49 & 1.8 \\
& $\pi^0$-decay \hi & 1 & 1 & 1.62 & 1.21 & 1.74 \\
& \hspace{32.5pt}CO & 1 & 1 & 1.42 & 1.4 & 0.3 \\
Pulsars & & & & & &  \\
index-scaled & IC & 1.71 & 1.71 & 1.6 & & \\
& $\pi^0$-decay \hi & (0.5,0.29,0.14)\tablenotemark{1} & (0.5,0.29,0.14) & (0.32,0.29,0.14) & & \\
& \hspace{32.5pt}CO & (0.22,0.30,0.30) & (0.22,0.30,0.30) & (0.37,0.30,0.30) & & \\
OBstars & & & & & &  \\
intensity-scaled & IC & 4.15 & 4.15 & 1.48 & 1.13 & 1 \\
& $\pi^0$-decay \hi & 3.7 & 3.7 & 1.2 & 1.19 & 1.41 \\
& \hspace{32.5pt}CO & 1.2 & 0.8 & 1.3 & 1.37 & 0.69 \\
OBstars & & & & & &  \\
index-scaled & IC & 2.21 & 2.21 & 1.48 & & \\
& $\pi^0$-decay \hi & (1,0.17,0.17) & (1,0.4,0.4) & (0.67,0.17,0.17) & & \\
& \hspace{32.5pt}CO & 1 & 1 & (0.17,0.41,0.06) & & \\ 
\enddata
\tablenotetext{1}{Tuple entries refer to parameters for Eq.~\ref{eq:breakfn}: $(f_0, \gamma_1, \gamma_2)$}
\end{deluxetable}

\begin{deluxetable}{llccccc}[tb]
\tablecolumns{7}
\tablecaption{Scaled fluxes\tablenotemark{1} $>1$~GeV per annuli for each IEM over the $15^\circ \times 15^\circ$ region. 
\label{table:fluxes}}
\tablehead{
\colhead{Model} &
\colhead{Process} &  
\colhead{Annulus~2} &
\colhead{Annulus~3} &
\colhead{Annulus~4} &
\colhead{Annulus~5} &
\colhead{Annulus~6}
}
\startdata
Pulsars & & & & & &\\
intensity-scaled & IC & 9.82\tablenotemark{2} & & 17.38 & 1.34 & 0.89\\
& $\pi^0$-decay \hi & 2.65 & 3.69 & 63.53 & 20.43 & 3.28\\
& \hspace{32.5pt}CO & 2.60 & 3.58 & 44.64 & 3.42 & 0.44\\
Pulsars & & & & & &\\
index-scaled & IC & 12.92 & & 17.38 & & \\
& $\pi^0$-decay \hi & 5.27 & 7.33 & 49.86 & & \\
& \hspace{32.5pt}CO & 2.43 & 3.34 & 49.38 & & \\
OB-stars & & & & & &\\
intensity-scaled & IC & 11.33 & & 16.53 & 1.19 & 0.40\\
& $\pi^0$-decay \hi & 6.28 & 9.47 & 45.10 & 20.36 & 2.39\\
& \hspace{32.5pt}CO & 4.05 & 3.75 & 45.46 & 3.49 & 1.14\\
OB-stars & & & & & &\\
index-scaled & IC & 6.04 & & 16.53 & &\\
& $\pi^0$-decay \hi & 3.82 & 17.59 & 56.85 & &\\
& \hspace{32.5pt}CO & 3.37 & 4.69 & 40.90 & &\\
\enddata
\tablenotetext{1}{Units: $10^{-8}$ ph cm$^{-2}$ s$^{-1}$.}
\tablenotetext{2}{IC flux for annuli 2 and 3 are combined.}
\end{deluxetable}

The statistical uncertainties on the scaling factors are
typically $\sim 10-20$\% per fitting region.
The scaling factors are held constant after fitting for individual angular 
ranges.
Subsequent fits do not propagate these statistical uncertainties so that 
there the scaling coefficients for different annuli are not cross-correlated.

\section{1FIG point source spectral parameters}
\label{appendix:spectral_parameters}

\begin{table}[H]
\tabletypesize{\tiny}
\hspace{-0.1cm}
\begin{deluxetable}{lcccccc}[htb]
\tablecolumns{6}
\tablewidth{0pt}
\tablecaption{1FIG point source spectral parameters.
\label{table:source_properties}}
\tablehead{
\colhead{Name} & 
\colhead{Type} &
\colhead{Pulsars} &
\colhead{Pulsars} &
\colhead{OBstars} & 
\colhead{OBstars} \\
\colhead{1FIG} & 
\colhead{} &
\colhead{Intensity-scaled} &
\colhead{Index-scaled} &
\colhead{Intensity-scaled} & 
\colhead{Index-scaled}}
\startdata
1FIG J1701.1-3004 & LP &  $(2.28\pm0.21,0.48\pm0.24,2.62)$\tablenotemark{1} & $(2.30\pm0.22,0.47\pm0.26,2.62)$ & $(2.23\pm0.22,0.51\pm0.26,2.62)$ & $(2.29\pm0.21,0.48\pm0.26,2.62)$ \\
1FIG J1717.5-3342 & LP &  $(2.52\pm0.23,0.63\pm0.27,0.35)$ & $(2.70\pm0.22,0.50\pm0.26,0.35)$ & $(2.59\pm0.21,0.57\pm0.25,0.35)$ & $(2.67\pm0.21,0.50\pm0.25,0.35)$ \\
1FIG J1718.0-3056 & PL &  $1.98\pm0.25$\tablenotemark{2} & $2.15\pm0.25$ & $2.05\pm0.23$ & $2.09\pm0.25$ \\
1FIG J1728.6-3433 & PL &  $2.54\pm0.19$ & $3.03\pm0.26$ & $2.48\pm0.18$ & $2.81\pm0.24$ \\
1FIG J1729.1-3502 & LP &  $(1.81\pm0.28,0.81\pm0.38,2.82)$ & $(2.33\pm0.27,0.66\pm0.49,2.82)$ & $(1.80\pm0.26,0.78\pm0.33,2.82)$ & $(2.12\pm0.28,0.66\pm0.41,2.82)$ \\
1FIG J1730.2-3351 & PL &  $2.67\pm0.20$ & $3.10\pm0.25$ & $2.61\pm0.19$ & $2.91\pm0.24$ \\
1FIG J1731.3-3235 & LP &  $(1.53\pm0.39,1.41\pm0.58,2.19)$ & $(2.33\pm0.31,1.12\pm0.57,2.19)$ & $(1.50\pm0.38,1.39\pm0.54,2.19)$ & $(2.11\pm0.31,1.10\pm0.52,2.19)$ \\
1FIG J1731.6-3001 & LP &  $(2.28\pm0.14,0.07\pm0.13,0.52)$ & $(2.41\pm0.14,0.00\pm0.00,0.52)$ & $(2.26\pm0.15,0.09\pm0.13,0.52)$ & $(2.32\pm0.14,0.04\pm0.12,0.52)$ \\
1FIG J1732.3-3131 & LP &  $(1.87\pm0.07,0.82\pm0.08,1.91)$ & $(2.03\pm0.07,0.76\pm0.08,1.91)$ & $(1.86\pm0.07,0.83\pm0.08,1.91)$ & $(1.99\pm0.07,0.77\pm0.08,1.91)$ \\
1FIG J1734.6-3228 & PL &  $2.09\pm0.20$ & $2.47\pm0.26$ & $2.00\pm0.20$ & $2.15\pm0.27$ \\
1FIG J1735.4-3030 & PL &  $2.62\pm0.19$ & $2.84\pm0.20$ & $2.58\pm0.19$ & $2.75\pm0.20$ \\
1FIG J1736.1-3150 & PL &  $1.90\pm0.19$ & $2.07\pm0.25$ & $1.85\pm0.19$ & $1.95\pm0.23$ \\
1FIG J1736.1-3422 & PL &  $2.48\pm0.18$ & $2.65\pm0.19$ & $2.49\pm0.16$ & $2.60\pm0.18$ \\
1FIG J1737.4-3144 & LP &  $(0.82\pm0.55,2.42\pm0.98,2.15)$ & $(1.63\pm0.52,2.48\pm1.34,2.15)$ & $(0.37\pm0.69,2.77\pm1.20,2.15)$ & $(1.00\pm0.70,3.01\pm1.85,2.15)$ \\
1FIG J1739.4-3010 & PL &  $2.71\pm0.20$ & $3.13\pm0.25$ & $2.59\pm0.24$ & $2.93\pm0.28$ \\
1FIG J1740.1-3057 & PL &  $2.27\pm0.17$ & $2.57\pm0.21$ & $2.13\pm0.19$ & $2.42\pm0.21$ \\
1FIG J1740.2-2834 & LP &  $(3.80\pm1.36,1.41\pm1.24,0.36)$ & $(4.84\pm1.12,1.77\pm0.89,0.36)$ & $(5.00\pm0.06,2.77\pm0.51,0.36)$ & $(5.00\pm0.01,2.22\pm0.42,0.36)$ \\
1FIG J1741.5-2538 & PL &  $1.66\pm0.27$ & $1.79\pm0.29$ & $1.63\pm0.27$ & $0.00\pm0.00$ \\
1FIG J1741.5-2054 & LP &  $(2.79\pm0.12,1.67\pm0.28,1.73)$ & $(2.86\pm0.13,1.62\pm0.29,1.73)$ & $(2.79\pm0.13,1.64\pm0.28,1.73)$ & $(2.83\pm0.13,1.63\pm0.28,1.73)$ \\
1FIG J1742.5-3318 & LP &  $(2.06\pm0.42,3.08\pm0.95,2.11)$ & $(2.53\pm0.37,2.61\pm0.83,2.11)$ & $(2.01\pm0.41,3.09\pm0.93,2.11)$ & $(2.33\pm0.38,2.76\pm0.87,2.11)$ \\
1FIG J1744.2-2930 & LP &  $(2.21\pm0.22,0.42\pm0.26,2.56)$ & $(2.87\pm0.22,0.29\pm0.29,2.56)$ & $(2.13\pm0.25,0.47\pm0.28,2.56)$ & $(2.49\pm0.22,0.34\pm0.27,2.56)$ \\
1FIG J1744.3-3051 & PL &  $2.25\pm0.18$ & $2.51\pm0.24$ & $2.15\pm0.21$ & $2.33\pm0.24$ \\
1FIG J1745.0-2905 & LP &  $(2.08\pm0.18,0.94\pm0.27,2.54)$ & $(2.23\pm0.24,1.07\pm0.36,2.54)$ & $(1.97\pm0.23,1.08\pm0.35,2.54)$ & $(2.28\pm0.20,0.85\pm0.29,2.54)$ \\
1FIG J1745.1-3012 & LP &  $(2.75\pm0.17,0.18\pm0.15,0.45)$ & $(3.10\pm0.25,0.08\pm0.19,0.45)$ & $(2.67\pm0.19,0.22\pm0.18,0.45)$ & $(2.90\pm0.20,0.11\pm0.17,0.45)$ \\
1FIG J1745.5-2859 & LP &  $(2.28\pm0.06,0.21\pm0.04,2.69)$ & $(2.33\pm0.07,0.22\pm0.05,2.69)$ & $(2.28\pm0.06,0.21\pm0.05,2.69)$ & $(2.30\pm0.07,0.21\pm0.05,2.69)$ \\
1FIG J1746.4-2843 & LP &  $(2.66\pm0.12,0.19\pm0.12,3.12)$ & $(2.73\pm0.14,0.22\pm0.14,3.12)$ & $(2.65\pm0.12,0.21\pm0.13,3.12)$ & $(2.69\pm0.13,0.20\pm0.13,3.12)$ \\
1FIG J1746.5-3240 & LP &  $(2.26\pm0.16,0.82\pm0.23,2.12)$ & $(2.49\pm0.16,0.71\pm0.24,2.12)$ & $(2.22\pm0.17,0.84\pm0.25,2.12)$ & $(2.40\pm0.16,0.73\pm0.24,2.12)$ \\
1FIG J1747.0-2826 & LP &  $(2.58\pm0.12,0.00\pm0.00,0.33)$ & $(2.83\pm0.15,0.00\pm0.00,0.33)$ & $(2.57\pm0.13,0.00\pm0.00,0.33)$ & $(2.71\pm0.14,0.00\pm0.00,0.33)$ \\
1FIG J1747.2-2959 & LP &  $(2.51\pm0.09,0.54\pm0.13,2.35)$ & $(2.65\pm0.10,0.57\pm0.15,2.35)$ & $(2.48\pm0.10,0.59\pm0.14,2.35)$ & $(2.67\pm0.09,0.50\pm0.13,2.35)$ \\
1FIG J1747.6-2442 & LP &  $(3.03\pm0.71,1.39\pm0.85,0.93)$ & $(3.06\pm0.62,0.87\pm0.56,0.93)$ & $(2.99\pm0.69,1.36\pm0.83,0.93)$ & $(3.05\pm0.68,1.19\pm0.75,0.93)$ \\
1FIG J1748.1-2449 & LP &  $(2.45\pm0.15,1.16\pm0.23,2.42)$ & $(2.64\pm0.14,0.86\pm0.22,2.42)$ & $(2.46\pm0.14,1.15\pm0.24,2.42)$ & $(2.58\pm0.14,0.97\pm0.23,2.42)$ \\
1FIG J1748.2-2856 & LP &  $(2.40\pm0.25,0.43\pm0.25,2.54)$ & $(2.65\pm0.24,0.39\pm0.28,2.54)$ & $(2.45\pm0.24,0.43\pm0.25,2.54)$ & $(2.59\pm0.24,0.39\pm0.26,2.54)$ \\
1FIG J1748.2-2816 & LP &  $(2.38\pm0.14,1.03\pm0.19,0.41)$ & $(2.57\pm0.15,0.89\pm0.20,0.41)$ & $(2.42\pm0.14,1.01\pm0.19,0.41)$ & $(2.49\pm0.14,0.91\pm0.19,0.41)$ \\
1FIG J1749.1-2917 & LP &  $(2.20\pm0.31,0.57\pm0.34,2.31)$ & $(2.47\pm0.31,0.56\pm0.40,2.31)$ & $(2.07\pm0.37,0.69\pm0.43,2.31)$ & $(2.25\pm0.36,0.68\pm0.47,2.31)$ \\
1FIG J1750.2-3705 & PL &  $2.53\pm0.21$ & $2.57\pm0.21$ & $2.51\pm0.20$ & $2.56\pm0.20$ \\
1FIG J1753.5-2931 & LP &  $(2.26\pm0.30,0.49\pm0.35,2.29)$ & $(2.53\pm0.28,0.40\pm0.36,2.29)$ & $(2.03\pm0.39,0.62\pm0.48,2.29)$ & $(2.28\pm0.34,0.49\pm0.42,2.29)$ \\
1FIG J1753.6-2539 & LP &  $(2.36\pm0.14,0.54\pm0.14,0.46)$ & $(2.59\pm0.13,0.35\pm0.13,0.46)$ & $(2.39\pm0.14,0.53\pm0.14,0.46)$ & $(2.51\pm0.14,0.41\pm0.14,0.46)$ \\
1FIG J1755.5-2511 & LP &  $(2.09\pm0.28,0.72\pm0.33,2.89)$ & $(2.72\pm0.29,0.56\pm0.41,2.89)$ & $(2.08\pm0.27,0.73\pm0.32,2.89)$ & $(2.47\pm0.28,0.58\pm0.35,2.89)$ \\
1FIG J1758.5-2405 & LP &  $(2.23\pm0.18,0.23\pm0.21,0.36)$ & $(2.48\pm0.17,0.17\pm0.20,0.36)$ & $(2.20\pm0.18,0.22\pm0.20,0.36)$ & $(2.41\pm0.18,0.14\pm0.19,0.36)$ \\
1FIG J1759.0-2345 & LP &  $(1.96\pm0.18,0.24\pm0.12,0.34)$ & $(2.19\pm0.19,0.17\pm0.14,0.34)$ & $(2.01\pm0.18,0.22\pm0.13,0.34)$ & $(2.03\pm0.22,0.22\pm0.15,0.34)$ \\
1FIG J1800.5-2359 & LP &  $(2.46\pm0.10,0.19\pm0.10,0.46)$ & $(2.57\pm0.11,0.12\pm0.10,0.46)$ & $(2.47\pm0.10,0.18\pm0.10,0.46)$ & $(2.50\pm0.11,0.14\pm0.10,0.46)$ \\
1FIG J1801.1-2313 & LP &  $(3.42\pm0.45,1.70\pm0.55,3.02)$ & $(4.13\pm1.01,2.24\pm1.03,3.02)$ & $(3.48\pm0.54,1.76\pm0.68,3.02)$ & $(3.82\pm0.75,2.02\pm0.82,3.02)$ \\
1FIG J1801.2-2451 & PL &  $2.43\pm0.18$ & $2.79\pm0.21$ & $2.48\pm0.16$ & $2.69\pm0.17$ \\
1FIG J1801.4-2330 & LP &  $(3.27\pm1.06,0.56\pm1.12,3.10)$ & $(2.89\pm0.19,0.00\pm0.02,3.10)$ & $(3.27\pm0.65,0.54\pm0.66,3.10)$ & $(2.90\pm0.21,0.04\pm0.23,3.10)$ \\
1FIG J1801.6-2358 & PL &  $2.11\pm0.24$ & $2.33\pm0.25$ & $2.11\pm0.22$ & $2.29\pm0.22$ \\
1FIG J1802.2-3043 & PL &  $1.91\pm0.22$ & $1.98\pm0.23$ & $1.85\pm0.22$ & $1.90\pm0.22$ \\
1FIG J1808.2-3358 & PL &  $2.60\pm0.23$ & $2.63\pm0.24$ & $2.51\pm0.24$ & $2.60\pm0.23$ \\
1FIG J1809.5-2332 & LP &  $(2.66\pm0.04,0.42\pm0.05,0.32)$ & $(2.69\pm0.04,0.40\pm0.05,0.32)$ & $(2.66\pm0.04,0.42\pm0.05,0.32)$ & $(2.68\pm0.04,0.40\pm0.05,0.32)$
\enddata
\tablenotetext{1}{Parameter tuple $\alpha, \beta, E_b$~(GeV) for spectral model $dN/dE \propto (E/E_b)^{-\alpha - \beta \log(E/E_b)}$.}
\tablenotetext{2}{Parameter $\alpha$ for spectral model $dN/dE \propto E^{-\alpha}$.}
\end{deluxetable}
\end{table}

\end{appendix}

\end{document}